\documentclass[twoside,floatfix,14pt]{mbook}
\large
\usepackage{graphicx}
\usepackage{14pt}
\usepackage{epsfig}
\usepackage{amssymb}
\usepackage{amsmath}
\usepackage{here}
\usepackage{cite}
  \newcommand{\be}{\begin{equation}}
  \newcommand{\ee}{\end{equation}}
  \newcommand{\nn}{\nonumber}
\begin{document}
   \pagestyle{empty}
     {\Large
      \ \vspace{-2cm}
      \begin{center}
       {\bf
        Jagellonian University
       }
      \end{center}
       \begin{center}
        {\bf
         Institute of Physics
        }
       \end{center}
      \vspace{2cm}
       \begin{center}
        {\Huge\bf
          Hadronic interaction of ${\bf \eta}$ and ${\bf \eta^{\prime}}$
          mesons with protons
        }
       \end{center}
   }
   \vspace{1cm}
     \begin{center}
       {\large\bf
         Pawe{\l} Moskal
       }
     \end{center}
    \large{
     \vspace{6cm}
    \hspace{-0.1cm}
    \parbox{13.0cm}{
    {\large\bf
     Habilitation thesis
     prepared at the Department of Nuclear Physics in the Institute of
     Physics of the Jagel\-lonian University
     and at the Institute of Nuclear Physics of the Research Centre J\"ulich,
     submitted to the Faculty of
     Physics, Astronomy and Applied Computer Science
     at the Jagel\-lonian University,
     for the postdoctoral lecture qualification.
   }
 
    \vspace{2cm}
      \begin{center}
       {\large\bf
           Cracow 2004
       }
      \end{center}
   }

\pagestyle{plain}
\chapter*{{\large{\mbox{} \hspace{4.7cm} Acknowledgement}}}
\pagestyle{myheadings}
\markboth{Acknowledgement}{Acknowledgement}
 
Accomplishment of the experiments presented
in this treatise was possible only thanks
to the joined effort
of friends from the COSY-11 group with whom
I had the luck to work during many years of my
involvement at the Jagellonian University
in Poland and the Research Centre J{\"u}lich in Germany.
I am greatly indebted to all of them for their help.
 
A person whom I own most and whom I revere and admire
is Professor Walter Oelert,
inventor of the COSY-11 facility and founder of the COSY-11 collaboration.
It is only due to his great open mindedness and understanding that I could devote
a lot of my time to study physics while working as a physicist.
I thank you from the bottom of my heart for the
time devoted, support and patience,
the latter so often tried to the utmost.
 
I am profoundly grateful to Professor Lucjan Jarczyk
and Professor Walter Oelert
for being my promotors in the truest sense of this word.
 
Wise and friendly advise, encouragement and direct
reprimands of Professor Lucjan Jarczyk
I value more highly than I can express in words.
 
At the last stage of completing this dissertation
I was pleased to work together with talented and zealous students:
Master of Science Rafa{\l}~Czy{\.z}ykiewicz,
Master of Science Micha{\l}~Janusz,
Master of Science Pawe{\l}~Klaja,
Master of Science Cezary~Piskor-Ignatowicz, and
Master of Science  Joanna~Przerwa.
When introducing you to the experimental technique I certainly gained more
than yourselves and for an instant I will not regret the time spent together.
I am especially thankful to Joanna who kindly agreed to defend our recent experimental proposal
and who with her unbelievable charm and serenity made our daily work pleasurable.
 
One of the many factors which motivated me to conduct this work was
the written wish of Master of Science Peter Winter
who had wanted to adress me in the acknowledgements of his PhD thesis
without a negation in front of the title he had granted me in advance.
Although it is probably too late, I did do my best.
 
For the perusal of parts of this manuscript I owe a great debt to
Prof. Lucjan  Jarczyk, Prof. Walter  Oelert,
Dr. Thomas  Sefzick,
Dr. Dieter Grzonka,
Dr.~hab.~Jerzy~Smyrski,
Dr.~Magnus~Wolke,
Mgr.~Rafa{\l} Czy{\.z}ykiewicz,
Prof.~Bogus{\l}aw~Kamys,
Dr.~Christoph~Hanhart,
Dr.~Vadim Baru, and
Mrs. Joanna Czekotowska.
This treatise benefited immeasurably from your emendations and critical remarks
concerning the earlier version. Still it does not make me any less
responsible for any mistake
I may have made.
 
For the support of my activites in the Department of Nuclear Physics
of the Jagellonian University and for the
approval of my work I am grateful to Prof. Reinhard Kulessa,
and for the inestimable support of my work in the Nuclear Physics Institute
of the Research Centre J{\"u}lich I am thankful to Prof. Kurt Kilian.\\
 
Last but not least, I do appreciate very much the support, patience and love
of {\.Z}aneta, Ines and Gabriel.
 
\clearpage
\newpage

   \normalsize
     \def\contentsname{\Large \mbox{} \hspace{5cm} Contents}
     \tableofcontents
     \pagestyle{myheadings}
     \markboth{Contents}{Contents}
     \newpage
     \clearpage
   \normalsize

 
\chapter*{\centerline{\Large{Preface}}}
\thispagestyle{empty}
   \addcontentsline{toc}{chapter}{\protect\numberline{}{Preface\dotfill}}
    \pagestyle{plain}
\large
 
\begin{center}
\parbox{0.8\textwidth}{
\hspace{0.5cm} Experimental results constituting the basis for this dissertation
have been published in twelve articles~\cite{moskal025203,khoukaz0401011,moskalf0a0,review,
winter251,moskal448,moskal295,moskal416,smyrski182,swave,baru445,moskal3202}
and eleven conference proceedings~\cite{qnpmoskal,panicmoskal,moskal1777,moskal2277,moskal3091,
moskal367,aipmoskal,moskal0110001,hadronmoskal,meson2002moskal,ecoolmoskal}.
The research has been realized at the cooler synchrotron COSY by means of the COSY--11 facility~\cite{Proposalwalter},
which was build to a large extent at the Jagellonian University.
 
\hspace{0.5cm} Ideas of experiments connected with this treatise, elaborated and formulated as
experimental proposals~\cite{proposaljoanna,proposal123,proposalrafal,proposalc11,proposal11_7,proposal11_5,proposal11_2},
have been  positively judged and
approved for realization by the Programme Advisory Committee of the COSY accelerator,
and the measurements have been carried out by the COSY-11 collaboration.
 
\hspace{0.5cm} To render the  reading of this dissertation more pleasant
I have attempted to adduce at the beginnig of each section
a few sentences from  books of philosophers and scientists whom I immensely admire.\\
}
\end{center}


   \normalsize
   \def\chaptername{Introduction}

\pagestyle{plain}
\chapter{Introduction}
\thispagestyle{empty}
\mbox{}

\pagestyle{myheadings}
  \markboth{Hadronic interaction of $\eta$ and $\eta'$ mesons with protons}{1. Introduction}
        
\large
\begin{flushright}
\parbox{0.7\textwidth}{
 {\em
  It is extremely advantageous to be able to bring a number of investigations
  under a formula of a single problem. For in this manner, we not only facilitate 
  our own labour, inasmuch as we define it clearly to ourselves, but also render 
  it more easy for others to decide whether we have done justice to our undertaking~\cite{kantcritique}.\\
 }
 \protect \mbox{} \hfill  Immanuel Kant \protect\\
 }
\end{flushright}
Pseudoscalar mesons $(\pi,K,\eta,\eta^{\prime})$ constitute
the lightest (basic) nonet of particles built out of quarks and antiquarks,
and the investigation of their interaction with nucleons is one of the key issues
in hadron physics. 

The interaction of $\pi$ and $K$  mesons with nucleons has been deduced from
experiments realized by means of charged pion or kaon beams.
Such experiments are, however,  generally not feasible  in the case of flavour neutral
mesons due to  their too short lifetime.
Therefore, although $\eta$ and $\eta^{\prime}$ mesons were discovered over 
fourty years ago~\cite{pevsner421,kalb64,gold64}
their hadronic interaction with nucleons has  not been established. The scattering length 
--~the very basic quantity describing the low energy interaction potential~-- in the case of the $\eta$ meson
is poorly estimated 
and in the case of the $\eta^{\prime}$ meson it is entirely unknown.
Estimated values of the real part of the proton-$\eta$ scattering length varies from 0.20~fm to 1.05~fm 
depending on the approach employed
for its determination~\cite{greenR2167,N.Kaiser-II,green053,green035208}.

It is the primordial purpose of this treatise to present
methods developed in order to study these interactions, to explain the applied experimental techniques
and to discuss the progress achieved.   We will address also the issue of the 
quark-gluon stucture of these mesons 
since in analogy to the connection between electromagnetic and Van der Vaals potentials,  
we may perceive the hadronic force 
as a residuum of the strong interaction that occurs between quarks and gluons -- the 
constituents of hadrons. 

In the quark model the $\eta$ and $\eta^{\prime}$ mesons are regarded as the mixture
of the singlet ($\eta_1~=~\frac{1}{\sqrt{3}}(u\bar{u}+d\bar{d}+s\bar{s})$)
and octet ($\eta_8~=~\frac{1}{\sqrt{6}}(u\bar{u}+d\bar{d}-2s\bar{s})$) states of the SU(3)-flavour pseudoscalar nonet.
A small mixing angle $\Theta\!\!\!\!~=~\!\!\!\!-15.5^{\circ}$~\cite{bramon271} implies
that the percentage amount of strange and non-strange quarkonium in both mesons is almost the same:
\begin{equation}
 \nonumber
 \eta~=~ 0.77 \frac{1}{\sqrt{2}} (u\bar{u}+d\bar{d}) - 0.63 s\bar{s},  \  \ \ \
 \eta^{\prime}~=~ 0.63 \frac{1}{\sqrt{2}} (u\bar{u}+d\bar{d}) + 0.77 s\bar{s}. 
\end{equation}
Although,  the $\eta$ and $\eta^{\prime}$ mesons possess similar quark structure,
their physical properties are unexpectedly different. Let us adduce three examples.
The $\eta^{\prime}$(958) meson mass
is almost two times larger than the mass of the $\eta$(547) meson. 
The branching ratios for the decays of $B$ and $D_s$ mesons
into the $\eta^{\prime}$ meson exceed significantly those into the $\eta$ meson
and the standard model predictions~\cite{jessop052002,branderburg3804}.
There exist excited states of nucleons
which decay via emission of the $\eta$ meson, yet none of the observed baryon resonances decays via  the
emission of the $\eta^{\prime}$ meson~\cite{PDG}. Thus, it is natural to expect 
that the production mechanisms of these mesons in the collision of nucleons as well as their 
hadronic interactions with nucleons should also differ from each other.

In spite of the fact that the strange and non-strange quarkonium content of $\eta$ and $\eta^{\prime}$ mesons
is similar, 
the $\eta$ meson remains predominantly the SU(3)-flavor octet
and the $\eta^{\prime}$  meson the SU(3)-flavor singlet. This indicates that the $\eta^{\prime}$ meson
may be mixed with pure gluon states  to a much larger extent than the $\eta$  and all other pseudoscalar and vector mesons.
The anomalously large mass of the $\eta^{\prime}$ meson~\cite{bass286} and the analysis of the decays of pseudoscalar,
vector, and J/$\psi$ mesons indicate~\cite{li335}, that the $\eta^{\prime}$ meson 
is indeed a mixture of quarkonium and gluonium.
The gluonic admixture of $\eta^{\prime}$ will influence the $\eta^{\prime}$-nucleon interaction
via the U(1) anomaly~\cite{bass286}. 
The range of the glue induced $\eta^{\prime}$-nucleon interaction,
if determined by the two-gluon effective potential, would be in the order of 0.3~fm~\cite{baru445}.
This range is large enough
to be important in the threshold production of the $\eta^{\prime}$ meson e.g. via the $pp\to pp\eta^{\prime}$ reaction
which occurs at distances of the colliding nucleons in the order of 0.2~fm. At such small distances the quark-gluon
degrees of freedom may play a significant role in the production dynamics of the $\eta$ and $\eta^{\prime}$ mesons.
It is essential to realize that the investigation of the interaction between hadrons is inseparably 
connected with the study of their
structure and  the dynamics of the processes in which they are created. Therefore, in this treatise
we present our study of the interaction between protons and $\eta$ or $\eta^{\prime}$  mesons 
in strict connection with 
the investigation of their structure and production mechanism.

The fact that fourty years after the discovery of the $\eta$ and $\eta^{\prime}$ mesons, 
their interaction 
with nucleons remains so weakly established, in spite of its crucial importance for hadron physics,
indicates that it is rather challenging to conduct such research.

From the methodological point of view, the first difficulty needed to be solved 
was to find a way of how the interaction with nucleons of such  short-lived objects  like $\eta$ and $\eta^{\prime}$ mesons
can manifest itself in a measurable manner. 
For, even when moving close to the velocity of light, both these mesons disintegrate
on the average within a distance of tens of femtometers rendering their direct detection impossible. Not to mention,
that it is completely unfeasible to accomplish out of them a beam or a target.
However, the creation of these mesons in the vicinity 
of nucleons with low relative velocities appeared to us to be a promising tool which could facilitate 
the  study of meson-nucleon interactions. 
For this purpose the best suited is the  production
close to the kinematical threshold or in kinematical regions
where the outgoing particles indeed possess low relative velocities and hence
remain in distances of few Fermi long enough to experience the hadronic interaction.
Investigations were started with  measurements of the 
excitation function for the $pp \to pp\eta$ and $pp \to pp\eta^{\prime}$ reactions, searching
for its statistically significant distortion caused by the interacting ejectiles. 
We scanned the energy region close to the kinematical threshold  where the effect 
was supposed to be most significant and where the interpretation of the results is not burdened
by the complexity resulting from contributions of many angular momentum states of initial and final systems.
Indeed, we  found out that the interaction between protons augments
the total production cross section gradually with decreasing
volume of the phase space. At the threshold the enhancement is  most pronounced and 
corresponds to an enlargement of the cross section by
more than an order of magnitude and it vanishes where the kinetic energy shared by ejectiles
in the centre-of-mass system exceeds $\sim$~100~MeV. It was most fascinating to see that the interaction
of the $\eta$ meson with protons enhanced the cross section by a further 
factor of two. As it will be demonstrated in this dissertation, we were able to discern this from the effect
caused by the  overwhelming  proton-proton interaction in a model independent manner.
This was possible by making a comparison of the excitation functions of the 
$pp\to pp\eta$ and $pp \to pp\pi^0$ reactions.

We would like to stress that the revealed phenomenon 
--~viz the interaction of the created object with the outgoing nucleons
affects the probability of its very production~--
is of a purely quantum mechanical nature.
Our understanding of this peculiar effect will be presented in chapter~\ref{inertact}.
However, it is by far easier to imagine that the interaction between ejectiles modifies the distributions 
of their relative momenta. This appears much more 
intelligible even on the analogy  of the well known effect that the momentum distributions
of $\beta$ particles are different for positrons and for electrons 
due to their Coulomb interaction with the residual nucleus.
Therefore seeking for another  measurable manifestation of the interaction 
between the $\eta$ meson and the proton
we performed a measurement of the $pp \to pp\eta$ reaction with a statistics 
which permitted to derive distributions of a complete set of differential cross sections.
The determined spectra of the invariant mass distributions for the two-particle subsystems of the $pp\eta$ final state
revealed  strong deviations from the predictions based on the assumption that all kinematically allowed
momentum configurations of ejectiles are equally probable. 
Deviations at low relative proton-proton momenta can be explained
satisfactorily well by the hadronic interaction between protons. Yet, a pronounced discrepancy observed in the region 
of large relative momentum between protons is by far larger than expected from an influence of the 
$\eta$-proton hadronic potential. This intricate observation will be examined in detail in section~\ref{dalitzsection}.
Its explanation appears to be quite a challenge due to the need for the description of a three-body system
subjected to the complex hadronic potentials.
Later on in section~\ref{searchforsection} we will outline  our ongoing investigations aiming 
at the determination of the hadronic interaction occuring between the $\eta^{\prime}$ meson and the proton.
In this case, contrary to the $\eta$ meson, 
the excitation function for the $pp\to pp\eta^{\prime}$ reaction did not reveal any statistically significant signal 
which could have been  assigned to 
the $\eta^{\prime}$-proton interaction. At present we are continuing 
these studies by determining
distributions of the invariant masses for two-particle subsystems of the $pp\eta^{\prime}$
final state which might deliver the first ever experimental evidence for this 
still unknown potential. Such investigations constitute a much more difficult experimental task
compared to the study of the $\eta$ meson. This is mostly due to the fact that the cross section 
for the $pp\to pp\eta^{\prime}$ reaction
is by about a factor of thirty smaller than the one of the $pp\to pp\eta$ reaction at the corresponding excess energy.

  Thorough  theoretical investigations have shown that in order to understand how the energy of motion
is converted into a meson, it is of utmost  importance to take into account
also an interaction between nucleons before the very collision. We will demonstrate in section~\ref{sectionfsiisi}
that the interaction between nucleons in the initial channel  reduces the total cross section drastically. For the 
$pp\to pp \eta$ reaction it is decreased 
by a factor of four and in the case of the $pp\to pp\eta^{\prime}$ reaction by a factor of three!

As an essence of the above part of the introduction we would like to emphasize that
the inferences about the hadronic interaction
of $\eta$ and $\eta^{\prime}$ with nucleons will be based on comparisons
of both the   differential cross sections and
the close-to-threshold excitation functions for the $pp\to pp\eta$ and $pp\to pp\eta^{\prime}$ reactions
with predictions based on the assumption
that the kinematically available phase space is  homogeneously populated.

It cannot be, however, a priori excluded that
possible deviations of the experimental distributions
from the phase space governed calculations may
also be due to other
physical effects caused by the specific production mechanism.
Therefore, the discussion of the dynamics of the considered reactions,
embodied in chapter~\ref{Dopsmp}, constitutes
a part of this thesis. 
We will show that at threshold the entire production dynamics manifests itself in a value of
a single constant which determines the magnitude of the total cross section.
Consequently, the measurements of one reaction channel (eg. $pp \to pp X$)
is by far not sufficient for a full understanding of the reaction mechanism,
since having  only one observable at disposal
it is impossible to establish  contributions from many
plausible production currents (as an example let us mention mesonic, nucleonic, or resonance currents).
Therefore,  an exploration of isospin and spin degrees of freedom
is mandatory if we like to corroborate or refute the hypotheses proposed. 
In addition, 
the question of the reaction mechanism is connected with the very basic problem of whether hadronic
or quark-gluon degrees of freedom are more appropriate for the description of phenomena occuring 
in the energy regime relevant to our studies.
We address this issue in chapter~\ref{Dopsmp}, which
comprises the description of currently considered
production models and gives account of present and planned future experimental studies 
of spin and isospin observables.
For example, we will discuss
to what extent one can infer a gluonic component
of the $\eta^{\prime}$ wave function from a comparison 
of its creation in  collisions of nucleons with different isospin combination.

Consideration of the quasi-free meson creation on virtual nucleons bound inside a nucleus
led us to the idea  of studying the dependence of the meson production cross section
on the virtuality of colliding baryons. 
The idea will be elucidated in section~\ref{virtuality},
and a method for the measurement of the close-to-threshold production 
of mesons in the quasi-free proton-neutron 
and even neutron-neutron collisions will be demonstrated in chapter~\ref{experiment}. 

From the experimental point of view, measuring the energy dependence of the total cross section
for the production of mesons close to the kinematical threshold was quite a challenge. 
First of all, because these cross sections
are  by more than seven orders
of magnitude smaller compared to the total yield of pp reactions, and secondly because they vary by few orders of magnitude
in a few MeV range of the excess energy. The afore mentioned relations are  demonstrated 
in figure~\ref{ppstrangetotal}. 
It shows 
recently determined excitation functions for the close-to-threshold 
production of some of the pseudoscalar mesons, and in particular 
it shows the data for the $\eta$ and $\eta^{\prime}$ mesons,
which constitute an experimental basis of this treatise. 
\begin{figure}[H]
\vspace{-0.1cm}
 \parbox{0.6\textwidth}{\epsfig{file=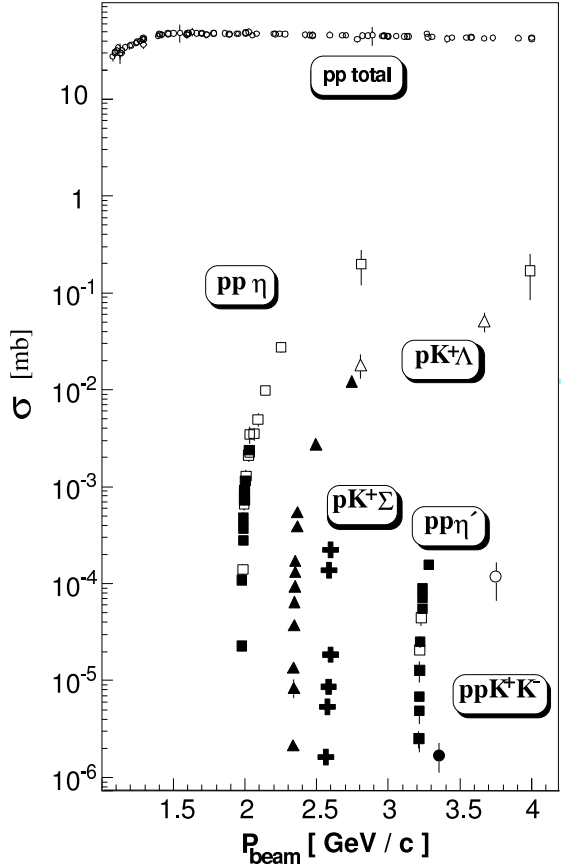,width=0.57\textwidth}} 
\hfill 
 \parbox{0.394\textwidth}{
   \caption{\label{ppstrangetotal}
       Close-to-threshold cross section for the proton-proton interaction leading to the
       production of mesons whose wave function comprises a significant amount of strangeness.
       For comparison, the total cross section of proton-proton collisions is also shown.
       The filled points depict data taken at 
       COSY~\cite{smyrski182,moskal367,balewski211,sewerin682,bilger217,moskal3202,moskal416,quentmeier276},
       and open symbols show results from
       other 
       laboratories~\cite{calen39,chiavassa270,bergdoltR2969,hibou41,pickup329,bodini475,fickinger2082,louttit1465,bierman922,balestra7,groom1}. 
     }
 }
\end{figure}
Conclusive inferences concerning the interaction 
between ejectiles could be drawn only due to the unprecedented experimental precision
obtained using proton beams available at storage ring facilities and in particular at the cooler synchrotron COSY. 
These beams are characterized by low emittance, small momentum spread, and well defined absolute momentum.
Experiments which will be discussed in this dissertation have been carried out
by means of the COSY-11 facility~\cite{Proposalwalter,brauksiepe397} installed at COSY~\cite{prasuhn167}.
Details of the experimental method are elucidated in chapter~\ref{experiment}.
In short, the technique is based on the determination of the four momentum vectors of
colliding and outgoing nucleons, and on the derivation of the mass of the produced meson or group of mesons
employing the principle of energy and momentum conservation.
The registered particles are identified by independent measurements of their velocity and momentum,
where the former is
determined from the time-of-flight between scintillation detectors and the latter
is deduced from the curvature of the particle trajectories in a magnetic field.
Figure~\ref{mmetap} presents an example of the mass spectrum determined 
for a system X produced via the  $pp\to ppX$ reaction
at an excess energy of 1.5~MeV with respect to the kinematical threshold 
for the production of the $\eta^{\prime}$ meson.
A signal from the creation of the $\eta^{\prime}$ meson can be well recognized and its width of
0.7~MeV (FWHM) indicates the precision achieved using the stochastically cooled proton beam of COSY
and the COSY-11 detection setup.  When examining the error bars, it is evident that if the accuracy was
twice worse the signal would be disputable. 
The method invented to  monitor the beam  geometry
at internal target facilities~\cite{moskal448,ecoolmoskal}, 
which is described in section~\ref{monitoring}, 
enabled us to improve 
the accuracy of mass determination
by correcting for variations of the avarage beam characteristics.
On the cognitive side, its application  led to the understanding of to what extent  the 
stochastic cooling improves the quality of the beam.
\begin{figure}[H]
\parbox{0.5\textwidth}{\epsfig{file=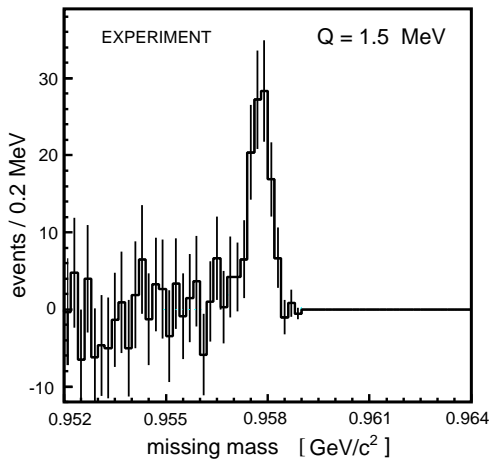,width=0.5\textwidth}} \hfill
\parbox{0.45\textwidth}{
   \vspace{-0.9cm}
   \caption{\label{mmetap}
       Missing mass spectrum for the $pp\to pp X$ reaction measured by means of the COSY-11 detection
       facility at the cooler synchrotron COSY. The measurement was performed  at a beam momentum corresponding to an 
       excess energy of 1.5~MeV for the $pp\to pp\eta^{\prime}$ reaction. 
       More details concerning the evaluation of the data will be explained in section~\ref{sectionmeasurement}.
       It is worth noting that the attained experimental mass resolution 
       is comparable with the natural width of the $\eta^{\prime}$ meson
       ($\Gamma$$_{\eta^{\prime}}$~=~0.202~MeV~\cite{PDG}).
     }
}
\end{figure}
\vspace{-0.3cm}
Necessary corrections for a non-ideal detection geometry were performed in an
utterly model independent manner. To arrive at this aim we expressed the experimental acceptance as a function 
of a complete set of mutually orthogonal variables describing the reaction, and to facilitate
calculations we digitized the phase space volume into a few thousand partitions in which the acceptance 
could be safely regarded as constant. 
In the analysis, we assigned to each measured event a weight equal to the inverse of the acceptance 
according to the phase space bin to which it belonged. 
This allowed us to elaborate the differential distributions independently of the reaction model used in simulations.
Employing the missing mass technique it is in principle impossible to discern between 
the multi-pion and $\eta$ or $\eta^{\prime}$ meson production on the event-by-event basis.
Therefore,  to derive the differential distributions we divide the range of the studied variable into bins  with a
width corresponding to the achieved experimental resolution
and separate the multi-pion background from the $\eta$ or $\eta^{\prime}$ meson signal for each bin.
The effort was undertaken in order to determine background- and model-free
distributions of differential
cross sections which in turn enable to test various hypotheses without 
an obscurity due to unnecessary assumptions. Hence we can benefit thoroughly from
simplifications  gained due to  the threshold kinematics.
The most crucial facilitation and attractiveness when interpreting the meson production 
at the vicinity of the threshold is the fact that the relative angular momenta larger than
$l = 0\,\hbar$ play no role due to the short range of the strong
interaction and small relative momenta of the produced particles.
It can be inferred from parity and angular momentum conservation laws that
the production of a nucleon-nucleon-meson system (for pseudoscalar or vector
mesons) in relative S-waves may only occur if the nucleons collide in P-wave.
Thus at threshold the transition $\mbox{P} \rightarrow \mbox{S}\mbox{s}$ is
the only possible one, with capital letters denoting the angular momentum
between nucleons and the small letter describing the meson angular momentum with
respect to the pair of nucleons. Section~\ref{partialwaves} will be devoted to the circumstantial
discussion of this issue. At this stage we would only like to note,
that the dominance of the $\mbox{P} \rightarrow \mbox{S}\mbox{s}$ transition
at threshold will not be unquestioningly assumed but rather the energy range of its applicability
--~which varies strongly with the mass of the created meson~--  will 
be established in section~\ref{range}. 

The present work is divided into seven chapters. After this introduction we will elucidate our understanding
of the manifestation of the hadronic interaction via the excitation function and the phase space abundance of
the studied reactions. Further, in the third chapter we will introduce the definitions of observables used
throughout this work. The fourth chapter comprises a discussion of the experimental evidence concerning 
the interaction between the pseudoscalar isosinglet mesons ($\eta$, $\eta^{\prime}$)  and the proton. 
In the fifth chapter we explain the reaction dynamics and discuss the relevance for the production process 
of hadronic and quark-gluon degrees of freedom.  
The final conclusion is preceded by the sixth chapter which contains the detailed description of the experimental method.


     \def\chaptername{Chapter}
     \cleardoublepage


\newpage
\clearpage
\pagestyle{plain}
\chapter{Manifestation of hadronic interactions} 
\thispagestyle{empty}
\pagestyle{myheadings}
\markboth{Hadronic interaction of $\eta$ and $\eta'$ mesons with protons}
         {2. Manifestation of hadronic interactions}
\label{inertact}                          

\begin{flushright}
\parbox{0.8\textwidth}{
 {\em ... what it can  mean to ask what is the probability  of past event given a future event!
      There is no essential problem with  this, however. Imagine the entire history of the universe
      mapped out in space-time. To find the probability of $p$ occurring, given that $q$
      happens, we imagine examining all occurrences of $q$ and counting up the fraction 
      of these which are accompanied by $p$. This is the required probability. It does not 
      matter whether $q$ is the kind of event that would normally occur later or earlier
     in time than $p$~\cite{penrose}.\\
 }
 \protect \mbox{} \hfill Roger Penrose
 }
\end{flushright}

The interaction of hadrons -- being the reflection of the strong force acting 
between their constituents -- provides information about their 
structure and the strong interaction itself. 
In the frame of the optical potential model the hadronic interaction can be 
expressed in terms of phase-shifts, which in the zero energy limit are described by the 
scattering length and effective range parameters. 
These are quite well known for the low-energy nucleon-nucleon 
interaction~\cite{machleidtR69,machleidt024001}, yet they are still poorly 
established in the case of meson-nucleon or  meson-meson interactions. 
This is partly due to the absorption of mesons when scattering on 
a baryon. To account for this effect the scattering length becomes a complex 
quantity where the imaginary part -- e.g.\ in case of the nucleon-$\eta$ 
interaction -- describes the $\eta N \rightarrow \pi N$ and $\eta N \rightarrow 
multi\mbox{-}\pi\,N$ processes. 
Moreover, the short life-time of all {\em neutral} ground state mesons 
prohibits their direct utilization as secondary beams and therefore the study of 
their interaction with hadrons is accessible only via 
their influence on the cross section of the reactions in which they were produced 
(eg. $NN \rightarrow NN\, Meson$).  
When created close to the kinematical threshold with the relative kinetic 
energy being in the order of a few MeV, the final state particles remain much 
longer in the range of the strong interaction than the typical life-time of 
$N^{*}$ or $\Delta$ baryon resonances with $10^{-23}\,\mbox{s}$, and hence 
they can easily experience a mutual interaction before escaping the area of an 
influence of the hadronic force. 
This -- as introduced by Watson~\cite{watson1163} -- final state interaction 
(FSI) may significantly modify both the original distributions of relative 
momenta of the outgoing reaction products and the magnitude of the production 
cross section. 
Although it is easy to understand that the FSI changes the distributions of the 
differential cross sections it is rather difficult to cope with the influences 
on the magnitude of the total reaction rate, since one tends to separate the 
primary production from the final state interaction in space as well as in 
time~\cite{watson1163}.
The influence of the FSI on the absolute scale of the cross section is a pure
quantum mechanical effect and 
our classical imagination must inevitably fail in this respect. 
However even when 
considering the primary production as a separate process it is well worth 
trying to understand the phenomenon qualitatively. 
If there were no final state interactions the total cross section would be 
fully determined by the kinematically available phase space volume $V_{ps}$, 
where each interval is populated with a probability governed by the primary 
production amplitude only:
\be
\sigma = \frac{1}{\mbox{F}} \int dV_{ps} \, |M|^2 \ \approx \ 
  const. \cdot V_{ps}.
\ee
The approximation in the equation results from the assumption that 
$|M|^2 = \mbox{constant}$ in a few MeV range above the production 
threshold~\cite{moalem445,bernard259,gedalin471}. 
F denotes the flux factor of the colliding particles.\vspace{1ex}

In the classical picture we might imagine that the reaction particles are 
created together with their appropriate force field and when escaping the 
interaction region they acquire a potential energy increasing or decreasing 
their kinetic energy depending whether the interaction is repulsive or 
attractive. 
For an attractive interaction they could be created also in those phase space 
partitions which are not available for non-interacting particles and 
subsequently be ``pulled down'' to the energetically allowed regions by final 
state interaction. 
The temporary growth of the primary production phase space would then 
increase the reaction rate. 
Contrary, in the case of a repulsive interaction the particles must be produced in 
the lower phase space volume since leaving the interaction area they will 
acquire additional kinetic energy.
The reduction of the total cross section, for example in case of the repulsive 
Coulomb force, can easily be understood when considering the production in a 
coordinate space of point-like objects. 
Here -- in contrast to non-interacting particles -- a strongly repulsive 
object cannot be produced at appropriately small distances since their later 
acceleration would violate energy conservation and thus 
the space available to primary production is reduced.

 Another more commonly anticipated presupposition of changes of the reaction yield due to the 
final state interaction is based on the time invariance principle. It asserts
that the cross section for the production of  repulsive particles is smaller
in comparison to  non-interacting particles, all other features  being equal,
because in the time inversed process 
the repulsive interaction hinders the slow particles 
from reaching distances at which a particular reaction may occur.
\begin{figure}[H]
\vspace{-0.4cm}
\parbox{0.50\textwidth}
  {\epsfig{figure=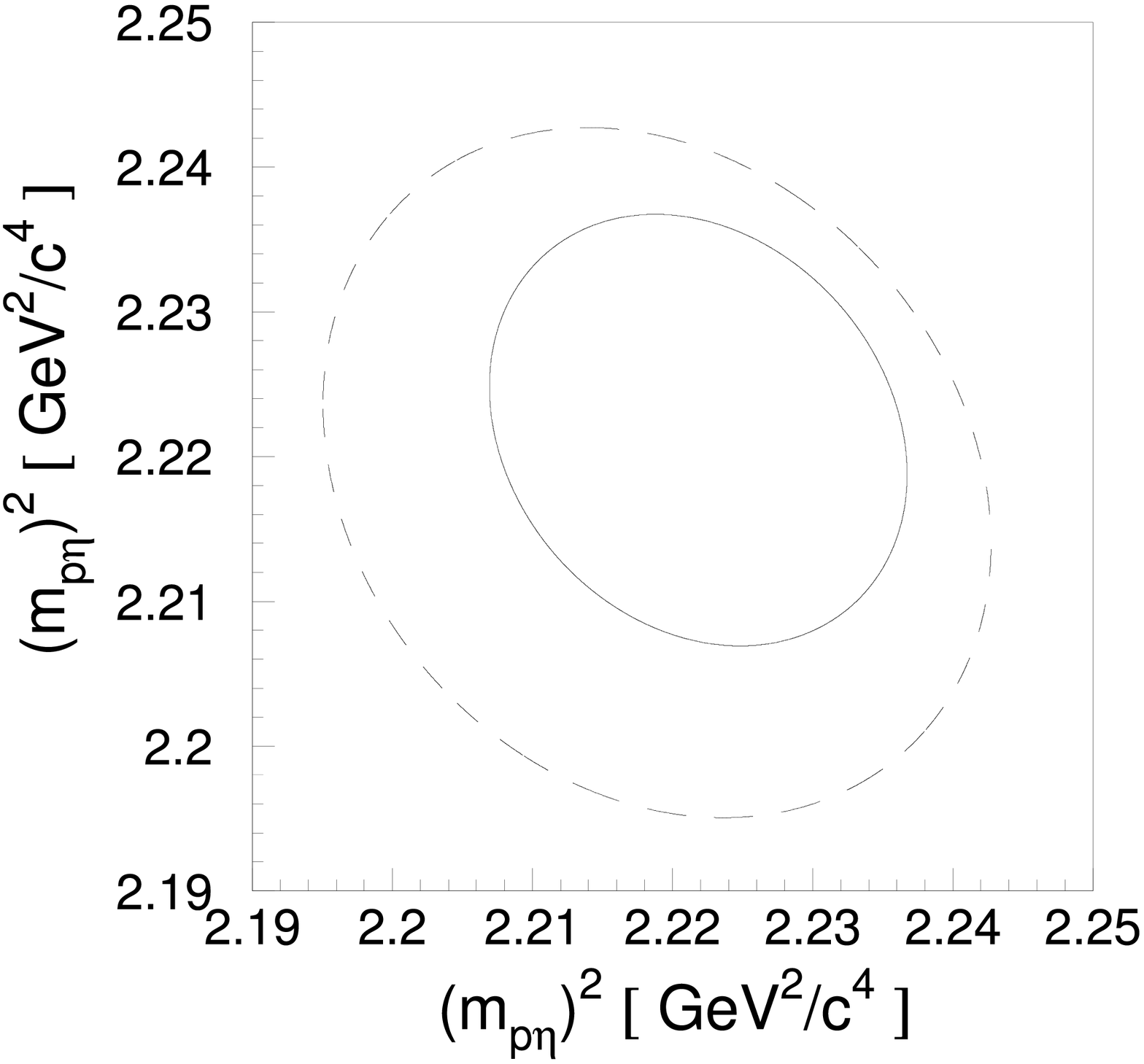,width=0.47\textwidth}} \hfill
\parbox{0.45\textwidth}
  {\epsfig{figure=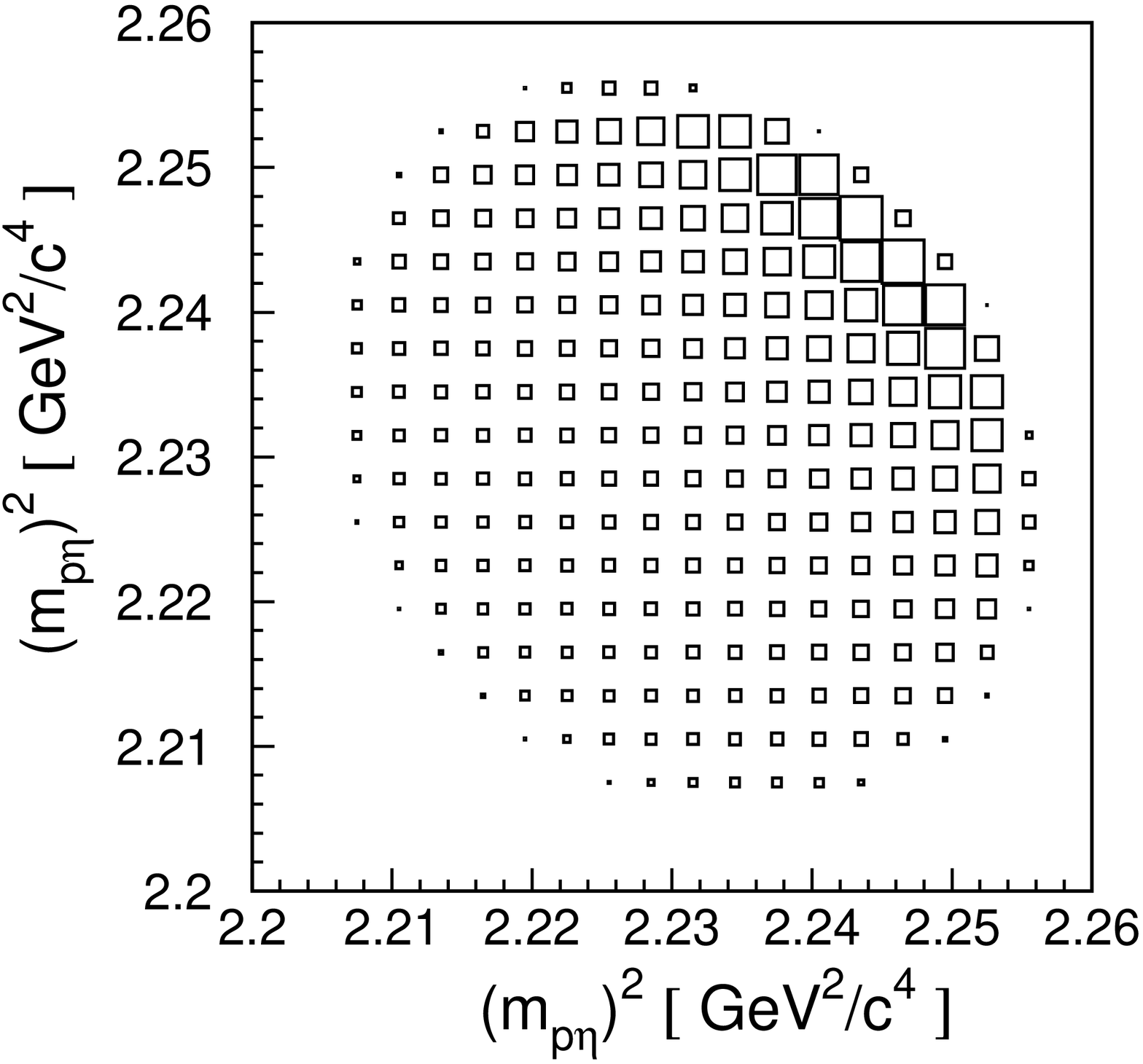,width=0.48\textwidth}}

\vspace{-0.3cm}
\caption{\label{dalitz_phasespace} ({\bf left}) The solid line indicates the 
kinematically available area for the $pp \eta$ system with a total 
centre-of-mass energy $\sqrt{\mbox{s}} = 2433.8\,\mbox{MeV}$ 
(excess energy Q~$\approx$~10~MeV).
The dashed 
line depicts the range assuming a reduction of the proton and $\eta$-meson 
masses by $2\,\mbox{MeV}$. The phase space volume results from an integral 
over the closed area: $V_{ps} = \frac{\pi^{2}}{4\,\mbox{\scriptsize s}}
\int\int d\,\mbox{m}^2_{p_{1}\eta}\,d\,\mbox{m}^2_{p_{2}\eta}$. 
(Terms 
of $(2\pi )^n$ are skipped here and will be included 
in the flux factor F 
according to the convention introduced by Byckling and 
Kajantie~\cite{bycklingkajantie}).\protect\\
({\bf right}) Distribution of the phase space for the $pp\eta$ system modified by 
the proton-proton interaction and calculated for an excess energy of 
$\mbox{Q} = 16\,\mbox{MeV}$. 
The area of the squares is proportional to the number 
of entries and is shown in a linear scale.}
\end{figure}
\vspace{-0.2cm}
One can also argue, that relativistically the primary mechanism creates the 
particles off the mass shell and subsequently they are lifted onto the mass 
shell by the final state interaction. 
The solid line in figure~\ref{dalitz_phasespace} (left) depicts the boundary 
of the Dalitz plot in case of the $pp \rightarrow pp \eta$ reaction calculated 
at the total centre-of-mass energy $\sqrt{\mbox{s}} = 2433.8\,\mbox{MeV}$ 
exceeding the threshold energy by $10\,\mbox{MeV}$. 
The area surrounded by that curve is a direct measure of the kinematically 
available phase space volume. 
The dashed line shows the corresponding plot at the moment of the primary 
creation if the mass of each particle was reduced by $2\,\mbox{MeV}$, 
demonstrating that now the available phase space grows significantly. 
Indeed, as shall be inferred from the experimental results presented in 
subsequent sections, at excess energies of a few MeV above threshold, the 
mutual interaction among the outgoing particles enhances drastically --~by 
more than an order of magnitude~-- the total cross section and modifies 
appreciably the occupation of the phase space. 
Figure~\ref{dalitz_phasespace} (right) indicates the phase space distribution 
expected for the $pp \eta$ system at an excess energy of $\mbox{Q} = 
\sqrt{\mbox{s}} - 2\,\mbox{m}_{p} - \mbox{m}_{\eta} = 16\,\mbox{MeV}$, 
assuming a homogeneous primary production and taking into account the S-wave 
interaction between the protons. 
The proton-proton FSI modifies the homogeneous Dalitz plot distribution of 
``non-interacting particles'', enhancing its population at a region where the 
protons have small relative momenta. 
The interaction of the proton-$\eta$ system would manifest itself at low 
invariant masses $\mbox{m}^2_{p\eta}$ corresponding to small relative momenta 
between the proton and the $\eta$ meson.
Such effects observed in the experiments are presented in section~\ref{dalitzsection}.

The effect of the nucleon-nucleon FSI diminishes with increasing excess 
energy since it significantly influences only that partition of the 
phase space at which the nucleons have small relative momenta. 
While this fraction stays constant, the full volume of the phase space 
grows rapidly: 
An increase in the excess energy from $\mbox{Q} = 0.5\,\mbox{MeV}$ to 
$\mbox{Q} = 30\,\mbox{MeV}$ corresponds to a growth of $V_{ps}$ by more than 
three orders of magnitude. 
As a result the S-wave nucleon-nucleon FSI is of less importance for higher 
excess energies where it affects a small fraction of the available 
phase space volume only.
A more quantitative discussion about the influence of the FSI 
and ISI (initial state interaction) effects 
at close-to-threshold production cross sections will  be presented in 
section~\ref{range}.

In this section we would only like to state that 
an interaction of the nucleons in the entrance channel,
 similarly to that among ejectiles, 
 influences the 
production process noticeably~\cite{hanhart176}.
For example, for $\eta$ production in the $pp \to pp \eta$ reaction the 
initial state interaction reduces the total cross section by a 
factor of about 3--5~\cite{batinic321,hanhart176} due to the repulsive proton-proton 
$^3\mbox{P}_{0}$-wave potential. 
This factor remains rather constant in the range of a few tens of MeV~\cite{batinic321} 
and hence, does not influence the energy dependence of the total cross section 
of the meson production, which is mostly determined by the final 
state interaction. 
In particular, the dominant S-wave nucleon-nucleon final state interaction 
is by far stronger than any of the low-energy meson-nucleon ones.
The reduction of the cross section by the initial state 
interaction we may intuitively conceive as a loss of the initial flux in favour 
of elastic, and a plethora of other --~than the studied one~-- inelastic reactions, 
which may occur when the colliding nucleons approach each other.

The above considerations point out that the hadronic interaction between 
nucleons and the short-lived mesons like $\eta$ and $\eta^{\prime}$
may be investigated
by the observation of the excitation 
function  for the production of these mesons off nucleons,
as well as by the studies of  the phase space abundance. 


\newpage
\clearpage
\pagestyle{plain}
\chapter{Definitions of observables}
\thispagestyle{empty}
\pagestyle{myheadings}
 
{

\markboth{Hadronic interaction of $\eta$ and $\eta'$ mesons with protons}
         {3. Definitions of observables} 
\label{appendfree}
\begin{flushright}
\parbox{0.69\textwidth}{
 {\em  
\hspace{-0.5cm}{\bf Hermogenes.} Suppose that we make Socrates a party to the argument?

\hspace{-0.5cm}{\bf Cratylus.} If you please. 

... 

\hspace{-0.5cm}{\bf Socrates.} Then the argument would lead us to infer the names
 ought to be given according to a natural process, and with a proper
 instrument, and not at our pleasure: in this and no other way shall
 we name with success. 

\hspace{-0.5cm}{\bf Hermogenes.} I agree. 

\hspace{-0.5cm}{\bf Socrates.} But again, that which has to be cut has to be cut with something? 

\hspace{-0.5cm}{\bf Hermogenes.} Yes. 

\hspace{-0.5cm}{\bf Socrates.} And that which has to be woven or pierced has to be woven
      or pierced with something? 

\hspace{-0.5cm}{\bf Hermogenes.}  Certainly. 

\hspace{-0.5cm}{\bf Socrates.} And that which has to be named has to be named with something? 

\hspace{-0.5cm}{\bf Hermogenes.} True~\cite{cratylus}. \\
 }
 \protect \mbox{} \hfill Plato
 }
\end{flushright}

\section{Cross section}
\label{Siv}
\begin{flushright}
\parbox{0.73\textwidth}{
 {\em
 Why is it that, in most cases, the definitions which satisfy scientists
 mean nothing at all to children?~\cite{poincare}\\
 }
 \protect \mbox{} \hfill  Henri Poincar$\acute{\mbox{e}}$ \protect\\
 }
\end{flushright}
\vspace{-0.0cm}
Investigations of the production of mesons and their interactions with 
nucleons are based on measurements determining the total and differential 
production cross sections and their dependence on the energy of the 
interacting nucleons.
Therefore, to enable a quantitative discussion on the mechanisms leading to 
the transformation of the energy-of-motion of nucleons into matter in the 
form of mesons let us recall the formula of the reaction cross section. 
In the case of the $NN \rightarrow NN \,Meson$ process -- with the four-momenta 
of the colliding nucleons denoted by $\mathbb{P}_a$ and $\mathbb{P}_b$ and 
with $n = 3$ particles in the exit channel -- it reads:

{{
\be 
\nonumber
\sigma = \frac{1}{\mbox{F}} \int dV_{ps} |M_{ab\,\rightarrow\,123}|^2 = 
\ee 
\be 
\label{phasespacegeneral}
 = \frac{1}{\mbox{F}} \int \prod_{i=1}^n d^{4}\mathbb{P}_{i} 
    \cdot \delta(\mathbb{P}_{i}^2 - \mbox{m}_{i}^2)
    \cdot \Theta(\mathbb{P}_{i}^2) 
    \cdot \delta^{4}(\mathbb{P}_a + \mathbb{P}_b - \sum_{j=1}^{n}\mathbb{P}_{j})
    \cdot |M_{ab\,\rightarrow\,123}|^2,
\ee
}}
where $|M_{ab\,\rightarrow\,123}|^2$ denotes the square of the 
Lorentz-invariant spin averaged matrix element describing the probability to 
create two nucleons and a meson with four-momenta of 
$\mathbb{P}_{i} = (\mbox{E}_i,\vec{\mbox{p}}_i)$ and $i = 1..n$, respectively.
The energy and momentum conservation as well as the on-shellness of the 
created particles is ensured by the Dirac-$\delta$ and the 
Heaviside-$\Theta$ functions.
The formula holds also for $n \neq 3$.\vspace{1ex}

The total cross section is then defined as an integral of the probabilities to 
populate a given phase space interval over the whole kinematically available 
range --~determined by energy and momentum conservation~-- normalized to the 
flux factor F of the colliding nucleons. 
In the case of non-interacting final state particles the matrix element close 
to threshold $|M_{ab\,\rightarrow\,123}|^2$ is nearly 
constant~\cite{moalem445,bernard259,gedalin471} and hence the allowed 
phase space volume, $V_{ps}$, is the decisive quantity which governs the 
growth of the cross section with increasing excess energy Q.
The latter -- defined as the total kinetic energy -- is shared among the 
outgoing particles in the reaction centre-of-mass 
frame~\footnote{\mbox{} Traditionally by the centre-of-mass system we understand a frame 
in which the momenta of all particles add to zero, called sometimes more 
explicitly centre-of-momentum frame.}:
\be
\mbox{Q} = \sqrt{\mbox{s}} - \sum_{i=1}^{n} \mbox{m}_{i}, 
\ee
where $\mbox{s} = |\mathbb{P}_a + \mathbb{P}_b|^2 = 
|\sum_{i=1}^{n} \mathbb{P}_{i}|^2$ denotes the square of the total 
centre-of-mass energy.  
Exactly at threshold, where the particles' kinetic energy in the centre-of-mass 
system equals  zero ($\mbox{Q} = 0\,\mbox{MeV}$), the total reaction 
energy $\sqrt{\mbox{s}}$ amounts to: 
$\sqrt{\mbox{s}_{th}} = \sum_{i=1}^{n} \mbox{m}_{i}$.
Before writing explicitly the formula for $V_{ps}$ let us
introduce the kinematical triangle function $\lambda$
defined as~\cite{bycklingkajantie}:
\be
\label{kaellen}
\lambda(x, y, z) = x^2 + y^2 + z^2 - 2xy - 2yz - 2zx.
\ee
It enables to formulate many useful kinematical variables in a very 
compact and Lorentz-invariant form.
In particular, the momenta of particles $i$ and $j$ in their centre-of-mass frame
may be expressed as follows:
\be
\label{momentum_kaellen}
\mbox{p}_{i}^{*} = \mbox{p}_{j}^{*} = 
 \frac{\sqrt{\lambda(\mbox{s}_{ij},\mbox{m}_{i}^2,\mbox{m}_{j}^{2})}}
      {2\sqrt{\mbox{s}_{ij}}},
\ee
where $\mbox{s}_{ij} = |\mathbb{P}_i + \mathbb{P}_j|^2$ stands for the square 
of the invariant mass of the $ij$ system considered as one quasi-particle. 
The above relation gives an expression for the flux factor F from 
equation~\eqref{phasespacegeneral} in terms of the colliding masses of the 
nucleons and the total energy s only, namely:
\be
\label{fluxfactor}
\mbox{F} = 4\,\sqrt{\mbox{s}} \,(2\pi)^{3n-4}\;\mbox{p}_{a}^{*} = 
  2\,(2\pi)^{3n-4}\;\sqrt{\lambda(\mbox{s},\mbox{m}_a^2,\mbox{m}_b^2)},
\ee
where we have chosen the convention introduced by Byckling and 
Kajantie~\cite{bycklingkajantie} and included the $(2\pi)$ factors for the 
phase space $(2\pi)^{3n}$ and for the matrix element $(2\pi)^{-4}$ into the 
definition of F. 
It is important to note that at threshold, for an excess energy range of a few 
tens of MeV, the small fractional changes of total energy 
$(\frac{\mbox{\scriptsize Q}}{\sqrt{\mbox{\scriptsize s}}} = 
\frac{\mbox{\scriptsize Q}}{\mbox{\scriptsize m}_1\,+\,
\mbox{\scriptsize m}_2\,+\,\mbox{\scriptsize m}_3\,+\,\mbox{\scriptsize Q}})$ 
causes weak variations of the flux factor and influences the 
shape of the energy dependence of the total cross section only slightly.\vspace{1ex}

For a two-particle final state $ab \rightarrow 12$ (for instance for the 
reaction $pn \rightarrow d \eta$) the phase space integral defined in 
equation~\eqref{phasespacegeneral} reduces to $V_{ps} := \int dV_{ps} = 
\frac{\pi}{\sqrt{\mbox{\scriptsize s}}}\,\mbox{p}_1^* = 
\frac{\pi}{2\,\mbox{\scriptsize s}}
\sqrt{\lambda(\mbox{s},\mbox{m}_1^2,\mbox{m}_2^2)}$ and the total cross 
section for such reactions (when neglecting variations due to dynamical 
effects $(|M_{ab\,\rightarrow\,12}| = const.)$) should increase linearly with 
the centre-of-mass momentum of the produced meson in the vicinity of the 
reaction threshold. 
The total cross section can be expressed analytically as a function of the 
masses of the particles participating in the reaction and the square of the total 
reaction energy s:
\be
\label{Vps_two_body}
\sigma_{ab\,\rightarrow\,12} = const \cdot \frac{V_{ps}}{\mbox{F}} = 
 \frac{const}{16 \pi\,\mbox{s}} \frac{\mbox{p}_1^*}{\mbox{p}_a^*} = 
 \frac{1}{16 \pi\,\mbox{s}} 
 \frac{\sqrt{\lambda(\mbox{s},\mbox{m}_1^2,\mbox{m}_2^2)}}
      {\sqrt{\lambda(\mbox{s},\mbox{m}_a^2,\mbox{m}_b^2)}}.
\ee
Near threshold, at a given excess energy $\mbox{Q} = \sqrt{\mbox{s}} - 
\mbox{m}_1 - \mbox{m}_2$, the emission of the reaction products in the 
centre-of-mass frame is isotropic and the whole dynamics of the process 
manifests itself in the absolute value of the transition matrix element 
$|M_{ab\,\rightarrow\,12}|$.
The underlying production mechanisms can also be extracted from the deviations 
of the total cross section energy dependence following the prediction of 
relation~\eqref{Vps_two_body}. 
A visualization of possible differences in the dynamics of the production of 
various mesons can be obtained by comparing the total cross sections of the 
studied reactions at the same value of the phase space volume normalized to 
the flux factor ($V_{ps}/\mbox{F}$). 
It is important to recognize that the comparison at the same centre-of-mass 
meson momentum ($\mbox{p}_2^*$) gives direct information about the amplitude 
differences only if the production of mesons with the same masses is concerned 
(for example $pp \rightarrow d \pi^+$ and $nn \rightarrow d \pi^-$), yet the 
reactions $pn \rightarrow d \eta$ and $pn \rightarrow d \pi^0$ have by far 
different $V_{ps}$ at the corresponding $\mbox{p}_2^*$.

Prior to the comparative analysis of the transition amplitudes for 
the $pp\to pp\eta$,  $pp\to pp\eta^{\prime}$, and $pp\to pp\pi^{0}$ reactions,
a corresponding formula of the phase space volume $V_{ps}$ 
for the three particle final state will be 
introduced in section~\ref{phasespacepop}. 

\section{Partial waves~--~selection rules}
\label{partialwaves}
\begin{flushright}
\parbox{0.70\textwidth}{
 {\em
 Not for nothing do we call the laws of nature 'laws': the more they 
 prohibit the more they say~\cite{popperlogic}.\\
 }
 \protect \mbox{} \hfill  Karl Raimund Popper \protect\\
 }
\end{flushright}

If one is interested in the decomposition of the production amplitude 
according to the angular momenta of the final state particles then indeed the
appropriate variable for a qualitative comparison even for particles with different 
masses is the meson momentum in the reaction centre-of-mass frame.
Correspondingly, for the more than two-body final state, the adequate variable 
is the maximum meson momentum, since it is directly connected with the maximum 
angular momentum by the interaction range. 
In the case of  a three-body exit channel ($ab \rightarrow 123$) the 
meson ($\mbox{m}_3$) possesses maximum momentum ($\mbox{q}_{max}$) when the remaining two 
particles are at rest relative to each other. Hence, employing 
definition~\eqref{momentum_kaellen} one obtains:

\be
\mbox{q}_{max} = 
  \frac{\sqrt{\lambda(\mbox{s},(\mbox{m}_1 + \mbox{m}_2)^2,\mbox{m}_3^2)}}
       {2\sqrt{\mbox{s}}}.
\ee

Contrary to the two-body final state, at a fixed excess energy the dynamics 
of the meson production associated with two or more particles reflects itself 
not only in the absolute value of the square of the matrix element but also in 
distributions of variables determining the final state kinematics.
Usually, in non-relativistic calculations of the total cross section, one 
takes the Jacobi momenta, choosing as independent variables the 'q'-meson 
momentum in the reaction centre-of-mass frame and 'k'-momentum of either 
nucleon in the rest frame of the nucleon-nucleon subsystem. 
By means of the $\lambda$ function they can be expressed as:
\be
\mbox{q} = 
 \frac{\sqrt{\lambda(\mbox{s},\mbox{s}_{12},\mbox{m}_3^2)}}{2\,\sqrt{\mbox{s}}},
 \;\;\;\mbox{and}\;\;\;
\mbox{k} = 
 \frac{\sqrt{\lambda(\mbox{s}_{12},\mbox{m}_1^2,\mbox{m}_2^2)}}
      {2\,\sqrt{\mbox{s}_{12}}},
\ee
with $\mbox{s}_{12} = |\mathbb{P}_1+\mathbb{P}_2|^2$ denoting the square of 
the invariant mass of the nucleon-nucleon subsystem.
In a non-relativistic approximation the expression of the total cross section 
defined by formula~\eqref{phasespacegeneral} for a meson production reaction 
in nucleon-nucleon interactions of the type $NN \rightarrow NN\,Meson$ 
simplifies to:
\be
\label{eq:sigma}
\sigma \propto \int_0^{\mbox{\scriptsize q}_{max}} 
  \mbox{k}\,\mbox{q}^2 \,|M_{ab\,\rightarrow\,123}|^2\,d\mbox{q}. 
\ee
Denoting by $L$ and $l$ the relative angular momenta of the nucleon-nucleon 
pair and of the meson relative to the $NN$ system, respectively, and 
approximating the final state particles by 
plane waves (case of 
non-interacting objects), whose radial parts $\psi_l(\mbox{q},\mbox{r})$ are 
given by the spherical Bessel functions:
\be
\psi_l(\mbox{q},\mbox{r}) \propto j_l(\mbox{qr}) 
  \stackrel{\mbox{\scriptsize qr }\rightarrow 0}{\longrightarrow} 
  \frac{(\mbox{qr})^l}{(2l+1)!},
\ee
an expansion of the transition amplitude $|M_{Ll}|^2$ for the $Ll$ partial 
wave around $\mbox{qr} = 0$ leads to
\be
|M_{Ll}|^2 \propto |M_{Ll}^0|^2  \ \mbox{k}^{2L} \ \mbox{q}^{2l} ,
\ee
where, $M_{Ll}^0$ denotes the matrix elements responsible for the primary production
with the angular momenta $L$ and $l$.
Thus, the total cross section can be expressed as the following sum 
of the partial cross sections $\sigma_{Ll}$: 
\be
  \sigma = \sum_{L,l} \sigma_{Ll} 
   \propto \sum_{L,l} \int_0^{\mbox{\scriptsize q}_{max}}
   |M_{Ll}^0|^2 \ \mbox{k}^{2L+1} \ \mbox{q}^{2l+2} \ d\mbox{q} 
\ee

Furthermore, if the strength of the the primary production  partial amplitudes $|M_{Ll}^0|$
were constant over the phase space then the energy dependence of the partial 
cross sections, obtained by solving the above equation, 
would be given by:
\be
\label{sigmaLl}
\sigma_{Ll} \propto \mbox{q}_{\,max}^{\,2L+2l+4} \propto \eta_{\,M}^{\,2L+2l+4},
\ee
where $\eta_{M} = \mbox{q}_{max}/\mbox{m}_M$ with $\mbox{m}_M$ denoting the 
mass of the created meson~\footnote{\mbox{} In former works~\cite{Rosenfeld,gellmann219} 
dealing only with pions this parameter is denoted by $\eta$, here in order to 
avoid ambiguities with the abbreviation for the eta-meson, we introduce an 
additional suffix $M$.}.
Thus -- at threshold -- for the Ss partial wave the cross section for the 
$NN \rightarrow NN \,Meson$ reaction should increase 
with the fourth 
power of $\eta_{M}$.
The dimensionless parameter $\eta_{M}$ was introduced by 
Rosenfeld~\cite{Rosenfeld} as a variable allowing for qualitative 
estimations of the final state partial waves involved in pion production. 
He argued that if the phenomenon of pion production takes place at a 
characteristic distance R from the centre of the collision, with R in the 
order of $\hbar/\mbox{Mc}$, then the angular momentum of the produced meson is 
equal to $l =\mbox{R}\,\mbox{q} = \hbar\,\mbox{q}/\mbox{Mc}$. 
Hence, $\eta_M$  denotes the classically calculated  maximum angular momentum of the 
meson relative to the centre-of-reaction. 
The same arguments one finds in the work of Gell-Mann and Watson~\cite{gellmann219}, 
where the authors do not expect the range of interaction to be larger than 
the Compton wavelength of the produced meson $\left(\hbar/\mbox{Mc}\right)$. 
However, it is rather a momentum transfer $\Delta\mbox{p}$ between the 
colliding nucleons which determines the distance to the centre of the collision 
$\mbox{R} \approx \hbar/\Delta\mbox{p}$ at which the production occurs. 
Based on indications from data, it is emphasized in the original article of 
Rosenfeld~\cite{Rosenfeld} that R is slightly less than $\hbar/(2\,\mbox{Mc})$, 
which is numerically close to the value of $\hbar/\Delta\mbox{p}$. 
Directly at threshold, where all ejectiles are at rest in the centre-of-mass 
frame, $\Delta\mbox{p}$ is equal to the centre-of-mass momentum of the 
interacting nucleons and hence, exploring equation~\eqref{momentum_kaellen} it 
can be expressed as:
\be
\label{momtranseq}
\Delta\mbox{p}_{th} = 
  \frac{\sqrt{\lambda(\mbox{s}_{th},\mbox{m}_{a}^2,\mbox{m}_{b}^{2})}}
       {2\,\sqrt{\mbox{s}_{th}}} 
  \stackrel{\mbox{\scriptsize for}\,pp \to ppX}{=\!=\!=\!=\!=\!=\!=\!=} 
  \sqrt{\mbox{m}_p \mbox{m}_X + \frac{\mbox{m}_X^2}{4}}.
\ee
Though the present considerations are limited to the spin averaged 
production only, it is worth noting that very close-to-threshold --~due to 
the conservation laws and the Pauli excluding principle~--
for many reactions there is only one possible angular momentum and spin 
orientation for the incoming and outgoing particles.
The Pauli 
  principle for the nucleon-nucleon system implies that 
  \begin{equation}
     (-1)^{{\scriptsize L}+{\scriptsize S}+{\scriptsize T}} = -1, 
  \end{equation}
  where $L$, $S$, and $T$ denote angular momentum, spin, and isospin of the nucleon 
  pair, respectively. For example if the nucleon-nucleon wave function is 
  symmetric in the configuration-space (${L} = 0$) as well as in the spin-space (${S} = 1$) then it must be 
  antisymmetric in the isospin-space (${T} = 0$) to be totally antisymmetric.

In the conventional notation~\cite{meyer064002,nakayama0302061} 
the transition between angular momentum parts of the initial and final  
states of the $NN\to NNX$ reactions are described in the following way:
\begin{equation}
        ^{2S^i+1}L_{J^i}^i \rightarrow ^{2S+1}\!\!L_{J},l
\end{equation}
where, superscript $i$ indicates the initial state quantities,
$l$ denotes the angular momentum of the  produced meson with respect
to the pair of nucleons, and $J$ stands for the total angular momentum of the N-N system.
The values of angular momenta 
are commonly being expressed using the spectroscopic notation 
($L$~=~S,P,D,... and $l$~=~s,p,d,...).
The parity conservation implies that $L^{i}+L+l$ must be an even number
if the parity of the produced meson 
is positive or odd if this parity is negative.
Employing additionally the conservation of the total angular momentum and Pauli principle
one can deduce that at threshold the $NN\to NNX$ reactions will be 
dominated by the transitions listed in table~\ref{partialtransitions}.
\vspace{-0.0cm}
\begin{table}[H]
\caption{\label{partialtransitions}Partial wave transitions for the $pp 
\rightarrow pp\,Meson$ and $nn \rightarrow nn\,Meson$ reactions at threshold}
\vskip -0.25cm
\tabskip=1em plus2em minus.5em
\halign to \hsize{\hfil#&\hfil#\hfil&\hfill#\hfill&\hfill#\hfill \cr
\noalign{\hrulefill}
type & meson  & spin and parity & transition  \cr
\noalign{\vskip -0.2cm}
\noalign{\hrulefill}
\noalign{\vskip -0.5cm}
\noalign{\hrulefill}
pseudoscalar & $\pi,\eta,\eta^{\prime}$ & $0^-$ & 
 $^3\mbox{P}_0 \,\rightarrow\, ^1\mbox{S}_0\,\mbox{s}$   \cr
vector       & $\rho,\omega,\phi $      & $1^-$ & 
 $^3\mbox{P}_1 \,\rightarrow\, ^1\mbox{S}_0\,\mbox{s}$   \cr
scalar       & $a_0, f_0$               & $0^+$ & 
 $^1\mbox{S}_0 \,\rightarrow\, ^1\mbox{S}_0\,\mbox{s}$   \cr
\noalign{\hrulefill} }
\end{table}
\vspace{-0.0cm}
For example, the production of neutral mesons with negative parity -- as 
pseudoscalar or vector mesons -- may proceed in the proton-proton collision 
near threshold only via the transition between $^3\mbox{P}_{0}$ and 
$^1\mbox{S}_0\mbox{s}$ partial waves.
This means that only the collision of protons with relative angular momentum equal 
to $1\,\hbar$ 
may lead to the production of such mesons.
Moreover, an inspection of the corresponding Clebsch-Gordan coefficients reveals that it is four times
more probable that the orientation of spins of the colliding nucleons is parallel than anti-parallel.\\
In the case of the production of neutral scalar mesons, the protons or 
neutrons must collide with anti-parallel spin orientations which remains 
unchanged after the reaction.
These simple considerations imply that in  close-to-threshold measurements
with  polarized beam and target one should see a drastic effect in the 
reaction yield depending whether the polarization of the reacting protons is 
parallel or anti-parallel.
Indeed, in reality, strong differences in the 
yield for various combinations of beam and target polarization have been 
determined in the pioneering measurements of the reaction
$\vec{p}\vec{p} \to pp\pi^{0}$ at the IUCF facility~\cite{HOM}.

Although close-to-threshold higher partial waves are not expected,
detailed studies of meson-nucleon interaction 
require a careful determination of their contributions, 
in order to avoid false assignment of the observed signals.
If we allow for the production of protons with 
the relative angular momentum $L~=~0$ or 1
and confine the $\eta$  meson production to the $l~=~0$ only,
then the $pp\to pp\eta$ reaction may proceed via
three possible transitions listed in table~\ref{tablehigher}.
\vspace{-0.0cm}
\begin{table}[H]
\caption{\label{tablehigher} Transitions for the $pp\to pp Meson(0^{-})$ reaction 
in the case where $l$~=~0 and $L$~=~0~or~1.
More comprehensive list of allowed transitions for the low partial waves in the
$NN\to NNX$ reaction can be found in references~\cite{christoph,deloff} }
\vskip -0.25cm
\tabskip=1em plus2em minus.5em
\halign to \hsize{\hfil#\hfil \cr
\noalign{\hrulefill}
 $^{2S^i+1}L_{J}^i  \ \ \ \rightarrow \ \ \ ^{2S+1}L_{J},l$ \cr
\noalign{\vskip -0.2cm}
\noalign{\hrulefill}
\noalign{\vskip -0.5cm}
\noalign{\hrulefill}
 $^3\mbox{P}_0 \, \ \ \rightarrow\, \ \ ^1\mbox{S}_0\,\mbox{s}$   \cr
 $^1\mbox{S}_0 \, \ \ \rightarrow\, \ \ ^3\mbox{P}_0\,\mbox{s}$   \cr
 $^1\mbox{D}_2 \, \ \ \rightarrow\, \ \ ^3\mbox{P}_2\,\mbox{s}$   \cr
\noalign{\hrulefill} }
\end{table}
\vspace{-0.0cm}
It is worth to note that the conservation rules mentioned above 
forbid  the production of the $\eta$ meson with $l$~=~1 if the 
protons are in the S-wave ($^1S_0 p $). In other words, the 
$\eta$ meson can be produced in the p-wave only if $L$ is larger 
than zero. This suggests that with an increase of the excess energy 
the higher partial waves in the pp$\eta$ system should first
appear in the  proton-proton subsystem. In fact such an effect was already observed
in experiments at the  SATURN laboratory in the case of the $\phi$ meson production
via the $pp\to pp\phi$ reaction~\cite{jim}.
Since the $\eta$ production 
is dominated by the formation and deexcitation of the S-wave
baryonic  resonance ($S_{11}(1535)$) this supposition 
becomes even more plausible. 

Another important feature 
of the $pp\to ppX$ and also $nn\to nnX$ reactions,
pointed out in reference~\cite{deloff}, is the fact that 
the interference terms between the transitions with odd and even 
values of the angular momentum $L$ of the final state baryons are
bound to vanish. This characteristic is due to the invariance of all observables under 
the exchange of identical nucleons in the final state. 
Thus, for instance, out of three possible interference terms
for transitions listed
in table~\ref{tablehigher} only one 
--~namely the interference between the $^3\mbox{P}_0\,\mbox{s}$ and $^3\mbox{P}_2\,\mbox{s}$ final states~--
will contribute to the production process  
significantly simplifying an interpretation of the experimental data.
In general, 
conservation of the basic quantum numbers leads
--~as derived in reference~\cite{christoph}~-- to the following selection rule
for the $NN\to NN X$ reaction:
\begin{equation}
  (-1)^{(\Delta S + \Delta T)}= \pi_{X} \ \ (-1)^{l},
\end{equation}
where $\pi_{X}$ describes the intrinsic parity of meson X,
$\Delta S$ denotes the change in the spin,
and $\Delta T$ in isospin, between the initial and final $NN$ systems. 
  A possible non-zero angular momentum
  between the outgoing particles can manifest itself in an unisotropic 
  population of the appropriate angles.
 In particular, the angular  distribution of the $\eta$ meson 
 observed in the overall centre-of-mass frame 
 should reflect a possible non-zero value of $l$.
 However, for the study of the angular momentum $L$ we need to use 
 another reference system, since  for a three body final state
 the beam axis is not a good reference direction
 to look for the angular distributions
 relevant for the relative angular momenta of the two particles~\cite{grzonkakilian}.
 This is because for the fixed relative settings of the two protons
 the angle between the beam and the  vector of relative protons momentum
 may acquire  any value.
 Searching for an appropriate variable let us consider a two body scattering.
 In this case, the beam line,
 which is at the same
 time the line along which the centre-of-mass system is moving,
 constitutes a reference frame for the angular distributions.
 Therefore,  by analogy to the two body system, an instructive reference axis
 for angular distributions in the proton-proton subsystem
 is now the momentum of the recoil $\eta$ meson, since the direction of that
 meson is identical to the direction of the movement of the proton-proton
 centre-of-mass subsystem. 
\vspace{-0.0cm}
\begin{figure}[h]
\centerline{
\parbox{0.99\textwidth}{\epsfig{file=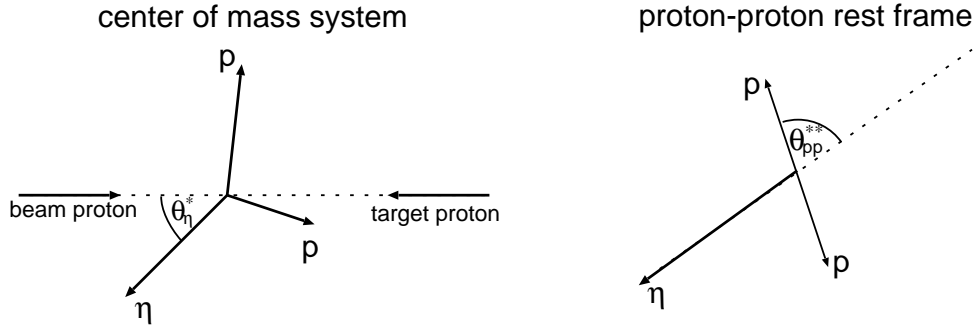,width=0.99\textwidth}}}
\vspace{0.4cm}
   \caption{\label{starstar}
     Definition of the angles used in the text.
     Throughout the whole work the angles in the overall
     centre-of-mass
     frame will be denoted by one asterisk, those in the two-particle 
     subsystems by two asterisks, and the angle in laboratory system
     will be left without any superscript.
     }
\end{figure}
\vspace{-0.0cm}
 Figure~\ref{starstar} visualises both above 
 introduced angles $\theta_{\eta}^{*}$ and $\theta_{pp}^{**}$, the latter
 being defined as the
 angle between the relative proton-proton momentum
 and the recoil particle~($\eta$) seen from the
 di-proton rest system~\cite{grzonkakilian}.
\vspace{-0.1cm}
\section{Phase space population}
\label{phasespacepop}
\begin{flushright}
\parbox{0.70\textwidth}{
 {\em
 It is essential to science that its matter should be in a space, but the space in which 
 it is cannot be exactly the space we see or feel~\cite{russell}.\\
 }
 \protect \mbox{} \hfill  Bertrand Russell \protect\\
 }
\end{flushright}
\vspace{-0.2cm}
Due to both the strong nucleon-nucleon low-energy interaction in the 
$^1\mbox{S}_0$ state and the meson-nucleon forces in the exit channel, the 
assumption of non-interacting plane waves leading to 
equation~\eqref{sigmaLl} failed when confronted with the experimental data. 
However, these deviations offer the possibility to determine still 
poorly known nucleon-meson interactions.
The interaction between particles depends on their relative momenta. 
Consequently, for investigations of final state interactions, more instructive 
coordinates than the q and k momenta are the squared invariant masses of the 
two-body subsystems~\cite{grzonkakilian}.
These are the coordinates of the Dalitz plot. In the original 
paper~\cite{dalitz} Dalitz has proposed a representation for the energy 
partitions of three bodies in an equilateral triangle whose sides are the axes 
of the centre-of-mass energies. He took advantage of the fact that the sum of 
distances from a point within the triangle to its sides is a constant equal to 
the height. Therefore, the height of the triangle measures the total energy 
$\sqrt{\mbox{s}} = \mbox{E}_1^* + \mbox{E}_2^* + \mbox{E}_3^*$ and interior 
points -- fulfilling four-momentum conservation -- represent energy 
partitions. For a constant $\sqrt{\mbox{s}}$, due to the energy conservation, 
without loosing any information, it is enough to consider the projection on any 
of the $\mbox{E}_i\,\mbox{E}_j$ planes. The linear relation between 
$\mbox{E}_i^*$ and $\mbox{s}_{jk}$ ($\mbox{s}_{jk} = \mbox{s} + \mbox{m}_i^2 - 
2\,\sqrt{\mbox{s}} \mbox{E}_i^*$) allows to use $\mbox{s}_{jk}\,\mbox{s}_{ik}$ 
or $\mbox{E}_i\,\mbox{E}_j$ coordinates equivalently, with the following 
relation between the phase space intervals: $d\mbox{E}_i^*\,d\mbox{E}_j^* = 
\frac{1}{4\mbox{\small s}}\;d\mbox{s}_{jk}\,d\mbox{s}_{ki}$. 
According to Kilian's geometrical representation~\cite{grzonkakilian} 
a Dalitz plot lies on a plane in the three dimensional space ($\mbox{s}_{12},\mbox{s}_{13},
\mbox{s}_{23}$) orthogonal to the space diagonal. 
The plane including a Dalitz plot corresponding to a fixed total energy $\sqrt{s}$
is then given by the following scalar product~\cite{grzonkakilian}:
\be
\label{scalarproduct}
(1,1,1)(\mbox{s}_{12},\mbox{s}_{13},\mbox{s}_{23}) = 
 \mbox{s}_{12} + \mbox{s}_{13} + \mbox{s}_{23} = 
 \mbox{s} + \mbox{m}_1^2 + \mbox{m}_2^2 + \mbox{m}_3^2.
\ee
The second equality of equation~\eqref{scalarproduct} means that there are only 
two independent invariant masses of the three subsystems and therefore a 
projection onto any of the $\mbox{s}_{ij}\,\mbox{s}_{jk}$ planes still 
comprises the whole principally accessible information about the final state 
interaction of the three-particle system.
In the case of no dynamics whatsoever and the absence of any final state 
interaction the occupation of the Dalitz plot would be fully homogeneous since 
the creation in each phase space interval would be equally probable. 
The final state interaction would then appear as a structure in that area.
\vspace{-0.1cm}
\begin{figure}[H]
\hspace{1cm}
\parbox{0.49\textwidth}{\epsfig{file=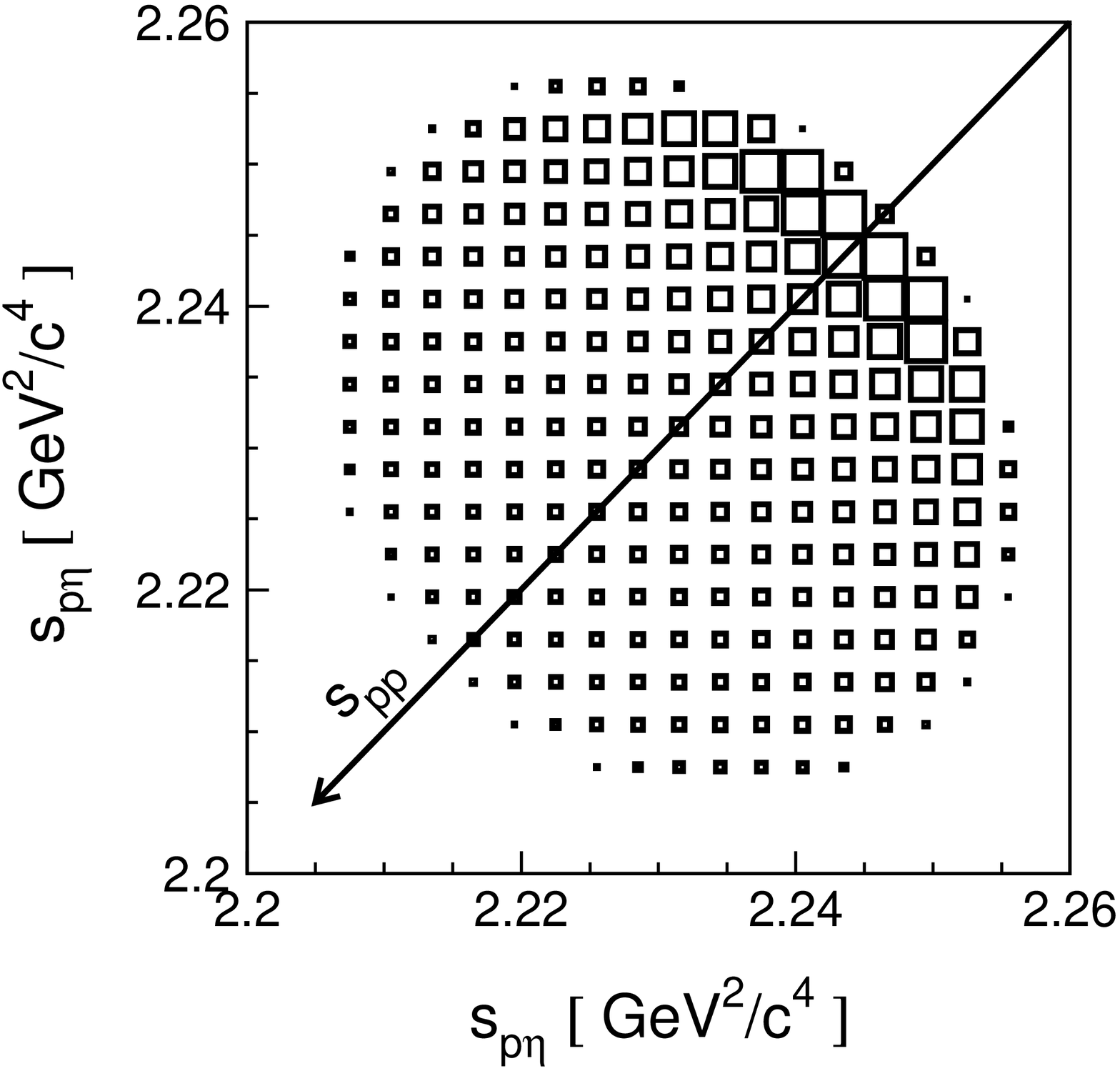,width=0.45\textwidth}}
\hfill
\parbox{0.49\textwidth}{\epsfig{file=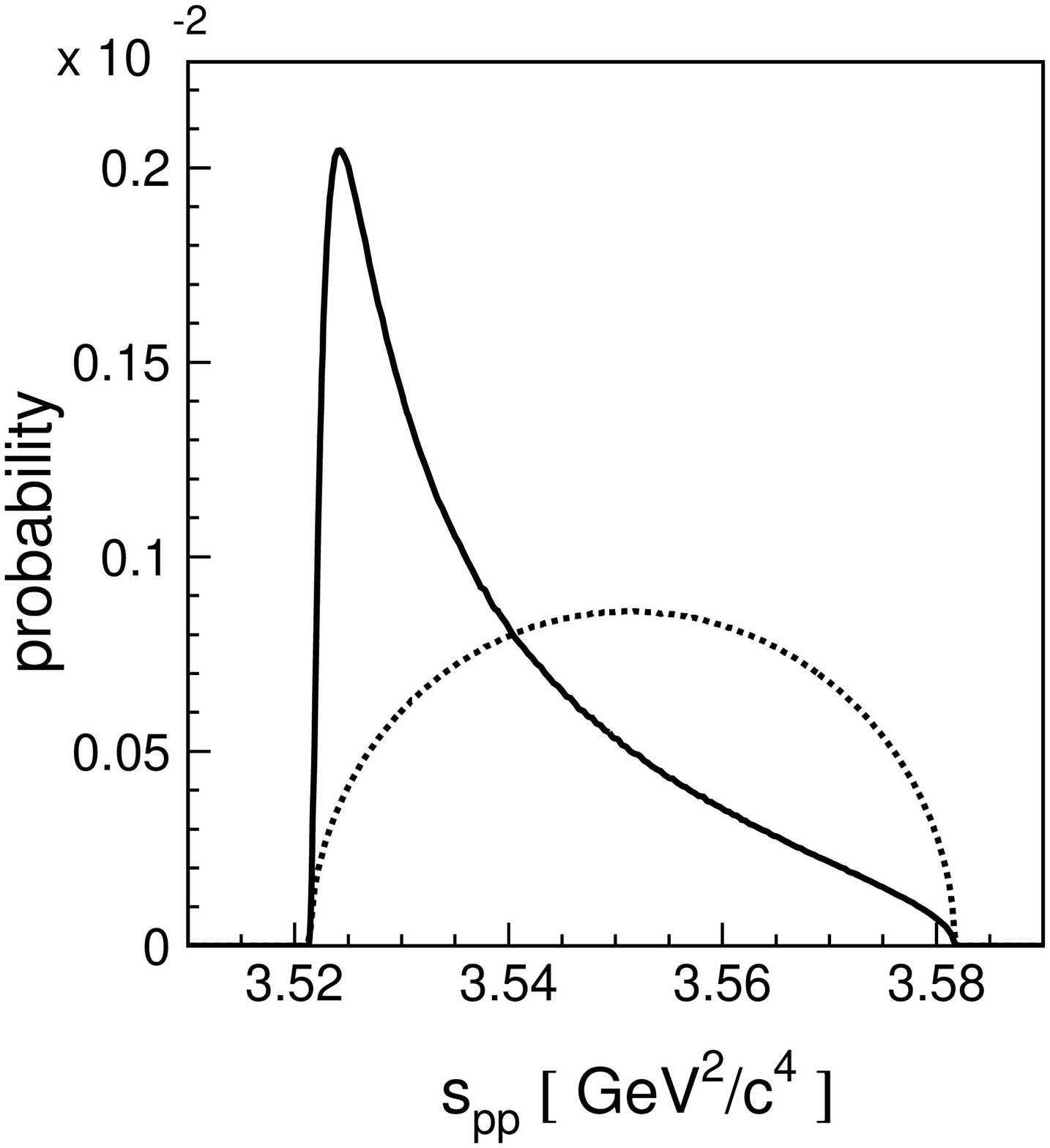,width=0.45\textwidth}}

\parbox{0.43\textwidth}{\raisebox{1ex}[0ex][0ex]{\mbox{}}}\hfill
\parbox{0.50\textwidth}{\raisebox{1ex}[0ex][0ex]{\large a)}}\hfill
\parbox{0.03\textwidth}{\raisebox{1ex}[0ex][0ex]{\large b)}}

\hspace{1cm}
\parbox{0.49\textwidth}{\epsfig{file=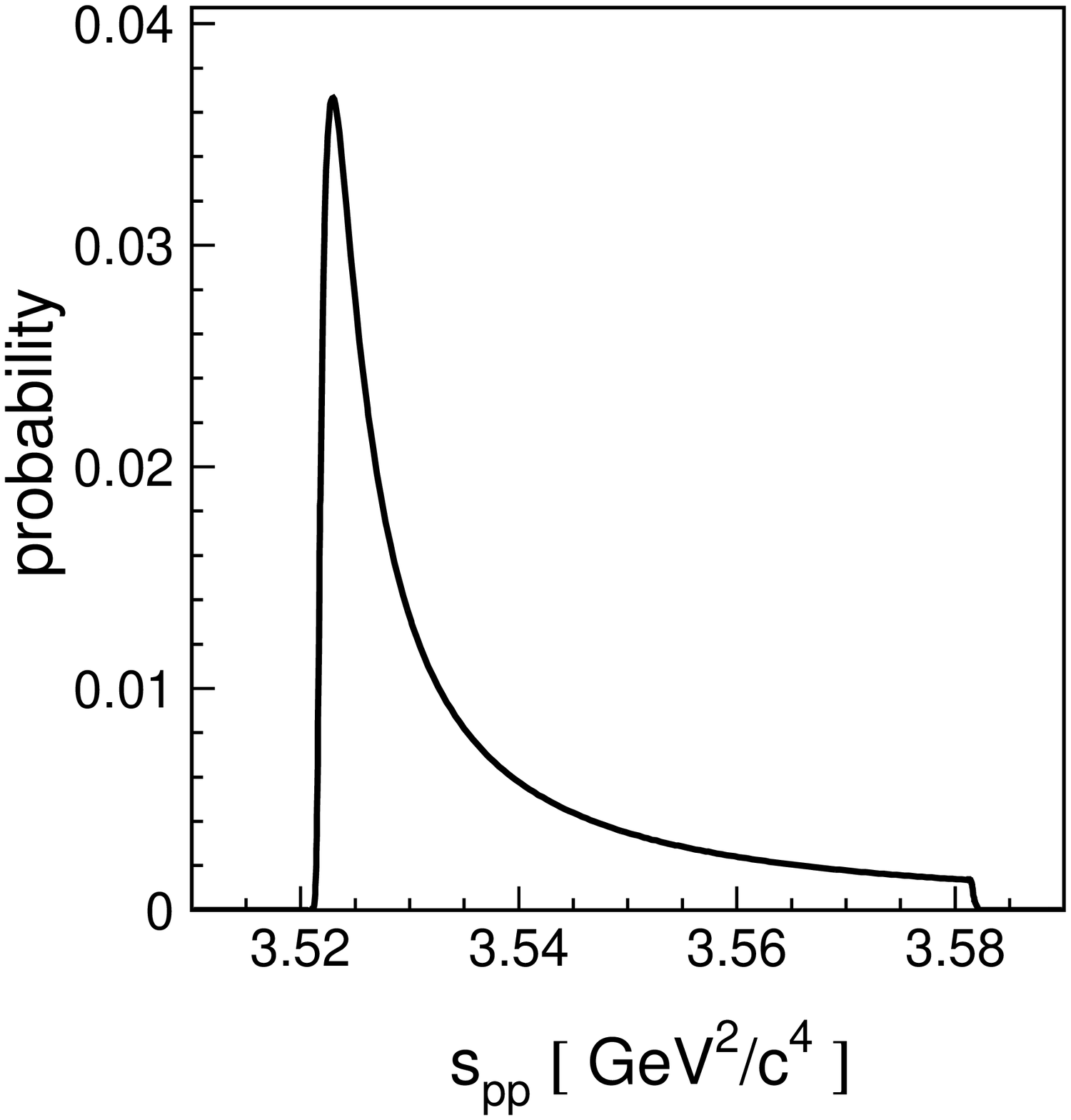,width=0.45\textwidth}}
\hfill
\parbox{0.49\textwidth}{\epsfig{file=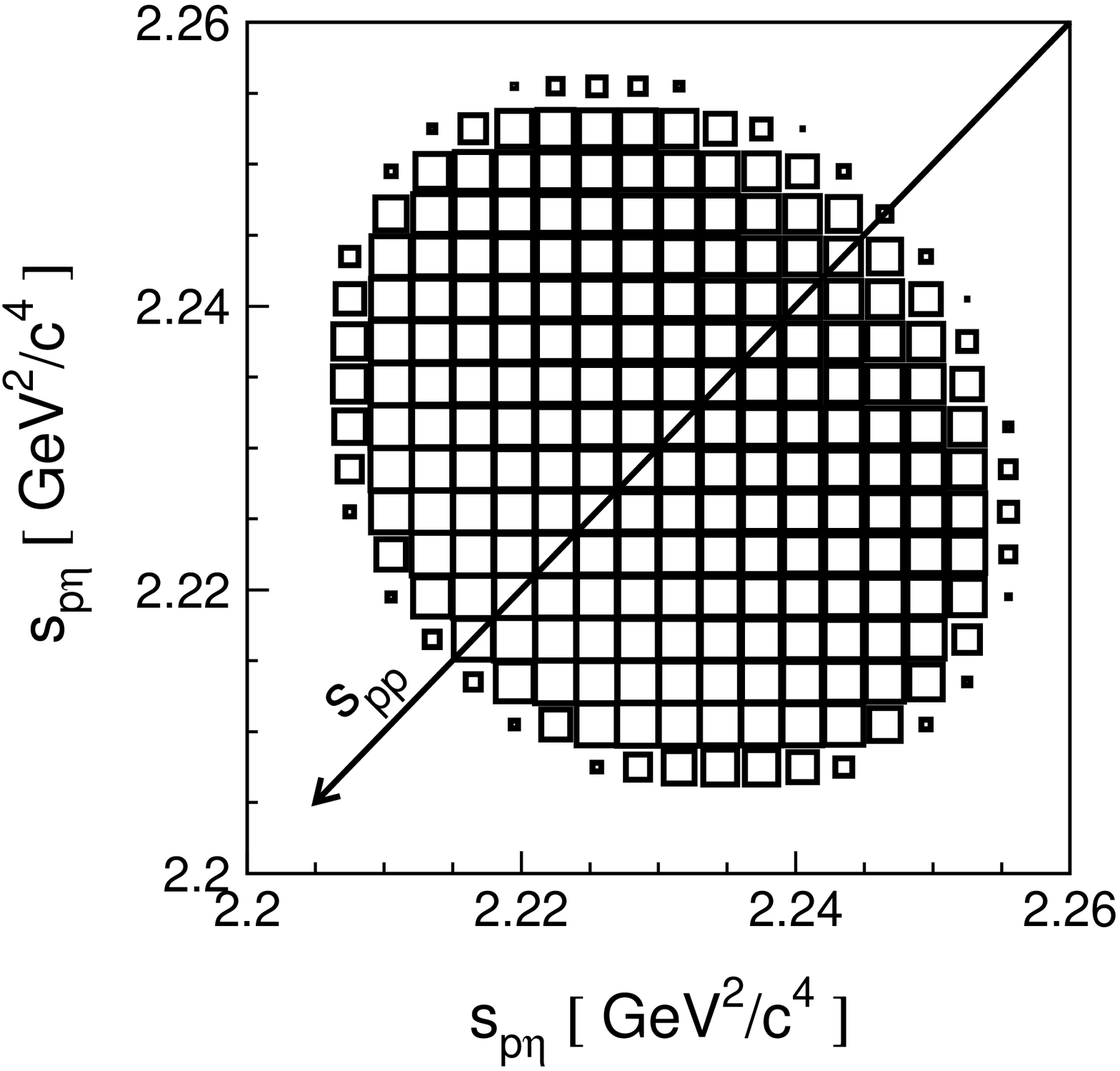,width=0.45\textwidth}}

\parbox{0.43\textwidth}{\raisebox{1ex}[0ex][0ex]{\mbox{}}}\hfill
\parbox{0.50\textwidth}{\raisebox{1ex}[0ex][0ex]{\large c)}}\hfill
\parbox{0.03\textwidth}{\raisebox{1ex}[0ex][0ex]{\large d)}}
\vspace{-0.2cm}
\caption{\label{dalitz_examples} 
    Monte-Carlo simulations: 
(a) Phase-space distribution for the $pp \rightarrow pp \eta$ reaction at 
    $\mbox{Q} = 16\,\mbox{MeV}$ modified by the proton-proton final state 
    interaction. \
(b) The dotted line shows the projection of the pure phase space density 
    distribution onto the $\mbox{s}_{pp}$ axis and the solid curve presents 
    its modification by the proton-proton FSI. \
(c) The square of the scattering amplitude for the $pp \rightarrow pp$ elastic 
    scattering as a function of the proton-proton invariant mass in the range 
    $2\,\mbox{m}_p < \sqrt{\mbox{s}_{pp}} < 2\,\mbox{m}_p + \mbox{Q}$, 
    calculated according to formula~\eqref{Mpppp} of section~\ref{influencesection}. \
(d) Phase-space density distribution modified by the proton-$\eta$ 
    interaction, with a scattering length equal to $\mbox{a}_{p\eta} = 
    0.7\,\mbox{fm} + i\,0.3\,\mbox{fm}$. 
     The proton-$\eta$ scattering amplitude has been calculated according
     to equation~\eqref{Mpeta}.  
     A detailed discussion of the nucleon-nucleon and nucleon-meson
     interaction will be presented in section~\ref{influencesection}.
     }
\end{figure}
\vspace{-0.2cm}
Figure~\ref{dalitz_examples}a shows --~for the example of the $pp \rightarrow 
pp \eta$ reaction~-- how the uniformly populated phase space density is 
modified by the S-wave ($^1\mbox{S}_0$) interaction between outgoing protons. 
An enhancement in the range corresponding to low relative momenta between 
protons is clearly visible. 
A steep decrease of the occupation density with increasing invariant mass of 
the proton-proton subsystem is even better seen in 
figure~\ref{dalitz_examples}b presenting the projection of the phase space 
distribution onto the $\mbox{s}_{pp}$ axis indicated by an arrow in 
figure~\ref{dalitz_examples}a.
This is a direct reflection of the shape of the proton-proton ($^1\mbox{S}_0$) 
partial wave amplitude shown in figure~\ref{dalitz_examples}c.
Figure~\ref{dalitz_examples}d shows the Dalitz plot distribution simulated when 
switching off the proton-proton interaction but accounting for the interaction 
between the $\eta$-meson and the proton. 
Due to the lower strength of this interaction the expected deviations from the 
uniform distributions are by about two orders of magnitude smaller, 
but still one 
recognizes a slight enhancement of the density in the range of low invariant 
masses of proton-$\eta$ subsystems.  
However, due to weak variations of the proton-$\eta$ scattering amplitude the 
enhancement originating from the $\eta$-meson interaction with one proton is 
not separated from the $\eta$-meson interaction with the second proton. 
Therefore an overlapping of broad structures occurs. 
It is observed that the occupation density grows slowly with increasing 
$\mbox{s}_{pp}$ opposite to the effects caused by the S-wave proton-proton 
interaction, yet similar to the modifications expected for the P-wave 
one~\cite{dyringPHD}.
From the above example it is obvious that only in high statistics experiments 
signals from the meson-nucleon interaction can appear over the overwhelming 
nucleon-nucleon final state interaction. 
It is worth noting, however, that the Dalitz plot does not reflect any possible 
correlations between the entrance and exit channels~\cite{grzonkakilian}. 

The Dalitz plot representation allows also for a simple interpretation of the 
kinematically available phase space volume as an area of that plot. 
Namely, equation~\eqref{phasespacegeneral} becomes:
\be
\label{Vpsdalitz}
\sigma = \frac{1}{\mbox{F}} \frac{\pi^{2}}{4\,\mbox{s}}
 \int\limits_{(\mbox{\scriptsize m}_1+\mbox{\scriptsize m}_2)^2}^{
   (\sqrt{\mbox{\scriptsize s}}-\mbox{\scriptsize m}_3)^2} d\,\mbox{s}_{12}
 \int\limits_{\mbox{\scriptsize s}_{23}^{min}(\mbox{\scriptsize s}_{12})}^{
   \mbox{\scriptsize s}_{23}^{max}(\mbox{\scriptsize s}_{12})} 
   d\,\mbox{s}_{23}\; |M_{ab\,\rightarrow\,123}|^2,
\ee
where the limits of integrations defining the boundaries of the Dalitz plot can 
be expressed as~\cite{bycklingkajantie}:
{\small{
\be
\mbox{s}_{23}^{max}(\mbox{s}_{12}) = 
  \mbox{m}_2^2 + \mbox{m}_3^2 - 
  \frac{(\mbox{s}_{12} - \mbox{s} + \mbox{m}_3^2) 
         (\mbox{s}_{12} + \mbox{m}_2^2 - \mbox{m}_1^2) - 
         \sqrt{\lambda(\mbox{s}_{12},\mbox{s},\mbox{m}_3^2) 
               \lambda(\mbox{s}_{12},\mbox{m}_2^2,\mbox{m}_1^2)}}
          {2\,\mbox{s}_{12}} \nn \\
\ee
\be
\mbox{s}_{23}^{min}(\mbox{s}_{12}) = 
  \mbox{m}_2^2 + \mbox{m}_3^2 - 
  \frac{(\mbox{s}_{12} - \mbox{s} + \mbox{m}_3^2) 
         (\mbox{s}_{12} + \mbox{m}_2^2 - \mbox{m}_1^2) + 
         \sqrt{\lambda(\mbox{s}_{12},\mbox{s},\mbox{m}_3^2) 
               \lambda(\mbox{s}_{12},\mbox{m}_2^2,\mbox{m}_1^2)}}{2\,\mbox{s}_{12}}.
     \nn
\ee
}}
Thus, the phase space volume kinematically available for the three-body final 
state can be written by means of only one integral:

\be
V_{ps} = \int dV_{ps} = 
 \frac{\pi^{2}}{4\,\mbox{s}}
 \int\limits_{(\mbox{\scriptsize m}_1+\mbox{\scriptsize m}_2)^2}^{
   (\sqrt{\mbox{\scriptsize s}}-\mbox{\scriptsize m}_3)^2} d\,\mbox{s}_{12}
 \int\limits_{\mbox{\scriptsize s}_{23}^{min}(\mbox{\scriptsize s}_{12})}^{
   \mbox{\scriptsize s}_{23}^{max}(\mbox{\scriptsize s}_{12})} 
   d\,\mbox{s}_{23}\; = 
  \nn 
\ee
\be
\label{Vps_relativistic}
=\;\frac{\pi^{2}}{4\,\mbox{s}}
 \int\limits_{(\mbox{\scriptsize m}_1+\mbox{\scriptsize m}_2)^2}^{
  (\sqrt{\mbox{\scriptsize s}}-\mbox{\scriptsize m}_3)^2}\;
 \frac{d\,\mbox{s}_{12}}{\mbox{s}_{12}}
 \sqrt{\lambda(\mbox{s}_{12},\mbox{s},\mbox{m}_3^2) 
       \lambda(\mbox{s}_{12},\mbox{m}_2^2,\mbox{m}_1^2)}, 
\ee
whose solution leads, in general, to elliptic functions~\cite{bycklingkajantie}.
However, in the nonrelativistic approximation it has the following closed form:
\be
\label{Vps_nonrelativistic}
V_{ps} = \frac{\pi^{3}}{2} 
  \frac{\sqrt{\mbox{m}_1\,\mbox{m}_2\,\mbox{m}_3}}
       {(\mbox{m}_1 + \mbox{m}_2 + \mbox{m}_3)^\frac{3}{2}} \; \mbox{Q}^2,
\ee
where the substitution of the non-relativistic relation between $\eta_{m_3}$ 
and~Q
\begin{equation*}
\mbox{Q} = 
  \frac{\mbox{m}_3^2 + 2\,\mbox{m}_3^2\,(\mbox{m}_1 + \mbox{m}_2)}
       {2\,(\mbox{m}_1 + \mbox{m}_2)} \; \eta_{m_3}^2 
\end{equation*}
gives the $\mbox{S}\,\mbox{s}$ partial cross section of 
equation~\eqref{sigmaLl}.
On the basis of formula~\eqref{Vps_nonrelativistic} the kinematically 
available phase space volume ($V_{ps}$) can be as easily calculated as the 
excess energy Q. 
Close-to-threshold -- in the range of  few tens of MeV -- the 
non-relativistic approximation differs only by a few per cent from the full 
solution given in equation~\eqref{Vps_relativistic}, which in fact with an 
up-to-date computer can be solved numerically with little effort. 
Therefore, in the following chapters we will describe the data as a function of 
$V_{ps}$ as well as of Q or $\eta_{M}$, if it is found to be appropriate.

\section{Orientation of the emission plane}
\label{choiceofobservables}

\begin{flushright}
\parbox{0.5\textwidth}{
 {\em Angles are the invention of Man;\protect\\
      God uses vectors!~\cite{wilkinmail}\protect\\
 }
 \protect \mbox{} \hfill Colin Wilkin
 }
\end{flushright}

For the full description of the three particle system five independent variables are required.
In the center-of-mass frame,
due to the momentum conservation,
the momentum vectors of the particles are lying in one  plane often referred 
to as the emission-,
reaction-, or decay plane. In this plane~(depicted in figure~\ref{figvariables})
a relative movement of the particles can be described by two variables only. 
As anticipated in the previous section,
the square of the invariant masses of the di-proton and proton-$\eta$
system denoted as $s_{pp}$ and $s_{p\eta}$, respectively,
constitute a natural choice for the study
of the interaction within  the pp$\eta$ system.
This is because in the case of non-interacting objects
the surface spanned by these variables is homogeneously populated. 
The interaction among
the particles modifies that occupation density
and in consequence facilitates an easy
qualitative  interpretation of the 
experimental results.
\begin{figure}[H]
    \parbox{1.0\textwidth}{
    \includegraphics[width=0.98\textwidth]{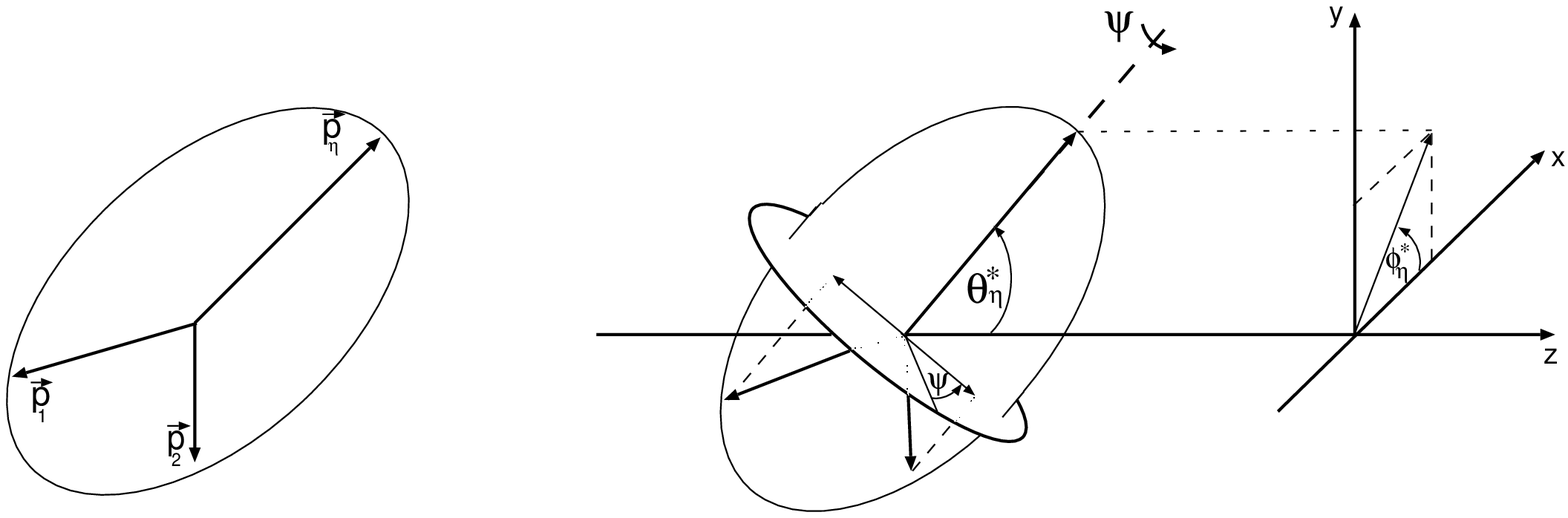}}
    \parbox{1.0\textwidth}{ \vspace{0.5cm}
    \includegraphics[width=0.98\textwidth]{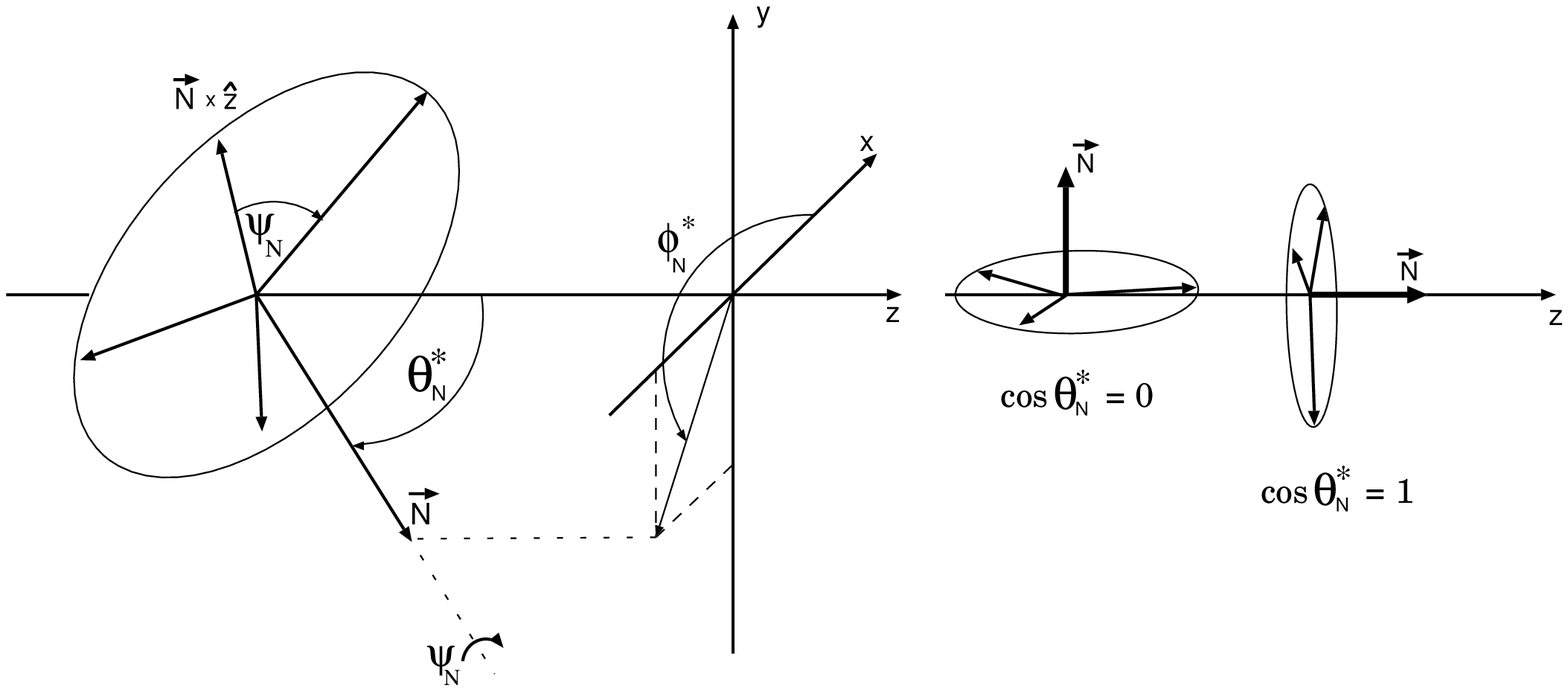}}
    \vspace{-0.3cm}
   \caption{ \label{figvariables}
     Definition of the centre-of-mass kinematical variables used 
     for the description of the pp$\eta$ system.
     In the centre-of-mass frame, the momenta of ejectiles lie in an emission plane.
     Within this plane the relative movement of the particles
     is fixed by the square of the invariant masses $s_{pp}$ and $s_{p\eta}$.
     As  the remaining three variables needed to define the system uniquely
     we use 
     either $\phi^{*}_{\eta}$, $\theta^{*}_{\eta}$, and $\psi$
     shown in the upper panel or $\phi^{*}_{N}$, $\theta^{*}_{N}$, and $\psi_{N}$
     defined in the lower panel.
     $\vec{N}$ is a vector normal to the emission plane, which can be calculated 
     as the vector product of the centre-of-mass momentum vectors of the outgoing protons.
     As an example two extreme orientations
     of the emission plane are shown in the right-lower panel.
     For  further descriptions see the text.
   }
\end{figure}
The remaining three variables must define an absolute
orientation of the emission plane in the distinguished coordinate system.
This may be attained --~for example~-- by defining the orientation
for the momentum of the arbitrarily chosen particle
in the center-of-mass frame
and the angle which describes the rotation around the direction fixed by that particle.
In one of our choices (following references~\cite{balestra092001,jim})
the corresponding variables are the polar and the azimuthal
angle of the $\eta$ momentum vector, depicted in figure~\ref{figvariables}
as $\phi^*_{\eta}$ and $\theta^{*}_{\eta}$, respectively,  and the angle $\psi$ describing
the rotation around the direction established by the momentum of the $\eta$ meson.
Figure~\ref{figvariables} demonstrates that such rotation neither affects the $\eta$ meson
momentum nor  changes the configuration
of the momenta in the emission plane.
For experiments with  unpolarized beams and targets the only favoured direction
is the one of the beam. Therefore, as a  zero value of the $\psi$ angle we have chosen the
projection of the beam direction
on the plane perpendicular to the momentum vector of the $\eta$
meson. 
Note that we identify the z-axis with the beam direction.
The $\phi^*_{\eta}$, $\theta^{*}_{\eta}$, and $\psi$ variables
can also be interpreted as Euler angles allowing  for the
rotation of the emission plane into a xz-plane.
The angle $\psi$ may be calculated as an angle between the emission plane
and the plane  containing
momentum vectors of the $\eta$ meson and the beam proton, or correspondingly as
an angle between the vectors normal to these planes. The angle $\psi$ is equal to zero when 
these two normals are parallel.
 
As a second possibility we will describe an orientation of the emission plane by the
azimuthal and polar angle of the vector normal to that plane~\cite{roderburg2299}.
These angles are
shown in figure~\ref{figvariables} as $\phi^{*}_{N}$ and $\theta^{*}_{N}$,
respectively.  Further  the absolute orientation
of the particles momenta in the emission plane will be described by $\psi_{N}$,
the angle between the $\eta$ meson and the vector product of the beam momentum
and the vector $\vec{N}$.
 
Obviously, the interaction between particles does not depend on the orientation 
of the emission plane, and therefore, it will fully manifest itself in the 
occupation density of the Dalitz plot which in our case will be represented in terms
of the square of the invariant masses of the two particle subsystems.
Yet, the distribution of the orientation of the emission plane will reflect
the correlation between the initial and final channels and hence its determination
should be helpful for the investigation of the production mechanism.


\newpage
\clearpage
\thispagestyle{empty}
\pagestyle{plain}
\chapter{Low energy interaction within the $pp\eta$ and $pp\eta^{\prime}$ systems}
\thispagestyle{empty}
\pagestyle{myheadings}
\markboth{Hadronic interaction of $\eta$ and $\eta'$ mesons with protons}
         {4. Low energy interactions within the $pp\eta$ and $pp\eta^{\prime}$ systems}
\label{Hinertact}                          
\vspace{-0.3cm}
\begin{flushright}
\parbox{0.71\textwidth}{
 {\em
   It frequently happens that, when particles are produced in a nuclear or elementary
   particle reaction, some of these interact among themselves so strongly that they influence
   appreciably the properties of the reaction cross 
   section.\!\! {\bf\em These interactions we shall call ``final state interaction''}~\cite{watson1163}.\\
 }
 \protect \mbox{} \hfill  Kenneth Watson \protect\\
 }
\end{flushright}
\vspace{-0.4cm}
In general, for a three-body exit channel one expects an energy dependence of the 
total cross section which can be described by the linear 
combination of partial cross sections from equation~\eqref{sigmaLl}. 
Therefore, to  extract  information about the final state 
interaction of the outgoing particles 
the contributions originating from different 
partial waves have to be known precisely. 
Appropriately, close-to-threshold there is only one important combination 
of angular momenta of emitted particles (Ss) and in this region the energy 
dependence of the total cross section is uniquely determined and
hence the interpretation of results is significantly simplified.
This is one of the most important  advantages of the meson creation at its production threshold,
however, in order to exploit this fact properly it is essential to determine the range of its applicability,
which --~as will be demonstrated in the next section~-- changes
significantly with the mass of the produced meson.
\vspace{-0.3cm}
\section{Range of the dominance of the $^3P_{0}\to ^1\!\!\!S_{0}s$ transition}
\label{range}
\begin{flushright}
\parbox{0.73\textwidth}{
 {\em
  ...~the most exact of sciences are those which are most concerned
  with primary considerations: for sciences based on few assumptions are more exact than those
  which employ additional assumptions~...~\cite{arystotelesmetaphysics}.\\
 }
 \protect \mbox{} \hfill  Aristotle \protect\\
 }
\end{flushright}
\vspace{-0.4cm}
Investigations with polarized beams and targets~\cite{meyer064002,meyer5439} 
of the $\vec{p}\vec{p} \to pp\pi^{0}$ reaction
allowed to deduce that the Ss partial-wave accounts for more than $95\,\%$ of 
the total cross section up to $\eta_M \approx 0.4$, as can be seen in 
figure~\ref{1s03p0}a, where the Ss contribution is indicated by the 
dotted line. 
The Ss contribution was inferred assuming the $\eta_M^6$ and $\eta_M^8$ 
dependence for Ps and Pp partial waves, respectively.
These are power-laws taken from proportionality~\eqref{sigmaLl}, which was 
derived under the assumption of non-interacting particles.
Relatively small values of $^3\mbox{P}_0$-wave 
nucleon-nucleon phase-shifts 
at low energies (compared to $^1\mbox{S}_0$ phase-shifts in 
figure~\ref{1s03p0}b) and similarly weak low-energy interactions of 
P-wave protons in other spin combinations~\cite{bergervoet1435} justify this 
assumption. 
\vspace{-1.0cm}
\begin{figure}[H]
       \hspace{-0.2cm}
       \parbox{0.45\textwidth}{\epsfig{file=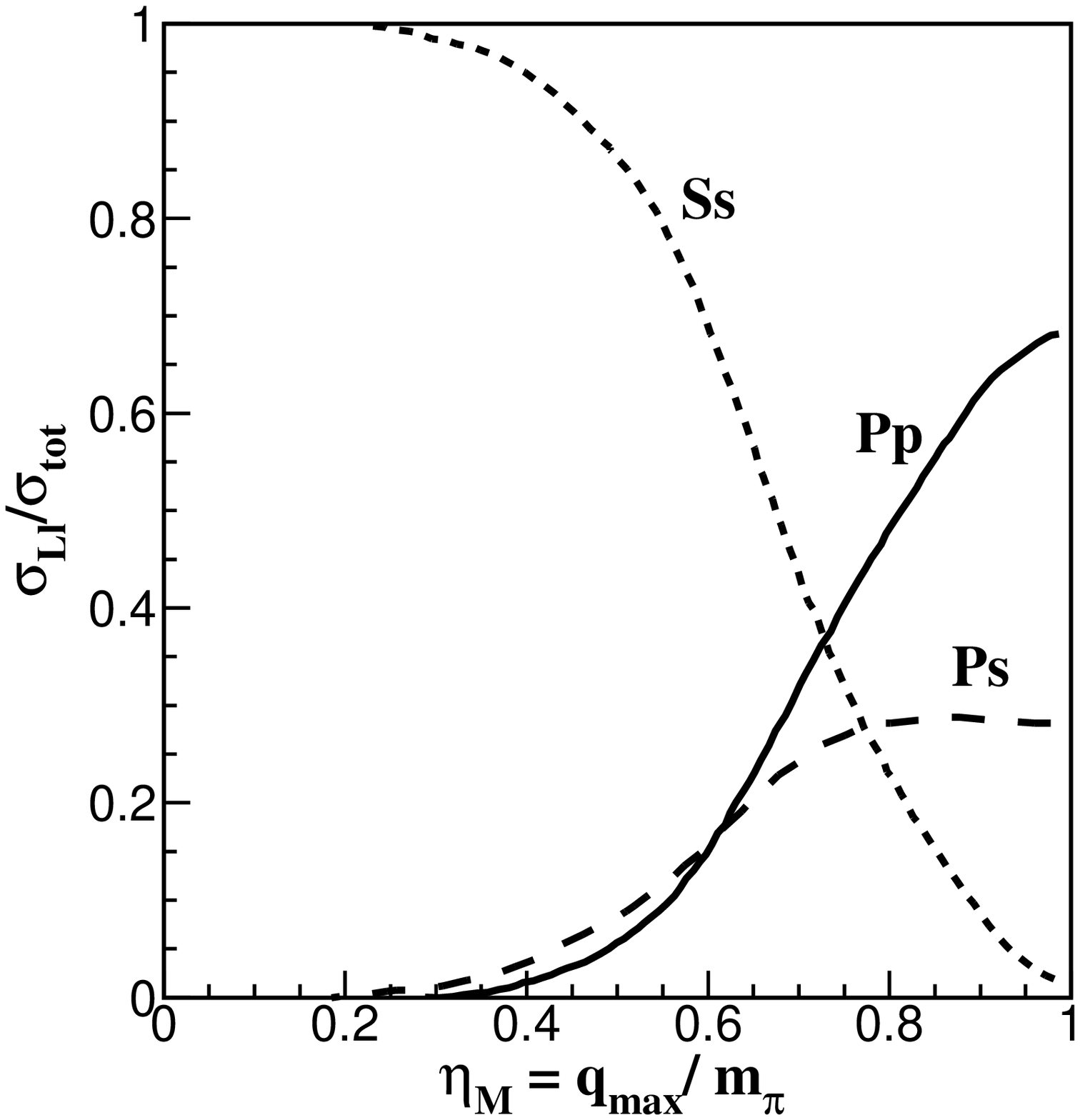,width=0.49\textwidth}}
       \parbox{0.55\textwidth}{\vspace{0.6cm}\epsfig{file=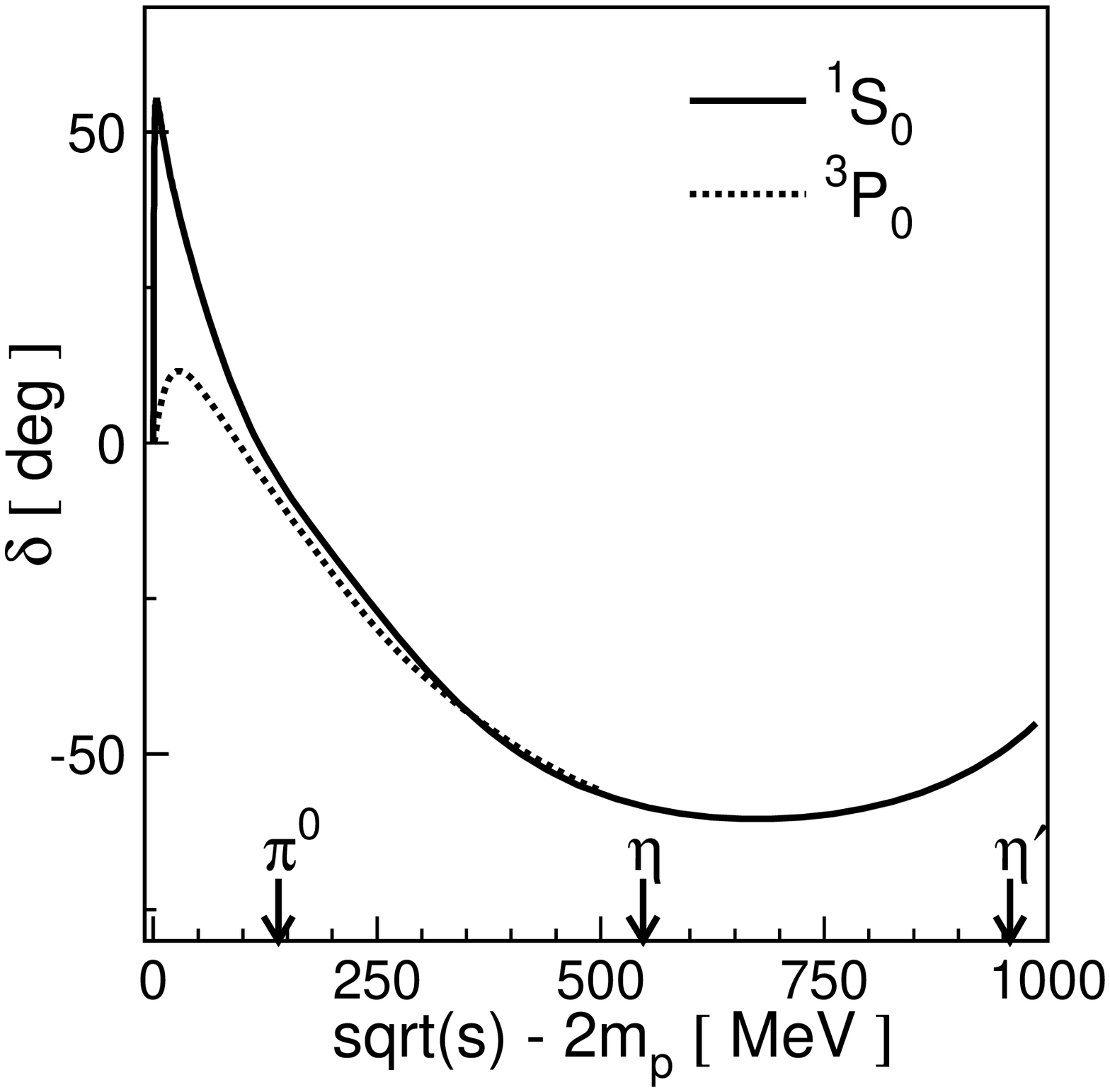,width=0.539\textwidth}}
 \vspace{-0.4cm}
 \parbox{0.44\textwidth}{\raisebox{2ex}[0ex][0ex]{\mbox{}}} \hfill
 \parbox{0.48\textwidth}{\raisebox{2ex}[0ex][0ex]{{\large a)}}} \hfill
 \parbox{0.04\textwidth}{\raisebox{2ex}[0ex][0ex]{{\large b)}}} 
 \vspace{-0.4cm}
 \caption{\label{1s03p0} 
(a) Decomposition of the total cross section of the $pp \rightarrow pp \pi^0$ 
reaction into Ss, Ps, and Pp final state angular momenta. The dashed and solid 
lines represent the $\eta_M^6$ and $\eta_M^8$ dependence of Ps and Pp partial 
cross sections, respectively. The remainder is indicated as the dotted line. 
The Sp partial wave is forbidden by the conservation laws and the Pauli 
excluding principle. Note that at $\eta_M = 1$ the Pp and Ps partial waves 
seem to dominate. However, the analysis of the differential cross section 
measured at CELSIUS at $\eta_M = 0.449$~\cite{zlomanczuk251,bilger633} showed 
that also a d-wave pion production --~due to the interference between Ss and 
Sd states~-- constitutes $7\,\%$ of the total cross section, when a 
meson-exchange model is assumed. The figure has been adapted 
from~\cite{meyer064002}.
(b) The $^1\mbox{S}_0$ and $^3\mbox{P}_0$ 
 phase-shifts of the nucleon-nucleon potential shown versus the 
 centre-of-mass kinetic energy available in the proton-proton system. The 
 values have been extracted from the SAID data base~\cite{arndt3005} (solution 
 SM97). For higher energies the S- and P-wave phase-shifts are nearly the 
 same. This is because the collision parameter required to yield the angular 
 momentum of $1\,\hbar$ diminishes significantly below $1\,\mbox{fm}$ with 
 increasing energy and consequently the interaction of nucleons --~objects of 
 about $1\,\mbox{fm}$ size~-- becomes almost central. 
 }
\end{figure}
\vspace{-0.3cm}
In accordance with the phenomenology of Gell-Mann and Watson~\cite{gellmann219} 
described in section~\ref{partialwaves} one expects that also in the case of heavier 
mesons the Ss partial wave combination will constitute the overwhelming 
fraction of the total production cross section for $\eta_M$ smaller than 0.4. 
This implies --~as can be deduced from the relation between $\eta_M$ and Q 
illustrated in figure~\ref{eta_Q_Vps}a~-- that mesons heavier than the 
pion are produced exclusively via the Ss state in a much larger excess energy 
range and hence larger phase space volume (see figure~\ref{eta_Q_Vps}b).
Thus, whereas for $\pi^0$ production the onset of higher partial waves is observed 
at Q around $10\,\mbox{MeV}$ it is expected only above $100\,\mbox{MeV}$ and 
above  $\approx 40\,\mbox{MeV}$ for $\eta^{\prime}$ and $\eta$ mesons, 
respectively. 
\vspace{-0.4cm}
\begin{figure}[H]
       \parbox{0.5\textwidth}{\epsfig{file=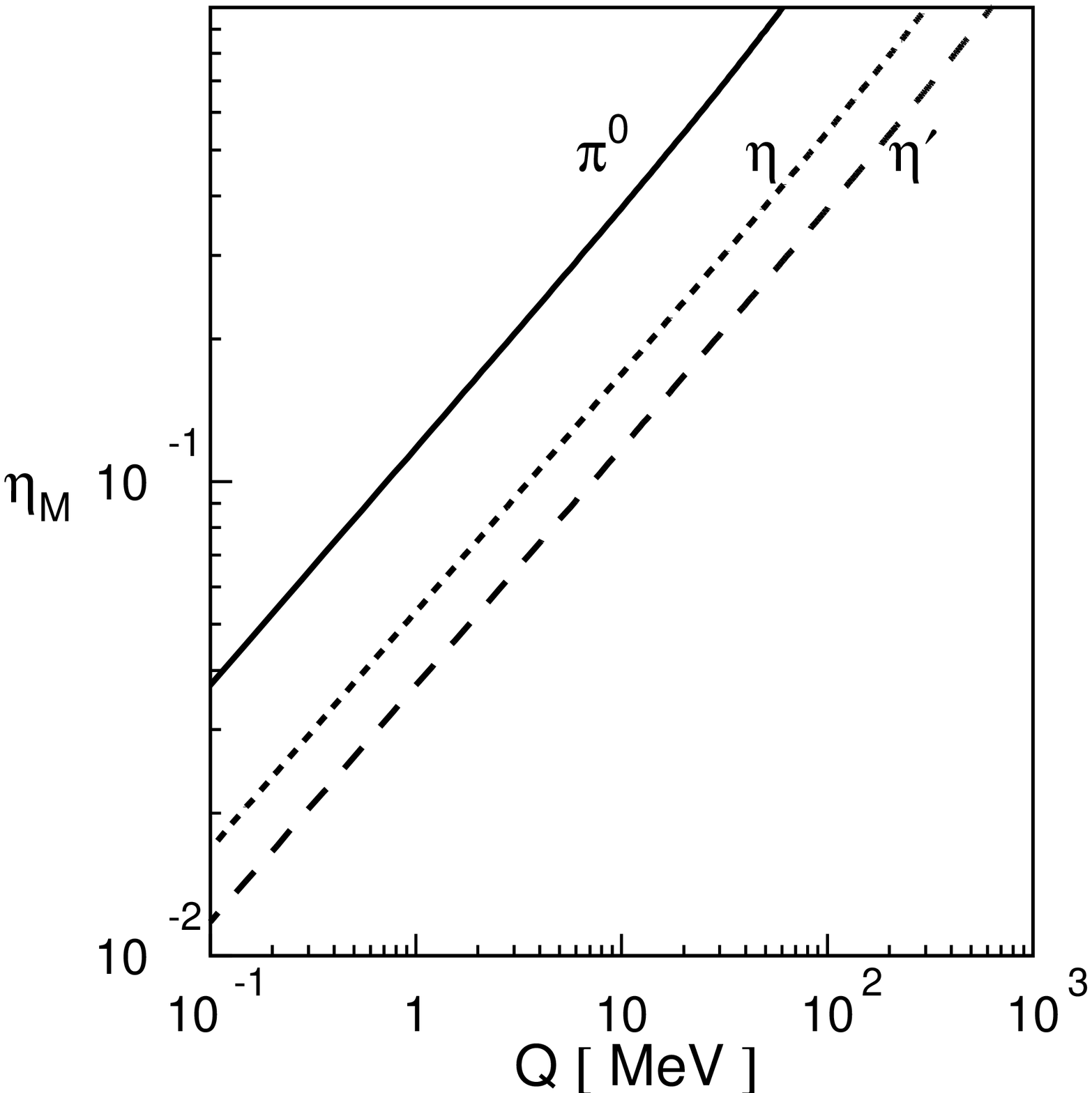,width=0.45\textwidth}}
       \parbox{0.5\textwidth}{\epsfig{file=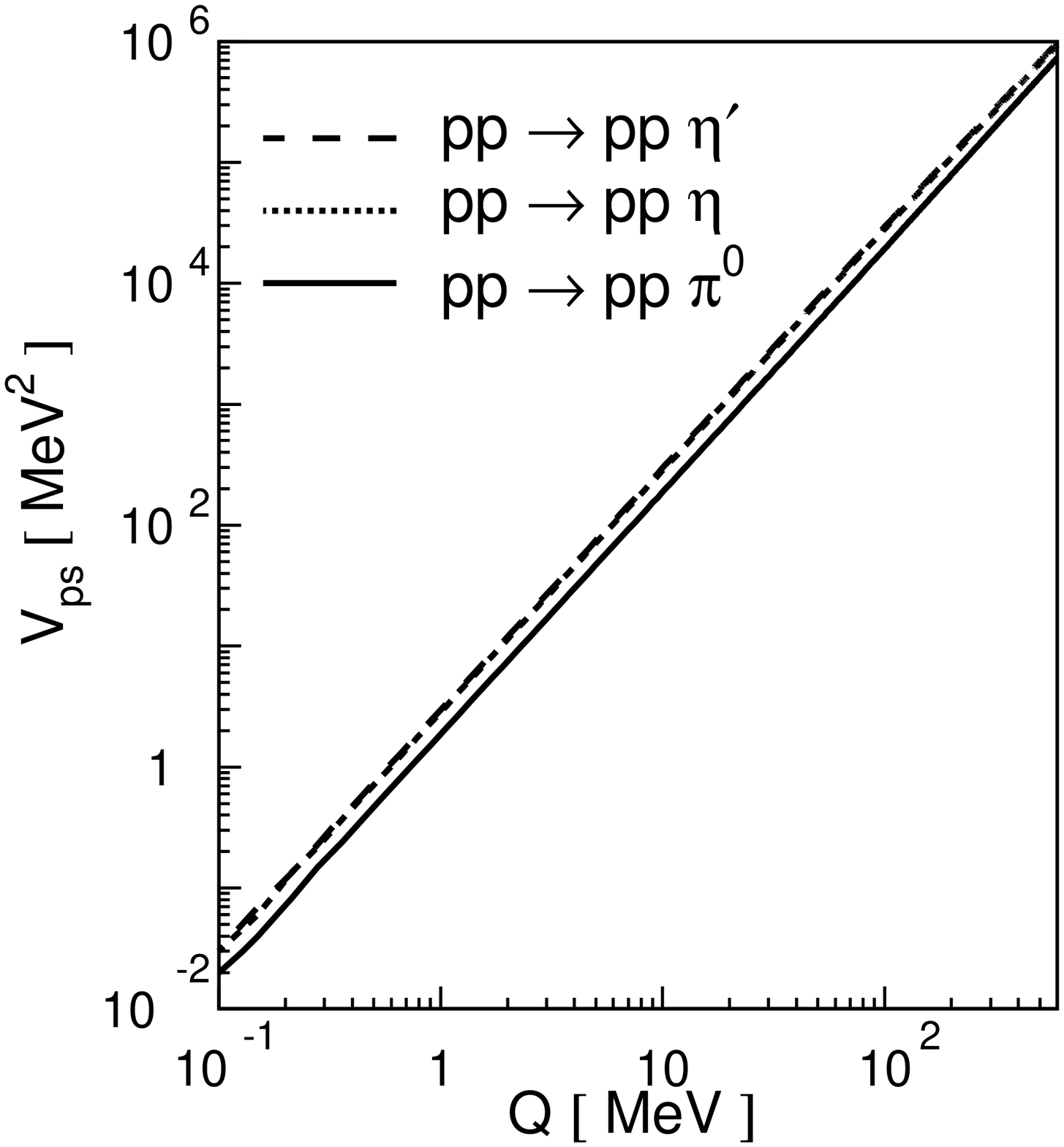,width=0.45\textwidth}}
 \parbox{0.43\textwidth}{\raisebox{2ex}[0ex][0ex]{\mbox{}}} \hfill
 \parbox{0.47\textwidth}{\raisebox{2ex}[0ex][0ex]{\large a)}} \hfill
 \parbox{0.05\textwidth}{\raisebox{2ex}[0ex][0ex]{\large b)}}
 \vspace{-0.4cm}
\caption{\label{eta_Q_Vps} 
(a) The variable $\eta_M$ as a function of the excess energy for $\pi^0 $, 
$\eta$, and $\eta^{\prime}$ mesons produced via $pp \to pp\,Meson$ 
reactions. 
(b) The phase space volume $V_{ps}$ (defined by 
equation~\eqref{Vps_relativistic}) versus the excess energy Q. The picture 
indicates that for the production of ``heavy mesons'' in the nucleon-nucleon 
interaction at a given Q value there is only a slight difference of the 
available phase space volume on the produced meson mass, which is larger 
than that of $\pi^0$ production by about $30\,\%$ only. Therefore, for the 
comparative studies of the production dynamics of different mesons, Q is as 
much a suitable variable as $V_{ps}$. Note, that the dashed and dotted lines
are almost undistinguishable.}
\end{figure}
\vspace{-0.6cm}
\begin{figure}[H]
\vspace{-1.2cm}
\parbox{0.5\textwidth}
  {\epsfig{file=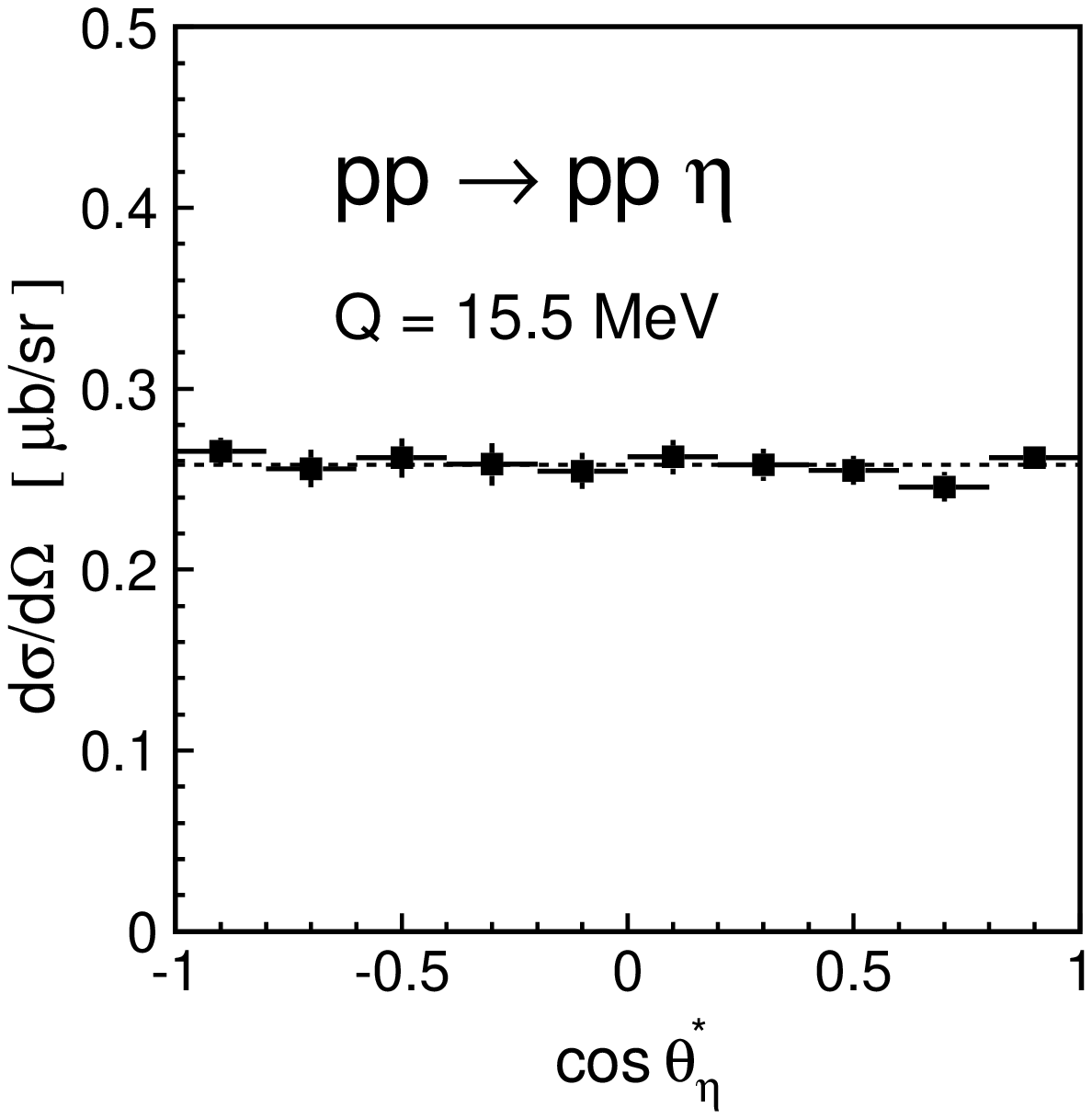,width=0.54\textwidth}}\hfill
\parbox{0.5\textwidth}
  {\epsfig{file=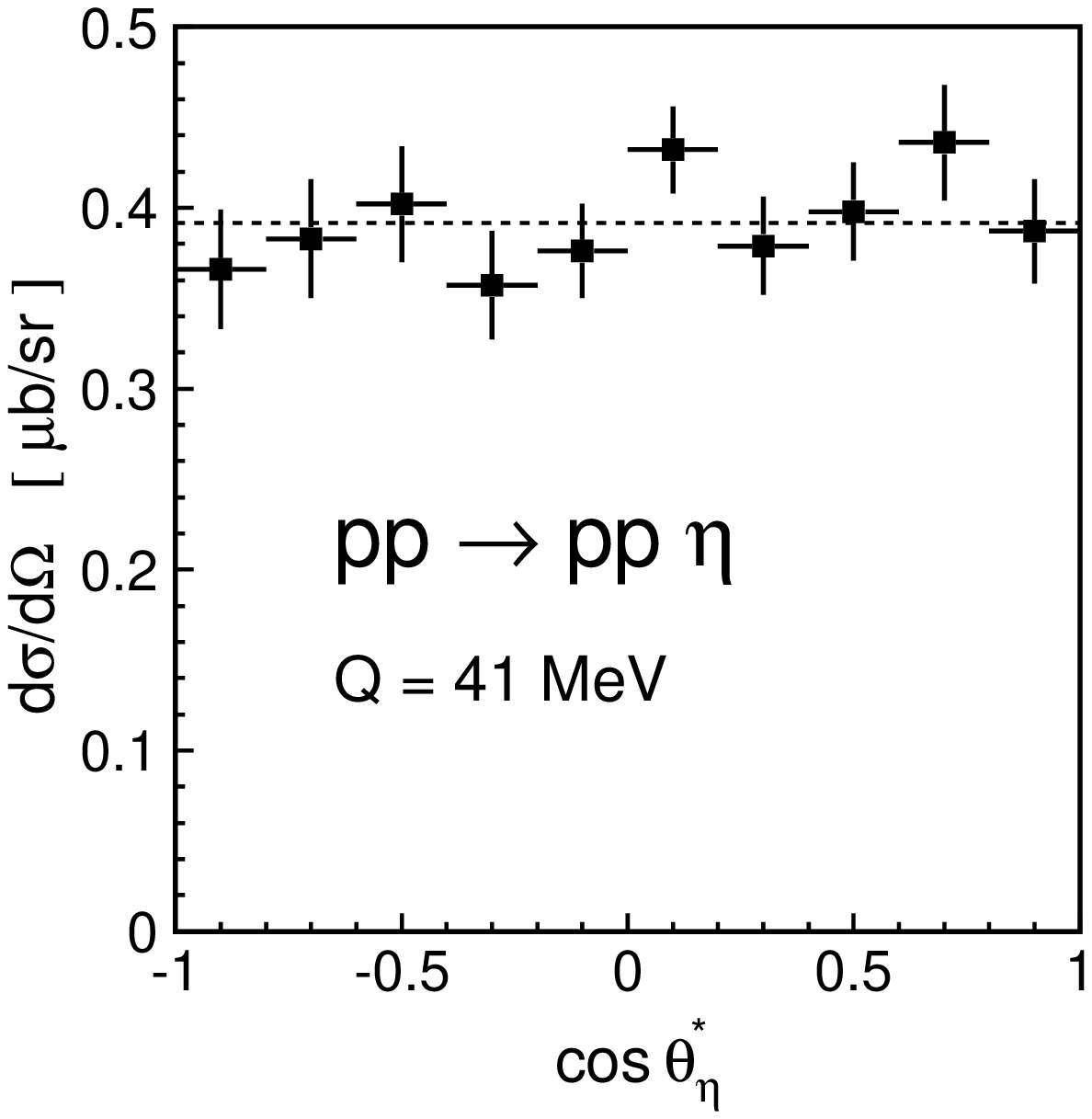,width=0.54\textwidth}}

\vspace{-0.4cm}
 \parbox{0.44\textwidth}{\raisebox{1ex}[0ex][0ex]{\mbox{}}} \hfill
\parbox{0.47\textwidth}{\raisebox{1ex}[0ex][0ex]{\large a)}} \hfill
\parbox{0.07\textwidth}{\raisebox{1ex}[0ex][0ex]{\large b)}}
\vspace{-0.8cm}
\caption{\label{differentialeta} Differential cross section of the $pp 
\rightarrow pp\eta$ reaction as a function of the meson centre-of-mass 
polar angle. Dashed lines indicate the isotropic distribution. Shown are 
results of measurements for
$Q~=~15.5\,\mbox{MeV}$~\cite{moskal025203,moskal367} (a) and $\mbox{Q} = 
41\,\mbox{MeV}$~\cite{TOFeta} (b). Only 
statistical errors are plotted, which in figure (a) are smaller than the size 
of symbols. The distribution presented in picture (a) is consistent with a 
measurement performed at an excess energy of $\mbox{Q} = 16\,\mbox{MeV}$ by 
means of the PROMICE/WASA detector~\cite{calen190}, whereas the data at 
$\mbox{Q} = 37\,\mbox{MeV}$ also from reference~\cite{calen190} deviate 
significantly from isotropy. However, data shown in picture (b) have been 
taken with a detector of much higher angular acceptance.}
\end{figure}
Figures~\ref{differentialeta}a and~\ref{differentialeta}b present the angular 
distributions of the created $\eta$ meson in proton collisions.
It is evident that at $\mbox{Q} = 15.5\,\mbox{MeV}$ and still at $\mbox{Q} = 
41\,\mbox{MeV}$ the production of the $\eta$ meson is completely isotropic 
within the shown statistical errors. 
Although at $\mbox{Q} = 41\,\mbox{MeV}$ the accuracy of the data does not 
exclude a few per cent of contributions originating from higher partial 
waves, the dominance of the s-wave creation is evident.
Similarly, the measurements of the differential cross section 
(figure~\ref{differentialetap}) for the $pp \rightarrow pp \eta^{\prime}$ reaction 
performed at SATURNE~\cite{balestra29} at $\mbox{Q} = 143.8\,\mbox{MeV}$ and at COSY~\cite{khoukaz0401011} 
at $\mbox{Q} = 46.6\,\mbox{MeV}$ are 
still consistent with pure Ss-wave production, though the relatively large 
error bars would allow for other contributions on a few per cent level 
($\approx 10\,\%$).
The observation, reported in section~\ref{detailedsection}, that distributions of the 
orientation of the emission plane~(fig.~\ref{costhetaetatof}) 
and  angle $\psi$~(fig.~\ref{rozkladpsi})
are  anisotropic,
also suggests a small contribution of higher order partial waves.
\begin{figure}[h]
\vspace{-1.1cm}
\parbox{0.5\textwidth}
  {\epsfig{file=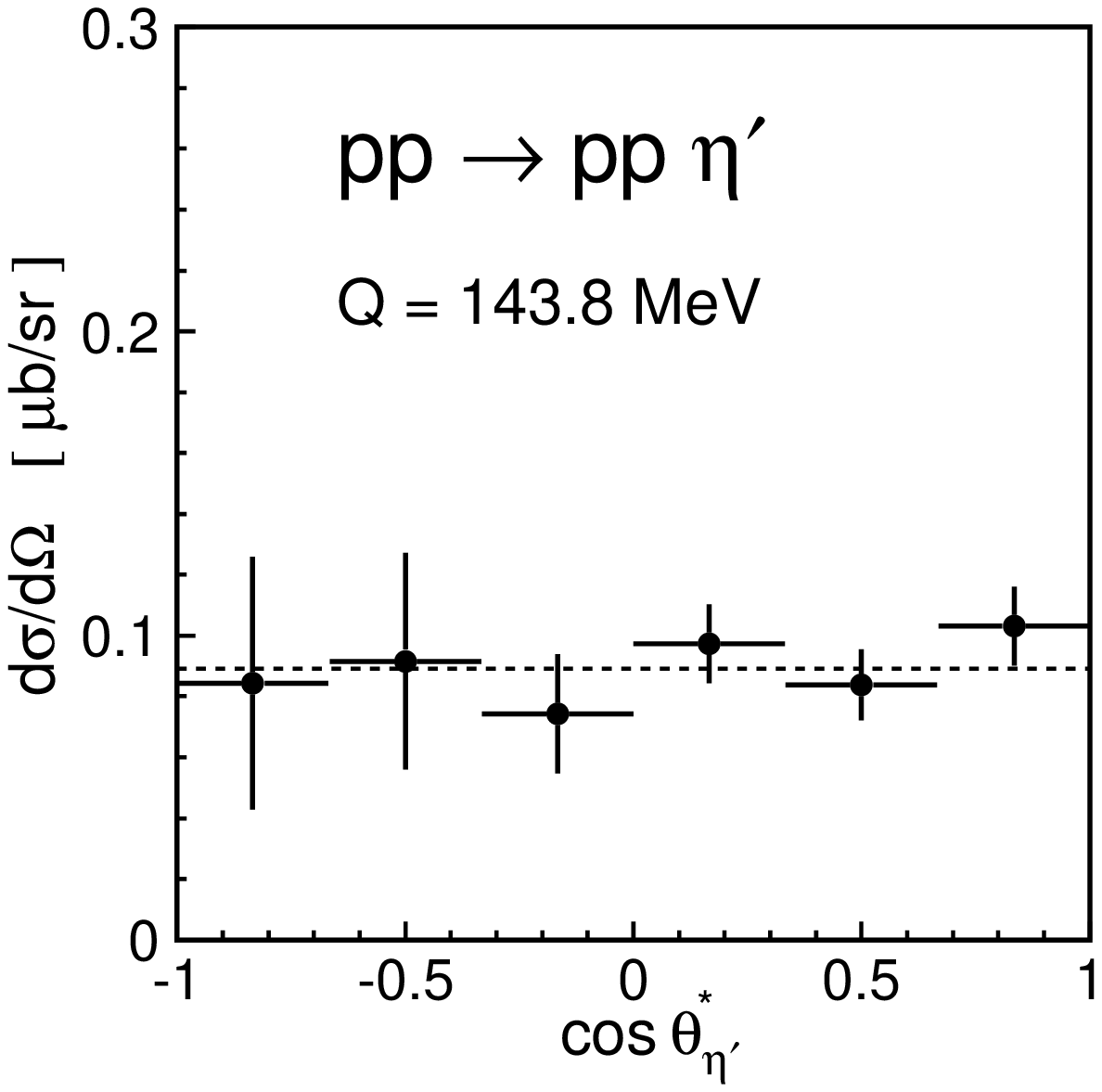,width=0.54\textwidth}}
\parbox{0.5\textwidth}
  {\epsfig{file=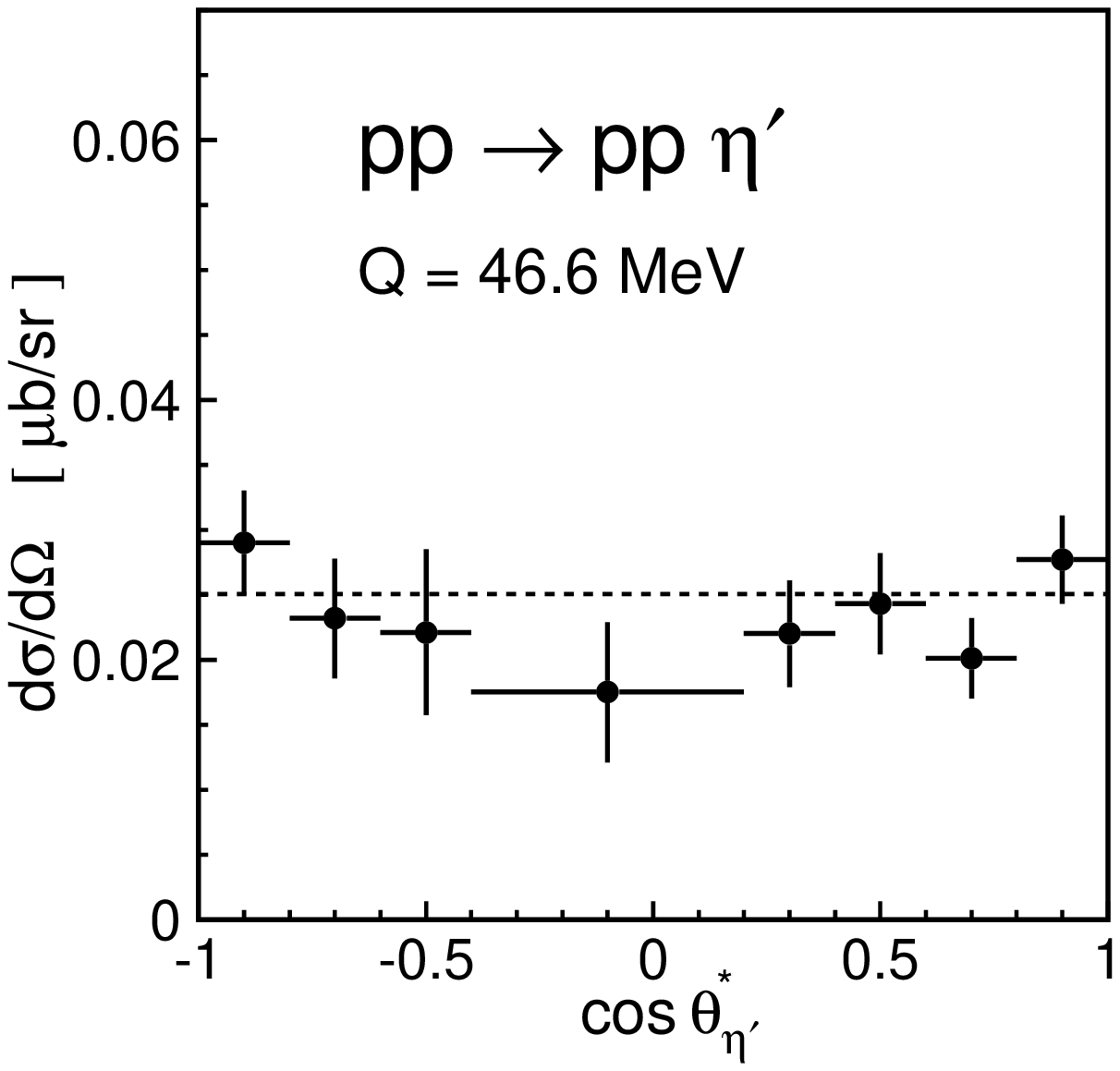,width=0.54\textwidth}}

 \vspace{-0.4cm}
 \parbox{0.44\textwidth}{\raisebox{1ex}[0ex][0ex]{\mbox{}}} \hfill
\parbox{0.47\textwidth}{\raisebox{1ex}[0ex][0ex]{\large a)}} \hfill
\parbox{0.07\textwidth}{\raisebox{1ex}[0ex][0ex]{\large b)}}
\caption{\label{differentialetap} Differential cross section of the $pp 
\rightarrow pp\eta^{\prime}$ reaction as a function of the meson centre-of-mass 
polar angle. Dashed lines indicate the isotropic distribution. Shown are 
results of measurements for the $pp \rightarrow pp \eta^{\prime}$ reaction 
taken at $\mbox{Q} = 143.8\,\mbox{MeV}$~\cite{balestra29} (a) and at 
$Q~=~46.6~\mbox{MeV}$~\cite{khoukaz0401011} (b).
}
\end{figure}
Specifically, the modulus of the cosine of the polar angle of the vector
normal to the emission plane, whose distribution is found to be anisotropic,
is a function of the sixth power of the final 
momenta
({\small $|cos(\theta_{N}^{*})| 
  \sim  \sqrt{|( \vec{p}_{proton_{1}}~\times~\vec{p}_{proton_2} )\cdot \vec{p}_{beam}|^2}$
}),
and even more peculiar the sine of the angle $\psi$,  which is a function of the eighth power 
of the final momenta
({\small $|sin(\psi)| 
  \sim  \sqrt{|( \vec{p}_{proton_{1}}~\times~\vec{p}_{proton_2} )\cdot 
                     (\vec{p}_{\eta}~\times~\vec{p}_{beam})|^2}$
}),
is found to be anisotropic too.
This could be a sign of an interference between partial waves  higher than 
Ss-wave~\footnote{\mbox{} We are grateful to Colin Wilkin for sharing with us this astonishment.}.

\section{Influence of the pp$\eta$ and pp$\eta^{\prime}$ interaction 
             on
            the excitation function of the $pp\to pp\eta (\eta^{\prime})$ reaction}
            
\label{influencesection}
\begin{flushright}
\parbox{0.73\textwidth}{
 {\em
 The only proof capable of being given that an object is visible,
 is that people actually see it. The only proof that a sound is audible,
 is that people hear it: and so of the other sources of our experience~\cite{mill}.\\
 }
 \protect \mbox{} \hfill  John Stuart Mill \protect\\
 }
\end{flushright}

Let us now consider to what extent the energy dependence of the total cross 
section in the estimated range of the dominance of the Ss partial waves can be 
understood in terms of the phase space variation and the interaction between 
the particles participating in the reaction.
Watson~\cite{watson1163} and Migdal~\cite{migdal2} proposed the factorization 
of the amplitude when the production is of short- and the 
interaction among the outgoing particles  of long range.
This requirement is well fulfilled for the close-to-threshold 
meson production due to the large momentum transfer ($\Delta\mbox{p}$) between 
the interacting nucleons needed to create the considered mesons ($\pi, \eta, 
\ldots, \phi$). 
\begin{table}[H]
\caption{\label{momtranstable} Momentum transfer $\Delta\mbox{p}$ calculated 
according to equation~\ref{momtranseq} and the corresponding distance 
$\mbox{R} \approx \hbar/\Delta\mbox{p}$ probed by the $NN \rightarrow 
NN\,Boson$ reaction at the kinematical threshold for different particles 
produced. The table has been adapted from~\cite{nak01}.}
\vskip -0.25cm
\tabskip=1em plus2em minus.5em
\halign to \hsize{\hfil#&\hfil#\hfil&\hfil#\hfil&\hfil#\hfil\cr
\noalign{\hrulefill}
particle  & mass [MeV] & $\Delta\mbox{p}$ [$\mbox{fm}^{-1}$] & R [fm] \cr
\noalign{\vskip -0.20cm}
\noalign{\hrulefill}
\noalign{\vskip -0.5cm}
\noalign{\hrulefill}
$\gamma$        &   0   &  0.0  &  $\infty$ \cr
$\pi$           & 140   &  1.9  &  0.53 \cr
$\eta$          & 550   &  3.9  &  0.26 \cr
$\rho , \omega$ & 780   &  4.8  &  0.21 \cr
$\eta^\prime$   & 960   &  5.4  &  0.19 \cr
$\phi$          &1020   &  5.6  &  0.18 \cr
\noalign{\hrulefill} }
\end{table}
According to the Heisenberg uncertainty relation the large momentum transfer 
brings about a small space in which the primary creation of the meson takes 
place.
In table~\ref{momtranstable} the distance probed by the $NN \rightarrow 
NN\,Meson$ reaction at threshold is listed for particular mesons. 
It ranges from $0.53\,\mbox{fm}$ for pion production to $0.18\,\mbox{fm}$ for 
the $\phi$ meson, whereas the typical range of the strong nucleon-nucleon 
interaction at low energies determined by the pion exchange may exceed a 
distance of a few Fermi and hence is by one order of magnitude larger than the 
values listed in table~\ref{momtranstable}.
Thus in analogy to the Watson-Migdal approximation for two-body 
processes~\cite{watson1163} the complete transition matrix element of 
equation~\eqref{phasespacegeneral} may be factorized approximately 
as~\footnote{\mbox{} For a comprehensive discussion of the FSI and ISI issue including 
a historical overview and a criticism of various approaches the reader is 
referred to~\cite{kleefeld51}.}
\be
\label{M0FSIISI}
|M_{pp \rightarrow pp X}|^2 \approx |M_{FSI}|^2 \cdot |M_0|^2 \cdot F_{ISI},
\ee
where $M_{0}$ represents the total short range production amplitude, 
$M_{FSI}$ describes the elastic interaction among particles in the exit 
channel and $F_{ISI}$ denotes the reduction factor accounting for the 
interaction of the colliding protons.
Further, in the first order approximation one assumes that the particles are 
produced on their mass shell and that the created meson does not interact with 
nucleons.
This assumption implies that the $|M_{FSI}|^2$ term can be substituted by the 
square of the on-shell amplitude of the nucleon-nucleon elastic scattering:
\be
\label{FSI_elastic}
|M_{FSI}|^2 = |M_{NN \rightarrow NN}|^2.
\ee
Effects of this rather bold assumption will be considered later, when
comparing the estimation with the experimental data.

In the frame of the optical potential model the scattering amplitude is 
determined by phase-shifts.
Particularly, the $^{1}\mbox{S}_0$ proton-proton partial wave -- relevant 
for further considerations -- can be expressed explicitly as 
follows~\cite{morton825}:
\be
\label{amppp}
M_{pp \rightarrow pp} = 
  \frac{e^{-i\delta_{pp}({^{1}\mbox{\scriptsize S}_{0}})} \cdot 
        \sin{\delta_{pp}({^{1}\mbox{S}_0})}}
       {C \cdot \mbox{k}},
\ee
where $C$ denotes the square root of the Coulomb penetration factor.
$C^2$ determines the ratio of the probability of finding two particles close 
together to the probability of finding two uncharged particles together, all 
other things being equal~\cite{jackson77} and can be expressed 
as~\cite{bethe38}:
\begin{equation} 
\label{penetrationfactor}
  C^{2} = \frac{2\pi\eta_c}{e^{2\pi\eta_c} - 1},
\end{equation}
where
$\eta_c$ is the relativistic Coulomb parameter, which for the collision of 
particles $i,j$ reads:
\begin{equation*}
\eta_c = \frac{\mbox{q}_i\,\mbox{q}_j \,\alpha}{\mbox{v}} = 
  \mbox{q}_i\,\mbox{q}_j \,\alpha 
  \frac{\mbox{s}_{ij} - \mbox{m}_i^2 - \mbox{m}_j^2}
       {\sqrt{\lambda(\mbox{s}_{ij},\mbox{m}_i^2,\mbox{m}_j^2)}},
\end{equation*}
with the fine structure constant $\alpha$, the relative velocity v of the 
colliding particles and with $\mbox{q}_i$, $\mbox{q}_j$ denoting their 
charges~\footnote{\mbox{} For collisions at an angular momentum $l$ larger than 
$0\,\hbar$ the $C^2$ of equation~\eqref{penetrationfactor} needs to be 
multiplied by a factor of $\prod_{n=1}^{l} 
\left(1 + (\eta_c/n)^2 \right)$~\cite{wong1866,arndt1002}.}.

The variable k in equation~\ref{amppp} stands for either proton momentum in the 
proton-proton rest frame and the phase-shift is indicated by $\delta_{pp}$. 
The phase-shifts $\delta_{pp}({^{1}\mbox{S}_0})$ can be extracted from the 
SAID data base (see fig.~\ref{1s03p0}b) or, alternatively, can be 
calculated according to the modified Cini-Fubini-Stanghellini formula 
including the Wong-Noyes Coulomb 
correction~\cite{naisse506,noyes995,noyes465}\\[-0.3cm]
\be
\label{CFS}
C^2 \;\mbox{k}\; ctg(\delta_{pp}) \;+\; 2\,\mbox{k}\,\eta_c\,h(\eta_c) = 
  - \frac{1}{a_{pp}} + \frac{b_{pp}\,\mbox{k}^2}{2} 
  - \frac{P_{pp}\,\mbox{k}^4}{1 + Q_{pp}\,\mbox{k}^2},
\ee
where $h(\eta_c) = -ln(\eta_c) - 0.57721 + \eta_c^2\,
\sum_{n=1}^{\infty}\frac{1}{n \cdot (n^2 + \eta_c^2)}$~\cite{jackson77}.

The phenomenological quantities $a_{pp} = -7.83\,\mbox{fm}$ and $b_{pp} = 
2.8\,\mbox{fm}$ denote the scattering length and effective 
range~\cite{naisse506}, respectively. 
The parameters $P_{pp} = 0.73\,\mbox{fm}^3$ and $Q_{pp} = 3.35\,\mbox{fm}^2$ 
are related to the detailed shape of the nuclear potential and derived from a 
one-pion-exchange model~\cite{naisse506}.
Substituting equation~\eqref{CFS} into equation~\eqref{amppp} allows to 
calculate the low-energy amplitude for the proton-proton elastic 
scattering~\footnote{\mbox{} In principle the formula is valid for
k$ \leq 133~MeV/c$~\cite{noyes995}.  
$|M_{pp \rightarrow pp}|^2$ is taken to be constant for larger 
values of k.}: \\[-0.4cm]
\be
\label{Mpppp}
|M_{pp \rightarrow pp}|^2 \;=\; 
 \frac{C^2}
  {C^4\,\mbox{k}^2 \;+\; 
  \left(- \frac{1}{a_{pp}}\,+\,\frac{b_{pp}\,\mbox{\scriptsize k}^2}{2}\, 
 -\,\frac{P_{pp}\,\mbox{\scriptsize k}^4}{1 + Q_{pp}\,\mbox{\scriptsize k}^2}\, 
 -\,2\,\mbox{k}\,\eta_c\,h(\eta_c)\right)^2}.
\ee
The result is presented as a solid line in figure~\ref{Mpppp_cross_pi}a and is 
in good agreement with the values obtained from the phase-shifts of the VPI 
partial wave analysis~\cite{arndt3005}, shown as solid circles and with the 
phase-shifts of the Nijmegen analysis~\cite{nijmpsa}, shown as open squares.
The factor $C^2$ is always less than unity due to the Coulomb repulsion 
between protons. 
At higher energies, where $C^2$ is close to unity, the nuclear scattering will 
be predominant and for the very low energies the Coulomb and nuclear 
interactions are competing. 
The Coulomb scattering dominates approximately up to about $0.8\,\mbox{MeV}$ 
of the proton energy in the rest frame of the other proton, where $C^2$ equals 
to one-half~\cite{jackson77}. 
In the case of the $pp \rightarrow pp\, Meson$ 
reaction the maximum possible 
energy of a proton seen from another proton is equal to $0.8\,\mbox{MeV}$ 
already at an excess energy of about $\mbox{Q} = 0.4\,\mbox{MeV}$. 
Therefore, a significant influence of the Coulomb repulsion on the energy 
dependence of the total production cross section is expected only at very low 
excess energies, i.e.\ conservatively for $\mbox{Q} \le 2\,\mbox{MeV}$.
\vspace{-0.5cm}
\begin{figure}[H]
\parbox{0.45\textwidth}{\vspace{-0.3cm}\epsfig{file=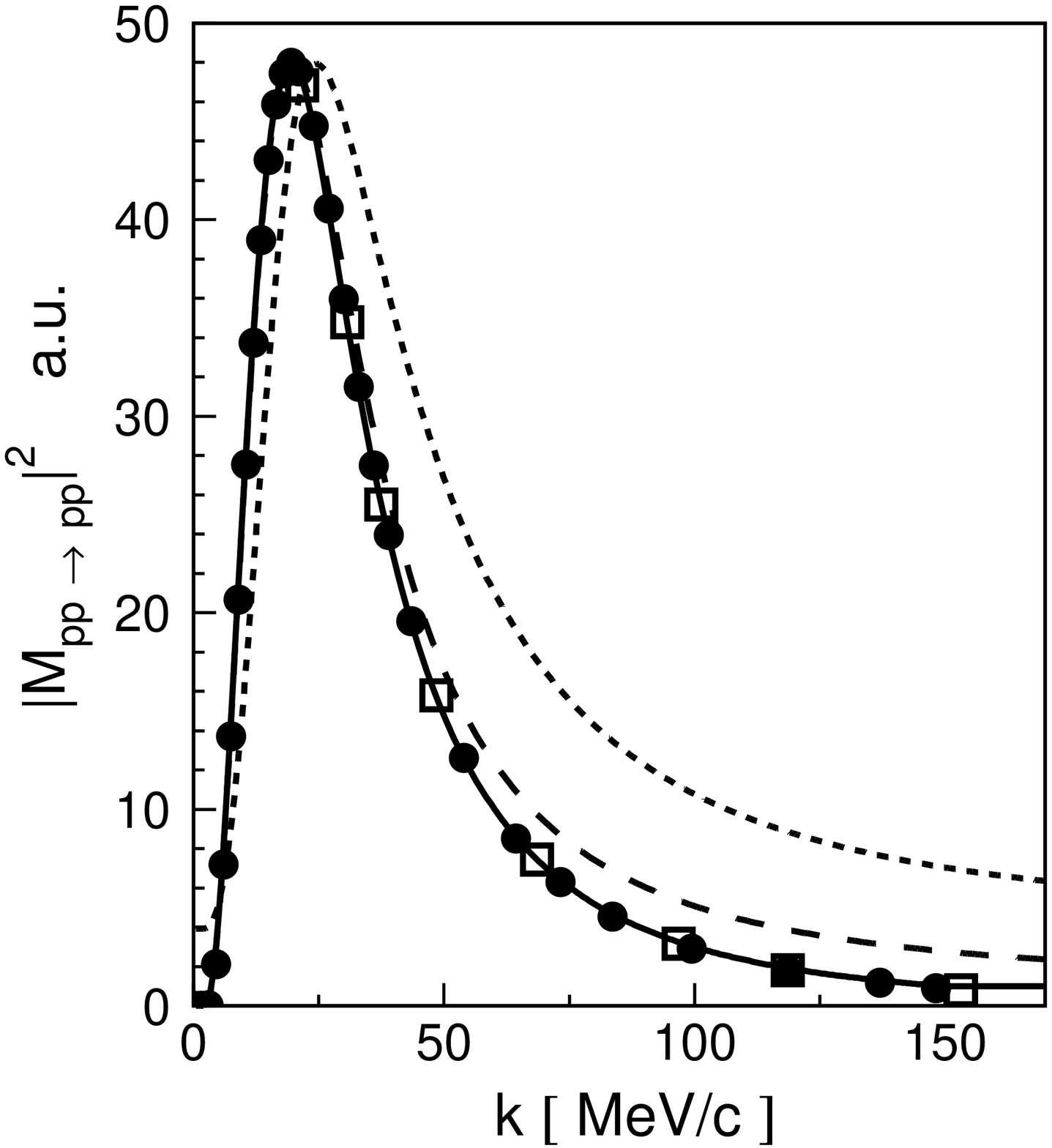,width=0.49\textwidth}} \hfill
\parbox{0.50\textwidth}
  {\epsfig{file=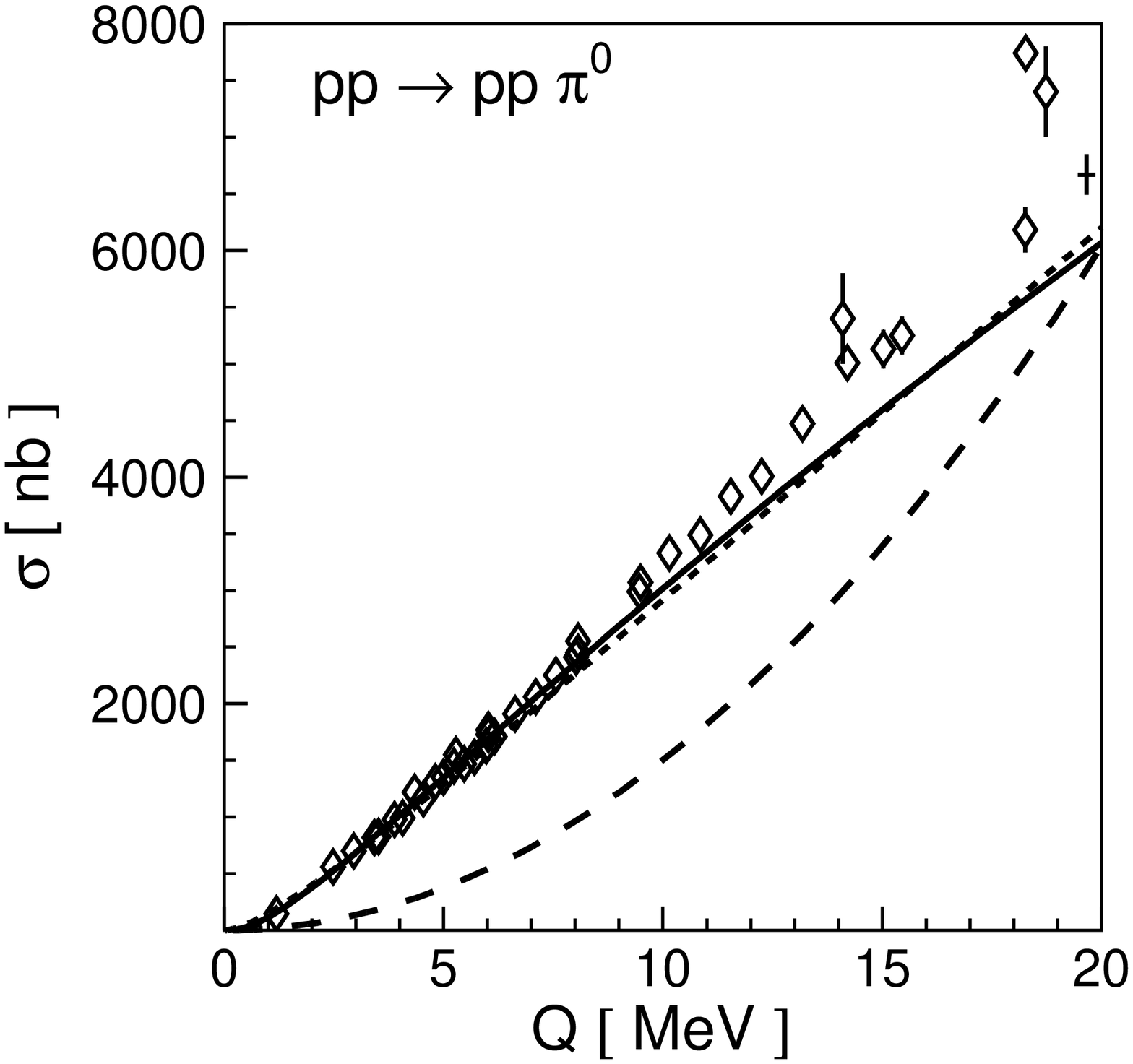,width=0.55\textwidth}}

\vspace{-0.3cm}
\parbox{0.38\textwidth}{\raisebox{1ex}[0ex][0ex]{\mbox{}}} \hfill
\parbox{0.49\textwidth}{\raisebox{1ex}[0ex][0ex]{\large a)}} \hfill
\parbox{0.03\textwidth}{\raisebox{1ex}[0ex][0ex]{\large b)}}

\vspace{-0.3cm}
\caption{\label{Mpppp_cross_pi} (a) Square of the proton-proton scattering 
amplitude versus k, the proton momentum in the proton-proton subsystem 
from~\cite{morton825,naisse506}~(solid line), \cite{druzhinin}~(dashed line), 
and \cite{shyammosel,goldbergerwatson}~(dotted line). The filled circles are 
extracted from~\cite{arndt3005} and the opened squares from~\cite{nijmpsa}.
The curves and symbols have been arbitrarily normalized to be equal at maximum to 
the result from reference~\cite{druzhinin}, shown as the dashed line. 
(b) Total cross section for the $pp \rightarrow pp \pi^0$ reaction as a 
function of the centre-of-mass excess energy Q. Data are from  
refs.~\cite{bondar8,meyer633,stanislausR1913,bilger633}. The dashed line 
indicates a phase space integral normalized arbitrarily. The phase space 
distribution with inclusion of proton-proton strong and Coulomb interactions 
fitted to the data at low excess energies is shown as the solid line. The 
dotted line indicates the parametrization of reference~\cite{faldt209} written 
explicitly in equation~\eqref{faldtwilkin}, with $\epsilon = 0.3$.}
\end{figure}
\vspace{-0.2cm}
Assuming that the on-shell proton-proton amplitude determines exclusively
the phase space population one can obtain the total cross section energy 
dependence substituting equation~\eqref{Mpppp} into 
formula~(\ref{Vpsdalitz}).
The solid line in figure~\ref{Mpppp_cross_pi}b 
represents the determined dependence for the $pp \rightarrow pp \pi^0$ 
reaction.
The absolute scale of the calculations has been fixed by  normalizing to 
the data.
One recognizes the good agreement with the experimental points in the excess 
energy range up to $\mbox{Q} \approx 10\,\mbox{MeV}$ 
in agreement with the previous conclusions based on the polarisation observables.

The dotted line in figures~\ref{Mpppp_cross_pi}b 
which is practically indistinguishable from 
the solid line, presents the excess energy dependence of the total cross 
section taking into account the proton-proton FSI effects according to the 
model developed by F{\"a}ldt and Wilkin~\cite{faldt209,faldt2067}.
Representing the scattering wave function in terms of a bound state wave 
function the authors derived a closed formula which describes the effects of 
the nucleon-nucleon FSI as a function of the excess energy Q only. 
This approach is specifically useful for the description of the spin-triplet 
proton-neutron FSI, due to the existence of a bound state (deuteron) with the 
same quantum numbers.
Though a bound state of the proton-proton system does not exist, the model 
allows to express the total cross section energy dependence for a $pp 
\rightarrow pp\,Meson$ reaction by a simple and easily utilizable formula: 
\be
\label{faldtwilkin}
\nonumber
\sigma \;=\; 
  const \cdot \frac{V_{ps}}{\mbox{F}} \cdot 
  \frac{1}
    {\left(1\;+\;\sqrt{1\,+\,\frac{\mbox{\scriptsize Q}}{{\displaystyle \epsilon}}}\right)^2} =
\ee
\be
{\mbox{}}\ \ =\; const^{\prime} \cdot 
  \frac{\mbox{Q}^{2}}{ \sqrt{\lambda(\mbox{s},\mbox{m}_p^2,\mbox{m}_p^2)}} 
  \cdot 
  \frac{1}
    {\left(1\;+\;\sqrt{1\,+\,\frac{\mbox{\scriptsize Q}}{{\displaystyle \epsilon}}}\right)^2},
\ee
where the parameter $\epsilon$ has to be settled from the data.
The flux factor F and the phase space volume $V_{ps}$ are given by 
equations~\eqref{fluxfactor} and~\eqref{Vps_nonrelativistic}, respectively.
The normalization can be determined from the fit of the data which must be 
performed for each reaction separately.
  
The determined energy dependences of the total cross section
  for $\eta^{\prime}$~\cite{balestra29,wurzinger283,moskal416,moskal3202,khoukaz0401011,hibou41} and
  $\eta$~\cite{hibou41,bergdoltR2969,chiavassa270,calen39,calen2642,smyrski182}
  mesons production in  proton-proton collisions
  are presented in figure~\ref{cross_eta_etap}.
  Comparing the data to the arbitrarily normalized phase space integrals (dashed lines)
  reveals that the proton-proton FSI enhances the total cross section by more than an order
  of magnitude for low excess energies.
  
  One recognizes also that
  in the case of the $\eta^{\prime}$
  the data are described very well (solid line)
    assuming that the on-shell proton-proton amplitude
    exclusively determines the phase space population.
 This indicates that the  proton-$\eta^{\prime}$ interaction is too small to manifest itself
  in the excitation function within the presently achievable accuracy.
In the case of $\eta$ meson production the interaction between nucleons is 
evidently not sufficient to describe the increase of the total cross section 
for very low and very high excess energies, as can be concluded from the comparison 
of the data and the upper solid line in figure~\ref{cross_eta_etap}.
This line was normalized to the data at an excess energy 
range between $15\,\mbox{MeV}$ and $40\,\mbox{MeV}$. 
The enhancement of the total cross section for higher energies can be assigned to 
the outset of higher partial waves.
As expected from the previous considerations, this is indeed seen at 
$\mbox{Q} \approx 40\,\mbox{MeV}$ where the energy dependence of the total 
cross section starts to change its shape.
On the contrary, the close-to-threshold enhancement --~being by about 
a factor of two larger than in the case of the $\pi^0$ and $\eta^{\prime}$ 
mesons~-- can be assigned neither to the contribution from other than Ss 
partial waves nor to the variation of the primary production amplitude 
$|M_{0}|$. 
The latter is expected to change at the most by a few per cent for excess energies 
below $20\,\mbox{MeV}$~\cite{moalem649}.
Instead, this discrepancy can be plausibly explained by the influence of the 
attractive interaction between the $\eta$ meson and the proton.
\vspace{-0.6cm}
 \begin{figure}[H]
    \centerline{\parbox{0.6\textwidth}{
    \includegraphics[width=0.58\textwidth]{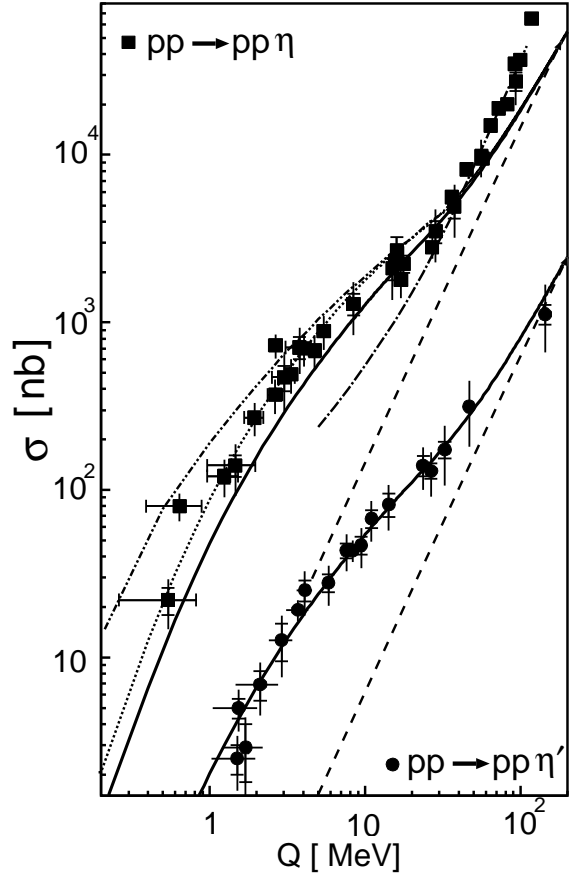}}} 
    \parbox{1.00\textwidth}{
    \vspace{0.0cm}
    \caption{\label{cross_eta_etap}
      Total cross section for the reactions
      $pp \rightarrow pp \eta^{\prime}$~(circles) and
      $pp \rightarrow pp \eta$~(squares)
      as a function of the centre-of-mass
      excess energy Q. Data are from
      refs.~\cite{balestra29,wurzinger283,moskal416,moskal3202,
                  khoukaz0401011,hibou41,bergdoltR2969,chiavassa270,calen39,calen2642,smyrski182}.
      The dashed lines
      indicate a phase space integral normalized arbitrarily. \protect\\
      The solid lines
      show the phase space distribution with inclusion of the
      $^1S_0$ proton-proton strong
      and Coulomb interactions.
    In case of the $pp\to pp\eta$ reaction the solid line
    was fitted to the data in the excess energy range
    between 15 and $40\,\mbox{MeV}$. Additional
    inclusion of the proton-$\eta$ interaction is indicated by the dotted line.
    The scattering length of $a_{p\eta} = 0.7\,\mbox{fm} + i\,0.4\,\mbox{fm}$ and the
    effective range parameter $b_{p\eta} = -1.50\,\mbox{fm} -
    i\,0.24\,\mbox{fm}$~\cite{greenR2167} have been chosen arbitrarily.
    The dashed-dotted line represents the energy dependence taking into account the
    contribution from the
    $^3P_{0}\to ^1\!\!\!S_{0}s$, $^1S_{0}\to ^3\!\!\!P_{0}s$ and
    $^1D_2\to ^3\!\!\!P_2 s$ transitions~\cite{nakayama0302061}.
    Preliminary results for the $^3P_{0}\to ^1\!\!\!S_{0}s$
    transition with full treatment of three-body effects
    are shown as a dashed-double-dotted line~\cite{Fixprivate,Fixprivate1}.
    The absolute scale of dashed-double-dotted line
    was arbitrary fitted to demonstrate
    the energy dependence only.
    }}
  \end{figure}
\vspace{-0.3cm}
Note, that the real part of the scattering length of the $\eta$-proton 
potential -- depending on the analysis method and the studied reaction -- is 
3 to 10 times~\cite{green053,sibirtsev086} larger than the scattering length 
for $\pi^0$-proton scattering ($a_{p\pi} \approx 
0.13\,\mbox{fm}$)~\cite{sigg269}.
Hence, the modifications of the total cross section energy dependence due to 
$\pi^0$-proton and $\eta^{\prime}$-proton interaction 
are too weak to be observed within the 
up-to-date accuracy of measurements and calculations. 
However, the influence of the $\eta$-proton interaction presented in 
figure~\ref{cross_eta_etap} is evident and hereafter it will be considered 
whether it may serve for the estimation of the $\eta$-proton scattering 
parameters.

A strict quantitative calculation requires the evaluation of the 
three-body Faddeev equation which is out of the scope 
of the present work\footnote{\mbox{} An exact derivation of the Faddeev 
equation can be found for example in~\cite{gloeckle107}. Presently few theory groups 
are carrying on the corresponding 
calculations~\cite{Fixprivate,Fixprivate1,garcilazopriv,wycechpriv,deloff}.}.
Here, we will rather present a simple phenomenological treatment which shall 
lead to the qualitative understanding how the mutual interaction among three 
outgoing particles affects the excitation function.
One of the simplest possibilities based on the naive probabilistic 
interpretation of the incoherent pairwise interaction would be to factorize 
the overall enhancement factor into corresponding pair 
interactions~\cite{schuberthPHD,bernard259}:
\be
\label{pairinteraction} 
|M_{FSI}|^2 \;=\; |M_{12 \rightarrow 12}|^2 \cdot |M_{13 \rightarrow 13}|^2 
  \cdot |M_{23 \rightarrow 23}|^2,
\ee
where $|M_{ij \rightarrow ij}|^2$ denotes the square of the elastic scattering 
amplitude of particles $i$ and $j$. 
The $|M_{pp\to pp}|^2$ term can be evaluated according to the formula~\eqref{Mpppp}, 
which for the s-wave $\eta$-proton scattering, after substitution of $C^2 
= 1$ and $\eta_c = P_{pp} = 0$, reduces to~\footnote{\mbox{} Note that the sign of the 
term $-1/a$ from equation~\eqref{Mpppp} was changed because the imaginary part 
of the proton-$\eta$ scattering length is positive~\cite{greenR2167}, as we 
also adopted here, whereas in the majority of works concerning 
nucleon-nucleon interaction, the scattering length is 
negative~\cite{machleidtR69}. We are grateful for this remark to A. Gasparyan.}:
\be
\label{Mpeta}
|M_{p\eta \rightarrow p\eta}|^2 \;=\; 
  \left| \frac{1}
    {\frac{1}{a_{p\eta}}\,+\,
     \frac{b_{p\eta}\,\mbox{\scriptsize k}_{p\eta}^2}{2} - i\,\mbox{k}_{p\eta}}
  \right|^2,
\ee
where
\begin{equation*}
\mbox{k}_{p\eta} \;=\;
  \frac{\sqrt{\lambda(\mbox{s}_{p\eta},\mbox{m}_{\eta}^2,\mbox{m}_p^2)}}
       {2\,\sqrt{\mbox{s}_{p\eta}}} 
\end{equation*}
denotes the $\eta$ momentum in the proton-$\eta$ rest frame. 
The scattering length $a_{p\eta}$ and effective range $b_{p\eta}$ are complex 
variables with the imaginary part responsible e.g.\ for the $p\eta 
\rightarrow p \pi^0$ conversion.\vspace{1ex}

The factorization of both \hspace{1ex}i) the overall production matrix element 
(eq.~\eqref{M0FSIISI}) and \hspace{1ex} ii) the three particle final state 
interactions (eq.~\eqref{pairinteraction}) applied to 
formula~\eqref{Vpsdalitz} gives the following expression for the total cross 
section of the $pp \rightarrow pp \eta$ reaction:
\vspace{-0.3cm}
{\small{
\be
\label{cross_with_FSI}
\sigma\!\! \;=\;\!\! \frac{F_{ISI}\,|M_0|^2}{\mbox{F}}\,\frac{\pi^2}{4\,\mbox{s}} 
   \int\limits_{(\mbox{\scriptsize m}_p+\mbox{\scriptsize m}_p)^2}^{
               (\sqrt{\mbox{\scriptsize s}}-\mbox{\scriptsize m}_{\eta})^2} 
       \hspace{-0.5cm} d\,\mbox{s}_{pp}\; |M_{pp \rightarrow pp}(\mbox{s}_{pp})|^2 \hspace{-0.6cm}
  \int\limits_{\mbox{\scriptsize s}_{p_2 \eta}^{min} 
                 (\mbox{\scriptsize s}_{pp})}^{
               \mbox{\scriptsize s}_{p_2 \eta}^{max}
                 (\mbox{\scriptsize s}_{pp})} \hspace{-0.5cm}d\,\mbox{s}_{p_2 \eta} \; 
  |M_{p_1 \eta \rightarrow p_1 \eta} (\mbox{s}_{p_1 \eta})|^2    \cdot 
  |M_{p_2 \eta \rightarrow p_2 \eta} (\mbox{s}_{p_2 \eta})|^2,
\ee
}}
\vspace{-0.1cm}
where the protons are distinguished by subscripts.
Exploring formulae~\eqref{Mpppp} and~\eqref{Mpeta} gives the results shown as 
the dotted line in figure~\ref{cross_eta_etap}. 
Evidently, the inclusion of the proton-$\eta$ interaction enhances the total 
cross section close-to-threshold by about a factor of 1.5 and leads to a 
better description of the data.
  A similar effect close-to-threshold is also observed 
  in the data of photoproduction of $\eta$
  via  the $\gamma d\to pn\eta$ reaction~\cite{eta_photo} indicating
  to some extent that the phenomenon is independent of the production
  process but rather related to the interaction among the $\eta$ meson and
  nucleons in the S$_{11}$(1535) resonance region.
  However, although a simple phenomenological treatment~\cite{review,bernard259,schuberthPHD}
   --~based on the factorization of the
  transition amplitude into the constant primary production and the on-shell
  incoherent pairwise interaction among outgoing particles~-- describes 
  the enhancement close-to-threshold~(dotted line) very well,
  it fails to describe the invariant
  mass distribution of the proton-proton and proton-$\eta$ subsystems
  determined recently at Q~=~15~MeV by the COSY-TOF~\cite{TOFeta}
  and at Q~=~15.5~MeV by the COSY-11~\cite{moskal025203} collaborations. 
  Though the groups utilized entirely different experimental
  methods the obtained results agree very well with each other.
   The structure of these invariant mass distributions
  may indicate  a non-negligible contribution from the P-waves in the
  outgoing proton-proton subsystem~\cite{nakayama0302061},
  which can be produced for instance via  $^1S_{0}\to ^3\!\!P_{0}s$ or $^1D_2\to ^3\!\!P_2 s$
  transitions. This hypothesis encounters, however, difficulties in describing the
  excess energy dependence of the total cross section.
  The amount of the P-wave admixture  derived from the proton-proton invariant mass
  distribution  leads to a good description of the excitation function
  at higher excess energies
  while at the same time  it spoils significantly
  the agreement with  the data at low values of Q, as depicted by the
  dashed-dotted line in figure~\ref{cross_eta_etap}.
  However, these difficulties in reproducing the observed energy dependence
  might be due to the particular model used in reference~\cite{nakayama0302061},
  and thus higher partial wave contributions cannot be excluded a priori.

  In contrast to the P-wave contribution
  the three-body treatment~\cite{Fixprivate,Fixprivate1} of the $pp\eta$
  system~(dashed-double-dotted line)
  leads to even larger enhancement of the cross section near threshold
  than that based on
  the Ansatz of the factorization of the proton-proton and proton-$\eta$ interactions.
  It must be kept in mind, however, that too strong FSI effect predicted by the three-body
  model must be partially assigned to the neglect of the Coulomb repulsion in this
  preliminary calculations~\cite{Fixprivate,Fixprivate1}.
  The above considerations  illustrate that the
  simple phenomenological approach shown by the dotted line
  could fortuitously lead to the proper result, due to a mutual cancellation
  of the effects caused by the approximations assumed in the calculations and
  the neglect  of higher partial waves.
  They  show also unambiguously that for the complete understanding
  of the low energy $pp\eta$ dynamics, in  addition to the already discussed
  excitation function of the total cross section, a knowledge of the
  differential observables is necessary.
  These will help to disentangle effects
  caused by the proton-$\eta$  interaction and the  contributions
  from higher partial waves.
  This issue will be discussed further in section~\ref{dalitzsection},
  where we present distributions determined experimentally
  for two sets of orthogonal variables
  fully describing the pp$\eta$  system, which was produced at an excess energy of Q~=~15.5~MeV
  via the $pp\to pp\eta$ reaction using the COSY-11~\cite{brauksiepe397,moskal448}
  facility at COSY~\cite{prasuhn167}.

\section{Phenomenology of the proton-proton initial and final state interaction}
\label{sectionfsiisi}
\begin{flushright}
\parbox{0.7\textwidth}{
 {\em
  Thus, it is suggested that among created beings
  there must be some basic agent which will move things
  and bring them together~\cite{arystotelesmetaphysics}.\\
 }
 \protect \mbox{} \hfill  Aristotle \protect\\
 }
\end{flushright}
When reducing the proton-proton FSI effect to a multiplicative 
factor, one finds that it depends on the assumed nucleon-nucleon potential 
and on the produced meson mass~\cite{baru579}. 

This issue was recently vigorously investigated e.g.\ by authors of 
references~\cite{baru579,kleefeld51,hanhart176,niskanen107,nak01} and we shall 
briefly report this here as well.
Up to now we factorized the transition matrix element into a primary 
production of particles and its on-shell rescattering in the exit channel 
(eqs.~\eqref{M0FSIISI}\eqref{FSI_elastic}\eqref{pairinteraction}).
Though it is a crude approximation, neglecting the off-shell effects of the 
production process completely, it astoundingly leads to a good description of 
the energy dependence of the total cross section, as already demonstrated in 
figures~\ref{Mpppp_cross_pi}b and \ref{cross_eta_etap}.
The off-shell effects, as pointed out by Kleefeld~\cite{kleefeld51}, could have 
been safely neglected in case of the electromagnetic transitions in atoms or 
$\beta$ decays, where the excitation energy of the involved nucleons is by 
many orders of magnitude smaller than their masses and the initial and final 
states go hardly off-shell~\cite{kleefeld51}.
However, in the case of the $NN \rightarrow NN\,Meson$ process the large 
excitation energy of the colliding nucleons is comparable with the nucleon 
masses and the primary interaction may create the particles significantly far 
from their physical masses, so that a priori the off-shell effects cannot be
disregarded.
\vspace{-0.9cm}
\begin{figure}[H]
\begin{center}
\parbox{1.0\textwidth}
  {\epsfig{file=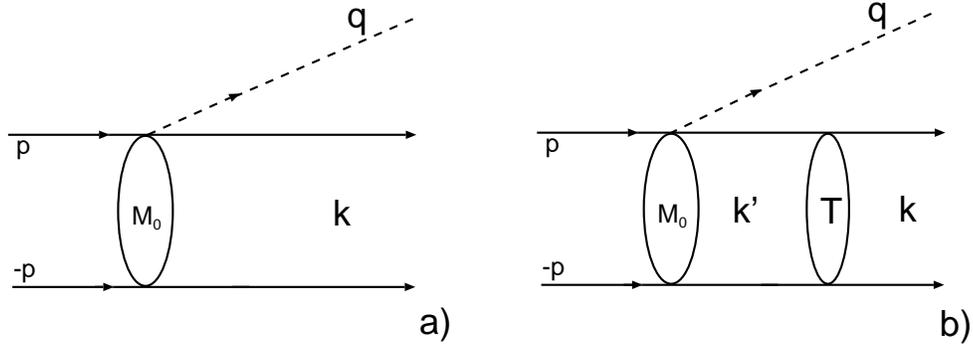,width=1.0\textwidth}}
\end{center}
\caption{\label{loop} Diagrammatic representation of the DWBA expressed by 
equation~\eqref{eqbaru1}. (a) The primary production term. (b) The loop 
diagram including the nucleon-nucleon FSI. $T(\mbox{k}^{\prime},\mbox{k})$ 
stands for the half-off-shell $T$ matrix with k and $\mbox{k}^{\prime}$ 
denoting the on- and off-shell centre-of-mass momentum in the 
nucleon-nucleon system, respectively. q indicates the momentum of the created 
meson in the reaction centre-of-mass system and p the momentum of the colliding 
nucleons.}
\end{figure}

Generally, the decomposition of the total production amplitude into the 
primary production and the subsequent nucleon-nucleon interaction visualized 
in figure~\ref{loop} is expressed by the formula:
\be
\label{eqbaru1}
M \;=\; M_0^{on} \:+\:  M_0^{off}\,G\,T_{NN},
\ee
where the second term of the equation represents the integration over the 
intermediate ($\mbox{k}^{\prime}$) momenta of the off-shell production 
amplitude and the half-off-shell nucleon-nucleon $T$ matrix~\cite{baru579}.
Assuming that the primary production occurs in such a way that one nucleon 
emits the meson which then re-scatters on the other nucleon and appears as a 
real particle the authors of reference~\cite{baru579} found that the 
enhancement of the cross section due to the nucleon-nucleon interaction 
depends strongly on the mass of the created meson. 
This is because with the increasing mass of the produced meson the distance 
probed by the nucleon-nucleon interaction decreases (see 
table~\ref{momtranstable}) and hence the relevant range of the off-shell 
momenta becomes larger.
The effect for the $pp \rightarrow pp\,Meson$ reactions is presented in 
figure~\ref{baruFSI}a, where one can see that, when utilizing the Bonn 
potential model for the nucleon-nucleon $T$ matrix, the enhancement in case 
of the $\pi^0$ production is by about a factor of four larger than for the 
$\eta$ or $\eta^{\prime}$ mesons.
A similar conclusion, but with the absolute values larger by about $40\,\%$, 
was drawn for the Paris $NN$ potential~\cite{baru579}.
\begin{figure}[H]
\hspace{-0.2cm}
\parbox{0.29\textwidth}
  {\epsfig{file=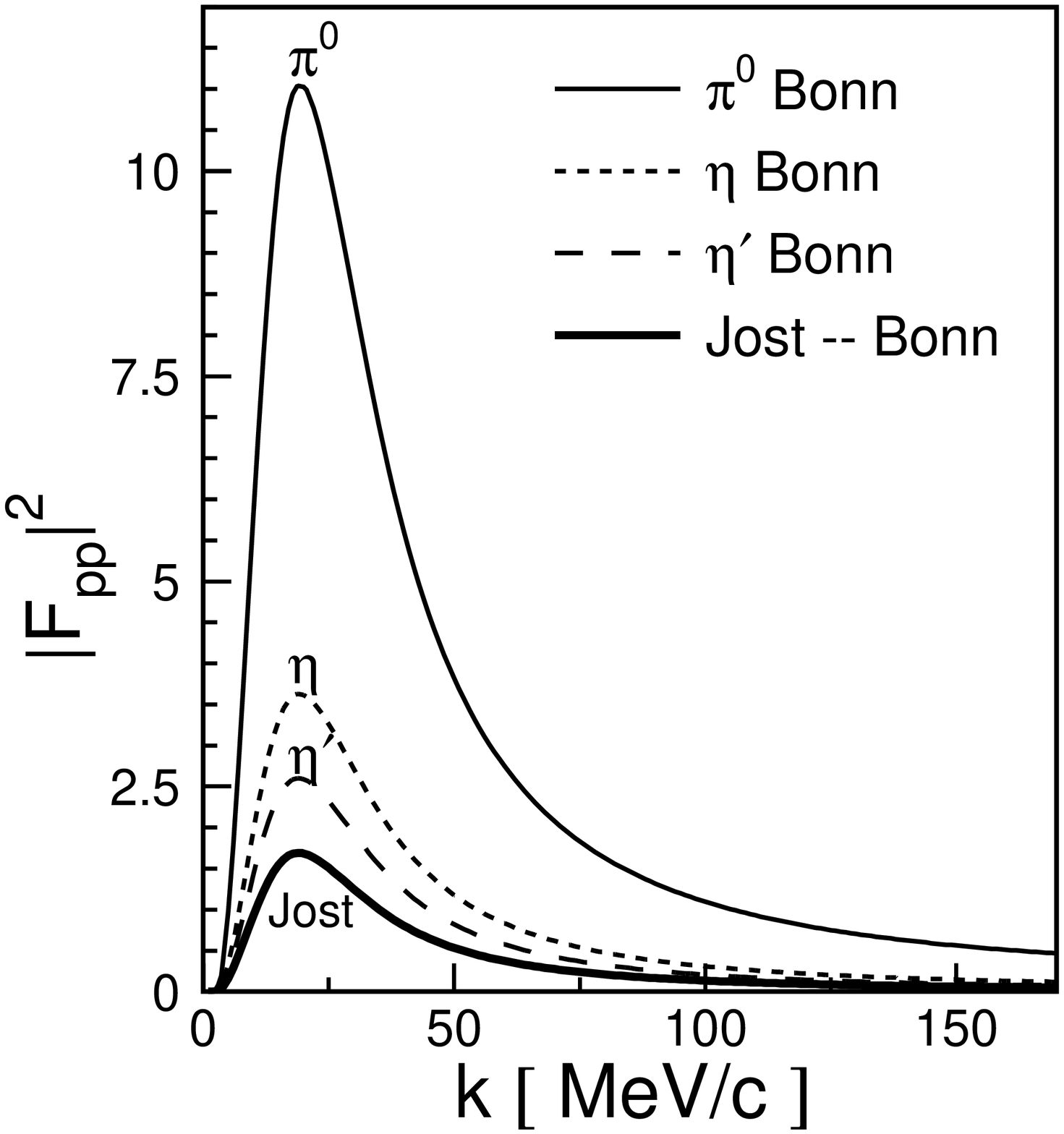,width=0.385\textwidth}} \hfill
\parbox{0.29\textwidth}
  {\hspace{-0.5cm}\epsfig{file=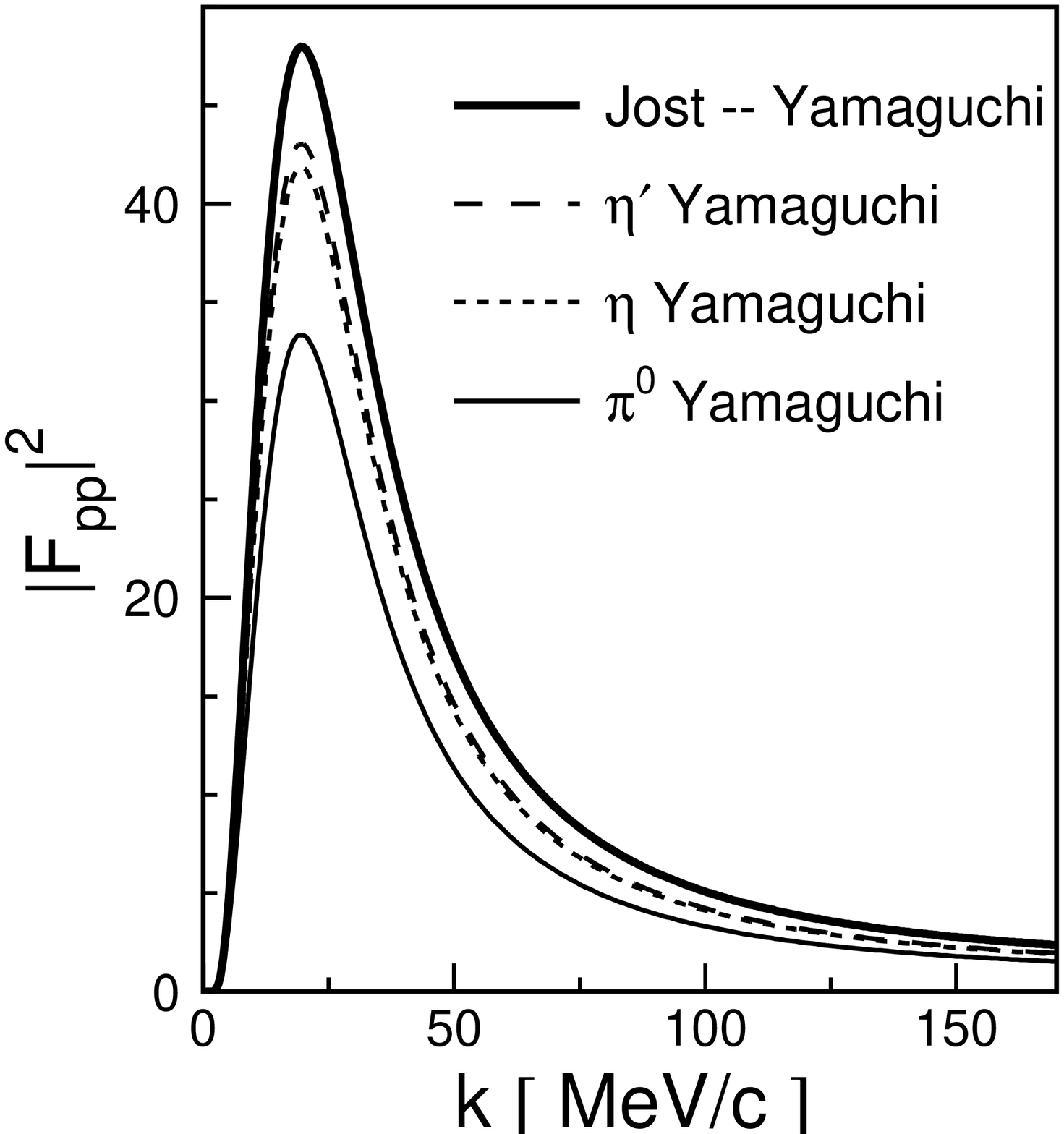,width=0.385\textwidth}} \hfill
\parbox{0.32\textwidth}
  {\hspace{-0.5cm}\epsfig{file=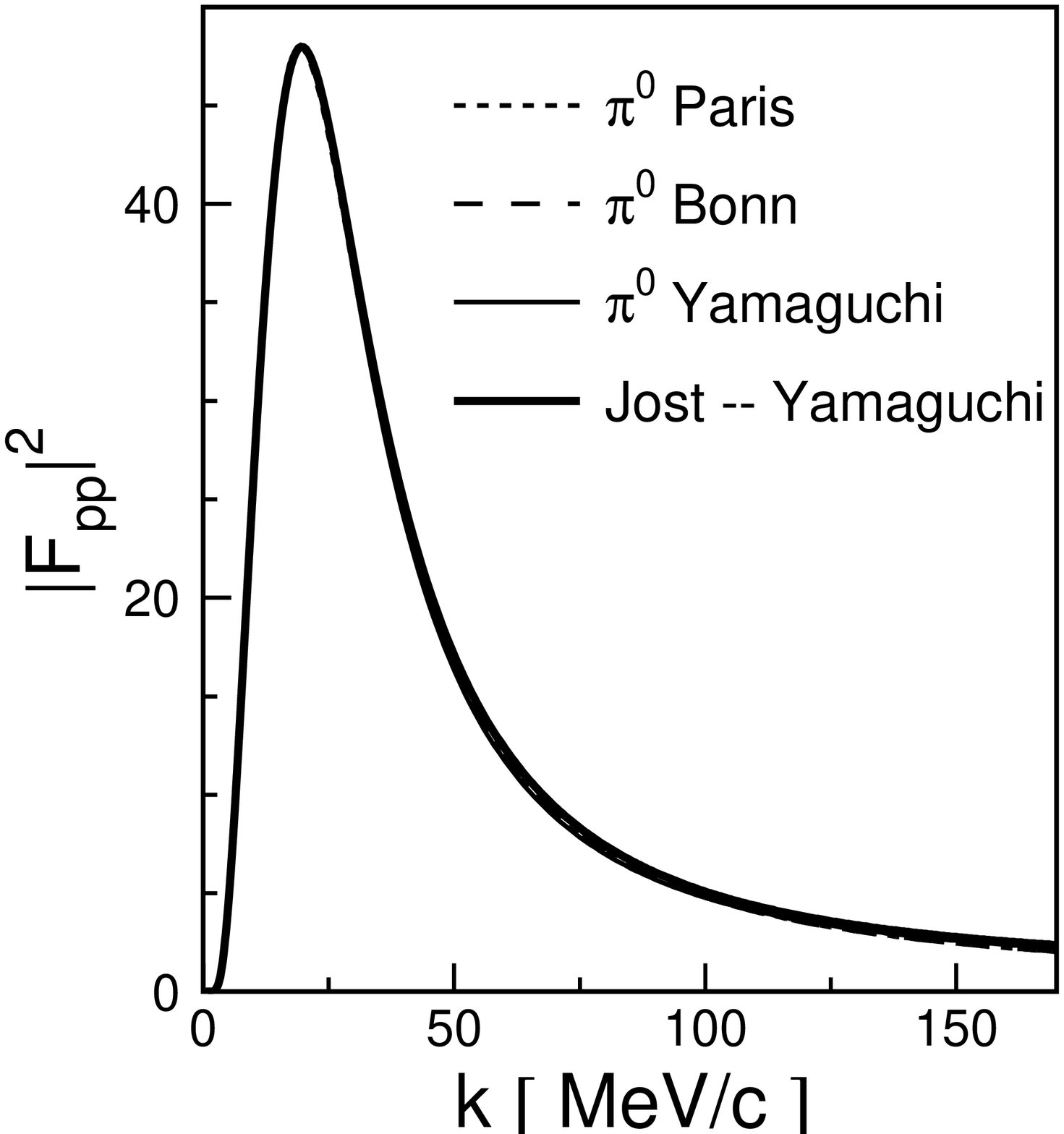,width=0.385\textwidth}} \hfill
\parbox{0.001\textwidth}{\mbox{}}
\parbox{0.31\textwidth}{\raisebox{0ex}[0ex][0ex]{\mbox{}}} \hfill
\parbox{0.31\textwidth}{\raisebox{0ex}[0ex][0ex]{\large a)}} \hfill
\parbox{0.325\textwidth}{\raisebox{0ex}[0ex][0ex]{\large b)}} \hfill
\parbox{0.025\textwidth}{\raisebox{0ex}[0ex][0ex]{\large c)}}
\caption{\label{baruFSI} The FSI factor for Bonn~\cite{haidenbauer2190} a) and 
Yamaguchi~\cite{haeringen355} b) potentials. The solid, dotted and dashed 
lines correspond to $\pi^0$, $\eta$, and $\eta^{\prime}$ meson production, 
respectively. Thick solid lines indicate the inverse of the squared Jost 
function ($|F_{pp}(\mbox{k})|^2 = |J(-\mbox{k})|^{-2}$). 
c) The FSI factors for $\pi^0$ production of Paris (dotted 
curve)~\cite{lacombe861}, Bonn (dashed curve), and Yamaguchi (solid curve) 
potentials normalized to be equal at maximum to the inverse of the squared 
Jost function of the Yamaguchi potential (thick solid line). The shapes 
stemming from different potentials are almost indistinguishable.
Note that the thick solid line corresponds to the dashed line in 
figure~\ref{Mpppp_cross_pi}a. The figures have been adapted from 
reference~\cite{baru579}.}
\end{figure}
On the  contrary, when applying the Yamaguchi potential into calculations the 
enhancement grows with the increasing mass of the meson, as shown in 
figure~\ref{baruFSI}b. 
The thick solid curves in figures~\ref{baruFSI}a and~\ref{baruFSI}b show the 
results of the frequently applied approximation of the nucleon-nucleon FSI 
effects:
\be
M \;=\; M_0^{on} \:+\:  M_0^{off}\,G\,T_{NN} \;\approx\; 
        M_0^{on} \cdot (1 \:+\: G\,T_{NN}) \;=\; 
\nn
\ee
\be
\label{eqbaru2}
 = M_0^{on}\,J^{-1}(-\mbox{k})
  \;\equiv\; M_0^{on}\,F_{NN}(k) ,
\ee
where the overall transition matrix element $M$ is factorized to the primary 
on-shell production and the $NN$ FSI  expressed as the inverse of the Jost 
function $J^{-1}(-\mbox{k})$~\cite{goldbergerwatson}.
As can be seen in figures~\ref{baruFSI}a and~\ref{baruFSI}b, the variation of 
the absolute values --~of such obtained enhancement factors~-- with the 
applied potential is significant. \\
Since the physical value of the total cross section cannot depend on the 
off-shell features of the potential used in calculations, which are in 
principle not measurable~\cite{fearing758}, the differences in the magnitude 
of the $|F_{pp}|$ factor must reflect itself in a corresponding dependence of 
the primary production amplitude on the potential used. 
Therefore, it is of vital importance to realize that the values of the 
threshold amplitudes $|M_0|$ are significant only in the context of the 
potential they were extracted from.
Figure~\ref{baruFSI}c demonstrates, however, that the shapes of the 
enhancement factors, with the meson exchange mechanism assumed for the primary 
production, are pretty much the same, independently of the applied $NN$ 
potential and correspond to the form of the Jost function inferred from the 
Yamaguchi potential.
This indicates that the energy dependence of the $NN$ FSI factors is 
predominantly determined by the on-shell $NN$ $T$ matrix.

In references~\cite{hanhart176,nak01,kleefeld51} the formula for the transition 
matrix element with explicit dependence on the considered off-shell features 
for the initial and final state interaction is derived:
{\small{
\be
M \:=\: \left\{1 \,\!\! + \,\!\! \frac{
  \left[\eta(\mbox{k}) e^{2i\delta(\mbox{\scriptsize k})} \,-\, 1 \right] 
  \cdot \left[ 1 \,+\, P_f(\mbox{p},\mbox{k})\right]}{2} \right\}  
  \: M_0 \: \left\{1 \!\!\,+\,\!\! \frac{
  \left[\eta(\mbox{p}) e^{2i\delta(\mbox{\scriptsize p})} \,-\, 1 \right]
  \cdot \left[ 1 \,+\, P_i(\mbox{p},\mbox{k})\right]}{2} \right\},
\ee
}}
\hspace{-0.2cm}where subscripts $i$ and $f$ indicate the initial and final state, 
respectively. 
The functions $P(\mbox{p},\mbox{k})$ exhibit all the off-shell effects of the 
$NN$ interaction and the primary production current~\cite{nak01} and $\delta$ 
and $\eta$ denote the phase-shift and inelasticity, correspondingly.
At threshold, the inelasticity in the exit channel is equal to unity 
($\eta(\mbox{k}) = 1$) due to the small relative momentum of the outgoing 
nucleons.
The last term of the formula expresses the influence of the initial state 
interaction on the production process. 
Due to the large relative momenta of the colliding protons needed to create a 
meson it is characterized by a weak energy dependence in the excess energy 
range of a few tens of MeV.
For example in figure~\ref{1s03p0}b one can see that the phase-shift 
variation of the $^3\mbox{P}_0$ partial wave (having predominant contribution 
to the threshold production of pseudoscalar mesons) in the vicinity of 
the threshold for mesons heavier than $\pi^0$ is indeed very weak.
Taking additionally into account that the initial state off-shell function 
$P_i(\mbox{p},\mbox{k})$ is small (as it is the case at least for meson exchange 
models~\cite{hanhart176}) one reduces the influence of the $NN$ initial state 
interaction to the reduction factor $F_{ISI}$ which can be estimated from the 
phase-shifts and inelasticities only:
\be
\label{F_ISI}
F_{ISI} \;=\; \frac{1}{4} \:\left| 
  \,\eta(\mbox{p}) \,e^{2i\delta(\mbox{\scriptsize p})} \,+\, 1 \:\right|^2. 
\ee

At the threshold for $\pi$ meson production this is close to unity since at 
this energy the inelasticity is still nearly 1 and the $^3\mbox{P}_0$ 
phase-shift is close to zero (see figure~\ref{1s03p0}b). 
However, at the $\eta$ threshold, where the phase-shift approaches its 
minimum, the proton-proton ISI diminishes the total cross section already by 
a factor of 0.2~\cite{hanhart176}. 
A similar result was obtained using a meson exchange model for $\eta$ 
production in the $pp \rightarrow pp \eta$ reaction and calculating the 
proton-proton distortion from the coupled-channel $\pi NN$ 
model~\cite{batinic321}.
The authors of reference~\cite{batinic321} concluded that the initial 
proton-proton distortion reduces the total cross section by about a factor of 
$\approx 0.26$, which keeps constant at least in the studied range of 
$100\,\mbox{MeV}$ in kinetic beam energy.
Hence, the closed formula~\eqref{F_ISI} disregarding the off-shell effects 
($P_i(\mbox{p},\mbox{k})$) permits to estimate the cross section reduction due 
to the initial state distortion with an accuracy of about $25\,\%$. 
The shape of the $^3\mbox{P}_0$ phase-shift shown in 
figure~\ref{1s03p0}b indicates that the effect is at most pronounced 
close to the $\eta$ production threshold, yet for the $\eta^{\prime}$ meson 
the formula~\eqref{F_ISI} leads to a factor $F_{ISI} = 
0.33$~\cite{nakayama024001}.
The primary production amplitude as well as the off-shell effects of the 
nucleon-nucleon FSI ($P_{f}(\mbox{p},\mbox{k})$) are also weakly energy 
dependent~\cite{nak01}, since they account for the short range creation 
mechanism, which shall be considered in chapter~\ref{Dopsmp}.

\section{Qualitative comparison between $pp\eta$, $pp\eta^{\prime}$, and $pp\pi^{0}$
         interactions}
\label{qualitativecomparison}
\begin{flushright}
\parbox{0.73\textwidth}{
 {\em
 But what we must aim at is not so much to ascertain 
 resemblances and differences, as to discover similarities
 hidden under apparent discrepancies~\cite{poincare}.\\
 }
 \protect \mbox{} \hfill  Henri Poincar$\acute{\mbox{e}}$ \protect\\
 }
\end{flushright}
The accordance of the experimental data with the simple factorization 
represented by solid lines in figures~\ref{Mpppp_cross_pi}b 
and~\ref{cross_eta_etap} fully confirms the suppositions considered in the previous sections
which imply that the energy dependence of the total cross section  
is in the first order determined by 
the on-shell scattering of the outgoing particles. 
However, since the distortion caused by the nucleons is by some orders of 
magnitude larger than that resulting from the meson-nucleon interaction, even small 
fractional inaccuracies in the description of nucleon-nucleon effects may 
obscure the inference of the meson-nucleon interaction.
The differences between the square of the on-shell proton-proton scattering 
amplitude and the Jost function prescription are presented in 
figure~\ref{Mpppp_cross_pi}a.
To minimize the ambiguities that may result from these discrepancies at least 
for the quantitative estimation of the effects of the unknown meson-nucleon 
interaction one can compare the spectra from the production of the meson under investigation to 
the spectra determined for the production of a meson whose interaction with 
nucleons is well established.
For instance, to estimate the strength of the $\eta pp$ and $\eta^{\prime} pp$ 
FSI one can compare the appropriate observables to those of the $\pi^0 pp$ 
system.
\vspace{-0.2cm}
\begin{figure}[H]
    \parbox{0.54\textwidth}{ 
       \includegraphics[width=0.54\textwidth]{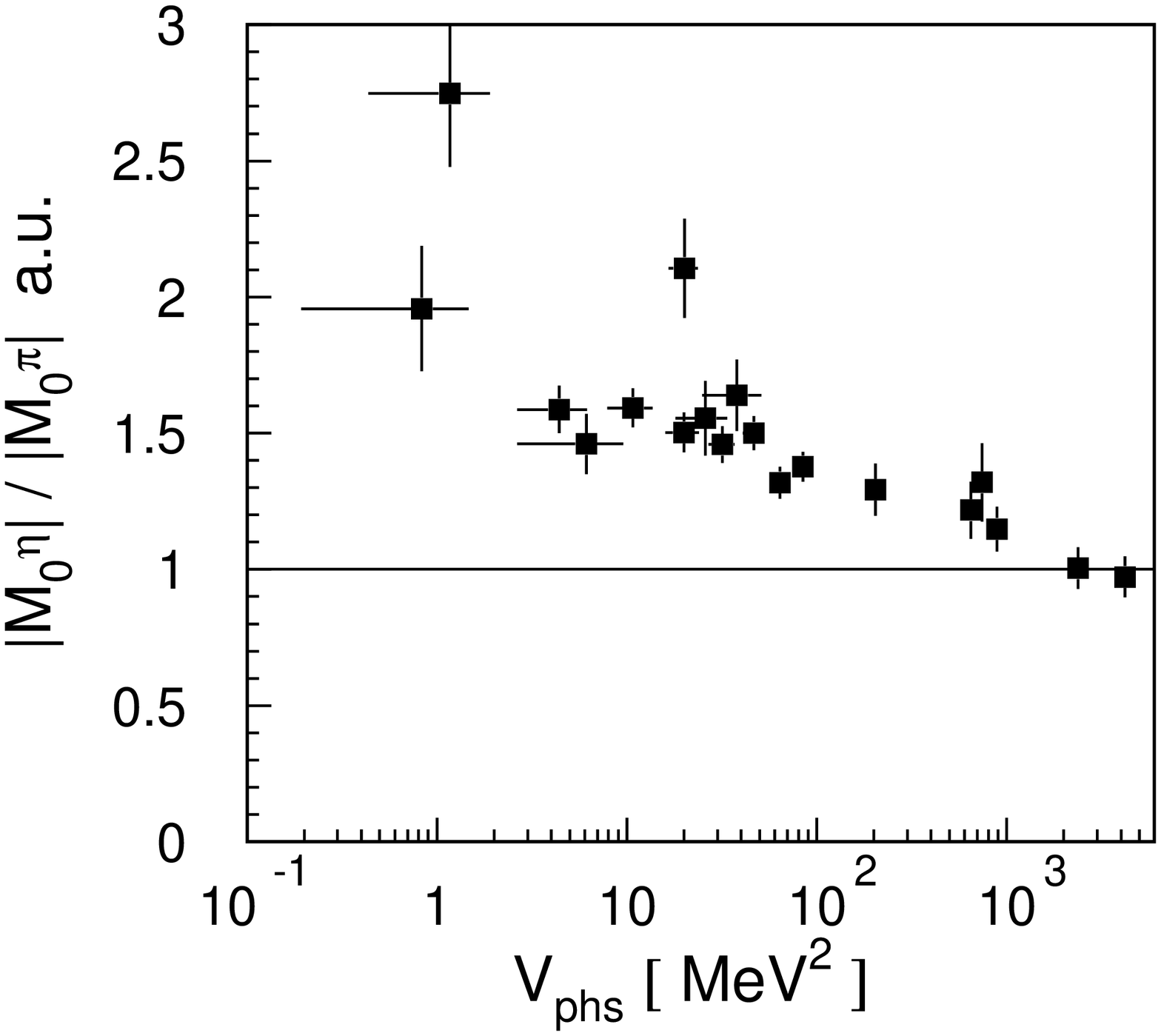}

       \vspace{-0.8cm}
       \includegraphics[width=0.54\textwidth]{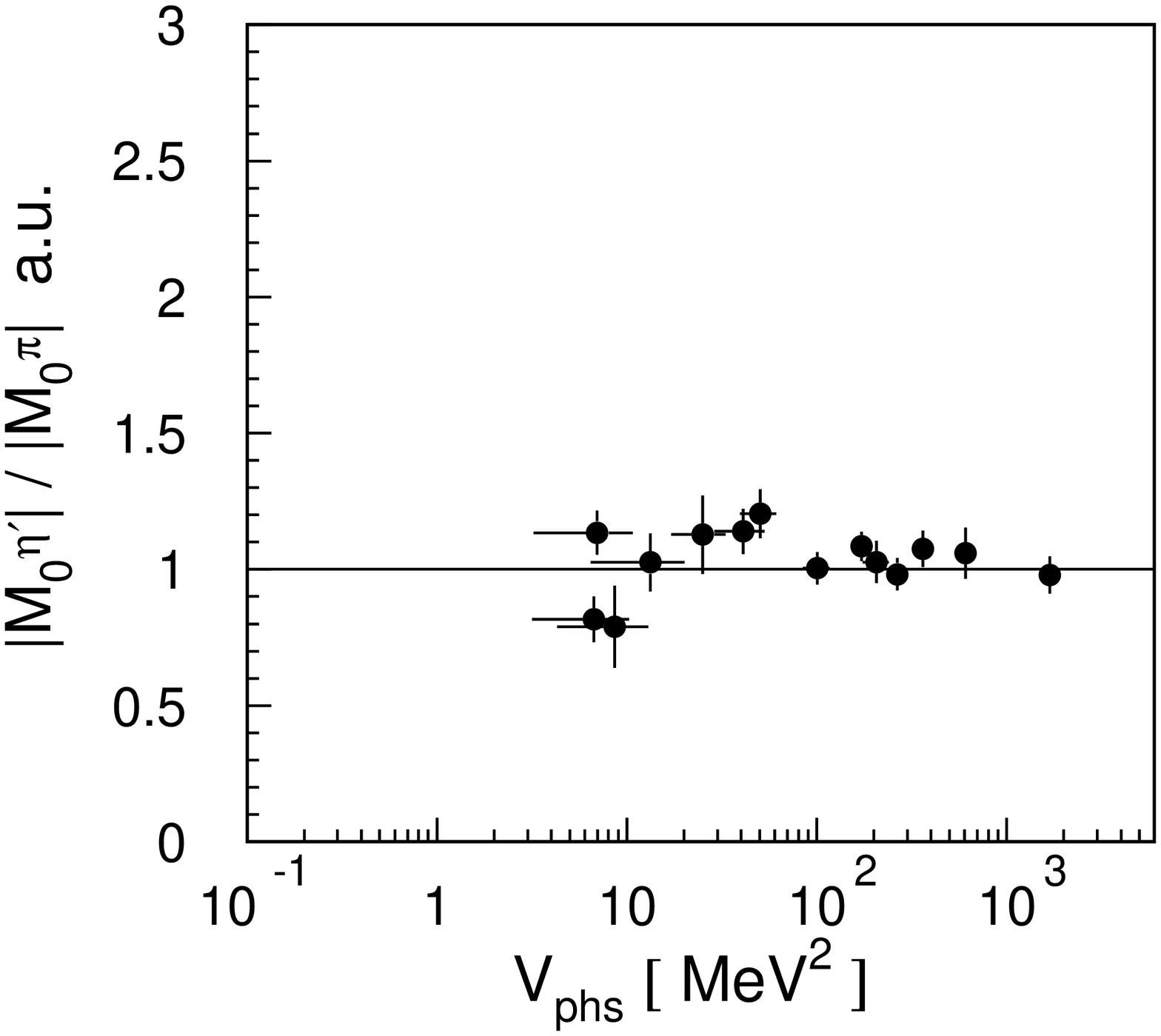}
    } \hfill
    \parbox{0.4\textwidth}{
        \caption{\label{mmnisk}\label{ratio_and_factors} The ratios of \hspace{1ex}
        $|M^{\eta}_0|/|M^{\pi^0}_0|$ (upper panel) and \hspace{1ex} 
        $|M^{\eta^{\prime}}_0|/|M^{\pi^0}_0|$ (lower panel)
        extracted from the data by means of 
        equation~\eqref{cross_with_FSI}, assuming a $pp$-FSI enhancement factor as 
        depicted by the dotted line in figure~\ref{Mpppp_cross_pi}a and neglecting 
        the proton-meson interaction~\cite{swave}.
    }}
\end{figure}
\vspace{-1.0cm}
Upper and lower panels of figure~\ref{ratio_and_factors} show the 
dependence of $|M_0|$ on the phase space volume for $\eta$ and 
$\eta^{\prime}$ production normalized to $|M_0^{\pi^0}|$.
The values of $|M_0|$ were extracted from the experimental data by means of 
equation~\eqref{cross_with_FSI} disregarding the proton-meson interaction 
($|M_{p \eta (\eta^{\prime}) \rightarrow p \eta (\eta^{\prime})}|$ was set 
to 1).
If the influence of the neglected interactions were the same in the cases of 
the $\eta (\eta^{\prime})$ and $\pi^0$ production
the points would be consistent with the 
solid line. 
This is the case for the $pp \rightarrow pp \eta^{\prime}$ reaction 
visualizing the weakness of the proton-$\eta^{\prime}$ interaction 
independently of the prescription used for the proton-proton FSI~\cite{swave}.
In case of the $\eta^{\prime}$ meson its low-energy interaction with the 
nucleons was expected to be very weak since there exists no baryonic resonance 
which would decay into $N \eta^{\prime}$~\cite{PDG}. 
Figure~\ref{ratio_and_factors}a shows --~independently of the model used for 
the correction of the proton-proton FSI~-- the strong effects of the 
$\eta pp$ FSI at low $V_{ps}$.

  \section{Dalitz plot occupation for the pp$\eta$ system} 
   \label{dalitzsection}
\vspace{-0.0cm}
\begin{flushright}
\parbox{0.67\textwidth}{
 {\em
   ...\! a knowledge of the effects is what leads to an investigation and discovery of causes~\cite{galileo}.\\
 }
 \protect \mbox{} \hfill  Galileo Galilei \protect\\
 }
\end{flushright}
\vspace{-0.0cm}

The strength of the interaction between particles depends on their relative momenta
or equivalently on the invariant masses of the two-particle subsystems.
Therefore it should show up as  a modification
of the phase space abundance in the kinematical regions where the outgoing
particles possess small relative velocities.
Only two  invariant masses of the three subsystems are independent (see eq. \ref{scalarproduct})
and therefore the entire  accessible information about the final
state interaction of the three-particle system can be presented in the
form of the Dalitz plot. Figure~\ref{dalitze}a
shows distribution of the events
as determined experimentally for the $pp\eta$
system at an excess energy of Q~=~15.5~MeV. 
This distribution originating from kinematically complete measurements 
comprises the whole experimentally available information about the interactions 
of the $pp\eta$-system. 
In this figure one easily recognizes
the growth of the population density at the region where the protons
have small relative momenta which can be assigned to the strong attractive
S-wave interaction between the two protons. This is qualitatively in agreement with the
expectation presented in figure~\ref{dalitze}b, which shows the result
of Monte-Carlo calculations  where the homogeneously populated
phase space was weighted by the square of the on-shell $^1S_{0}$
proton-proton scattering amplitude. However, already in this two dimensional
representation it is visible that
the experimentally determined distribution remains rather homogeneous
outside the region of the small proton-proton invariant masses,
whereas the simulated abundance decreases gradually with growing s$_{pp}$
(as indicated by the arrow).
Figure~\ref{dalitze}c shows the simulated  phase space density
distribution
disregarding the proton-proton interaction but accounting for the interaction
between the $\eta$-meson and the proton.
At  this excess  energy, corresponding  
to the small relative  momentum range ($\mbox{k}_{p \eta}^{max} \approx 105\,\mbox{MeV/c}$)
the variations of the 
proton-$\eta$ scattering amplitude are quite moderate 
(see figure~\ref{factors}b in the next section).
  \begin{figure}[H]
    \parbox{0.5\textwidth}{ \vspace{-1.cm}
    \includegraphics[width=0.57\textwidth]{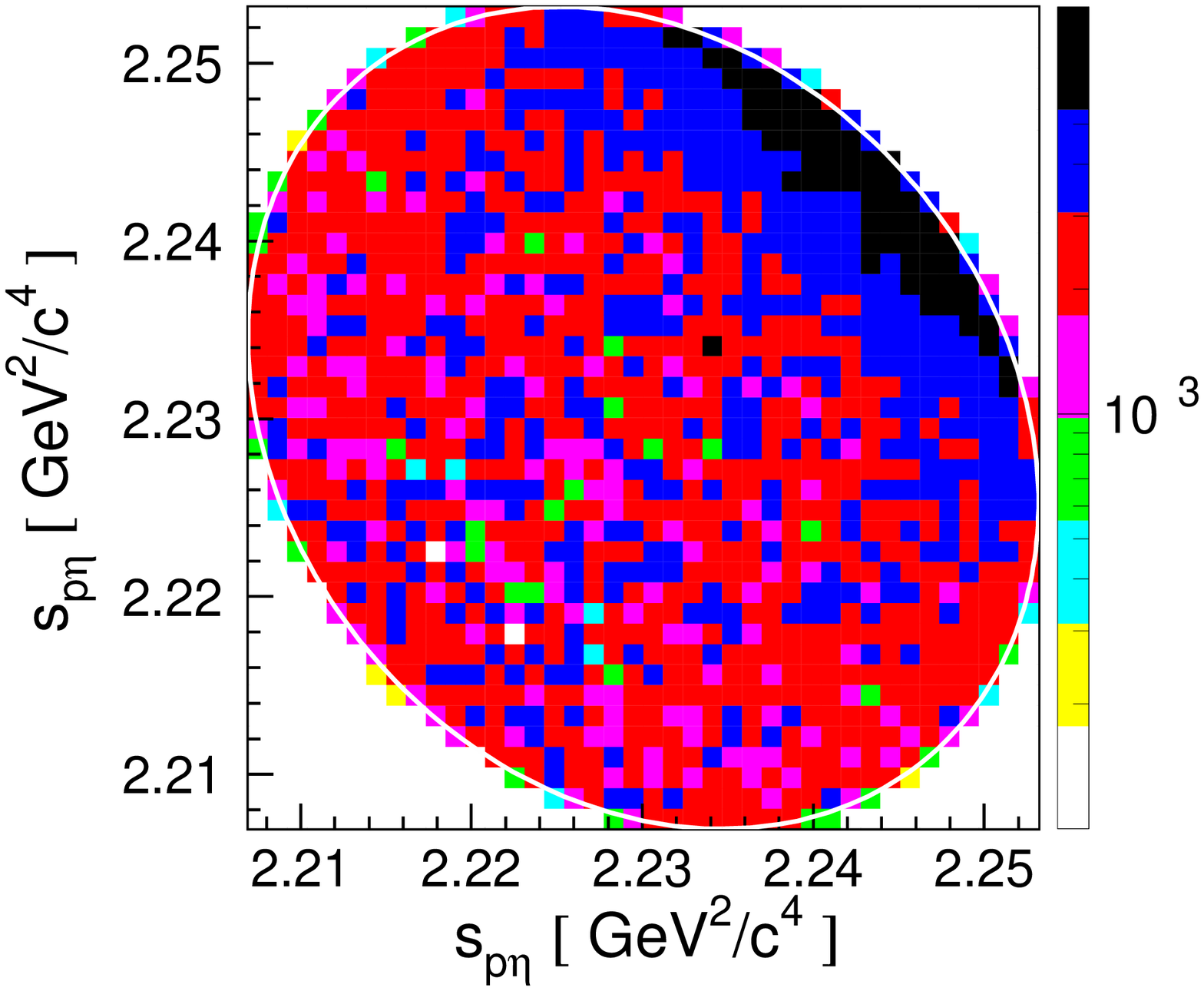}}
    \parbox{0.5\textwidth}{\vspace{-1.cm}
    \includegraphics[width=0.57\textwidth]{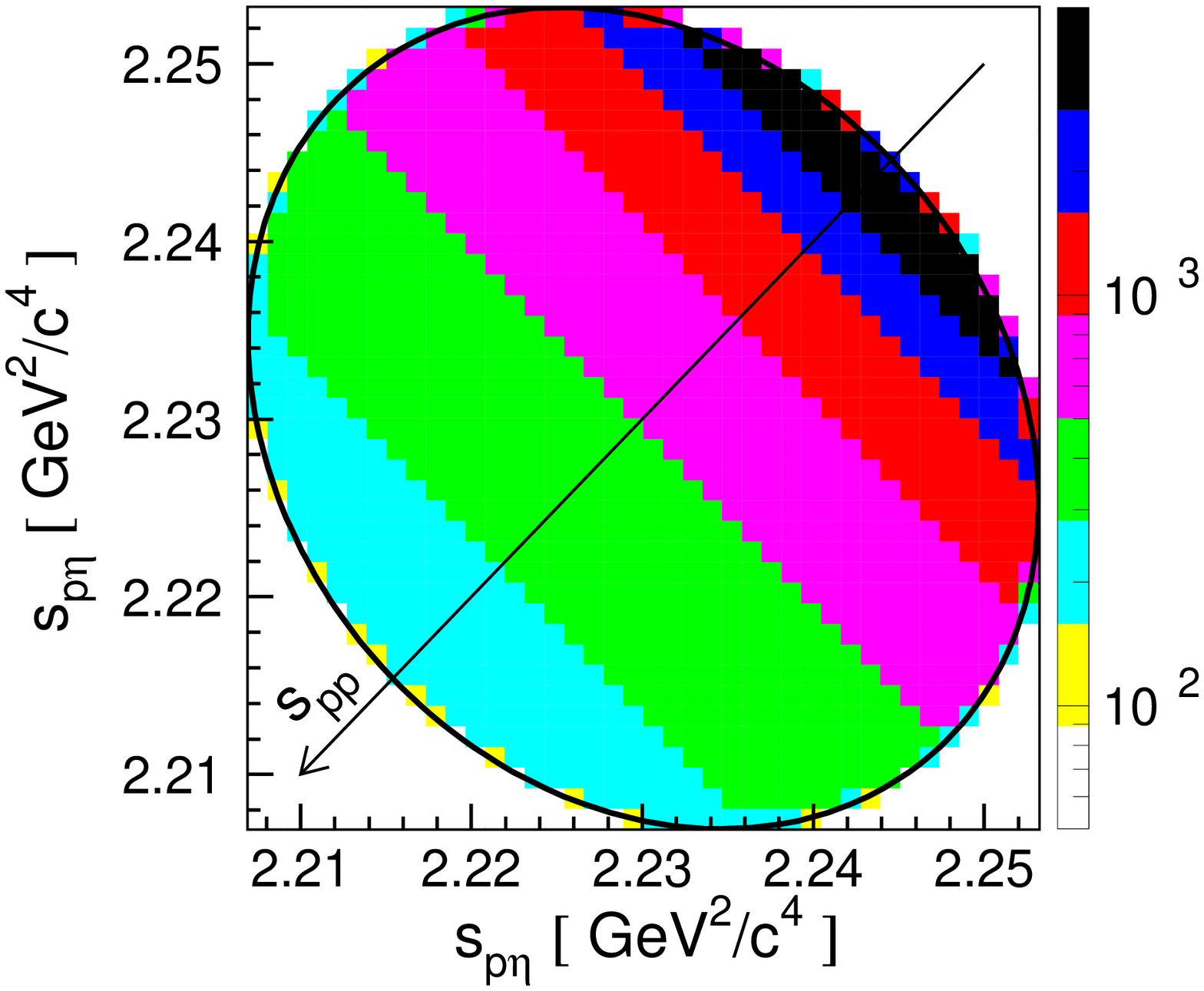}}
 
 \vspace{-1cm}
 \parbox{0.44\textwidth}{\raisebox{0ex}[0ex][0ex]{\mbox{}}} \hfill
 \parbox{0.48\textwidth}{\raisebox{0ex}[0ex][0ex]{\large a)}} \hfill
 \parbox{0.04\textwidth}{\raisebox{0ex}[0ex][0ex]{\large b)}}
    \parbox{0.5\textwidth}{\vspace{-0.2cm}
    \includegraphics[width=0.57\textwidth]{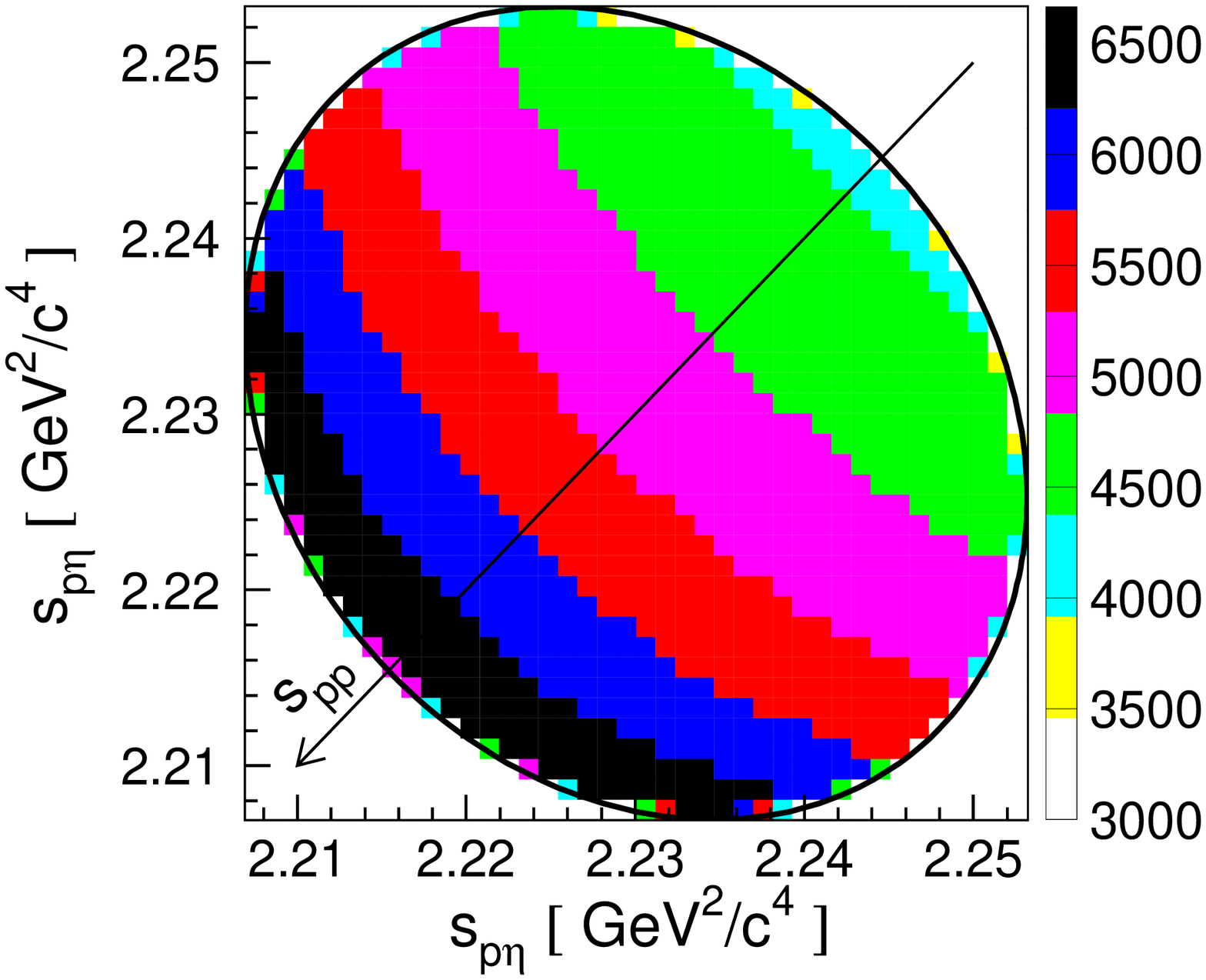}}\hfill
    \parbox{0.45\textwidth}{\vspace{-0.2cm}
    \includegraphics[width=0.43\textwidth]{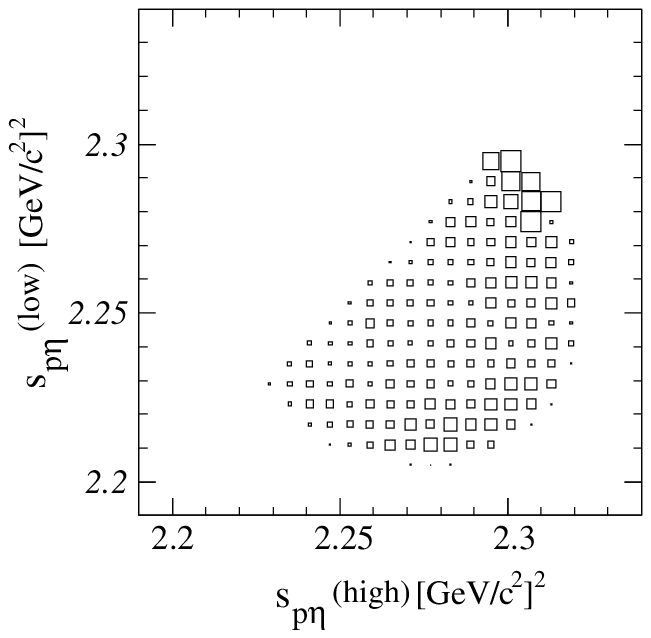}}
 
 \vspace{-1cm}
 \parbox{0.44\textwidth}{\raisebox{0ex}[0ex][0ex]{\mbox{}}} \hfill
 \parbox{0.48\textwidth}{\raisebox{0ex}[0ex][0ex]{\large c)}} \hfill
 \parbox{0.04\textwidth}{\raisebox{0ex}[0ex][0ex]{\large d)}}
    \vspace{0.2cm}
    \caption{\label{dalitze} Dalitz plots. 
      Lines surrounding the plots show the kinematical limits.\protect\\
     {\bf (a)} Experimental result
       determined for the $pp\to pp\eta$ reaction at Q~=~15.5~MeV.
      Data were corrected for the detection acceptance and efficiency. \protect\\
     {\bf (b)}
      Monte-Carlo simulations for the $pp \to pp\eta$ reaction at Q~=~15.5~MeV:
      Phase-space density distribution modified by the
      $^1S_{0}$ proton~-~proton final state interaction. \protect\\
     {\bf (c)}
     Simulated phase space density distribution modified by the proton-$\eta$ interaction
      with a scattering length of $a_{p\eta}$~=~0.7~fm~+~$i~0.3$~fm.
      Details of the calculations together with the discussion of the nucleon-nucleon
      and nucleon-meson final state interaction can be found in reference~\cite{review}.
      The scale in the figure is linear in contrast to the above panels.\protect\\
      {\bf (d)} 
          Dalitz plot distribution of the $pp \rightarrow pp\eta$ reaction at an
          excess energy of $\mbox{Q} = 37.6\,\mbox{MeV}$, corrected for the detection
          acceptance. Out of the two invariant masses corresponding to two $p-\eta$
          pairs the one being larger is plotted along the x-axis. The figure is taken
          from reference~\cite{calen39}. A similar spectrum for the $pp \pi^0$
          system~\cite{bondar8} reveals the influence of the proton-proton interaction,
          yet again, as for the total cross section energy dependence
         (fig.~\ref{Mpppp_cross_pi}b), the proton-$\pi^0$ interaction is too weak to
         affect the density distribution of the Dalitz plot observably.
    }
  \end{figure}
Due to the low strength of this interaction the expected deviations from a
uniform distribution is smaller --~from that caused by the proton-proton force~--
by about two orders of magnitude, yet
an enhancement of the density in the range of low invariant
masses of proton-$\eta$ subsystems is clearly visible. 
Note that the scale
in the lower panel of figure~\ref{dalitze} is linear whereas in the upper panel
it is logarithmic.
Due to weak variations of the proton-$\eta$ scattering amplitude the
enhancement originating from the $\eta$-meson interaction with one proton is
not separated from the $\eta$-meson interaction with the second proton.
Therefore an overlapping of broad structures occurs.
It is observed that the occupation density grows slowly with increasing
$\mbox{s}_{pp}$, opposite to the effects caused by the S-wave proton-proton
interaction, but similar to the modifications expected for the NN P-wave~\cite{dyringPHD}.
From the above example it is obvious that only
from experiments with high statistics,
signals of the meson-nucleon interaction can be observed  on top of  the overwhelming
nucleon-nucleon final state interaction.
The increase of the distribution 
density at regions of small invariant masses of the proton-proton and 
proton-$\eta$ subsystems is also visible in figure~\ref{dalitze}d showing
experimental results determined at $\mbox{Q} = 37.6\,\mbox{MeV}$ by the
WASA/PROMICE collaboration. 
At this excess energy these regions are quite 
well separated. 
However, since this is close to the energy where the advent of higher partial 
waves is awaited, the possible contribution from the P-wave proton-proton 
interaction cannot be a priori excluded. 
Specifically, the latter leads to the enhancement at large invariant masses of 
the proton-proton pair~\cite{dyringPHD} and hence affects the phase space 
region where the modification from the proton-$\eta$ interaction 
is expected.
 A deviation of the experimentally observed  population of the phase space
 from the expectation based on the mentioned assumptions
 is even better visible in figure~\ref{petaspp}.
 This figure presents the  projection
 of the phase space distribution onto the $s_{pp}$ axis corresponding
 to the axis indicated
 by the arrows in figures~\ref{dalitze}b and~\ref{dalitze}c.
 The superimposed lines in figure~\ref{petaspp} correspond to the calculations
 performed under the assumption that the production amplitude can be factorized
 into a primary production and final state interaction.
 The dotted lines result from calculations where only the proton-proton FSI was taken into account,
 whereas the  thick-solid lines represent results where the overall enhancement
 was factorized into the corresponding pair interactions of the $pp\eta$ system.
 This factorisation Ansatz is only valid if the different amplitudes
 are completely decoupled which is not the case here.
 Therefore,
 these calculations should be considered as a rough estimate of the effect introduced
 by the FSI in the different two body systems.
 As introduced in section~\ref{influencesection}, the enhancement factor accounting
 for the proton-proton FSI has been  calculated~\cite{swave,review}
 as the square of the on-shell proton-proton scattering amplitude
 derived according to the modified Cini-Fubini-Stanghellini  formula including
 the Wong-Noyes Coulomb corrections~\cite{noyes995}.
 The homogeneous phase space distributions (thin solid lines)
 deviate strongly from
 the experimentally determined spectra.
 The curves including the  proton-proton and proton-$\eta$
 FSI reflect the
 shape of the data for small invariant masses of the proton-proton system,
 yet they  deviate significantly for large $s_{pp}$ and small $s_{p\eta}$ values.
  An explanation of this discrepancy
  could be  a
  contribution of P-wave proton-proton
  interaction~\cite{nakayama0302061},
  or a
  possibly inadequate assumption
  that proton-$\eta$ and proton-proton interaction modify the
  phase space occupations only as
  incoherent weights~\cite{kleefeld51}.
  \vspace{-0.2cm}
  \begin{figure}[H]
    \parbox{0.5\textwidth}{\hspace{-0.7cm}
    \includegraphics[width=0.58\textwidth]{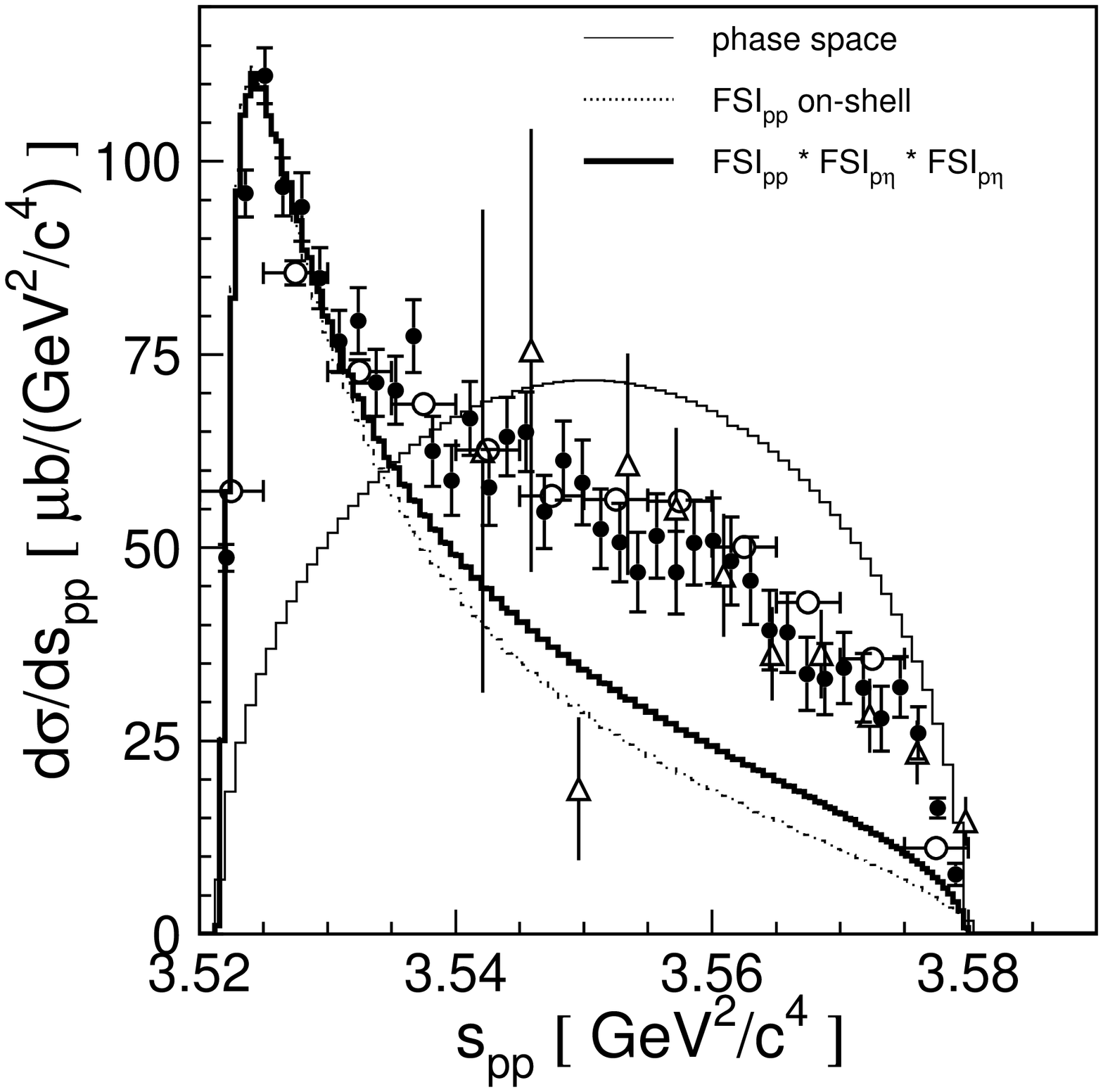}}
    \parbox{0.5\textwidth}{\hspace{-0.7cm}
    \includegraphics[width=0.58\textwidth]{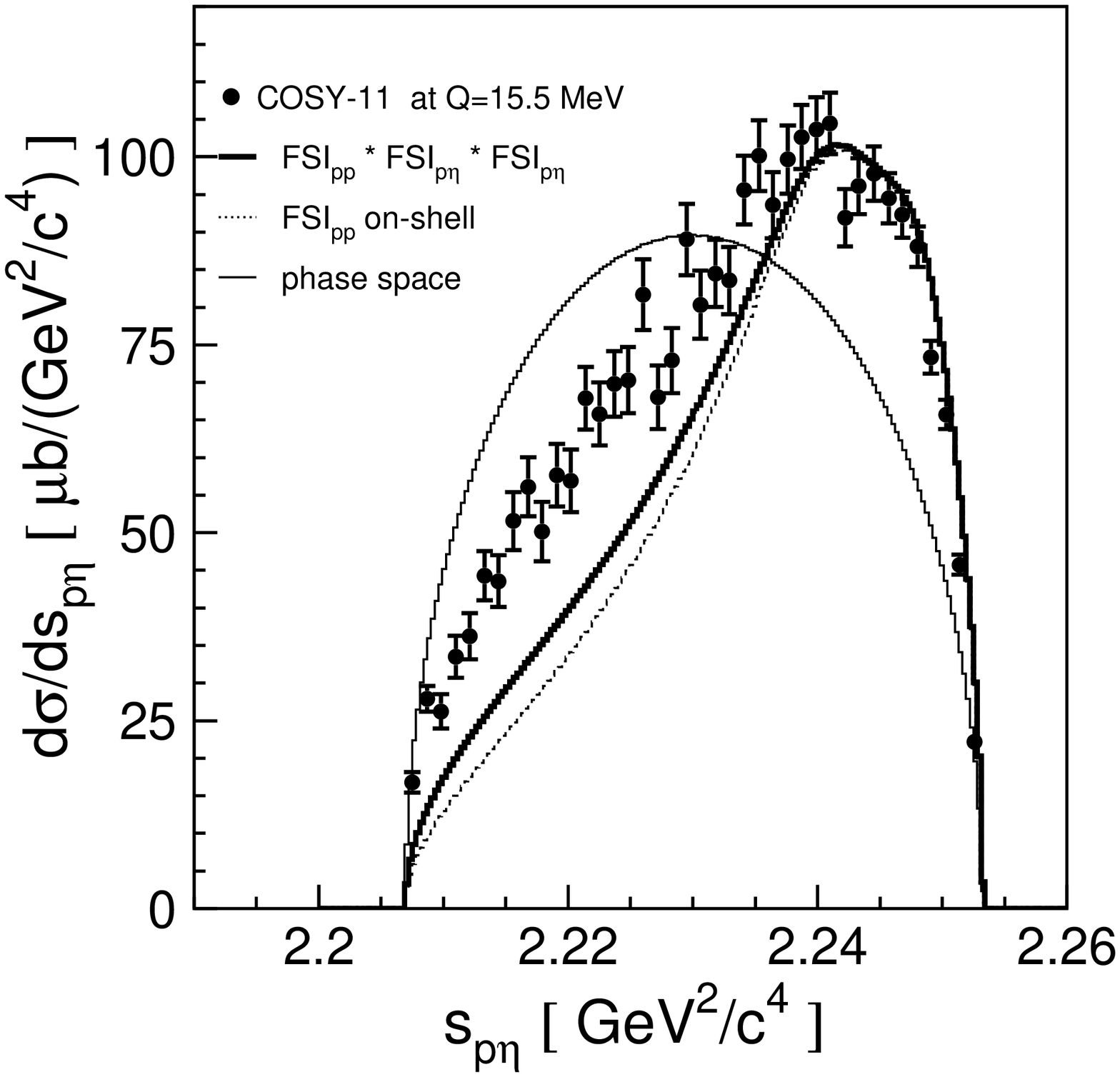}}
    \vspace{-0.6cm}
    \caption{\label{petaspp}
         Distributions of the square of the proton-proton~($s_{pp}$)
          and proton-$\eta$~($s_{p\eta}$)
          invariant masses
          determined experimentally
          for the $pp  \to pp\eta$ reaction at the excess energy of Q~=~15.5~MeV
          by the COSY-11 collaboration (closed circles),
          at Q~=~15~MeV by the TOF collaboration (open circles)~\cite{TOFeta},
          and at Q~=~16~MeV by PROMICE/WASA (open triangles)~\cite{calen190}.
          The TOF and PROMICE/WASA data have been normalized
          to those of  COSY-11, since these  measurements
          did not evaluate the luminosities but rather normalized
          the results to  reference~\cite{calen39}
          (see also comment~\cite{TOTAL}).
    The integrals of the phase space weighted by
    the square of the proton-proton on-shell
    scattering amplitude~FSI$_{pp}$(dotted lines), and by the product of FSI$_{pp}$ and
    the square of the proton-$\eta$ scattering amplitude~(thick solid lines),
    have been normalized arbitrarily at small values of $s_{pp}$.
    The thick solid line was obtained assuming a scattering length of
   $a_{p\eta}$~=~0.7~fm~+~$i$~0.4~fm.
    The expectations under the assumption of the homogeneously populated phase space
    are shown as thin solid curves.
    }
  \end{figure}
  A slightly better description is achieved when
  the  proton-proton interaction is accounted for by the
  realistic nucleon-nucleon potential.
  Figure~\ref{petaspp_kanzo}a depicts the results obtained using
  two different models for the  production process as well as for the
  NN interaction~\cite{vadim024002,vadimmpriv,nakayama0302061}.
 The calculations for the $^3P_{0} \to ^1\!\!S_{0}s$ transition differ slightly,
  but the differences between the models are, by far, smaller than the
  observed signal.
  Therefore we can safely claim that the
  discussed effect is rather too large to be caused
  by the particular assumptions used for the production operator and
  NN potential.
  \vspace{-0.2cm}
  \begin{figure}[H]
    \parbox{0.5\textwidth}{\hspace{-0.7cm}
    \includegraphics[width=0.58\textwidth]{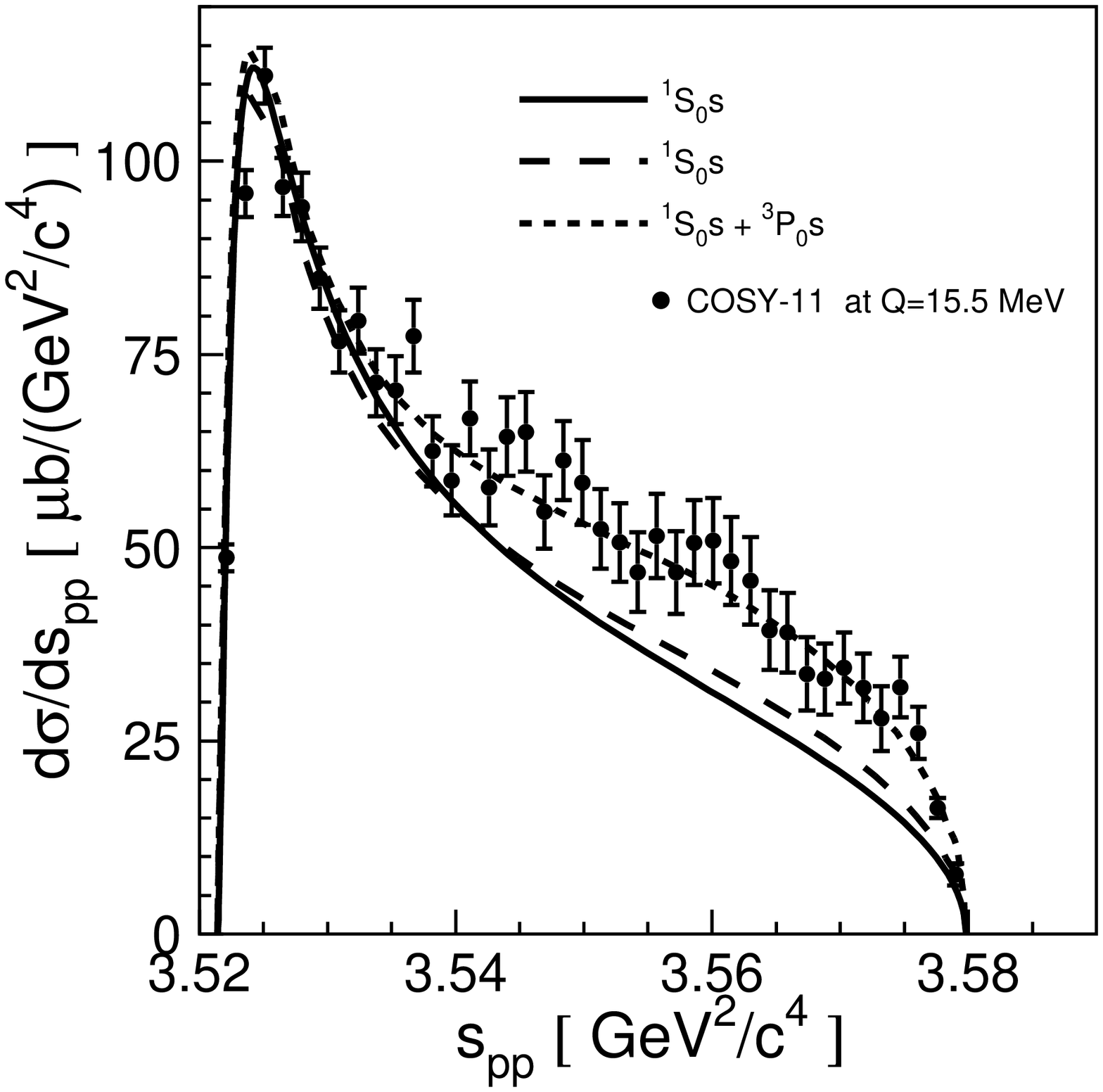}}
    \parbox{0.5\textwidth}{\hspace{-0.7cm}
    \includegraphics[width=0.58\textwidth]{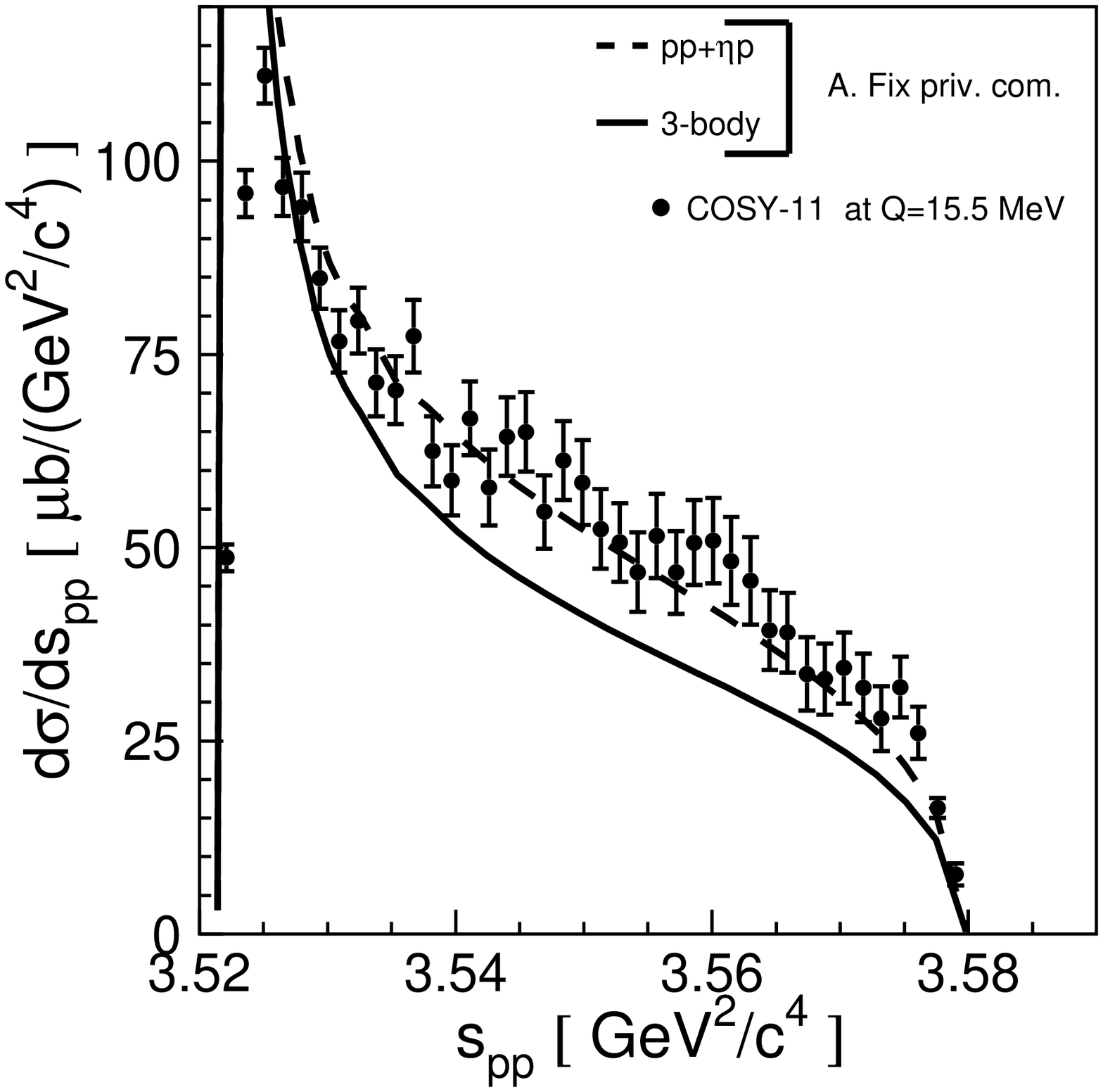}}

    \vspace{-0.7cm}
    \parbox{0.44\textwidth}{\raisebox{0ex}[0ex][0ex]{\mbox{}}} \hfill
    \parbox{0.48\textwidth}{\raisebox{0ex}[0ex][0ex]{\large a)}} \hfill
    \parbox{0.04\textwidth}{\raisebox{0ex}[0ex][0ex]{\large b)}}
    \caption{\label{petaspp_kanzo}
       {\bf (a)}
       Distribution of the square of the proton-proton~($s_{pp}$)
       invariant mass
       for the $pp\to pp\eta$ reaction at an excess energy of Q~=~15.5 MeV.
       Solid and  dashed lines correspond to the calculations under the assumption
       of the $^3P_0\to ^1\!\!S_0s$ transition according to the models described in references~
       \cite{vadim024002,vadimmpriv} and \cite{nakayama0302061}, respectively.
       The dotted curve shows the result with the inclusion
       of the  $^1S_0\to ^3\!\!P_0s$ contribution
       as suggested in reference~\cite{nakayama0302061}. \protect\\
       {\bf (b)} The same data as above but with
       curves denoting preliminary three-body calculations~\cite{Fixprivate,Fixprivate1} of the
       final $pp\eta$ system as described in~\cite{fix277}.
       At present only the dominant transition  $^3P_0\to ^1\!\!S_0s$
       is taken into account and
       the production mechanism is reduced to the excitation of the
       $S_{11}(1535)$ resonance via the exchange of the $\pi$ and $\eta$ mesons.
       The solid line was determined with the rigorous three-body approach~\cite{Fixprivate,Fixprivate1}
       where the proton-proton sector is described in terms of the separable
       Paris potential (PEST3)~\cite{haidenbauer1822}, and for the $\eta$-nucleon scattering
       amplitude an isobar model analogous to the one of
       reference~\cite{benhold625} is used
       with $a_{\eta N}~=~0.5$~fm~$+~i\cdot 0.32$~fm.
       The dashed line is obtained if only pairwise interactions ($pp+p\eta$) are allowed.
       The effect of proton-proton FSI at small $s_{pp}$ is overestimated due to neglect
       of Coulomb repulsion between the protons.
       The lines are normalized arbitrarily but their relative amplitude is fixed from the model.
    }
  \end{figure}
  \vspace{-0.0cm}
  As we will see in subsection~\ref{angularsection}  in figure~\ref{costhetaetatof}b,
  the experimental distribution of the $\eta$ polar angle in the center-of-mass frame
  is fully isotropic.
  This is the next evidence --~besides the shape of the excitation function
  and the kinematical arguments
  discussed in reference~\cite{review}~-- that
  at this excess energy (Q~=~15.5~MeV) the $\eta$ meson is produced
  in the center-of-mass frame predominantly with the angular momentum
  equal to zero. Similarly, the distribution determined for the
  polar angle of the relative
  proton-proton momentum with respect to the momentum of the $\eta$ meson
  as seen in the di-proton rest frame is also consistent with isotropy.
  Anyhow, even the isotropic distribution in this angle does not imply
  directly that the relative angular momentum between protons is equal to zero,
  because of their internal spin equal to $\frac{1}{2}$.
  Therefore,
  the contribution from  the $^3$P$_0$-wave
  produced via
  the $^1S_{0}\to ^3\!\!P_{0}s$ transition
  cannot be excluded. 
  Moreover, as pointed out in reference~\cite{nakayama0302061}, 
  the isotropic angular distribution,
  can principally also be achieved by the destructive interference between the
  transitions $^1S_{0} \to ^3\!\!P_{0}s$ and $^1D_{2} \to ^3\!\!P_{2}s$.
\vspace{-0.3cm}
\begin{figure}[H]
\parbox{0.48\textwidth}{\hspace{-0.4cm}
\includegraphics[width=0.54\textwidth]{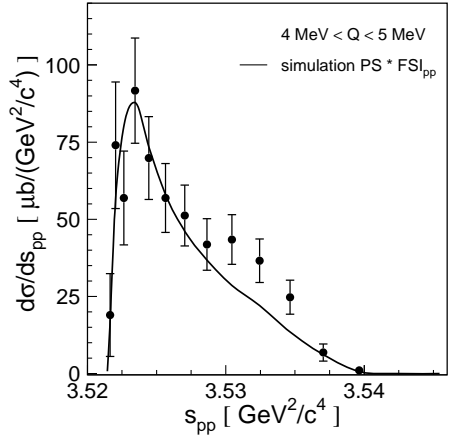}} \hfill
\parbox{0.50\textwidth}{ \vspace{-0.6cm}
\caption{\label{figurejurek} 
 Distribution of square of the proton-proton invariant mass from the $pp\to pp\eta$ reaction
  measured at COSY-11 for the excess energy range
   4~MeV~$\le$~Q~$\le$~5~MeV~\cite{jurekraport,smyrski182,moskal025203}.
  The superimposed line shows the result of simulations performed under the assumption that 
  the phase space population is determined exclusively by the on-shell interaction 
  between outgoing protons. 
  The ``tail'' at large $s_{pp}$ values is due to the smearing of the excess energy,
  since this former COSY-11 data have not been kinematically fitted.
  Additionally to the 1~MeV range of Q a smearing of about 0.3~MeV~($\sigma$) should be taken
  into account.
}}
\end{figure}
\vspace{-0.3cm}
  Since flat angular distributions do not preclude the occurrence of 
  higher partial-waves an effect of their contribution to the production process
  was proposed to 
  explain the structure observed in the invariant mass spectra~\cite{nakayama0302061}.
  In fact, as depicted by the dotted line in figure~\ref{petaspp_kanzo}a,
  an admixture of the $^1S_{0}\to ^3\!\!P_{0}s$ transition to the main 
  $^3P_{0}\to ^1\!\!S_{0}s$ one leads to the
  excellent agreement with the experimental points.
  However, at the same time, this conjecture leads
  to  strong discrepancies in the shape of the excitation function
  as can be deduced from the comparison of the dashed-dotted
  line and the data in figure~\ref{cross_eta_etap}.
  While it describes the data points
  in the excess energy range
  between 40~MeV  and 100~MeV it underestimates the total cross section
  below 20~MeV by a factor of 2.
  This observation entails that either the contribution of higher partial 
  waves, though conforming with invariant mass spectra and angular distributions,
  must be excluded or that the simplifications performed e.g. for the description 
  of the final state three-body system were too strong.
  Interestingly, the enhancement at large $s_{pp}$ is visible
  also at much lower excess energy. This can be concluded from figure~\ref{figurejurek}
  in which the COSY-11 data at  Q~$\approx$~4.5~MeV~\cite{moskal025203,jurekraport}
  are compared to the simulations
  based on the assumption that the phase space abundance is due to  
  the proton-proton FSI only.
   This observation could imply that the effect is caused by the 
  proton-$\eta$ interaction
  rather than by higher partial waves, since their contribution
  at such small energies is quite improbable~\cite{review}.
  However, as shown in the lower part of figure~\ref{petaspp_kanzo}
  the rigorous three-body treatment of the $pp\eta$ system leads
  at large values of $s_{pp}$ 
  to the reduction of the 
  cross section  in comparison to the calculation   
  taking into account only first order rescattering (pp+p$\eta$)~\cite{Fixprivate,Fixprivate1}.
  Here, both calculations include only the $^3P_{0}\to ^1\!\!S_{0}s$ transition.
  Although the presented curves are still preliminary, we can qualitatively assess  
  that the  rigorous three-body approach, in comparison to the present estimations,
  will on one hand enhance the total cross section 
  near threshold as shown in figure~\ref{cross_eta_etap}, while on the other hand
  it will
  decrease the differential cross section at large values of $s_{pp}$.
  This is just opposite to the influence  of P-waves
  in the proton-proton system.
 
  From the above presented considerations it is rather obvious 
  that the rigorous three-body treatment of the produced $pp\eta$ system
  and the exact  determination of the contributions from the 
  higher partial waves may result in a simultaneous explanation of  
  both observations:  the  near-threshold enhancement of the excitation function  
  of the total cross section, and the strong increase of
  the invariant mass distribution at large values of $s_{pp}$.
  For the unambiguous determination of the contributions
  from different partial waves spin dependent observables are required~\cite{nakayama0302061}.
  The first attempt has been already reported in~\cite{winter251},
  and more comprehensive investigations are in progress~\cite{proposalrafal}
  (see also section~\ref{spindegreessection}).
\vspace{-0.1cm}
\section{Quasi-bound state}
\begin{flushright}
\parbox{0.71\textwidth}{
 {\em
 ...~which exists either because of some impossible blip
    on the curve of probability or because the gods enjoy a joke as much as anyone~\cite{pratchett}.\\
 }
 \protect \mbox{} \hfill  Terry Pratchett \protect\\
 }
\parbox{0.71\textwidth}{
 {\em
 ...\!\!\! I am inclined to think that scientific discovery is impossible without
 faith in ideas which are of a purely speculative kind, 
 and sometimes even quite hazy ...~\cite{popperlogic}.\\
 }
 \protect \mbox{} \hfill  Karl Raimund Popper \protect\\
 }
\end{flushright}
\vspace{-0.3cm} 
With the up-to-date experimental accuracy, from all meson-$N$-$N$ systems the
$\eta NN$ one reveals by far the most interesting features.
Recently, the dynamics of the $\eta NN$ system has become a subject of
theoretical investigations in view of the possible existence of quasi-bound
or resonant states~\cite{fix119}.
A direct measure of the formation --~or non-formation~-- of an
$\eta$-nuclear quasi-bound state is the real part of the $\eta$-nucleon
scattering length~\cite{svarc024}.
The determined values of $\mbox{Re}\,(a_{\eta N}$) range between
$0.20\,\mbox{fm}$~\cite{N.Kaiser-II} and $1.05\,\mbox{fm}$~\cite{green035208}
depending on the analysis method and the reaction studied~\cite{green053},
and at present an univocal answer whether the attractive interaction between
the $\eta$ meson and nucleons is strong enough to form a quasi-bound state is
not possible.
\begin{figure}[H]
\hfill
\parbox{0.54\textwidth}
  {\epsfig{file=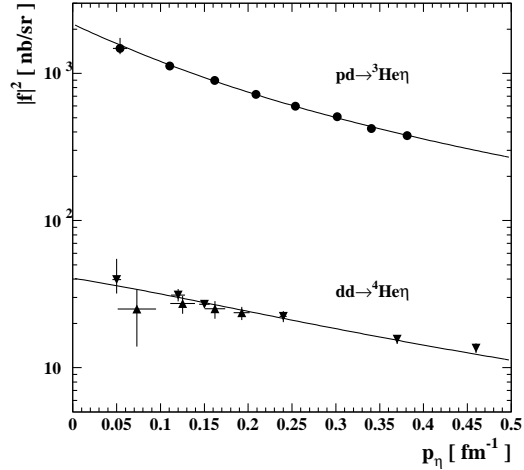,width=0.525\textwidth}} \hfill
\parbox{0.45\textwidth}
  {\vspace{-0.5cm} \caption{\label{he34} Averaged squared production amplitudes of the
  reactions $pd \rightarrow {}^3He\,\eta$ and $dd \rightarrow {}^4He\,\eta$ as
  function of the centre-of-mass $\eta$ momentum. 
  The data are taken from references~\cite{May96,Fra94,Wil97}.
  The amplitude $f_n$ is related to the unpolarized centre-of-mass
  cross section by the equation $\frac{d\sigma}{d\Omega}=\frac{p^*_\eta}{p^*_d}|f_n|^2$,
  where $p^*_\eta$ and $p^*_d$ denote the centre-of-mass momentum in $\eta ^3\!He$ and $pd$
  systems, respectively.
  Superimposed lines correspond to the fit of the s-wave scattering-length formula:
   $f(p_n) = \frac{constans}{1 - i p_\eta a_{\eta X}}$, where $a_{\eta X}$ 
   denotes the scattering length of the $\eta-^3\!He$ or $\eta-^4\!He$ potential.}} 
\end{figure}
The shape of the energy dependence of the $pd \rightarrow {^3\!He}\,\eta$ 
production amplitude (see figure~\ref{he34}) implies that either the real or imaginary part of the $\eta\;{^3\!He}$
scattering length has to be very large~\cite{wilkinR938}, which may be
associated with a bound $\eta\;{^3\!He}$ system.
Similarly encouraging are results of reference~\cite{shevchenko143}, where it
is argued that a three-body $\eta NN$ resonant state, which may be formed
close to the $\eta d$ threshold, may evolve into a quasi-bound state for
$\mbox{Re}\,(a_{\eta N}) \ge 0.733\,\mbox{fm}$.
Also the close-to-threshold enhancement of the total cross section of the
$pp \rightarrow pp \eta$ reaction was interpreted as being either a Borromean
(quasi-bound) or a resonant $\eta pp$ state~\cite{wycech2981} provided that
$\mbox{Re}\,(a_{\eta N}) \ge 0.7\,\mbox{fm}$.
Contrary, recent calculations performed within a three-body
formalism~\cite{fix119} indicate that a formation of a three-body $\eta NN$
resonance state is not possible, independently of the $\eta N$ scattering
parameters.
Moreover, the authors of reference~\cite{garcilazo021001} exclude the
possibility of the existence of an $\eta NN$ quasi-bound state.
Results of both calculations~\cite{fix119,garcilazo021001}, although performed
within a three-body formalism, were based on the assumption of a
separability of the two-body $\eta N$ and $NN$ interactions.
However, in the three-body system characterized by the pairwise attractive
interactions, the particles can be pulled together so that their two-body
potentials overlap, which may cause the appearance of qualitatively new
features in the $\eta NN$ system~\cite{fix119}.
Particularly interesting is the $\eta d$ final state where the pair of
nucleons alone is bound by the strong interaction.
\vspace{-0.2cm}
\begin{figure}[H]
    \parbox{0.5\textwidth}{\hspace{-0.8cm}
       \includegraphics[width=0.6\textwidth]{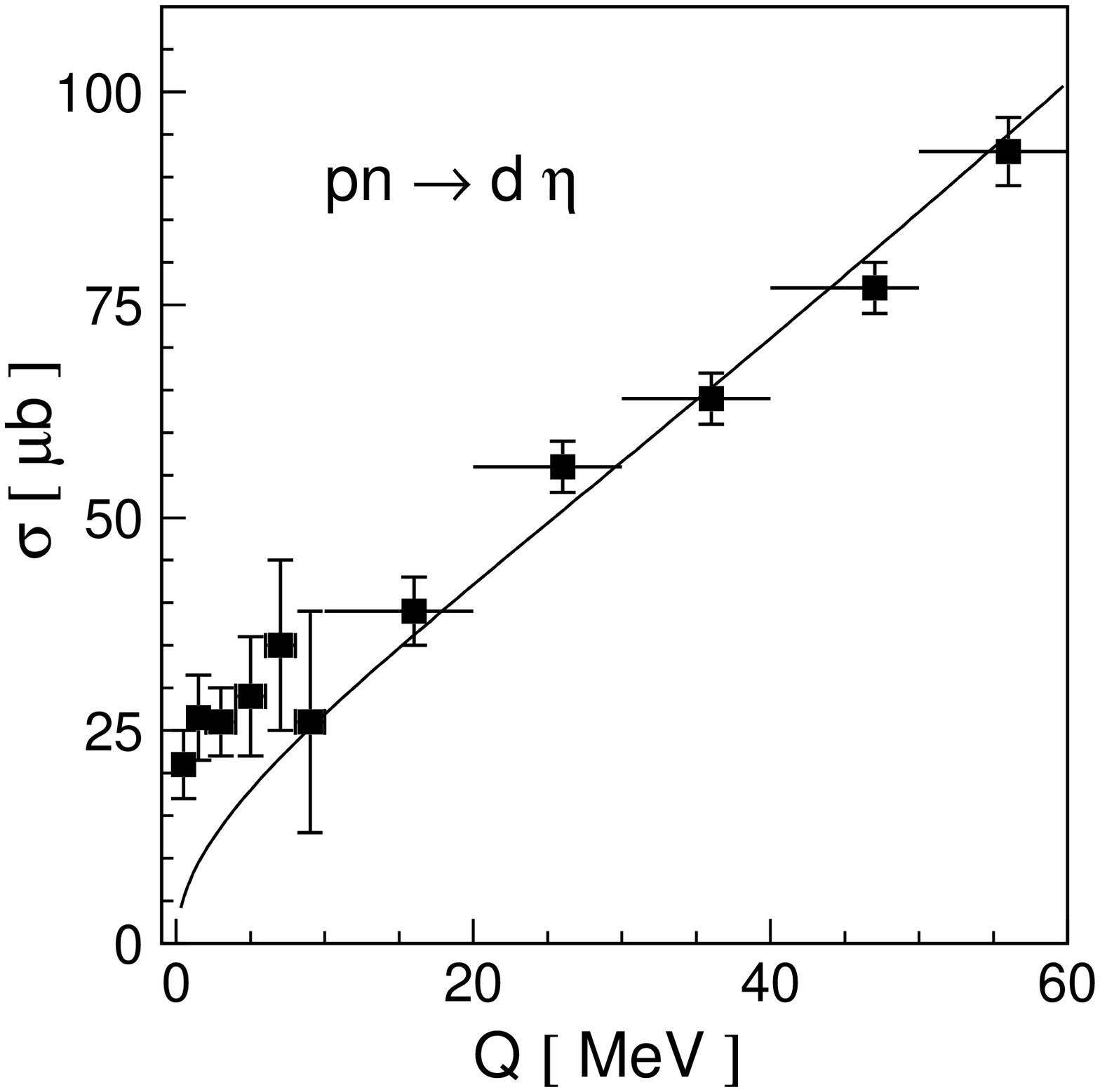}
    } 
    \parbox{0.5\textwidth}{ \vspace{-0.6cm}
       \includegraphics[width=0.49\textwidth]{factor.epsi}
    } 
    
\vspace{0.9cm}

    \parbox{0.43\textwidth}{\raisebox{0ex}[0ex][0ex]{\mbox{}}} \hfill
    \parbox{0.49\textwidth}{\vspace{-2.5cm}\raisebox{0ex}[-10ex][0ex]{\large a)}} \hfill
    \parbox{0.04\textwidth}{\vspace{-2.5cm}\raisebox{0ex}[-10ex][0ex]{\large b)}}

\vspace{-1.5cm}

     \caption{\label{factors}
     {\bf (a)} Total cross section of the quasi-free $pn \rightarrow d\eta$ reaction as a
       function of the excess energy~\cite{calen2069,calen2642}. The curve --~fitted in amplitude~--
       indicates
       the energy dependence proportional to the function $\sqrt{\mbox{Q}} \cdot
       \left(1 + \mbox{Q}/83.5 \right)$ which accounts for the s- and p-wave contribution.
       The ratio of the s- and p-waves magnitudes
       was taken the same as determined for the $np \rightarrow d \pi^0$
       reaction~\cite{bilger64,hutcheon176,hutcheon618}. It was assumed that the
       p-wave to s-wave ratio is the same for the $pn \rightarrow d \eta$ and $pn
       \rightarrow d \pi^0$ reactions at a corresponding value of $\eta_M$.
     {\bf (b)} Arbitrarily normalized enhancement factors for $pd$-, $d \eta$-, and
       $p \eta$-FSI. The $\eta$-proton factor is calculated according to
       equation~\eqref{Mpeta}, with $a_{p \eta} = 0.717\,\mbox{fm} +
       i\,0.263\,\mbox{fm}$~\cite{Batinic023} and $b_{p \eta} = -1.50\,\mbox{fm} -
       i\,0.24\,\mbox{fm}$~\cite{greenR2167}. The enhancement factor for
       $\eta$-deuteron has been extracted from the data of panel (a) parametrizing
       the ratio of the cross section to the phase space volume by the
       expression~\cite{smyrskic11proc}: $F_{d \eta} = 1 + 0.5/(0.5 + (\mbox{Q}/5)^2)$.
       The proton-deuteron FSI factor is calculated according to
       ref.~\cite{meyer2474}.
    }
\end{figure}
\vspace{-0.1cm}
Figure~\ref{factors}a shows the total cross section for the $pn
\rightarrow d \eta$ reaction measured close to the production threshold.
For excess energies below $10\,\mbox{MeV}$ the data are enhanced over the
energy dependence determined for the $pn \rightarrow d \pi^0$ reaction
indicated by the solid curve.
This is in qualitative agreement with the calculations of Ueda~\cite{ueda297}
for the three-body $\eta NN$-$\pi NN$ coupled system, which predict the
existence of an $\eta NN$ quasi-bound state with a width of $20\,\mbox{MeV}$.
Ueda pointed out that the binding of the $\eta NN$ system is due to the
$\mbox{S}_{11}$ $\eta N$ and $^{3}\mbox{S}_1$ $NN(d)$ interaction which is
characterized by no centrifugal repulsion.
Such repulsion makes the $\pi NN$ system, in spite of the strong
$\mbox{P}_{33}$ $\pi N$ attraction, hard to be bound~\cite{ueda68}.
Whether the observed cusp at the $pn \rightarrow d \eta$ threshold is large
enough to confirm the existence of the $\eta NN$ bound state has been recently
vigorously discussed~\cite{wycech045206,deloff024004}.
 
The enhancement factor for the deuteron-$\eta$ interaction inferred from the
data in figure~\ref{factors}a varies much stronger in comparison to
the proton-$\eta$ one, as demonstrated in figure~\ref{factors}b.
This suggests that the effects of this interaction should be even more
pronounced in the differential distributions of the cross section for the $pd
\rightarrow pd \eta$ reaction as those observed in case of the $pp \rightarrow
pp \eta$ process, especially because the ``screening'' from the
proton-deuteron interaction is by more than an order of magnitude smaller
compared to the proton-proton interaction as can be deduced from the
comparison of the solid lines in figures~\ref{factors}b
and~\ref{Mpppp_cross_pi}a.
Experimental investigations on that 
issue~\cite{hibou537,smyrskic11proc,cezary}
as well as searches for $\eta$-mesic nuclei~\cite{gillitzerTOF} by
measuring proton-deuteron induced reactions in the vicinity of the $\eta$
production threshold are under way.
Theoretically, the existence of mesic nuclei or 
quasi-bound meson-nucleus systems is not excluded, however, up to now a 
compelling experimental proof for the formation of such a state is still 
missing. 
The probability to create such states depends crucially on 
the sign and 
the strength of the meson-nucleus interaction. 
Since the $\pi$-nucleon and $K^+$-nucleon interaction have been found to be 
repulsive, it is unlikely that they will form quasi-bound states. 
Contrary, there is evidence for an attractive $K^-$-nucleon interaction~\cite{lau99},
however, corresponding experiments would suffer from the low  cross 
sections for the creation of $K^-$ mesons~\cite{quentmeier276}. 
Furthermore, the Coulomb interaction is expected to screen corresponding 
physical observables and might lead to the formation of mesonic atoms bound by 
the electromagnetic interactions.
In contradistinction to $K^{-}$, the $\eta$ meson is uncharged and the observation of an 
attractive $\eta$-nucleon interaction led to speculations concerning the 
existence of $\eta$-nuclear quasi-bound states. 
In the absence of $\eta$-meson beams such states bound by the strong 
interaction would offer a new possibility to study the $\eta$-nucleon 
interaction since the meson would be trapped for a relatively large time in 
the nuclear medium. 
In a very first prediction of such  states termed $\eta$-mesic nuclei 
Haider and Liu~\cite{Hai86,Hai86b}
in the framework of an optical model have estimated that it can be formed 
already for nuclei with atomic numbers of  $\mbox{A} \ge 12$. 
Further on, the topic was vigorously discussed~\cite{rakityanskyR2043,Wyc95,Aba96,wilkinR938,CHIANG738,Li87},
the limit was lowered considerably, and presently one considers seriously the existence 
of $\eta$-nucleus bound states even for $d, t, {}^3\!He$, or  $^4\!He$ nuclei~\cite{wilkinR938,rakityanskyR2043}. 

Experimental evidences for bound states have been found in the near-threshold 
production of $\eta$ mesons in the reaction channel $pd \rightarrow 
{}^3He\,\eta$. 
The unexpected large production amplitude as well as its rapid decrease with 
increasing energy (figure~\ref{he34}) is attributed to a strong s-wave 
final state interaction associated with a large $\eta$-$\,{}^3He$ scattering 
length ($a_{\eta-^3\!He} \sim (-2.31 + i\,2.57)\,\mbox{fm}$)~\cite{wilkinR938}.\newline
Assuming that $^3\!He$ and $\eta$ can form a quasi-bound state, it is 
interesting to compare $pd \rightarrow {}^3He\,\eta$ and $dd \rightarrow {}^4He\,\eta$
production data,
where for the latter, according to the 
higher mass number, the formation of a bound state should be even more 
probable. 
A comparison of corresponding near-threshold data is presented in 
figure~\ref{he34}.
Interestingly, the slope of the $f_{\eta}(p_\eta)$ function of the $dd\to ^4\!He\eta$ reaction
is smaller than the one of the reaction $pd\to ^3\!He\eta$. This observation suggests, though contra-intuition,
that the existence of the $\eta$-mesic nucleus is more probable in the case of $^3\!He$ than $^4\!He$.

 \section{Search for a signal from $\eta^{\prime}$-proton 
          interaction in the invariant mass  distributions}

\label{searchforsection}
\begin{flushright}
\parbox{0.7\textwidth}{
 {\em
   I am sure that everyone, even among those who follow the profession, will
   admit that everything  we know is almost nothing compared with what remains
   to be discovered...~\cite{descartes1}.\\
 }
 \protect \mbox{} \hfill  Ren\'{e} Descartes \protect\\
 }
\end{flushright}

A phenomenological analysis --~presented in section~\ref{influencesection}~--
of the determined excitation function for
the $pp\to pp\eta^{\prime}$ reaction (lower part of figure~\ref{cross_eta_etap}) revealed
no signal which could have been assigned to the proton-$\eta^{\prime}$ interaction.
Later on, in section~\ref{qualitativecomparison}, a comparison of the energy dependence of the
production amplitudes
for the  $pp \to pp\eta^{\prime}$,
$pp \to pp\eta$ and $pp \to pp\pi^0$ enabled to conclude, in a model-free way, that the proton-$\eta^{\prime}$
interaction is indeed much weaker than the proton-$\eta$ one, and a trial to determine quantitatively the scattering
length of the proton-$\eta^{\prime}$ potential resulted only in a very modest estimation of an upper limit
of its real part ($|Re~a_{p\eta^{\prime}}| < 0.8$~fm)~\cite{moskal416}.
Seeking for a visible manifestation of the proton-$\eta^{\prime}$ interaction
 we have performed a high statistics measurement of $pp\to pp\eta^{\prime}$
reaction in order to determine a distribution of events over the phase space~\cite{proposal123}.
We expect that the invariant mass spectra  for two
particle subsystems of the $pp\eta^{\prime}$ final state can give the first ever
experimental evidence for this still completely unknown interaction.
The data are presently analyzed~\cite{joannarap}.
    The missing mass spectrum determined on-line from 10\% of the data written on tapes
    is shown in  figure~\ref{missonline}. 
From a clear signal visible at the mass of the $\eta^{\prime}$ meson,
    one can estimate that the overall number of registered
    $pp \to pp \eta^{\prime}$ events amounts to about 13000.
The measurement  has been carried out at a beam momentum corresponding to
  the excess energy
  of Q~=~15.5~MeV for the $pp\to pp\eta^{\prime}$ reaction.
Such value of Q was chosen to enable a direct comparison with the invariant mass spectra
obtained for the $pp\to pp\eta$ reaction (see figure~\ref{petaspp}), without a need for a correction of kinematical factors.
\begin{figure}[H]
\vspace{-0.1cm}
\parbox{0.59\textwidth}{\hspace{-0.3cm}
\epsfig{file=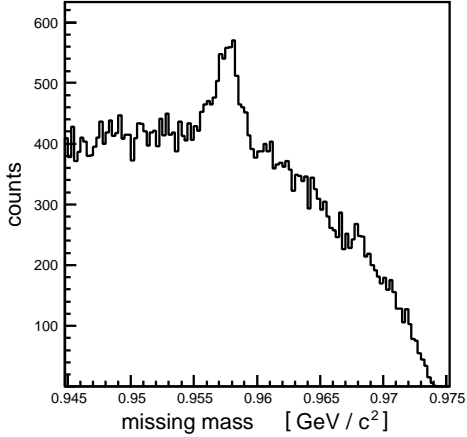,width=0.48\textwidth}}
\vspace{-1.0cm}
\parbox{0.4\textwidth}{
\caption{\label{missonline}
  On-line missing-mass distribution of the $pp\to ppX$ reaction
  measured  by means of the COSY-11 detection system in October 2003~\cite{joannarap}
  at
  the beam momentum of 3.257~GeV/c, which in the case of the
  $pp\to pp\eta^{\prime}$ reaction
  is equivalent to the excess energy
  of Q~=~15.5~MeV.
}}
\end{figure}
\vspace{0.6cm}
At present due to the need of the subtraction of the unavoidable  multi-pion background,
the available statistics allows only for the background-free determination of one-dimensional invariant mass distributions.
Yet it would be also desired to determine an occupation density over the Dalitz-plot since this two-dimensional distribution
comprises full empirically accessible information of the mutual interaction within a three particle system.
However, the present experimental conditions render it impossible,
unless the binning was made so large that the possible effect would be smeared out completely.
In spite of all, we still endeavour to measure this two-dimensional spectrum for both $pp\eta$ and $pp\eta^{\prime}$
systems free from the background.
A detection of protons  and gamma quanta from the decay of the meson would allow us to reach this aim.
Therefore,
when the WASA\cite{zabierowski159} detector will be
installed at the cooler synchrotron COSY~\cite{prasuhn167}
 we plan to seize the opportunity
 of  performing  pertinent investigations~\cite{LoImoskal}.
\clearpage


\newpage
\clearpage
\thispagestyle{empty}
\pagestyle{plain}
\chapter{Dynamics of the near threshold production of $\eta$ and $\eta^{\prime}$ 
         mesons 
         in collisions of nucleons}
\thispagestyle{empty}
\pagestyle{myheadings}
\markboth{Hadronic interaction of $\eta$ and $\eta'$ mesons with protons}
         {5. Dynamics of the  near threshold production of $\eta$ and $\eta^{\prime}$
         mesons ... }
\label{Dopsmp}    
\begin{flushright}
\parbox{0.7\textwidth}{
 {\em
   Everything outside of the dynamics is 
   just a verbal description of the table of
   data, and even then the data table
   probably yields more information than
   the verbal description can~\cite{heisenberg1}. \\
 }
 \protect \mbox{} \hfill Werner Karl Heisenberg \protect\\
 %
 }
\end{flushright}
\vspace{-0.4cm}
\section{Comparison of the production yields}
\begin{flushright}
\parbox{0.7\textwidth}{
 {\em
    All kinds of reasoning consist in nothing but a {\bf\em comparison},
    and a discovery of those relations, either constant or inconstant,
    which two or more objects bear to each other~\cite{humetreatise}.\\
 }
 \protect \mbox{} \hfill David Hume  \protect\\
 }
\end{flushright}
\vspace{-0.2cm}
Considerations presented in chapter~\ref{Hinertact} led to the conclusion, that close 
to the kinematical threshold, the energy dependence of the total cross section 
is in first approximation determined via the interaction among the 
outgoing particles and that the entire production dynamics manifests itself 
in a single constant,  which determines the absolute scale of the total cross 
section. We demonstrated also that the interaction among created particles increases
the production probability drastically, though in the naive time ordered approach
the final state interaction takes place after the production.  
Therefore, the observed increase of the total cross sections close to the kinematical
threshold, for the $pp\to pp\eta$ and $pp\to pp\eta^{\prime}$ reactions 
(see figure~\ref{cross_eta_etap}), can 
be considered as a pure quantum mechanical effect, which clearly demonstrates
that the production of particles and their mutual interaction must be treated
coherently for the proper appraisal of the absolute production cross section.
Thus, for the sake of completeness, we will devote one chapter to a brief description
of the  mechanisms responsible  for the production of $\eta$ and $\eta^{\prime}$ 
mesons in the collisions of nucleons.

As a first step towards the understanding of the creation mechanism underlying 
the production let us compare the total cross sections for these mesons
with their flavour-neutral pseudoscalar partner 
--~the meson $\pi^0$.

Since the masses of these mesons are significantly different\footnote{\mbox{} The 
$\pi^0$, $\eta$, and $\eta^{\prime}$ masses amount to 
$134.98\,\mbox{MeV}/\mbox{c}^2$, $547.30\,\mbox{MeV}/\mbox{c}^2$, and 
$957.78\,\mbox{MeV}/\mbox{c}^2$, respectively~\cite{PDG}.} the influence  
of the kinematical flux factor F on the total cross section and the 
suppression due to the initial state interaction $F_{ISI}$ depend 
substantially on the created meson. 
Therefore, for the comparison of the primary dynamics we will correct for 
these factors and instead of comparing the total cross section we will 
introduce -- according to reference~\cite{swave} -- a dimensionless quantity 
$\sigma \cdot \mbox{F}/F_{ISI}$, which depends only on the primary production 
amplitude $M_0$ and on the final state interaction among the produced 
particles.

Close-to-threshold the initial state interaction, which reduces the total 
cross section, is dominated by proton-proton scattering in the 
$^{3}\mbox{P}_0$ state which may be estimated in terms of phase-shifts and 
inelasticities by employing equation~\eqref{F_ISI}. 
The $F_{ISI}$ factor is close to unity for pion production and amounts to 
$\sim 0.2$~\cite{hanhart176} and $\sim 0.33$~\cite{nakayama024001} for the 
$\eta$, and $\eta^{\prime}$ meson, respectively, at threshold.  

A comparative study of the production of mesons with significantly different \
masses encounters the difficulty of finding a proper variable at which the 
observed yield can be compared. 
Seeking such quantity let us recall the meaning of the total cross section
which is defined 
as the integral over the available 
phase space volume of transition probabilities -- reflecting the dynamics of 
the process -- from the initial to the final state as written explicitly in 
equation~\ref{phasespacegeneral}.
Thus, if the dynamics of the production process of two different mesons were 
exactly the same then the above introduced yield would also be strictly the 
same for both mesons, provided it was extracted at the same value of the volume
of the phase space $V_{ps}$ (eq.~\ref{Vps_relativistic}). 
This inference would, however, not be valid if the production yields were 
compared at the $\eta_M$ or Q variables.
Therefore, the volume of the available phase space for the produced particles 
is the most suited quantity for the regarded comparison~\cite{swave}. 
This could also be the best choice for the investigation of isospin breaking 
where the cross sections for the production of particles with different masses 
need to be compared (e.g.\ $\pi^+ d \rightarrow pp \eta$ and $\pi^- d 
\rightarrow nn \eta$~\cite{tippens052001}).
Figure~\ref{cfisi_Vps} shows the yield of $\pi^0$, $\eta$, and 
$\eta^{\prime}$ mesons in the proton-proton interaction as a function of the 
available phase space volume. 
\vspace{-0.3cm}
\begin{figure}[H]
{\centerline{\parbox{0.68\textwidth}
  {\epsfig{figure=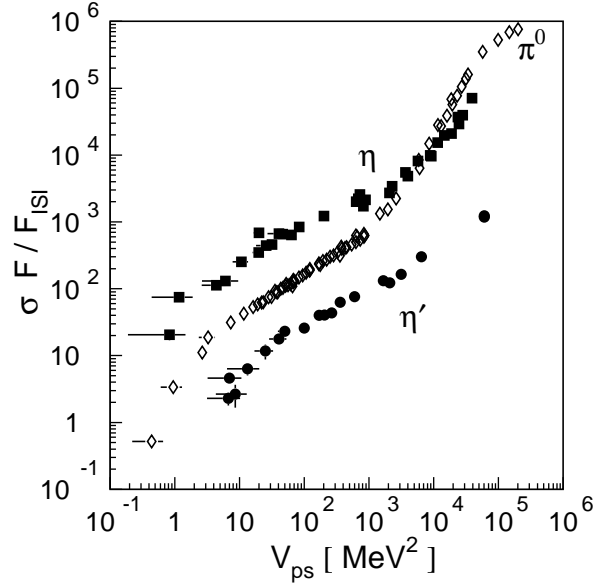,width=0.68\textwidth}}}}
\vspace{-0.3cm}
\caption{\label{cfisi_Vps} Total cross section ($\sigma$) multiplied by 
the flux factor F and divided by the initial state interaction reduction factor 
$F_{ISI}$ versus the available phase space volume for the reactions $pp 
\rightarrow pp \eta$ (squares~\cite{hibou41,bergdoltR2969,chiavassa270,calen39,
smyrski182,moskal367,calen2642}), $pp \rightarrow pp \pi^0$ 
(diamonds~\cite{bondar8,meyer633,stanislausR1913,bilger633,rappenecker763}), 
and $pp \rightarrow pp\eta^{\prime}$ (circles~\cite{hibou41,balestra29,
moskal3202,moskal416,khoukaz0401011}).
}
\end{figure}
\vspace{-0.3cm}
The onset of higher partial waves is seen for $\pi^0$ and $\eta$ mesons in the 
$V_{ps}$ range between $10^3$ and $10^4\,\mbox{MeV}^2$, whereas the whole 
range covered by the $\eta^{\prime}$ data seems to be consistent with the pure 
Ss production.
One can also recognize that the data for the Ss final state are grouped on 
parallel lines indicating a dependence according to the power 
law $\sigma \cdot \mbox{F}/F_{ISI} \approx 
\alpha \cdot V_{ps}^{0.61}$~\cite{raport48} and that over the relevant range 
of $V_{ps}$ the dynamics for $\eta^{\prime}$ meson production is about six 
times weaker than for the $\pi^0$ meson, which again is a further factor of 
six weaker than that of the $\eta$ meson.
This is an interesting observation, since the quark wave functions of $\eta$ 
and $\eta^{\prime}$ comprise a similar amount of strangeness 
($\approx 70\,\%$~\cite{moskalphd}) and hence, in the nucleon-nucleon 
collision one would expect both these mesons to be produced much less 
copiously than the meson $\pi^0$ being predominantly built out of {\em up} and 
{\em down} quarks. 
On the hadronic level, however, one can qualitatively argue that the $\eta$ 
meson --~contrary to $\pi^{0}$ and $\eta^{\prime}$~-- 
owes its rich creation to 
the existence of the baryonic resonance $N^*(1535)$ whose branching ratio into 
the $N \eta$ system amounts to 30--$55\,\%$~\cite{PDG}.
There is no such established resonance, which may decay into an s-wave 
$\eta^{\prime} N$ system~\cite{PDG}, and the $\pi^0$ meson production with 
the formation of the intermediate $\Delta(1232)$ state is strongly suppressed 
close-to-threshold, because of conservation laws.
 
\section{Mesonic degrees of freedom}
\begin{flushright}
\parbox{0.58\textwidth}{
 {\em
   For what is determinate cannot have innumerable explanations~\cite{kopernik}.\\
 }
 \protect \mbox{} \hfill  Nicholas Copernicus \protect\\
 }
\end{flushright}
One of the most crucial issue in the 
investigations of the dynamics of the close-to-threshold meson production
is the  determination of the relevant degrees of freedom for the 
description of the nucleon-nucleon interaction, especially in case when the 
nucleons are very close together\footnote{\mbox{} These investigations are listed as 
one of the key issues in hadronic physics~\cite{capstick238}.}. 
The transition region from the 
hadronic to constituent quark degrees of freedom does not have a well defined 
boundary and at present both approaches are evaluated in order to test 
their relevance in the description of close-to-threshold meson production in 
the collision of nucleons.
Simple geometrical considerations presented in figure~\ref{cartoon} indicate 
that at distances smaller than $2\,\mbox{fm}$ the internucleon potential 
should begin to be free of meson exchange effects and may be dominated by the 
residual colour forces~\cite{maltmanisgur}. 
\vspace{-0.4cm}
\begin{figure}[H]
\hfill
\parbox{0.45\textwidth}
  {\caption{\label{cartoon} A cartoon illustrating in naive geometrical terms 
   that for $\mbox{r} < 2\,\mbox{r}_N + 2\,\mbox{r}_M$ meson exchange is 
   unlikely to be appropriate for the description of the internucleon 
   potential. Figure and caption are taken from reference~\cite{maltmanisgur}.}} \hfill
\parbox{0.45\textwidth}
  {\epsfig{file=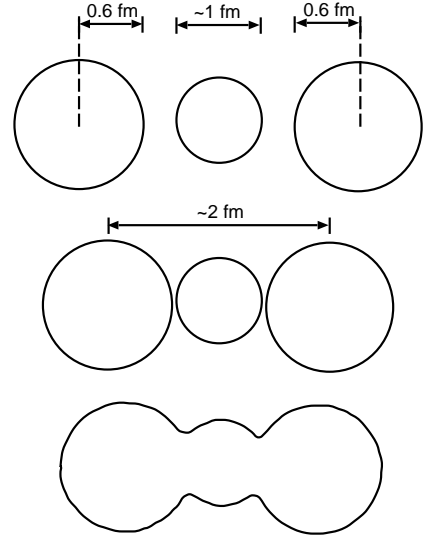,width=0.43\textwidth}} 
\end{figure}
\vspace{0.3cm}
In section~\ref{influencesection} it was shown, that the close-to-threshold production 
of mesons occurs when the colliding nucleons approach distances of about 
$0.5\,\mbox{fm}$ in case of $\pi^0$ and of about $0.18\,\mbox{fm}$ in case of 
$\phi$ production (see table~\ref{momtranstable}).
This distance is about one order of magnitude smaller than $2\,\mbox{fm}$ 
and it is rather difficult to imagine -- in coordinate space -- an exchange 
of mesons between nucleons as mechanism of the creation process.
Thus, the meson exchange for the threshold meson production can 
be understood as an effective 
description of the process occuring on the deeper level.
Such small collision parameters imply that the interacting nucleons --~objects 
of about 1~fm~-- overlap and their internal degrees of freedom may be of 
importance. On the other hand, as demonstrated by Hanhart~\cite{christoph},
if there is a meson exchange current possible at leading order 
it should dominate the creation process. This can be derived from the 
strong dependence of the production operator on the distance between the 
colliding nucleons. 
\vspace{-0.2cm}
\begin{figure}[H]
\centerline{\epsfig{file=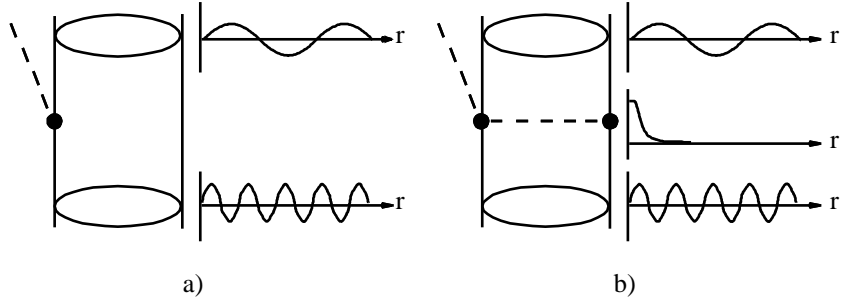,width=0.85\textwidth}}
\vspace{-0.5cm}
\caption{\label{figchristoph}
 Illustration of the momentum mismatch. The picture shows why one should
 a priori expect meson exchange currents 
 to be very important close to the threshold.
 The figure and title are adapted from reference~\cite{christoph}.
 By courtesy of C. Hanhart.
}
\end{figure}
\vspace{-0.8cm}
\begin{figure}[H]
\begin{center}
\epsfig{file=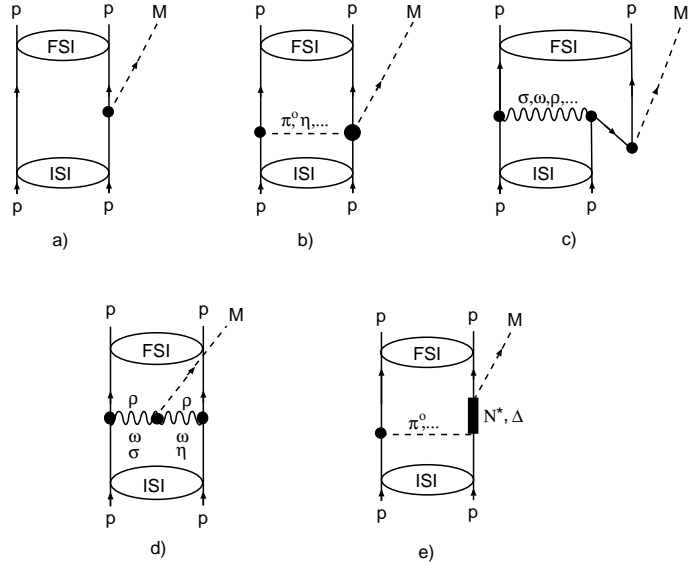,width=0.70\textwidth}
\end{center}
\vspace{-0.8cm}
\caption{\label{graph_mesonexchange} Diagrams for the $pp \rightarrow 
pp\,Meson$ reaction near threshold: 
(a) $Meson$-bremsstrahlung (nucleonic current) \hspace{1ex}
(b) ``rescattering'' term (nucleonic current) \hspace{1ex}
(c) production via heavy-meson-exchange \hspace{1ex}
(d) emission from virtual meson (mesonic current) \hspace{1ex}
(e) excitation of an intermediate resonance (nucleon resonance current).}
\end{figure}
This particular feature of the meson exchange
current will prevent the cancellation of the transition matrix element
being a convolution of the rapidly and mildly oscillating initial
and final state wave functions, as it is illustrated in figure~\ref{figchristoph}.
Consequently, during the last decade, the effective theory based on meson 
exchanges, which accounts for the size of the participating particles by 
introduction of the momentum transfer dependent form factors, has been 
extensively employed for the description of the creation process. 
Figure~\ref{graph_mesonexchange} represents the mechanisms in question. 
In the next sections we will report on the stage of understanding 
of the $\eta$ and $\eta^{\prime}$  mesons creation on the hadronic level,
and thereafter we shall briefly present first attempts to explain
the meson threshold production in the domain of quarks and gluons.

\section{Baryonic resonance -- a doorway state for the production of the meson $\eta$}
\label{doorway}
\begin{flushright}
\parbox{0.73\textwidth}{
 {\em
   ...\!\!\!\!\!\! although these excitations of the soul are often joined 
   with the passions  that are like them, they may also frequently be found 
   with others, and may even originate  from those that are in 
   opposition to them~\cite{passion}. \\
 }
 \protect \mbox{} \hfill Ren\'{e} Descartes \protect\\
 %
 }
\end{flushright}
It is at present rather well established~\cite{germond308,laget254,faldt427,
moalem445,moalem649,vetter153,nakayama012,alvaredo125,batinic321} that the $\eta$ 
meson is produced predominantly via the excitation of the $\mbox{S}_{11}$ 
baryonic resonance $N^*(1535)$ which subsequently decays into $\eta$ and 
nucleon, and whose creation is induced through the exchange of $\pi$, $\eta$, 
$\rho$, $\sigma$, and $\omega$ mesons, as shown in 
figure~\ref{graph_mesonexchange}e. 
Although all the quoted groups reproduce the magnitude of the total cross 
section, their models differ significantly as far as the relative contributions
from the $\pi$, $\eta$, and $\rho$ exchange mechanisms are concerned.
The discrepancies are due to the not well known strength of 
$Meson$-$N$-$\mbox{S}_{11}$ couplings and the $\eta N$ scattering potential.
For example, while the dominance of the $\rho$ meson exchange is anticipated 
by authors of references~\cite{germond308,laget254,faldt427,moalem445,
moalem649}, it is rather the exchange of the $\eta$ meson which dominates the 
production if one takes into account the effects of the off-shell $\eta N$ 
scattering~\cite{pena322}, or uses the multi-channel multi-resonance 
model~\cite{batinic321}.
In any case, the hitherto performed studies aiming to describe the total cross 
section show that the close-to-threshold production of $\eta$ mesons in 
nucleon-nucleon collisions is dominated by the intermediate virtual 
$\mbox{S}_{11}$ nucleon isobar whose width overlaps with the threshold. 
\vspace{-0.4cm}
\begin{figure}[H]
\hfill
\parbox{0.52\textwidth}
  {\epsfig{file=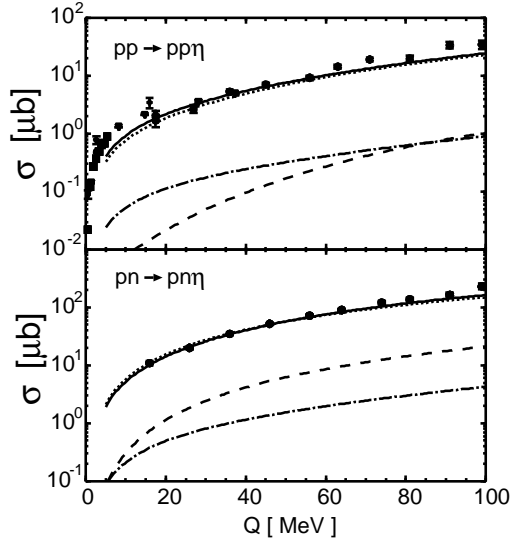,width=0.52\textwidth}} \hfill
\parbox{0.45\textwidth}
  {\caption{\label{nakayama_eta} Total cross sections for the $pp \rightarrow 
  pp\eta$ (upper panel) and $pn \rightarrow pn \eta$ (lower panel) reactions 
  as a function of excess energy. The dashed curves correspond to the 
  nucleonic current contribution and the dash-dotted curves to the mesonic 
  current consisting of $\eta\rho\rho$, $\eta\omega\omega$, and $\eta a_0 \pi$ 
  contributions. The resonance current presented by the dotted line consists 
  of the predominant $\mbox{S}_{11}(1535)$ and of the $\mbox{P}_{11}(1440)$ 
  and $\mbox{D}_{13}(1520)$ resonances excited via exchange of $\pi$, $\eta$, 
  $\rho$, and $\omega$ mesons. The solid curves are the total contribution. The 
  deviation from the data at low Q in the upper panel reflects the $\eta p$ 
  FSI which was not included in the calculations. The data are from 
  refs.~\cite{hibou41,bergdoltR2969,chiavassa270,calen39,smyrski182,calen2667}.
  The figure is adapted from~\cite{nak01}.}}
\end{figure}
In order to disentangle the various scenarios of the $\mbox{S}_{11}$ 
excitation a confrontation of the predictions with other observables is needed.
The interference between considered amplitudes causes a different behaviour 
--~depending on the assumed scenario~-- e.g.\ of the $\eta$ meson angular 
distributions. 
These differences, however, are too weak in the close-to-threshold region to 
judge between different models. 
Also the ratio of the $\eta$ meson production via the reactions $pp 
\rightarrow pp \eta$ and $pn \rightarrow pn \eta$ can be equally well 
described  by either assuming the $\rho$ meson exchange 
dominance~\cite{faldt427} or by taking pseudoscalar and vector mesons for 
exciting the $\mbox{S}_{11}$ resonance~\cite{nakayama012}. 
In the latter case, shown in figure~\ref{nakayama_eta}, the excitation of the 
resonance via the $\rho$ meson exchange was found to be negligible.
Yet, promisingly, the predictions of the analyzing power depend crucially on the assumed 
mechanism~\cite{faldt427,nakayama012}. 
This fact has already triggered  experimental investigations which aim to 
determine the spin observables. First results of tentative measurements
and perspectives of these studies
will be presented in section~\ref{spindegreessection}.
\clearpage
\newpage

\section{Spin degrees of freedom~--~a tool to study
         details of the $pp\to pp\eta$ reaction dynamics}
        
    \label{spindegreessection}

\begin{flushright}
       \parbox{0.73\textwidth}{
          {\em
            I thought ..., that scientific theories were not the digest
            of observation, but that they were inventions-conjectures
            boldly put forward for trial, to be eliminated if they clashed
            with observation; with observations which were rarely accidental 
            but as a rule 
           {\bf\em undertaken with definite intention of testing a theory 
                by obtaining, if possible, a decisive refutation}~\cite{humetreatise}.\\
          }
          \protect \mbox{} \hfill David Hume  \protect\\
       }
    \end{flushright}

  In spite of the precise measurements of the  total cross section
  for the creation of $\eta$ meson in  proton-proton~\cite{hibou41,bergdoltR2969,chiavassa270,calen39,
                                                           calen2642,smyrski182,moskal367}
  as well as  proton-neutron~\cite{calen2667} collisions
  there are still many ambiguities in the description of the
  mechanism underlying the production process. Calculations performed under different
  -~often mutually exclusive~- assumptions led to an equally satisfying description of the data.
  As we had already described in section~\ref{doorway},
  it is generally anticipated~\cite{batinic321,moalem445,germond308,laget254,vetter153,alvaredo125,
                                    pena322,kleefeld3059,ceci0301036} that the $\eta$ meson is produced
  predominantly via the excitation of the S$_{11}$ baryonic resonance $N^{*}(1535)$,
  whose creation is induced through the exchange of the virtual $\pi$, $\eta$, $\rho$,
  $\sigma$, and $\omega$ mesons, however, at present it has still not been established what 
  the relative contributions originating
  from a particular meson are. Measurements of the total cross
  section in  different isospin channels add some more limitations to the models,
  yet still the $\eta$ meson production in the $pp\to pp\eta$ and $pn\to pn\eta$ reactions
  can be equally well described by e.g. assuming the $\rho$ meson exchange dominance~\cite{faldt427}
  or by taking contributions from the pseudoscalar meson exchanges~\cite{nakayama012}.
  Therefore, for a full understanding of the production dynamics the determination  of
  more selective observables is mandatory. A~natural choice seems to be an extension of the 
  research to encompass the spin degrees of freedom.
  Figure~\ref{figay} shows  that predictions of the analyzing power $A_{y}$ are strongly sensitive
  to the type of the exchanged particle.
  A discrepancy  between  considered models is so pronounced that even a measurement of the
  beam analysing power with a precision of $\pm 0.05$ 
  {\bf  could exclude at least one of the above mentioned possibilities}
  with a statistical significance  better than three standard deviations.
  A tentative measurement of that quantity was conducted~\cite{winter251}
  and  points in figure~\ref{figay}b indicate the obtained result.
  Unfortunately, the present statistics  is 
  insufficient to allow discrimination between the alternatives considered,
  and for a conclusive inference a better accuracy of the data is required.
\vspace{0.1cm}
\begin{figure}[h]
\parbox{1.0\textwidth}{\centerline{\epsfig{file=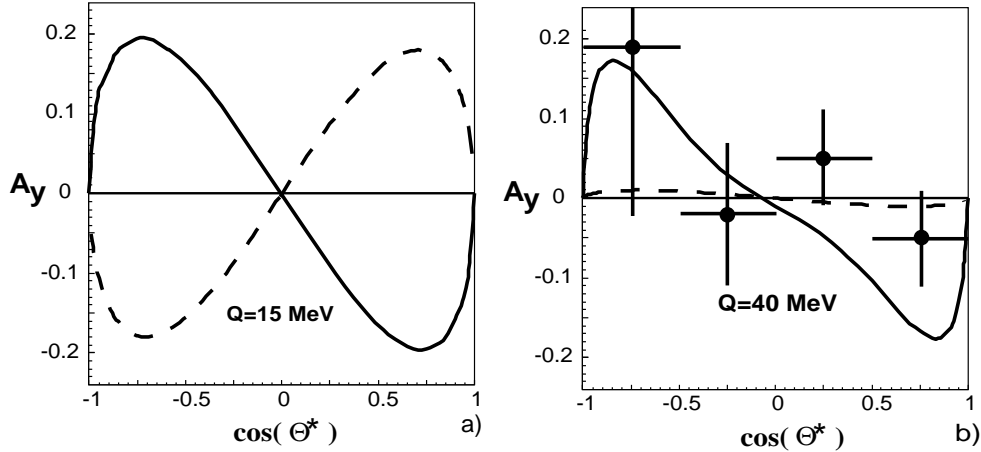,width=1.0\textwidth}}}
\caption{
Predictions for the angular dependence of the 
analysing power of the 
reaction $\vec{p}p\to pp\eta$ at $Q~=~15$ (a) and at $Q~=~37$\,MeV~(b)
compared to the data taken  at $Q=40\,$MeV~\cite{winter251}.
Solid lines present results of the  model~\cite{nakayama012}
characterized by the dominance of the exchange of the pseudoscalar mesons in the production process.
Dashed lines show expectation obtained under the assumption that  the production mechanism is predominated 
by the exchange of the $\rho$ meson~\cite{faldt427}.
\label{figay}
}
\end{figure}
\vspace{0.1cm}
  After successful  measurement of $A_{y}$ for the $\vec{p}p \to pp\eta$ reaction at Q~=~40~MeV,
  we continued the study at Q~=~37~MeV and at Q~=~10~MeV~\cite{proposalrafal,rafalraport,rafalphd}.
The measurement at Q~=~37~MeV was performed with 
proton beam polarisation amounting to approximately 70~\%,
a value significantly larger than in the first measurement.
Moreover, the luminosity integrated over the measurement period
was larger by about a factor of 1.5. 
These two 
factors together, improve the accuracy of the measurement more than twice, in the sense, 
that the errors corresponding to those in figure~\ref{figay}b 
are expected to be two times smaller. 
The second measurement was performed at Q~=~10~MeV since at  this excess energy 
the maximum of the discrepancy 
between the predictions of the regarded models~\cite{nakayama012,faldt427} is expected.
The data on the angular distributions of the analysing power
would also enable to determine the relative magnitudes  
--~or at least to set upper limits~-- 
for the contribution from the higher than
s-wave partial waves to the production dynamics.
This can be done with an accuracy by far better 
than the one resulting from the
measurements of the distributions of the spin averaged cross sections.
This is because the polarisation observables are sensitive to the interference 
terms between various partial amplitudes,  which may become measurable
even if one of the interfering terms alone appears to be insignificant in case 
of the spin averaged cross sections.
The analysis would be based on the formalism developed for the measurement 
of the $\pi^{0}$ meson production with a polarized proton beam and target~\cite{meyer064002},
which we have already used for the analysis of our first measurement with a polarized
proton beam~\cite{winter251,peterdipl,rafalhadron}.
This procedure  permits the derivation of  all available information contained
in the distributions of the cross sections and the analysing power.
Information about the partial wave decomposition is of crucial importance
for the interpretation of the production dynamics but also for the study 
of the mutual interaction of the produced $\eta$-proton-proton system. 
The strength of the proton-$\eta$ interaction cannot be derived
univocally 
from the differential cross section distributions without the knowledge 
of the contribution of  various partial waves.

Measurements with the polarized beam and target would allow  
determination of these shares.
As demonstrated  in references~\cite{meyer064002,meyer5439} the 
close-to-threshold contributions of the $Ps$ and $Pp$ partial waves can be
determined in the model-free way from the measurements of the spin dependent
total cross sections only. For example the strength of the $Ps$ final state
can be expressed as~\cite{meyer064002}:
      \begin{equation}
        \sigma(Ps) = \frac{1}{4}
                     \left( \sigma_{tot} +
                            \Delta{\sigma_T} +
                            \frac{1}{2}\Delta{\sigma_L}
                     \right)\!\!,
      \end{equation}
where $\sigma_{tot}$ denotes the total unpolarized cross section and
$\Delta{\sigma_T}$ and $\Delta{\sigma_L}$ stand for differences between the
total cross sections measured with anti-parallel and parallel beam and target
polarizations. Subscripts $T$ and $L$ associate the measurements with
the transverse and longitudinal polarizations, respectively.

At present none of the experimental facilities  enable
to investigate the $\vec{p}\vec{p} \to pp\eta$ reaction. However, an installation
of the WASA detector at COSY will offer possibility to accomplish such studies.
Corresponding measurements have been already proposed~\cite{LoIpeterrafal},
and the predictions of the spin correlation functions are 
given e.g. in references~\cite{nakayama0302061,rekalo2630}.
\clearpage
\newpage

\section{Possible mechanisms responsible for the creation of the $\eta^{\prime}$ meson}
\begin{flushright}

\parbox{0.67\textwidth}{
 {\em
   But in fact, even if all writers were honest and plain; even if they never passed off
   matters of doubt upon us as if they were truths, but set forth everything in good faith;
   nevertheless, since there is hardly anything that one of them says but someone else asserts 
   the contrary, we should be continually uncertain which side to believe. It would be no good
   to count heads, and then follow the opinion that has most authorities for it; for if the
   question that arises is a difficult one, it is more credible that the truth 
   of the matter may have been discovered by few men than by many~\cite{descartes2}. \\
 }
 \protect \mbox{} \hfill Ren\'{e} Descartes \protect\\
 } 
\end{flushright}

In case of the $\eta^{\prime}$  meson
the investigations of the mechanisms
underlying the production process are even less advanced than 
these of $\eta$ and $\pi$.
This is partly due to the fact that there is no well established 
baryonic resonance decaying into $\eta^{\prime} N$ channel,
which could constitute a doorway state for the $\eta^{\prime}$ 
production, and hence there is no mechanism whose strength could 
be regarded as dominant allowing for the neglect of other possibilities
at least in the first order. 
There is also not much known about the
$\eta^{\prime} NN$ and other relevant coupling constants, 
making impossible inferences about the relative strength 
of different mechanisms on the level achieved in the understanding
of the $\pi$ meson creation~\cite{christoph,machnerR231}.
It is also partly because the experimental database
is much poorer than this for $\eta$ and $\pi$ mesons.
This holds not only for the data on the $\eta^{\prime}$ production
in the collisions of nucleons, but also for the data at relevant energies
on nucleon-nucleon elastic scattering. 
The latter are needed for the estimation of the reduction  of the 
production cross section due to the interaction between nucleons 
in the initial state. It must be stated, however, 
that in the recent years the 
database 
--~relevant for the production of the $\eta^{\prime}$ meson~--
for the proton-proton reactions~\cite{altmeier,moskal416,hibou41,balestra29}
improved significantly, yet 
measurements with a corresponding accuracy for the proton-neutron channel 
are still missing.

\subsection{Meson exchange models}
\begin{flushright}
\parbox{0.73\textwidth}{
 {\em
\hspace{-0.5cm}{\bf Salviati.} But if, of many computations, not even two came out in agreement,
  what would you think of that?

(...)

\hspace{-0.5cm}{\bf Simplicio.}  If  that is how matters stand, it is truly a serious defect~\cite{galileo}. \\
 }
 \protect \mbox{} \hfill Galileo Galilei
 }
\end{flushright}

For the $\eta$ meson case a contribution from the mesonic 
current where the meson is created in the fusion of virtual e.g.\ 
$\rho$ or $\omega$ mesons emitted from both colliding nucleons
is by a factor of thirty weaker 
in comparison to the overwhelming strength of the resonance 
current~(figure~\ref{nakayama_eta}). 
In contrast, the mesonic current suffices to explain the magnitude of the close-to-threshold 
$\eta^{\prime}$ meson production in the proton-proton interaction, which is 
just by about a factor of thirty smaller compared to the $\eta$ meson.
Among the mesonic currents regarded by authors of 
reference~\cite{nakayama024001} the $\rho\rho\eta^{\prime}$ gives the dominant 
contribution, by a factor of five larger than the $\sigma\eta\eta^{\prime}$- 
and $\omega\omega\eta^{\prime}$-exchange. 
However, the understanding of the $\eta^{\prime}$ production on the hadronic 
level is far from being satisfactory. 
The magnitude of the total cross section was also well reproduced in the frame 
of a one-boson-exchange model (nucleonic current) where the virtual boson 
($B = \pi, \eta, \sigma, \rho, \omega, a_0$) created on one of the colliding 
protons converts to the $\eta^{\prime}$ on the other one~\cite{gedalin471}. 
Taking into account the off-shell effects of the $B p \rightarrow 
\eta^{\prime} p$ amplitude it was found that the short range $\sigma$ and 
$\rho$ meson exchanges dominate the creation process, whereas the $\pi$ 
exchange plays a minor role. 
On the contrary, other authors~\cite{sibirtsev333} reproduced the magnitude of 
the total cross section with $\pi$ exchange only and concluded that the 
$\eta$, $\rho$, and $\omega$ exchange currents either play no role or cancel 
each other. 
However, in both of the above quoted 
calculations~\cite{gedalin471,sibirtsev333} the initial state interaction 
between protons, which reduces the rate by a factor of about 3, was not taken 
into account and hence the obtained results would in any case reproduce only 
$30\,\%$ of the entire magnitude of the cross section and could be at least qualitatively 
reconciled with the mentioned result of reference~\cite{nakayama024001}, where 
the nucleonic current was found to be small.
Moreover, the choice of  another prescription for the form factors could 
reduce the one-pion-exchange contribution 
substantially~\cite{nakayama024001}. 
However, the picture that the $\eta^{\prime}$ meson is predominantly created 
through the mesonic current, 
remains at 
present unclear as well. 
This is because the magnitude of the total cross section could have also been 
described assuming that the production of $\eta^{\prime}$ is 
resonant~\cite{nakayama024001}. 
As possible intermediary resonances the recently reported~\cite{ploetzke555} 
$\mbox{S}_{11}(1897)$ and $\mbox{P}_{11}(1986)$ have been considered.
These resonances were deduced from $\eta^{\prime}$ photoproduction data, under 
the assumption that the close-to-threshold enhancement observed for the 
$\gamma p \rightarrow \eta^{\prime} p$ reaction can be utterly assigned to 
resonance production. 
Further, as well strong, assumptions have been made in the derivation of the 
$g_{N N^* \eta^{\prime}}$ and $g_{N N^* \pi}$ coupling 
constants~\cite{nakayama024001}.

Hence, it is rather fair to state that in the case of the close-to-threshold 
$\pi^0$ and $\eta$ meson production in nucleon-nucleon collisions the dynamics 
is roughly understood on the hadronic level, but the mechanisms leading to the 
$\eta^{\prime}$ creation are still relatively unknown.

Until now it has not been possible to satisfactorily estimate the relative 
contributions of the nucleonic, mesonic, and resonance current to the 
production process. 
In fact, model uncertainties allow that each one separately could describe the 
absolute values of the $pp \rightarrow pp \eta^{\prime}$ total cross section.
This rather pessimistic conclusion calls for further experimental and 
theoretical research.  
The understanding of the production dynamics of the $\eta^{\prime}$ meson on 
both hadronic and the quark-gluon level is particularly important since its 
wave function comprises a significant gluonic component~\cite{ball367}, 
distinguishing it from other mesons and hence the comprehension of the 
mechanism leading to its creation in collisions of hadrons may help 
to determine its quark-gluon structure. That, in turn will be helpful when
investigating possible glueball candidates~\cite{ball367}. 
Hereafter we will briefly report of the gluonic mechanisms which may 
--~along with the meson exchange processes discussed above~-- contribute to 
the $\eta^{\prime}$ production.\vspace{1ex}

\subsection{Approaches on the quark-gluon level}
\begin{flushright}
\parbox{0.73\textwidth}{
 {\em
   We must clearly acquire knowledge of factors that are primary. For we claim to know a thing only
   when we believe that we have discovered what primarily accounts for its being~\cite{arystotelesmetaphysics}.\\
 }
 \protect \mbox{} \hfill  Aristotle \protect\\
 }
\end{flushright}

As afore mentioned, the close-to-threshold production of $\eta$ and 
$\eta^{\prime}$ mesons in the nucleon-nucleon interaction requires a large 
momentum transfer between the nucleons and hence can occur only at distances 
of about $0.3\,\mbox{fm}$~(see table~\ref{momtranstable}). 
This suggests that the quark-gluon degrees of freedom may indeed play a 
significant role in the production dynamics of these mesons. 
A possibly large glue content of the $\eta^{\prime}$ and the dominant 
flavour-singlet combination of its quark wave function may cause that the 
dynamics of its production process in nucleon-nucleon collisions is 
significantly different from that responsible for the production of other 
mesons.
In particular, the $\eta^{\prime}$ meson can be efficiently created via a 
``contact interaction'' from the glue which is excited in the interaction 
region of the colliding nucleons~\cite{bass286}.
A gluon-induced contact interaction contributing to the close-to-threshold 
$pp \rightarrow pp \eta^{\prime}$ reaction derived in the frame of the 
U(1)-anomaly extended chiral Lagrangian is discussed in 
references~\cite{bass286,bass429,bass348}.
The strength of this contact term is related to the amount of spin carried by 
polarized gluons in a polarized proton~\cite{bass348,bass17}, thus making the 
study of the close-to-threshold $\eta^{\prime}$ meson production even more 
interesting.
\vspace{0.9cm}
\begin{figure}[H]
\vspace{-0.6cm}
\hfill
\parbox{0.6\textwidth}
  {\epsfig{file=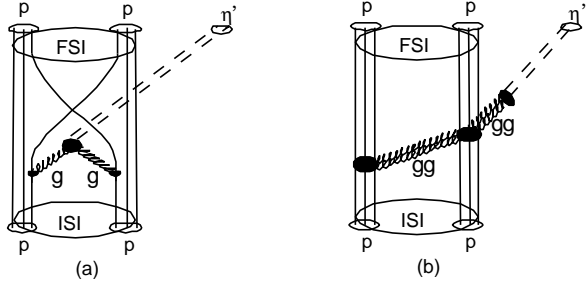,width=0.6\textwidth}} \hfill
\parbox{0.35\textwidth}
  {\caption{\label{graf_glue} Diagrams depicting possible quark-gluon 
  dynamics of the reaction $pp \rightarrow pp \eta^{\prime}$. 
  (a) production via a fusion of gluons~\cite{kolacosynews} with 
  rearrangement of quarks.
  (b) production via a rescattering of a ``low energy pomeron''.}}
\end{figure}
\vspace{1.2cm}
\begin{figure}[H]
\begin{center}
\epsfig{file=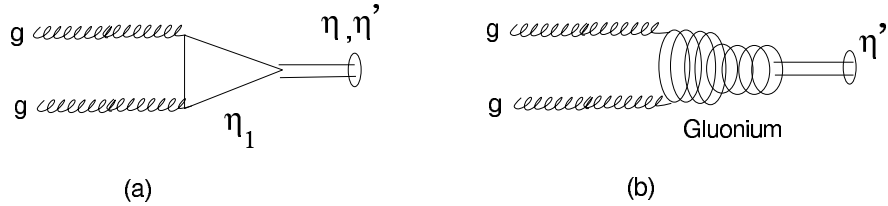,width=0.9\textwidth}
\end{center}
\vspace{-0.5cm}
\caption{\label{triangle} Coupling of $\eta$ and $\eta^{\prime}$ to two gluons 
through 
(a) quark and antiquark triangle loop and 
(b) gluonic admixture. 
$\eta_{1}$ denotes the flavour-singlet 
quark-antiquark state. This indicates that gluons may convert into the $\eta$ 
or $\eta^{\prime}$ meson via a triangle quark loop only by coupling through 
their flavour singlet part. The figure is taken from 
reference~\protect\cite{kou054027}.}
\end{figure}
\vspace{0.3cm}
Figure~\ref{graf_glue} depicts possible short-range mechanisms which may lead 
to the creation of the $\eta^{\prime}$ meson via a fusion of gluons emitted 
from the exchanged quarks of the colliding protons~\cite{kolacosynews} or via 
an exchange of a colour-singlet object made up from glue, which then 
re-scatters and converts into $\eta^{\prime}$~\cite{basspriv}.
The hadronization of gluons to the $\eta^{\prime}$ meson may proceed directly 
via its gluonic component or through its overwhelming flavour-singlet 
admixture~$\eta_1$ (see fig.~\ref{triangle}).
Contrary to the significant meson exchange mechanisms and the fusion of gluons 
of figure~\ref{graf_glue}~graph~a), the creation through the colour-singlet 
object proposed by S.D.~Bass (graph~\ref{graf_glue}b) is isospin independent, 
and hence should lead to the same production yield of the $\eta^{\prime}$ 
meson in both reactions ($pp \rightarrow pp \eta^{\prime}$ and $pn \rightarrow 
pn \eta^{\prime}$) because gluons do not distinguish between flavours.
This property should allow to test the relevance of a short range gluonic 
term~\cite{bassproc} by the experimental determination of the cross section 
ratio $R_{\,\eta^{\prime}} = \sigma(pn \rightarrow pn \eta^{\prime})/
\sigma(pp \rightarrow pp \eta^{\prime})$, which in that case should be close 
to unity after correcting for the final and initial state interaction between 
participating baryons.
The other extreme scenario -- assuming the dominance of the isovector meson 
exchange mechanism -- should result in the value of $R_{\,\eta^{\prime}}$ 
close to 6.5 as had already been established in the case of the $\eta$ meson~\cite{calen2667}.
Perspectives of the experimental investigations aiming to determine $R_{\,\eta^{\prime}}$
and the discussion to what extent these studies may help to establish
contributions from quarks and gluons in the $\eta^{\prime}$ meson
will be presented in section~\ref{isospindegrees}.
In addition to the 
investigation presented 
above~\cite{bass286,bass429}, the interesting features of the 
close-to-threshold meson production, in particular the large cross section 
for the $\eta$ meson in proton-proton interactions exceeding the one of the 
pion, or even more surprisingly large cross section of the $\eta$ 
production in proton-neutron collisions, encouraged also other authors to 
seek the underlying -- OZI rule violating -- creation mechanisms in the 
frame of microscopic models of QCD~\cite{koc00,dil02,kleefeld2867}.
The hitherto regarded processes are presented in figure~\ref{dillig_gluons}.
As indicated in the pictures the structure of participating baryons has been 
modeled as quark-diquark objects with harmonic confinement~\cite{dillig050}.
In the upper graphs the large momentum transfer is shared by the exchanged 
gluons with a subsequent interchange of quarks to provide a colourless object 
in the final state.
The lower graphs depict the two examples of instanton induced interactions 
with a 6-quark-antiquark (fig.~\ref{dillig_gluons}d) and two-gluon vertex 
(fig.~\ref{dillig_gluons}e).
Adjusting the normalization to the cross section of the $pp \rightarrow 
pp \pi^0$ reaction at a single energy point the model~\cite{dil02} accounts 
roughly for close-to-threshold cross sections of other pseudoscalar and 
vector mesons in proton-proton collisions. 
\begin{figure}[H]
\vspace{-0.6cm}
\begin{center}
\epsfig{file=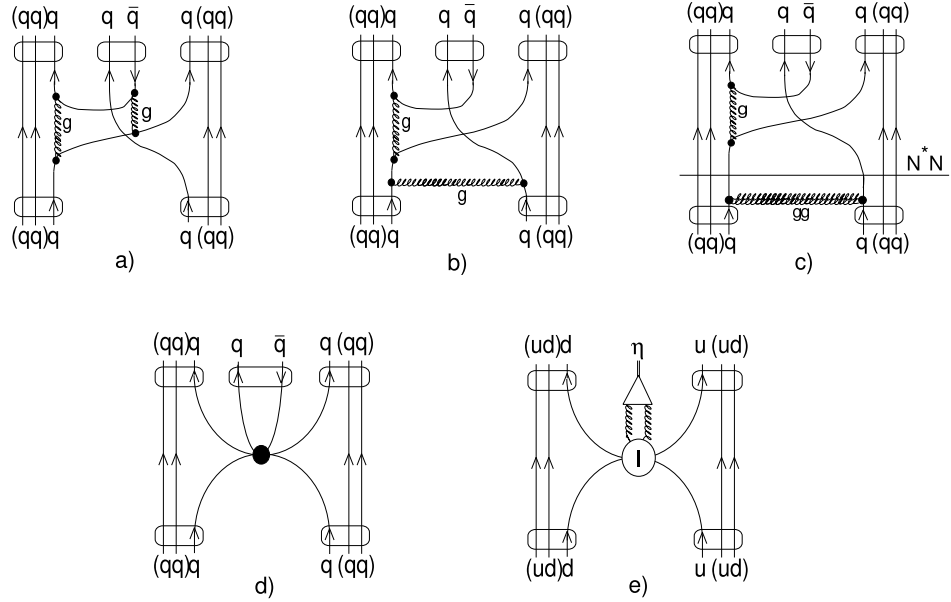,height=0.96\textwidth,angle=-90}
\end{center}
\vspace{-0.3cm}
\caption{\label{dillig_gluons} Diagrams for the $qq \rightarrow qq (q\bar{q})$ 
production operator: \hspace{1ex}
(a), (b) Two-gluon exchange and rescattering mechanism \hspace{1ex}
(c) Correlated, colourless two-gluon exchange. The dashed line indicates the 
excitation of an intermediate $N N^*$ system. \hspace{1ex}
(d) Instanton induced 6-quark interaction. Figures (a-d) according 
to~\cite{dil02}. \hspace{1ex}
(e) The instanton contribution to the $\eta$ meson production in the 
proton-neutron interaction. Figure copied from~\cite{koc00}.}
\end{figure}
Though this approach is characterized by a significantly smaller number of 
parameters than in the meson exchange models, their uncertainties allow for 
the description of the data equally well, either by the gluon exchange or by 
the instanton induced interactions, at least in case of the $\pi^0$, 
$\eta^{\prime}$, $\omega$, and $\phi$ mesons. 
Yet for the resonance dominated $\eta$ and $K^+$ meson production in 
proton-proton collisions it was found that the instanton induced interaction 
presented by graph~\ref{dillig_gluons}d is not sufficient.
Similarly, the authors of reference~\cite{koc00} argue that the instanton 
induced interaction with a quark-gluon vertex (graph~\ref{dillig_gluons}e) 
should be of no importance for the $\eta$ production in proton-proton 
collisions. 
This arises from the properties of the vertex which lead to the $\eta$ 
production only in case of an interaction between quarks of different flavours 
($ud \rightarrow udgg$) and in the quark-diquark model of baryons the proton 
consists of $ud$-diquark and $u$-quark and correspondingly the neutron is 
modeled as $ud$-diquark and $d$-quark.
Although negligible in case of proton-proton interactions, this mechanism may 
contribute significantly to the total cross section of the  $pn \rightarrow 
pn \eta$ reaction and indeed as shown in reference~\cite{koc00} it reproduces 
the data. 
However, the magnitude of the cross section is very sensitive to the size of 
the instanton in the QCD vacuum and -- similar to the uncertainty of 
coupling constants in case of the hadronic approach -- at present it makes 
precise predictions impossible.
Thus, despite the partial successes, at the present stage of developments both 
approaches --~on the quark-gluon and on the hadronic level~-- do not 
provide an unambiguous answer to the dynamics of the close-to-threshold 
meson production in the nucleon-nucleon interaction.

 \section{Exploration of isospin degrees of freedom}
   \label{isospindegrees}
\begin{flushright}
\parbox{0.73\textwidth}{
 {\em
   When a person is placed between two choices,  the person bends himself
   or herself toward the choice he or she desires~\cite{hildegarda}.\\ 
 }
 \protect \mbox{} \hfill Hildegard von Bingen \protect\\
 }
\end{flushright}
   Treating proton and neutron as  different states of nucleon
   distinguished only by the isospin projection, $+\frac{1}{2}$ for the
   proton and  $-\frac{1}{2}$  for the neutron, we may classify the
   $NN\to NN X$ reactions according to the total isospin of the nucleons pair
   in the initial and final state.  A total isospin of two nucleons
   equals 1 for proton-proton and
   neutron-neutron pairs,
   and  may acquire the value of 1 or 0 for the neutron-proton system.\\
   Since $\eta$ and $\eta^{\prime}$ mesons are isoscalars,
   there are only two pertinent transitions
   for the $NN\to NN X$ reaction, provided that it
   occurs via the isospin conserving interaction.
   These are $\sigma_{00}$ and $\sigma_{11}$, where the first and second subscript
   --~labeling the total cross section~--
   describe the total isospin of the nucleon pair
   in the initial and final state, correspondingly.

   It is thus enough to measure two reaction channels
   for an unambiguous determination of isospin 0 and 1 cross sections.
   In particular for the total cross section the following
   relations are satisfied:
   \begin{equation}
     \label{sigma11}
     \sigma_{pp\to pp \eta^{\prime}} = \sigma_{11},
   \end{equation}
   and
   \begin{equation}
     \label{sigma00}
     \sigma_{pn\to pn \eta^{\prime}} = \frac{1}{2}(\sigma_{00} + \sigma_{11}).
   \end{equation}
   In the case of the proton-neutron reaction the total cross sections
   add incoherently since the different isospin of the $NN$ pairs implies
   that either total spin or angular momentum  of these pairs must differ.
   This is because the $NN$ system is bound to satisfy the Pauli Principle.
   Therefore,  different isospin states
   do not interfere as far as the
   total cross section is concerned~\cite{christoph}.

   It is needless to mention, that
   experimental determination of cross sections for different
   isospin configurations allows for deeper insight into the production mechanism.
   In sections~\ref{Sitd} and~\ref{nnnnn} we will present how one can conduct the study
   of the meson production in the proton-neutron and neutron-neutron reactions
   and in the next section we will give an example of inferences
   which may be drawn from  the comparison
   of the total cross sections for the  $pp\to pp\eta^{\prime}$
   and $pn\to pn\eta^{\prime}$ reactions.

\subsection{Glue content of the $\eta^{\prime}$ meson}
\begin{flushright}
\parbox{0.73\textwidth}{
 {\em
   ...\!\! even God does not know his own nature~\cite{mageeconfess}. \\
 }
 \protect \mbox{} \hfill Bryan Magee \protect\\
 }
\end{flushright}
\begin{flushright}
\parbox{0.73\textwidth}{
 {\em
   Experience no doubt teaches us that this or that object is constituted in such and such
   a manner, but not that it could not possibly exist otherwise~\cite{kantcritique}. \\
 }
 \protect \mbox{} \hfill Immanuel Kant \protect\\
 }
\end{flushright}

The most remarkable feature --~in the frame of the quark model~--
distinguishing the $\eta^{\prime}$ meson from 
all other pseudscalar and vector ground state mesons, is the fact,
that the $\eta^{\prime}$ is predominantly 
a flavour-singlet combination  of quark-antiquark pairs
and therefore can mix with  purely gluonic states.
A comprehensive discussions on the issue of how one can experimentally 
determine the differences in the quark-gluon
structure between $\eta^{\prime}$ and other mesons can be found in the proceedings
of the International Workshop on the Structure of the $\eta^{\prime}$ Meson
held in Las Cruces~\cite{etaprimeworkshop}. Unfortunately, except for
the gedanken experiments,  no experimentally feasible test  
allowing for the model-independent conclusions was suggested.

Due to the short life-time of this meson it is impossible to use it as a beam 
or target. This fact entails that the study must be performed by measurements of the
reactions where the $\eta^{\prime}$ meson is created in the collisions or decays
of more stable particles, and preferentially the studied observables should
depend --~in a predictive manner~--
on properties of the $\eta^{\prime}$ meson.

In this section we will argue that 
a comparison of the close-to-threshold total cross sections for the $\eta^{\prime}$
meson production in both the $pp\to pp\eta^{\prime}$ and $pn \to pn \eta^{\prime}$ reactions 
should provide  insight into the flavour-singlet (perhaps also into gluonium)
content of the $\eta^{\prime}$ meson and the relevance of quark-gluon
or hadronic degrees of freedom in the creation process.
It is of great credit to S.~D.~Bass
for this idea to be shared with us~\cite{basspriv}.
However, prior to the formulation of the main conjectures we will recall the 
most important facts of the production mechanism 
necessary for the understanding of the further conclusions.

Close-to-threshold production of $\eta$ and $\eta^{\prime}$ mesons
in the nucleon-nucleon interaction requires a large momentum 
transfer between the nucleons and occurs at distances
in the order of $\sim$0.3~fm.  
This implies that the quark-gluon 
degrees of freedom may 
play a significant role in the production dynamics of these mesons.
Therefore, additionally to the mechanisms associated with meson
exchanges it is possible that the $\eta^{\prime}$ meson is created from excited glue
in the interaction region of the colliding nucleons~\cite{bass286,bass348}, 
which couple to the $\eta^{\prime}$ meson directly via its gluonic
component or through its SU(3)-flavour-singlet admixture. The production through the
colour-singlet object as suggested in reference~\cite{bass286} is isospin independent
and should lead to the same production yield of the
$\eta^{\prime}$ meson in the $pn\to pn\ gluons \to pn\eta^{\prime}$ 
and                          $pp\to pp\ gluons \to pp\eta^{\prime}$ reactions
after correcting for the final and initial state interaction between the nucleons.
 
Investigations of the $\eta$-meson production  in collisions of nucleons
allowed to conclude that, close to the kinematical threshold,
the creation of $\eta$ meson from isospin I~=~0 
exceeds the production with I~=~1
by about a factor of 12. This  was derived from the measured ratio of the 
total cross sections for the reactions $pn \to pn \eta$ and 
$pp \to pp\eta$ ($R_{\eta} = \frac{\sigma{(pn\to pn\eta})}{\sigma{(pp\to pp\eta)}}$), 
which was determined to be $R_{\eta}\approx 6.5$ in the excess energy 
range between 16~MeV and 109~MeV~\cite{calen2667}.
The large difference of the total cross section
between the isospin channels
suggests the dominance of
isovector meson ($\pi$ and $\rho$) exchange
in the creation of $\eta$ in nucleon-nucleon collisions~\cite{wilkinjohansson,calen2667}.
\vspace{0.0cm}
\begin{figure}[t]
  {\epsfig{file=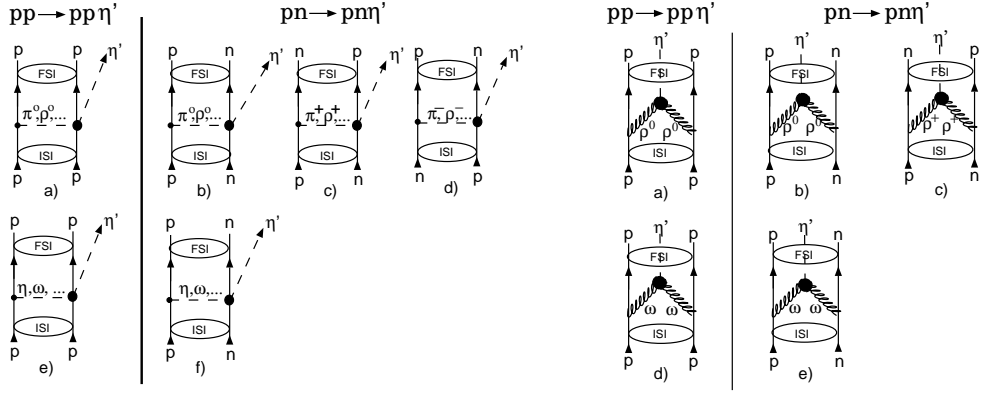,width=0.41\textwidth,angle=270}}  
  
  {\caption{\label{graf_isospin1} ({\bf left}) Example of diagrams 
  with the isovector (upper row)
  and isoscalar (lower row) 
  meson exchange leading to the creation of the meson
  $\eta^{\prime}$ in the proton-proton and proton-neutron collisions.
  ({\bf right})
  Fusion of the virtual $\omega$ (isoscalar) and $\rho$ (isovector) mesons
  emitted from the colliding nucleons. }}
\end{figure}
\vspace{0.0cm}
Since the quark structure of $\eta$ and $\eta^{\prime}$ mesons is very similar
we can --~by analogy to the $\eta$ meson production~-- 
expect that in the case of  
dominant isovector meson exchange the ratio $R_{\eta^{\prime}}$  
should also be about 6.5. If, however, the $\eta^{\prime}$ meson was produced via its 
flavour-blind gluonic component from the colour-singlet glue excited
in the interaction region, the ratio $R_{\eta^{\prime}}$ should approach unity
after corrections for the interactions between the participating baryons.

Figure~\ref{graf_isospin1} demonstrates  
qualitatively the fact that the production of mesons in the proton-neutron
collisions may be  more probable 
than in the proton-proton interaction if it
is driven by the isovector meson exchanges only. 
This is because in the case of the proton-neutron collisions there are always more possibilities
to realize an exchange or fusion of the
isovector mesons than in the case of the reaction of protons. 
More precisely, for the one meson exchange diagrams, 
when considering $\pi$, $\eta$, $\rho$, and $\omega$ mesons 
and neglecting initial and final state interaction,
the matrix element has the following structure~\cite{wilkinjohansson}:
\begin{equation}
  \nonumber
  |M(pn\to pn\eta)|^2 = \frac{1}{2}\left[ 
              |T_{\pi} + T_{\eta} - T_{\rho} - T_{\omega}|^2 
            + |3T_{\pi} - T_{\eta} + 3T_{\rho} - T_{\omega}|^2
            \right];
\end{equation}
\begin{equation}
  \label{isocolin}
  \hspace{-5.4cm}|M(pp\to pp\eta)|^2 = | T_{\pi} + T_{\eta} - T_{\rho} - T_{\omega}|^2 
\end{equation}
where the amplitudes are labeled  by the names of the exchanged mesons.
In general, when regarding the isospin structure of the exchange current,
as shown in reference~\cite{christoph}, if the production operator
accounts for the isoscalar meson exchange then $\sigma_{00}$~=~$\sigma_{11}$,
and in case of isovector exchange $\sigma_{00}$~=~$9 \sigma_{11}$.
This is also clear from the particular case represented by equations~\ref{isocolin},
when combined with formulae~\ref{sigma11} and~\ref{sigma00}.
Exploiting these  equations
we can estimate that, if the production of $\eta^{\prime}$ 
meson via the $pp\to pp\eta^{\prime}$ was governed only by the exchange of the
isovector mesons  the ratio $R_{\eta^{\prime}}$ would be equal to five,
and in the case of isoscalar mesons it would be equal to unity. 

A very important result
of the theoretical investigation,
relevant for further consideration,
is that regardless of whether it is a mesonic, nucleonic, or resonance  current
the contribution from the exchange of isovector mesons ($\rho$ or $\pi$) 
is much larger from that of isoscalar 
ones~($\omega$ or $\eta$)~\cite{nakayama024001,wilkinjohansson,vadim024002}.
This unequivocally entails that if the 
ratio $R_{\eta^{\prime}}$ --~corrected for FSI and ISI distortions~--
will be found to be close to unity we will have a clear indication
that the $\eta^{\prime}$ is produced directly by  gluons.
Gluons, as we discussed in chapter~\ref{Dopsmp},
may hadronize to $\eta^{\prime}$ either via its $SU(3)_{F}$ 
flavour-singlet component or via 
its gluonic content.  In order to disentangle these two effects 
a substantial theoretical input is required. The benefit of the invested effort will 
be the determination of the quark-gluon structure 
of the $\eta^{\prime}$ meson.
If, however, the measured ratio $R_{\eta^{\prime}}$ will not be equal to one,
a quantitative determination of the contribution from gluonic mechanism
to the production process will require a better understanding of the 
meson exchange currents. 
Then if the contribution of meson exchange currents 
is once understood we will also be able to infer the one from gluons.
The dynamics of the meson production 
in both hadro- and photo-production is at present vigorously
studied. The recent results are reported 
e.g. in references~\cite{kanzoheber,review,christoph,krusche399}.

The close-to-threshold 
excitation function for the $pp\to pp\eta^{\prime}$ 
reaction has been already determined~\cite{moskal3202,moskal416,hibou41,balestra29}
whereas the total cross section for 
the $\eta^{\prime}$ meson production in the proton-neutron 
interaction remains unknown. 
As a first step towards the determination of the value of $R_{\eta^{\prime}}$
the feasibility of the measurement of the $pn\to pn\eta^{\prime}$ 
reaction by means of the COSY-11 facility
was studied using a Monte-Carlo method~\cite{moskal0110001,rafalmgr}. As a second step, 
a test experiment of the $pn\to pn\eta$ 
reaction --~suspected  to have by at least a factor of  
thirty larger cross section than the one
for the $pn\to pn\eta^{\prime}$ reaction~--
was performed. 
  In this test measurement,
  using a beam of stochastically cooled protons 
  and a deuteron cluster target,
  we have proven the ability of the COSY-11 facility to study 
  the quasi-free creation of mesons via the $pn\to pn X$ reaction.
Appraisals of simulations~\cite{moskal0110001,rafalmgr} 
and preliminary  results of the tentative measurements 
of the quasi-free $pn\to pn\eta$
reaction performed using the newly extended COSY-11 
facility~\cite{hadronmoskal,proposalc11,proposaljoanna}
will be presented in section~\ref{testpnpneta}.

\vspace{-0.4cm}
\section{Transition probability as a function of particles' ``virtuality''}
    \label{virtuality}
    \begin{flushright}
       \parbox{0.73\textwidth}{
          {\em
            Es gibt keine freie Masse, genauso wie es kein leeres Vakuum gibt~\cite{kilian5k}.\\
          }
          \protect \mbox{} \hfill Kurt Kilian  \protect\\
       }
    \end{flushright}
     \vspace{-0.4cm}
     First of all the title of this section requires apology and the explanation
     what is meant under the notion of virtuality. 
     Usually we consider and measure the reactions of free particles
     for which the difference between the squared energy and momentum
     is equal to the square of its mass.
     This is not the case for the nucleons bound inside the nuclei,
     whose masses may differ from the mass of the free particles.
     In particular, in case of the deuteron,
     the entire difference (summed for both nucleons) amounts to at least
     a value equivalent to the binding energy.
     The inner motion of nucleons inside a nucleus makes this
     difference even larger. Thus principally the virtual nucleons inside
     a deuteron may not be identical with their free equivalent.
     By virtuality we would like to name the difference between
     the mass of the real particle and the mass calculated as a difference between
     squared energy and squared momentum of its virtual counterpart.
     When using a deuteron as a source of neutrons for the measurement of e.g.
     proton-neutron reactions
     this difference is usually regarded as an obstacle making difficult
     a direct comparison of free and quasi-free scattering.
     Here we would like
     to suggest how to take advantage of this fact.
     Namely we intend to make a systematic study of the dependence of the
     cross sections for a meson production in nucleon-nucleon collisions
     as a function of their virtuality, which in the first order, if a deuteron is
     used as a target,
     can  be deduced from the momentum of the nucleon
     which does not take part in the reaction.

     There is a difference between the free and bound nucleon since
     they have different masses. Hence it is natural to expect that they
     will also differ in other aspects. Recently a chance appeared to 
     realise such studies at the WASA facility~\cite{zabierowski159},
     which is planned to be installed at COSY in the near future, and the corresponding
     letter of intent has been already prepared~\cite{LoImoskal}.

\section{Natural width of mesons~--~does it depend on the momentum transfer?}
\begin{flushright}
\parbox{0.57\textwidth}{
 {\em
  Throughout the history of spectroscopy, improved resolution has led to discoveries of finer
  and finer structures~\cite{maglichhadron}.\\
 }
 \protect \mbox{} \hfill  Bogdan Castle Maglich \protect\\
 }
\end{flushright}

Although only loosly connected to the subject of this monograph, we would like to comment
on a very recent and highly interesting hypothesis announced
by Maglich at the HADRON'03 conference in Aschaffenburg~\cite{maglichhadron}.
Based on the experience gained from measurements of the $a_1$ and $a_2$ mesons
via the $\pi^{-} p \to p X^{-}$ reactions,
he postulated that the width of mesons may be a function of the momentum transfer $|t|$,
and in particular that the mesons width grows with decreasing value of $|t|$.
According to his idea the intrinsic widths of all meson states are narrow and can
be observed only at high momentum transfers $|t| > 0.2$. Since the production
cross sections are inversely proportional to $|t|$, the experimental
data samples are dominated by events corresponding to low values of the momentum transfer,
and therefore the observed widths of mesons appear 
to be broad~\cite{maglichhadron}~\footnote{
An alternate explanation proposed after completion of this treatise can be found in reference~\cite{maglichc11}.
}.

Nothing can be a priori excluded, however we would like to indicate that the broadening
of the width of mesons  may only be an apparent effect
caused by the error propagation through the formula exploited for the missing mass calculation.
Such effects have been observed at the missing mass spectra
from measurements of the $\eta$ meson at the COSY-11~\cite{moskal025203}
and TOF~\cite{TOFeta} facilities.
\begin{figure}[H]
\begin{center}
\vspace{-2.0cm}
\epsfig{file=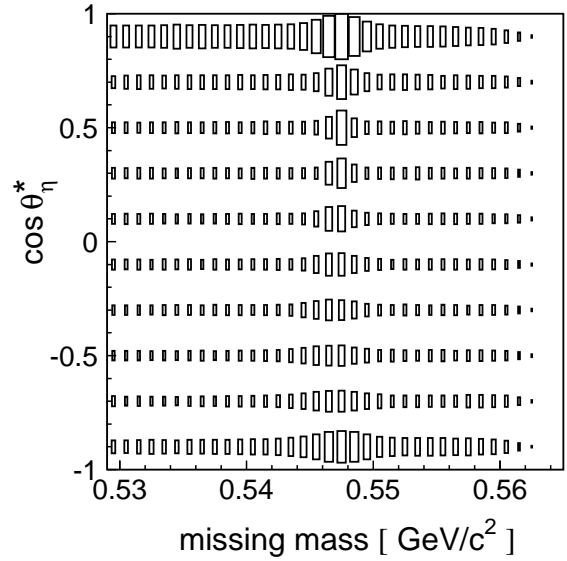,width=0.7\textwidth}
\vspace{-1.0cm}
\end{center}
\caption{\label{missodtheta}
  Distribution of the polar scattering angle of the X system created via  the $pp \to ppX$ reaction
  measured at Q~=~15.5~MeV above the threshold for the production of the $\eta$ meson~\cite{moskal367}.
}
\end{figure}
Figure~\ref{missodtheta} presents the dependence of the missing mass
distribution as a function of the polar angle of the emission of the produced system X.
At a mass value corresponding to the mass of the $\eta$ meson
one observes a strip of enhanced population-density over
a smooth multi-pion background (figure~\ref{missall} from section~\ref{detailedsection}
constitutes the projection of this plot onto a mass axis).
Here it is incontestably evident that the width of the signal originating from the reaction $pp\to pp\eta$ is not constant
but rather varies with the polar angle $\Theta^*_{\eta}$. Specifically, the signal from $\eta$ meson is much sharper for
$cos\Theta_{\eta}$~=~0.3 than for $cos\Theta^*_{\eta}$~=~-0.95.  The difference cannot be assigned to the
variation of the width of the $\eta$ meson, since this is three orders of magnitude smaller
($\Gamma_{\eta}$~=~1.18~$\pm$~0.11~KeV).
The effect is also seen by comparison of left and right panel in figure~\ref{invmiss}.

In general the ground-state pseudoscalar mesons are too narrow to study a possible variation of their
width via the missing mass method at any of the up-to-date missing mass spectrometers.
For this purpuse vector mesons would be much better suited,
and in particular mesons $\omega$ and $\phi$ possessing
width of 8.4~MeV and 4.3~MeV, respectively.
In figure~\ref{mmetap} we demonstrated that the COSY-11 detection setup
combined with the stochastically cooled proton beam~\cite{stockhorst} of COSY results in a mass resolution
of about 0.7~MeV (FWHM). In the case of $\phi$ or $\omega$ mesons, such a precision  would enable
to observe effects in the order of 10\%
without necessity of any sophisticated analysis.
Unfortunately the limited acceptance of COSY-11 setup allows for the reliable (model independent)
evaluation of the data only in the vicinity of the kinematical threshold, which in case of the
broad\footnote{\mbox{} For a comprehensive discussion of
a notion of the threshold and the definition of the total cross section in case of broad
resonances the reader is referred to reference~\cite{moskalf0a0}.}
mesons
makes the derivation of their spectral function rather complicated due the strong
variation of the phase space over a resonance range~\cite{moskalf0a0}.
A comparable resolution but with by far higher acceptance shall be obtainable
at the WASA detector. Its installation at cooler synchrotron COSY will allow for investigations
of all ground state pseudoscalar and vector flavour-neutral mesons and we consider to perform
such a study for the $\omega$ meson.

In case the bold hypothesis rised by Maglich~\cite{maglichhadron} were wrong
it can be easily falsified by the high statistics measurements
of the $\omega$ mesons at the upcoming facility WASA@COSY.
If confirmed it would be a sensation.

\newpage
\clearpage
\thispagestyle{empty}
\pagestyle{plain}
\chapter{Experiment}
\thispagestyle{empty}
\pagestyle{myheadings}
\markboth{Hadronic interaction of $\eta$ and $\eta'$ mesons with protons}
         {6. Experiment}
\label{experiment}                          
\begin{flushright}
\parbox{0.73\textwidth}{
 {\em 
   Science is the attempt to discover, by means of observation,
   and reasoning based upon it, first, particular facts about the world,
   and then laws connecting facts with one another and (in fortunate cases)
   making it possible to predict future occurrences~\cite{russelreligion}.\\
 }
 \protect \mbox{} \hfill Bertrand Russell \protect\\
 } 
\end{flushright}

Major part of the experimental results,
constituting the basis for the inferences considered
in this work, has been obtained using the COSY-11 facility.
This chapter will be devoted to the description of this detection system 
placing  emphasis on its use for the determination of the total cross section 
in the case of the
$pp\to pp\eta^{\prime}$ reaction and derivation of the differential cross sections
for the reaction $pp\to pp\eta$.
The description will be focused on the data analysis including the methods 
for the off-line monitoring of the beam and target parameters, model independent
multi-dimensional acceptance corrections,
the procedure of kinematical fitting and the subtraction of the background, especially
in case of
the derivation of the differential cross sections. 

For  details of the functioning of the employed detectors, the description 
methods of their calibration, as well as for the elucidation 
of the hardware event selection and the data acquisition the reader is referred to 
references~\cite{wolkephd,moskalphd}.

\section{Measurement of the total cross section on the example of the $pp\to pp\eta^{\prime}$ reaction}
\label{sectionmeasurement}
\begin{flushright}
\parbox{0.73\textwidth}{
 {\em
   I hold that most observations are more or less indirect, and that it is doubtful whether the distinction
   between directly observable incidents and whatever 
   is only indirectly observable leads us anywhere~\cite{popperconjecture}. \\
 }
 \protect \mbox{} \hfill Karl Raimund Popper \protect\\
 }
\end{flushright}
The COSY-11 facility enables exclusive measurements of the meson production occurring via interaction
of nucleons in the range of the kinematical threshold.
The collisions of protons are realized 
using the cooler synchrotron COSY-J{\"u}lich~\cite{prasuhn167,maie97}
and the $H_{2}$ cluster
target~\cite{dombrowski228,khou96} installed in front of one of the regular COSY dipole magnets,
as shown schematically in figure~\ref{detector}.
The target being  a beam 
of $H_{2}$ molecules grouped inside clusters of up to $10^{5}$ atoms,
crosses perpendicularly  the beam of $\sim 2\cdot 10^{10}$ protons circulating in the ring.
\vspace{-0.0cm}
\begin{figure}[H]
    \parbox{1.0\textwidth}{\vspace{-0.4cm}
    \includegraphics[width=0.94\textwidth]{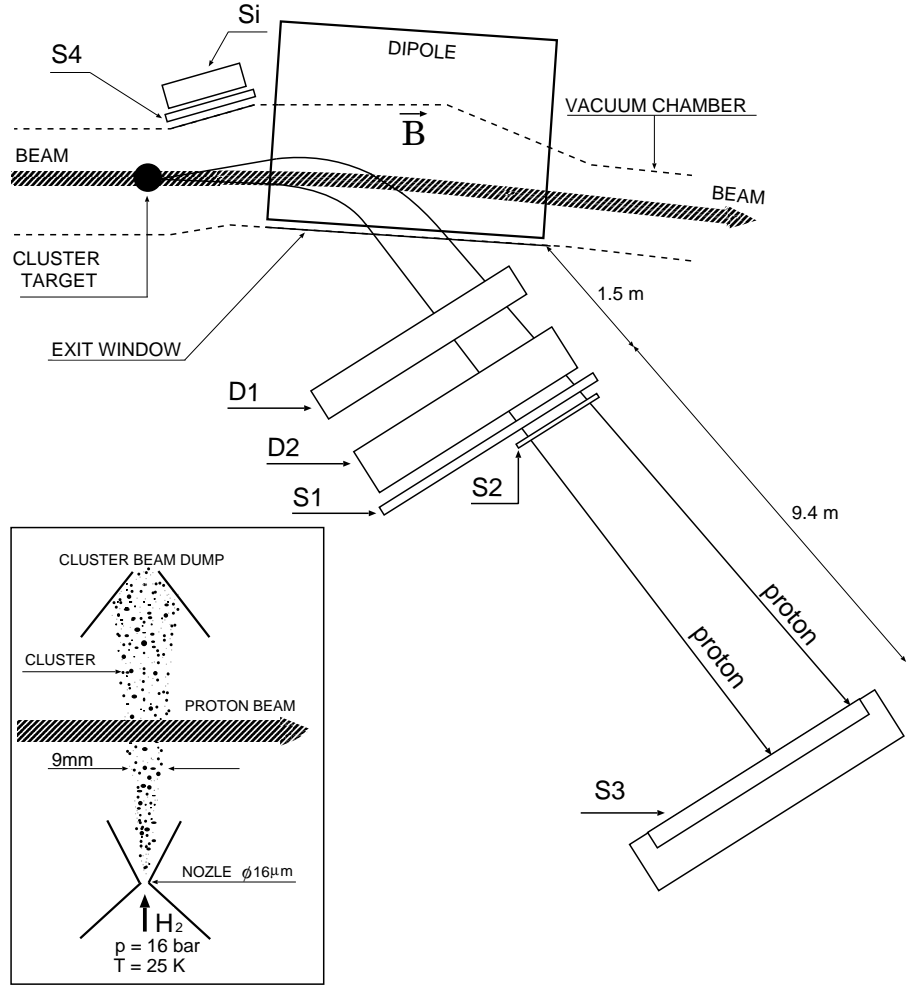}}
    \vspace{-0.3cm}
    \caption{\label{detector}
     Schematic view of the part of the COSY-11 detection setup~\cite{brauksiepe397}.
     The cluster target~\cite{dombrowski228} is located in front 
     of the accelerator dipole magnet. 
     Positively charged particles which leave the
     scattering chamber through the thin exit foil are detected
     in two drift chamber stacks D1, D2~\cite{gugulski} and in the scintillator hodoscopes
     S1, S2~\cite{moskaldiplom}, and S3~\cite{wolkediplom,anton631}.
    Scintillation detector S4 and the 
    position sensitive silicon pad detector Si~\cite{sili}
    are used in coincidence with the S1 counter for the registration  of the elastically
    scattered protons. Elastic scattering is  used for an absolute normalisation 
    of the cross sections of the 
    investigated reactions and for monitoring both the geometrical spread of the proton 
    beam and the position at which  the beam crosses the target~\cite{moskal448}. \ \ \ 
    In the left-lower corner  a schematic view of the interaction region
    is depicted.  The figure has been  adapted from reference~\cite{moskalphd}.
    }
    \vspace{-0.2cm}
\end{figure}  
\vspace{-0.2cm}
The beam of accelerated protons
is cooled stochastically during the measurement cycle~\cite{stockhorst}. 
Longitudinal and vertical cooling enables to keep the circulating beam 
practically without energy losses and without a spread of its dimensions
even during a 60 minutes cycle,
 when passing more than  $10^{6}$ times per second through the 
10$^{14}$~atoms/cm$^{2}$ thick target. 

If at the intersection point of the cluster beam with  the COSY proton beam the collision of protons
results  in the production of a  meson, then
the ejected protons
  -~having smaller momenta than the beam protons~- are separated
 from the circulating beam by the magnetic field.
Further they leave the vacuum chamber through a thin exit foil
 and are registered by the detection system consisting of drift chambers and scintillation
 counters~\cite{brauksiepe397,smyrski182}.
The hardware trigger, based on signals from scintillation detectors, 
was adjusted to register all events with at least two positively charged
particles. Tracing back trajectories from  drift chambers 
through the dipol magnetic field to the target point allowed 
for the determination of the particles momenta.
Having momentum and velocity, the latter measured using scintillation detectors,
it is possible to identify the mass of the particle. Figure~\ref{invmass}a 
shows the squared mass of two simultaneously detected particles.  
\begin{figure}[H]
\vspace{-0.4cm}
  \parbox{0.31\textwidth}{
    \includegraphics[width=0.32\textwidth]{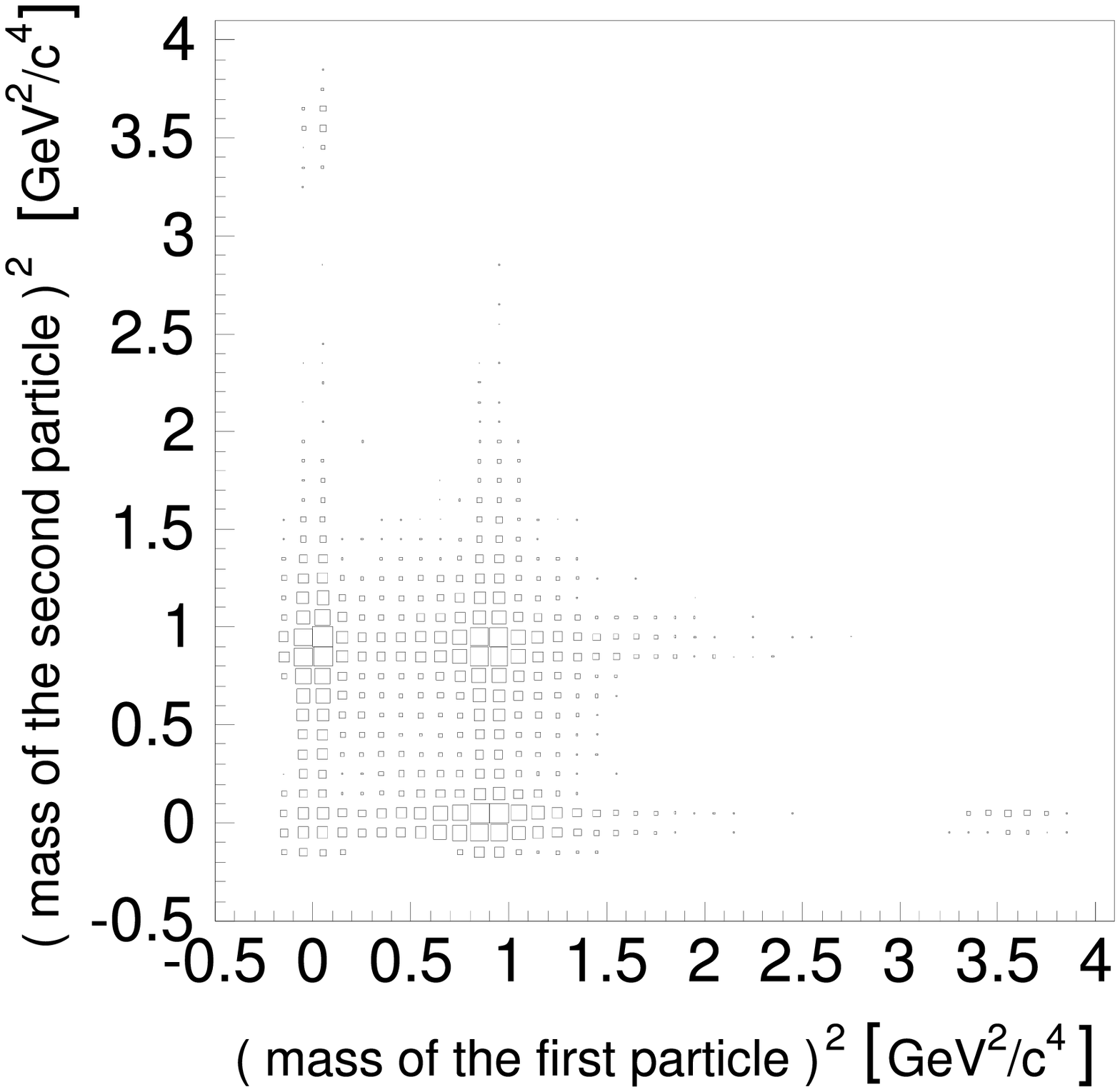}}
  \parbox{0.33\textwidth}{
    \includegraphics[width=0.32\textwidth]{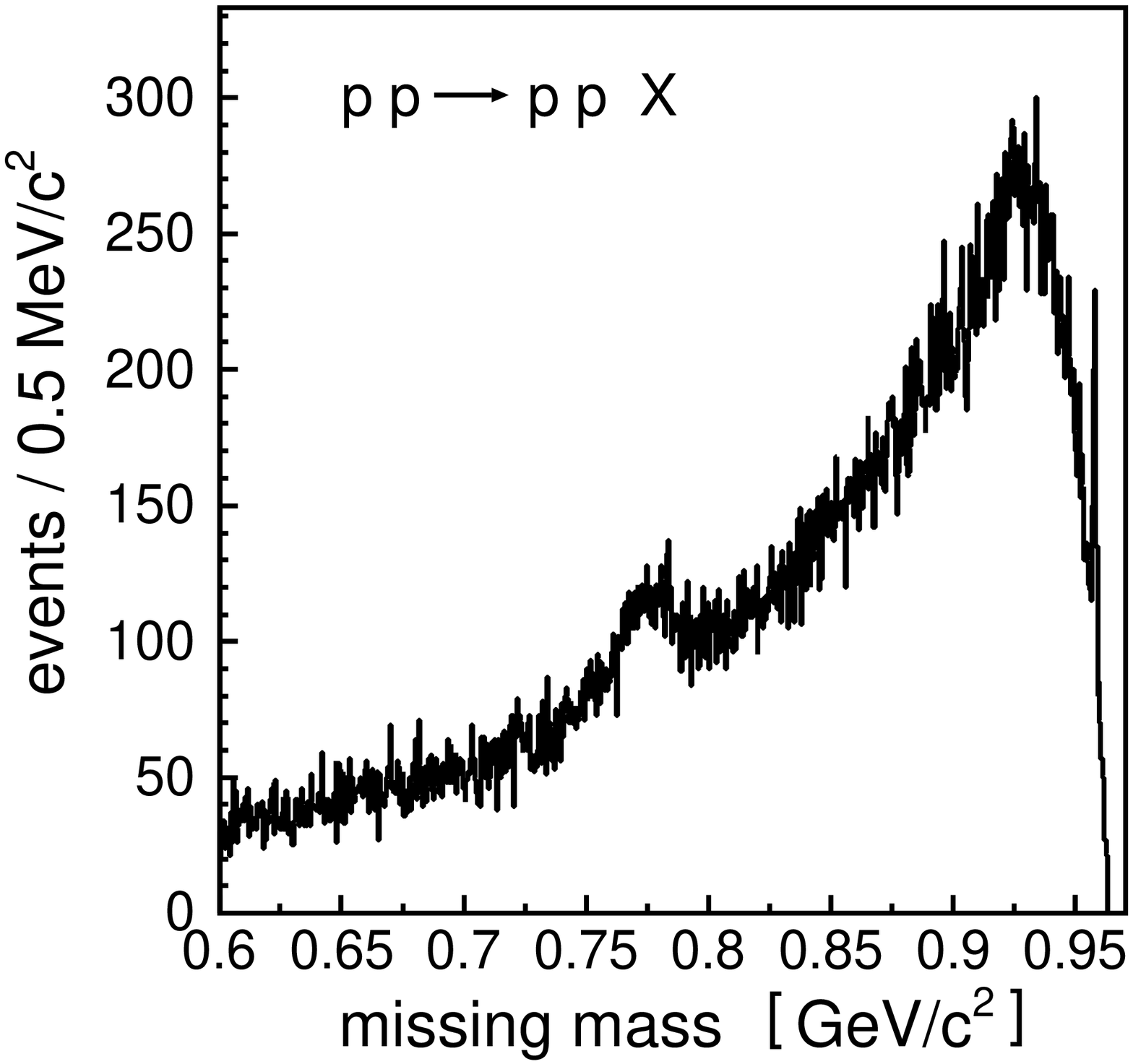}}
  \parbox{0.33\textwidth}{
    \includegraphics[width=0.34\textwidth]{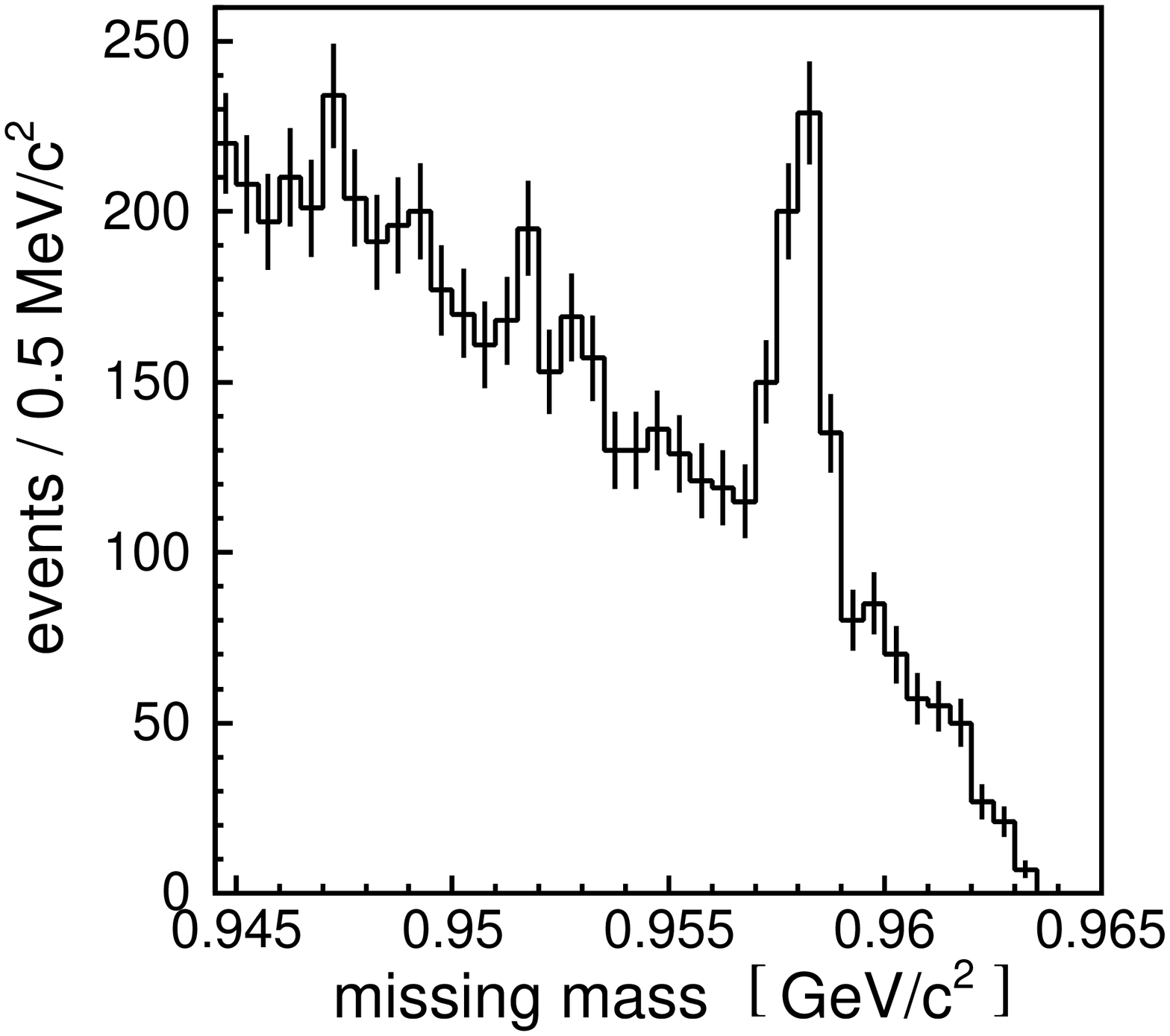}}

     \vspace{0.0cm}
    \parbox{0.270\textwidth}{\raisebox{0ex}[0ex][0ex]{\mbox{}}} \hfill
\parbox{0.240\textwidth}{\raisebox{0ex}[0ex][0ex]{\large a)}} \hfill
\parbox{0.29\textwidth}{\raisebox{0ex}[0ex][0ex]{\large b)}} \hfill
\parbox{0.011\textwidth}{\raisebox{0ex}[0ex][0ex]{\large c)}}
\vspace{-0.35cm}
\caption{
          (a) Squared masses of two positively charged particles measured in coincidence.
          Pronounced peaks are to be recognized when two protons, proton and pion, two pions,
          or pion and deuteron were registered. Note that the number of events is
          shown in logarithmic scale.
          (b)(c) Mass spectrum of the unobserved particle or system of particles
           in the $pp\rightarrow ppX$ reaction
           determined at Q~=~5.83~MeV above the $\eta^{\prime}$ production threshold.
        }
\label{invmass}
\end{figure}
\vspace{-0.3cm}
A~clear separation is seen into groups of events with two protons,
two pions, proton and pion, and also deuteron and pion. This spectrum
enables to select events with two registered protons.
The knowledge of the momenta of both protons before and
after the reaction allows to calculate the mass of an unobserved particle or system of particles
created in the reaction. 
The four-momentum conservation applied to the $pp\to ppX$ reaction
leads to the following expression:
\vspace{-0.2cm}
\begin{equation}
  \nn
  m_{x}^{2}  =  E_{x}^{2} - \vec{P}_{x}^{2} =
 (E_{beam}+E_{target}-E_{1}^{p}-E_{2}^{p})^{2}
  -  |\vec{P}_{beam}+\vec{P}_{target}-\vec{P}_{1}^{p}-\vec{P}_{2}^{p}|^{2},
\end{equation}
where the used notation is self-explanatory\footnote{\mbox{} The missing 
      mass technique was first applied by
      authors of reference~\cite{maglic185}.}.
Figure~\ref{invmass}b depicts the missing mass spectrum obtained for the
$pp \rightarrow pp X$ reaction at an excess-energy  of Q~=~5.8~MeV above the $\eta^{\prime}$
meson production threshold. 
Most of the entries in this spectrum originate 
from the multi-pion production~\cite{moskal3202,moskalphd},
forming a continuous background to the well distinguishable peaks accounting for the creation of $\omega$
and $\eta^{\prime}$ mesons, which can be seen at  mass values of 782~MeV/c$^{2}$
and 958~MeV/c$^{2}$, respectively.
The signal of the $pp\rightarrow pp\eta^{\prime}$ reaction
is better to be seen in the figure~\ref{invmass}c, where the missing mass distribution only in
the vicinity of the kinematical limit is presented. 
Figure~\ref{miss2}a shows the missing mass spectrum
for the measurement at Q~=~7.57  MeV (above kinematical threshold of the $pp\to pp\eta^{\prime}$ reaction)
together with the multi-pion background~(dotted line) as
combined  from the measurements at different excess-energies.
Subtraction of the background leads to the spectrum with a clear signal at the mass of the
$\eta^{\prime}$ meson as shown by the solid line in figure~\ref{miss2}b.
The dashed histogram
in this figure corresponds to the Monte-Carlo simulations where the beam and target conditions
were deduced from the measurements of  elastically scattered protons 
in the manner described in the next section.
The magnitude of the simulated distribution was  fitted to the data, but the consistency of the
widths is a measure of understanding of the detection system and the target-beam conditions.
Histograms from a measurement at Q~=~1.53~MeV shown in figures~\ref{miss2}c,d demonstrate
the achieved missing-mass resolution at the COSY-11 detection system, when using a stochastically cooled proton beam.
The width of the missing mass distribution (figure~\ref{miss2}d), 
which is now close to the natural width
of the $\eta^{\prime}$ meson ($\Gamma_{\eta^{\prime}}=0.202$~MeV~\cite{PDG}), 
is again well reproduced by the
Monte-Carlo simulations.
 The broadening of the width of the $\eta^{\prime}$ signal with increasing excess-energy
(compare figures~\ref{miss2}b and~\ref{miss2}d) is a kinematical effect discussed in more
detail in reference~\cite{smyrski182}. The decreasing of the signal-to-background ratio
with growing excess-energy is due to the broadening 
of the $\eta^{\prime}$ peak 
and the increasing  background
 when moving away from the kinematical limit~(see e.g. figure~\ref{invmass}c or~\ref{miss2}c).
 At the same time,  the shape of the background, determined by the convolution
of the detector acceptance and the distribution of the two- and three-pion production~\cite{moskalphd},
remains unchanged within the studied range 
of the beam momentum.
The signal-to-background ratio changes from
1.8 at Q~=~1.53~MeV to 0.17 at Q~=~23.64~MeV.
 The geometrical acceptance, being defined by the gap of a dipole yoke 
and the scintillation detector most distant from the target~\cite{brauksiepe397,smyrski182},
decreases from 50~$\%$ to 4~$\%$ within this excess-energy range. However, 
in the horizontal plane it is still 100~$\%$ and thus covers the whole phase space. 
This issue will be considered in details in subsection~\ref{multiaccept}.
\vspace{-0.0cm}
\begin{figure}[H]
\vspace{-0.0cm}
\hspace{0.0cm}
\parbox{0.49\textwidth}{\vspace{0.1cm}\epsfig{file=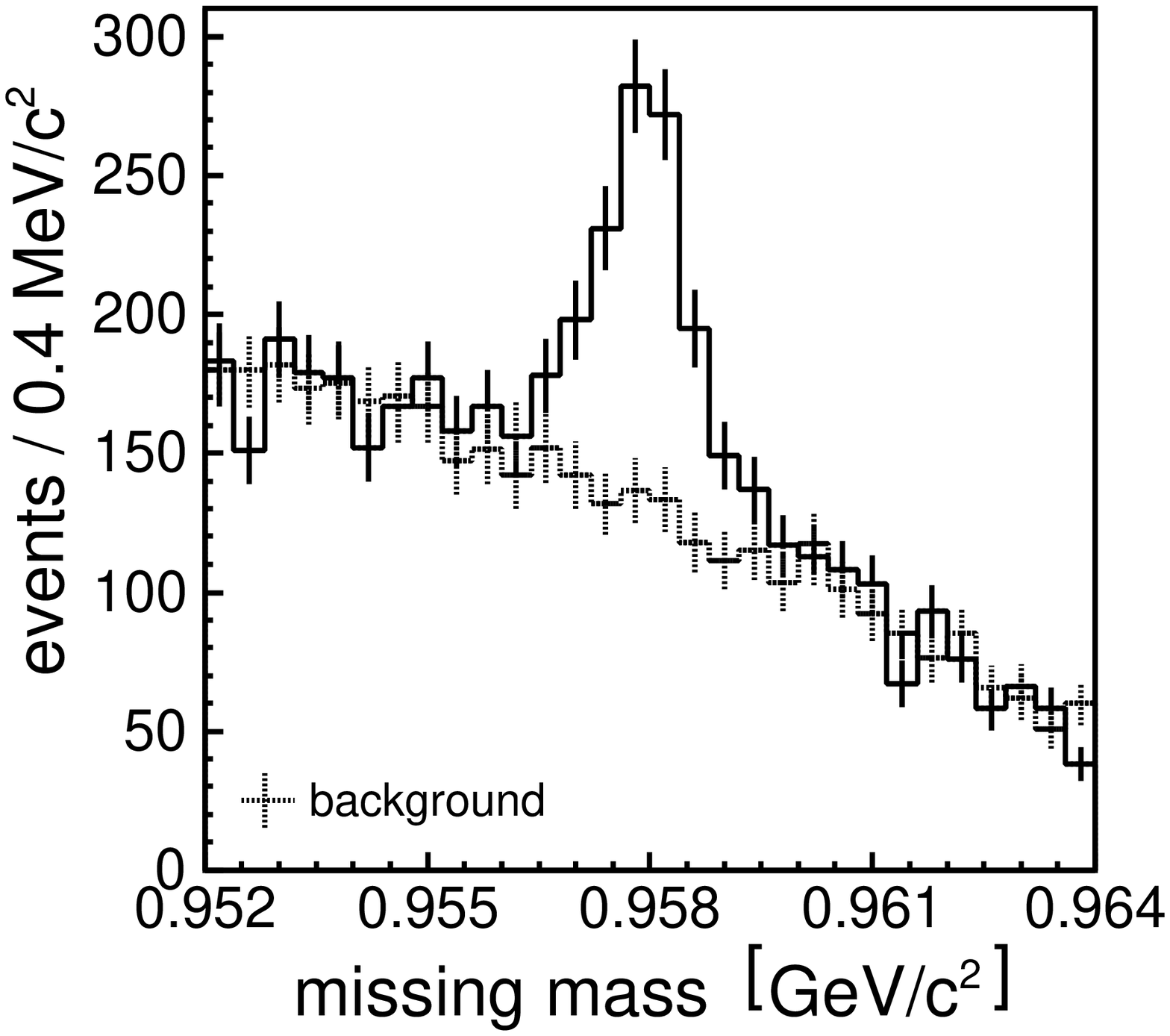,width=0.49\textwidth}}
\hfill
\parbox{0.49\textwidth}{\epsfig{file=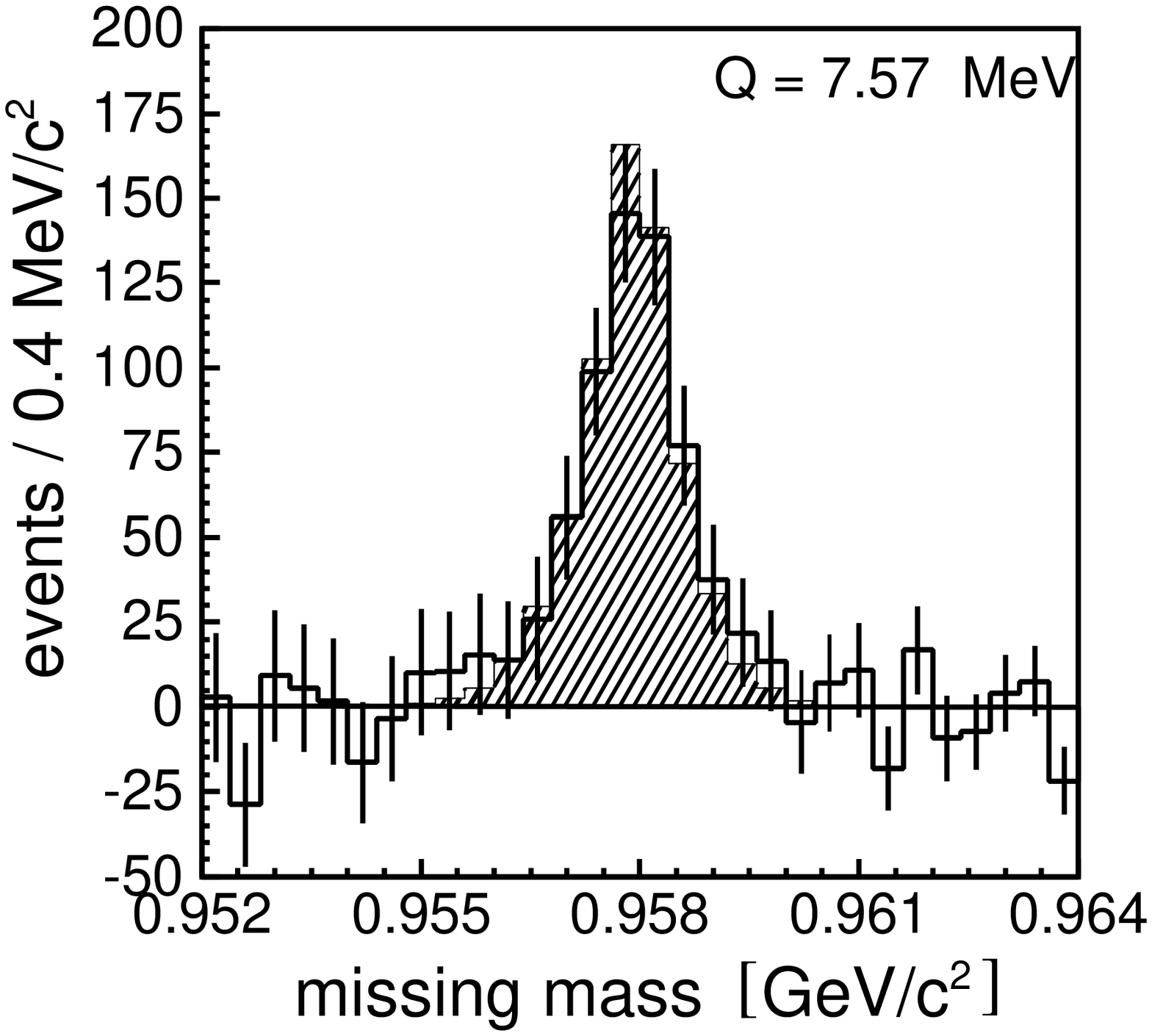,width=0.49\textwidth}}
 
\parbox{0.45\textwidth}{\raisebox{1ex}[0ex][0ex]{\mbox{}}}\hfill
\parbox{0.48\textwidth}{\raisebox{1ex}[0ex][0ex]{\large a)}}\hfill
\parbox{0.03\textwidth}{\raisebox{1ex}[0ex][0ex]{\large b)}}
 
\vspace{-0.1cm}
\hspace{0.0cm}
\parbox{0.49\textwidth}{\vspace{-0.1cm}\epsfig{file=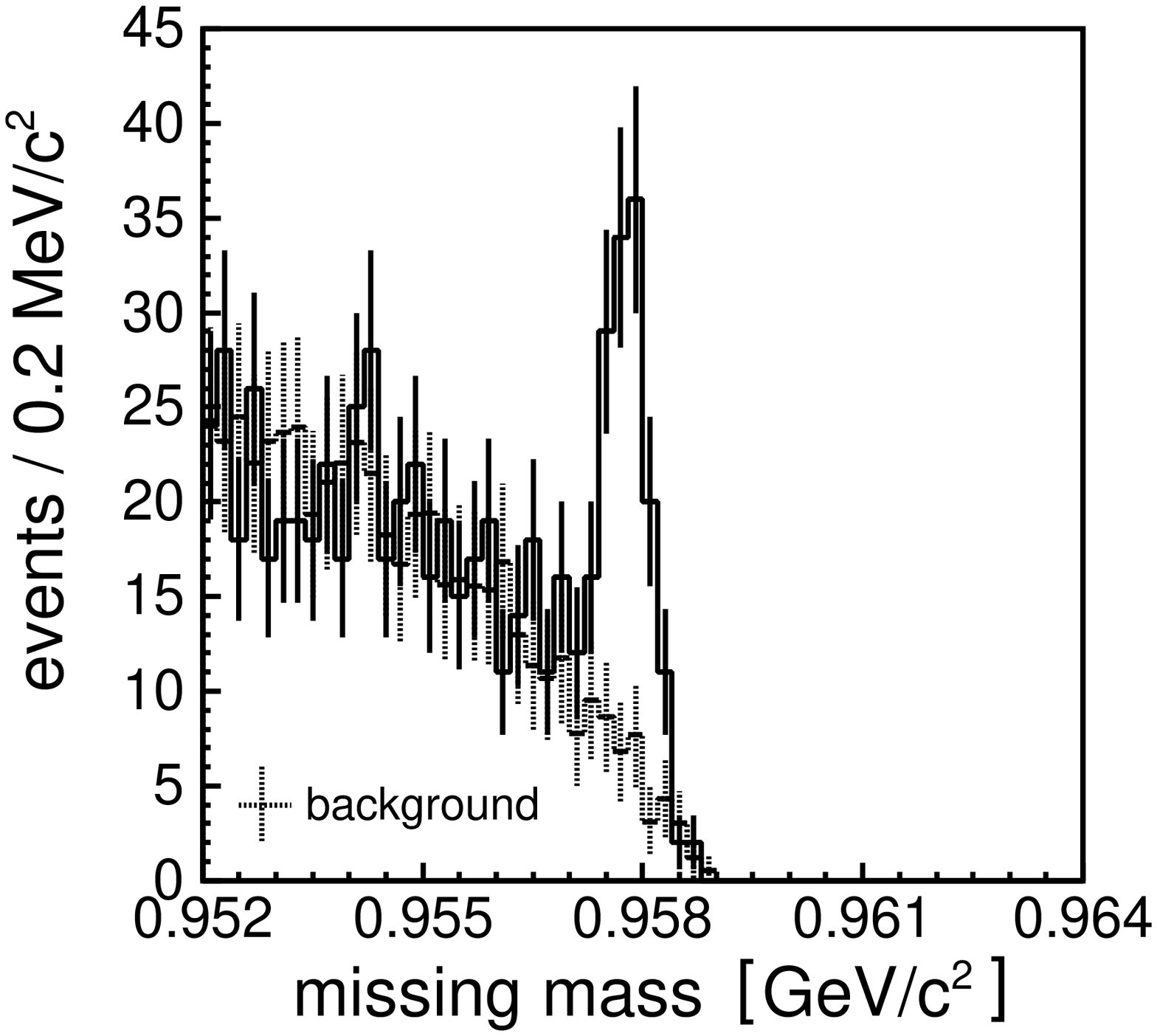,width=0.49\textwidth}}
\hfill
\parbox{0.49\textwidth}{\epsfig{file=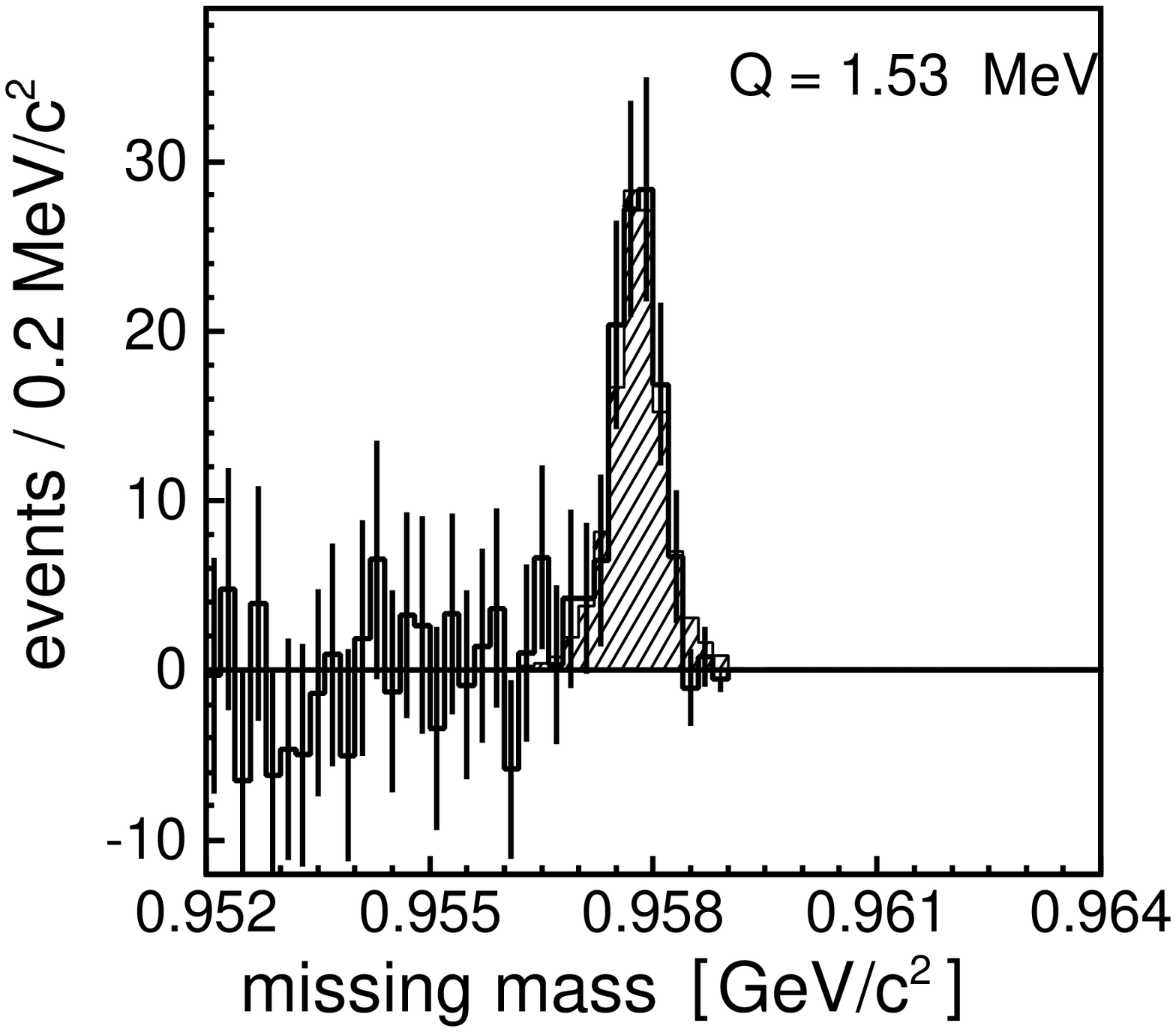,width=0.49\textwidth}}
 
\parbox{0.45\textwidth}{\raisebox{1ex}[0ex][0ex]{\mbox{}}}\hfill
\parbox{0.48\textwidth}{\raisebox{1ex}[0ex][0ex]{\large c)}}\hfill
\parbox{0.03\textwidth}{\raisebox{1ex}[0ex][0ex]{\large d)}}
\vspace{-0.3cm}
  \caption{
           Missing mass distribution with respect to the proton-proton system: \  \
           (a),(b)   measurements at Q~=~7.57~MeV above the threshold of the $pp\to pp\eta^{\prime}$ reaction 
           and  (c),(d) at Q~=~1.53~MeV.
           Background shown  as  dotted lines is combined from the measurements at different energies
           shifted to the appropriate kinematical limits and normalized to the solid-line histogram.
           Dashed histograms are obtained by means of the Monte-Carlo simulations.
        }
\label{miss2}
\end{figure}
\vspace{-0.0cm}
The absolute value of the excess-energy was determined from the position
of the $\eta^{\prime}$ peak in the missing mass spectrum, which should correspond to
the actual mass of the meson $\eta^{\prime}$. The systematic error of the excess-energy
established by this method equals to 0.44~MeV and constitutes of 0.14~MeV due to the uncertainty
of the $\eta^{\prime}$ meson mass~\cite{PDG} and of 0.3~MeV due to the
inaccuracy of the detection system geometry~\cite{moskalabsolut1}, whereby the largest effect
originates in the inexactness of relative settings of target, dipole, and drift chambers. 

\section{Monitoring of beam and target parameters}
\label{monitoring}
\begin{flushright}
\parbox{0.73\textwidth}{
 {\em
  Since we can never know anything for sure, it is simply not worth
  searching for certainty; but it is well worth searching for truth;
  and we do this chiefly by searching for mistakes, 
  so that we can correct them~\cite{popperbetter}.\\
 }
 \protect \mbox{} \hfill Karl Raimund  Popper  \protect\\
 }
\end{flushright}

 An exact extraction of absolute cross sections from the measured data demands
a reliable estimation of the acceptance of the detection system.
This in turn crucially depends on the accuracy of the determination of the
position and
dimensions of both the beam and the target.
In this section we will describe a method for estimating the dimensions of
the  proton beam, based on the momentum distribution of
elastically scattered protons, which can be measured
simultaneously with the investigated reaction.

As already pointed out,
the production of short-lived uncharged mesons 
($\eta$,
$\omega$, $\eta^{\prime}$, or $\phi$) 
is investigated 
 at the COSY-11 facility 
(see figures~\ref{detector} and ~\ref{elas_princip}b)
by means of the
missing mass technique via the $pp\rightarrow ppX$ 
reaction,
with the four-momenta of 
protons
being fully determined
experimentally.
The accuracy of the extracted missing mass value, which decides whether
the signal from the given meson is visible over a
background, depends on the precision of the momentum reconstruction of
the registered protons, which in turn depends on the detector
resolution and both the momentum- and the geometrical spread of the
accelerator
proton beam interacting with the internal cluster target beam.
The momentum reconstruction is performed by tracing back trajectories
from  drift chambers through the dipole magnetic field to the
target,
 which is ideally assumed
to be an infinitely thin vertical line.
In reality, however, the reactions take place in that region of finite
dimensions where beam and target overlap, as depicted in figure~\ref{elas_princip}a.
Consequently, assuming in the analysis an infinitesimal target implies a
smearing out of the momentum vectors and hence of the resolution of the
missing mass signal.
\begin{figure}[t]
       \parbox{0.4\textwidth}{\hspace{-2.8cm}\epsfig{file=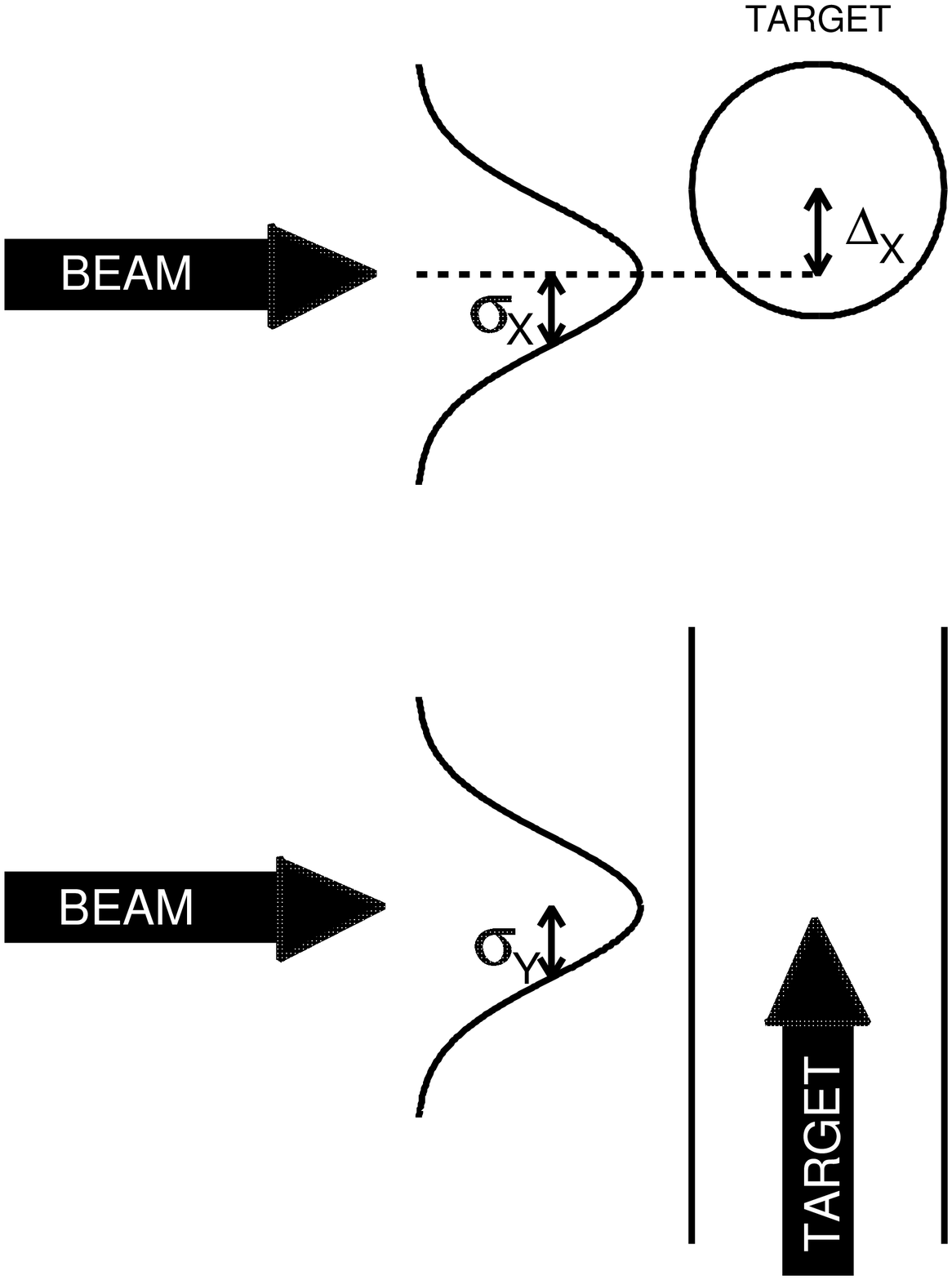,width=0.6\textwidth}}
       \parbox{0.6\textwidth}{\epsfig{file=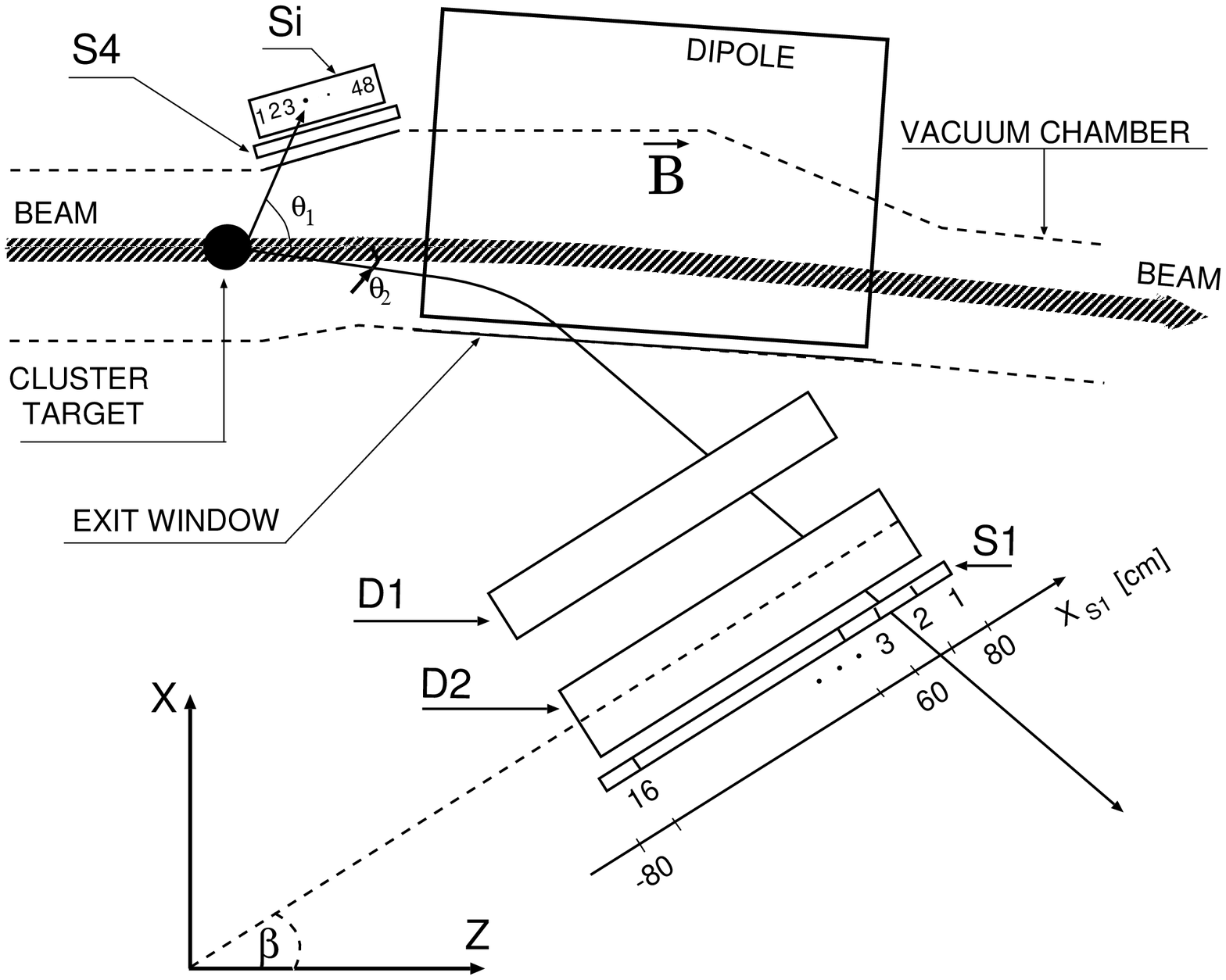,width=0.6\textwidth}}
 \parbox{0.30\textwidth}{\raisebox{0ex}[0ex][0ex]{\mbox{}}} \hfill
 \parbox{0.48\textwidth}{\raisebox{0ex}[0ex][0ex]{\large a)}} \hfill
 \parbox{0.04\textwidth}{\raisebox{0ex}[0ex][0ex]{\large b)}}
\caption{ (a) Schematic description of the relative beam and target
         setting. Seen from above (upper part), and from aside (lower
         part), $\sigma_{X}$ and $\sigma_{Y}$ denote the horizontal and
         vertical standard deviation of the assumed Gaussian distribution of
         the proton beam density, respectively. The distance between the
         centres of the proton  and the target beam is described as
         $\Delta_{X}$. \protect \\
 (b) Schematic view of the COSY-11 detection setup. Only
 detectors used for the measurement of elastically scattered protons
 are shown. Numbers, at the silicon pad detector (Si), and below the
 scintillator hodoscope (S1), indicate the order of segments. D1 and
 D2 denote drift chambers. The X$_{S1}$  axis is defined such that
 the first segment of the S1 ends at 80~cm and the sixteenth ends at
 -80~cm. The proton beam, depicted by a shaded line, circulates in the 
 ring and crosses each time the $H_{2}$ cluster target installed in front
 of one of the bending dipole magnets of the COSY accelerator.
 }
\label{elas_princip}
\end{figure}
The part of the COSY-11 detection setup used for the registration of
elastically scattered protons is shown in figure~\ref{elas_princip}b.
Trajectories of protons scattered in the forward direction are
measured by means of two drift chambers (D1 and D2) and a scintillator hodoscope~(S1),
whereas the recoil protons are registered in coincidence with the
forward ones using a silicon pad detector arrangement (Si) and a scintillation detector~(S4).
The two-body kinematics gives an unambiguous relation between the
scattering angles $\Theta_{1}$ and $\Theta_{2}$
of the recoiled and forward flying
protons.
Therefore, as seen in figure~\ref{correl}a, events of
elastically scattered protons 
  can be identified from the correlation
  line formed between the position in the silicon pad detector Si, and
  the scintillator hodoscope S1,
  the latter measured by the two drift chamber stacks.
For those
protons which are elastically scattered in forward direction and  are deflected
in the magnetic field of the dipole the momentum vector at the target point
can be determined.
According to two-body kinematics, momentum components parallel and
perpendicular to the beam axis form an ellipse, from which a
section is shown as a solid line in figure~\ref{correl}b,
superimposed on data selected according to the correlation
criterion from figure~\ref{correl}a for elastically scattered events.
\begin{figure}[H]
\vspace{-1.0cm}
  \parbox{0.31\textwidth}{\vspace{0.5cm}
    \includegraphics[width=0.367\textwidth]{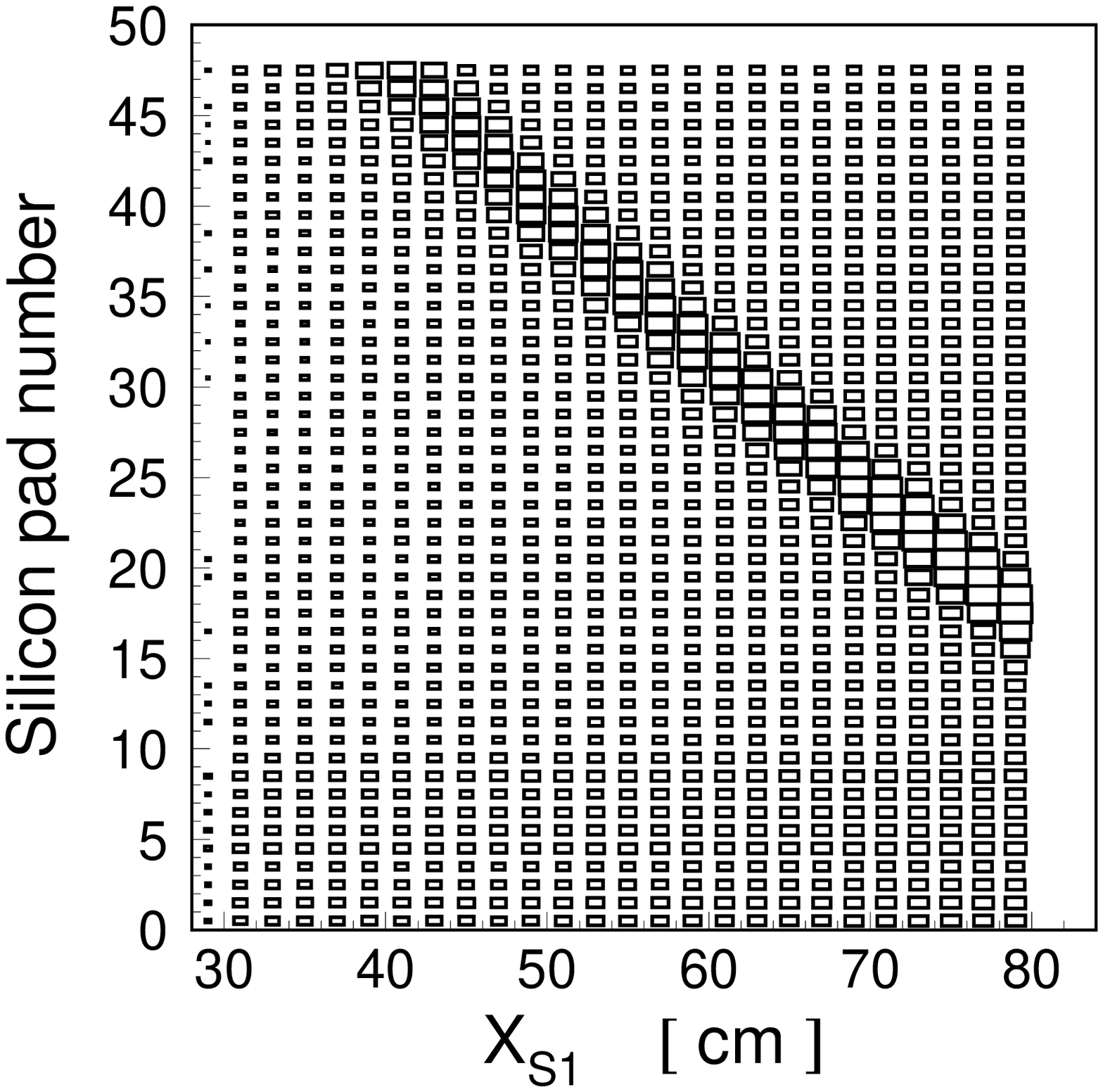}}
  \parbox{0.33\textwidth}{
    \includegraphics[width=0.409\textwidth]{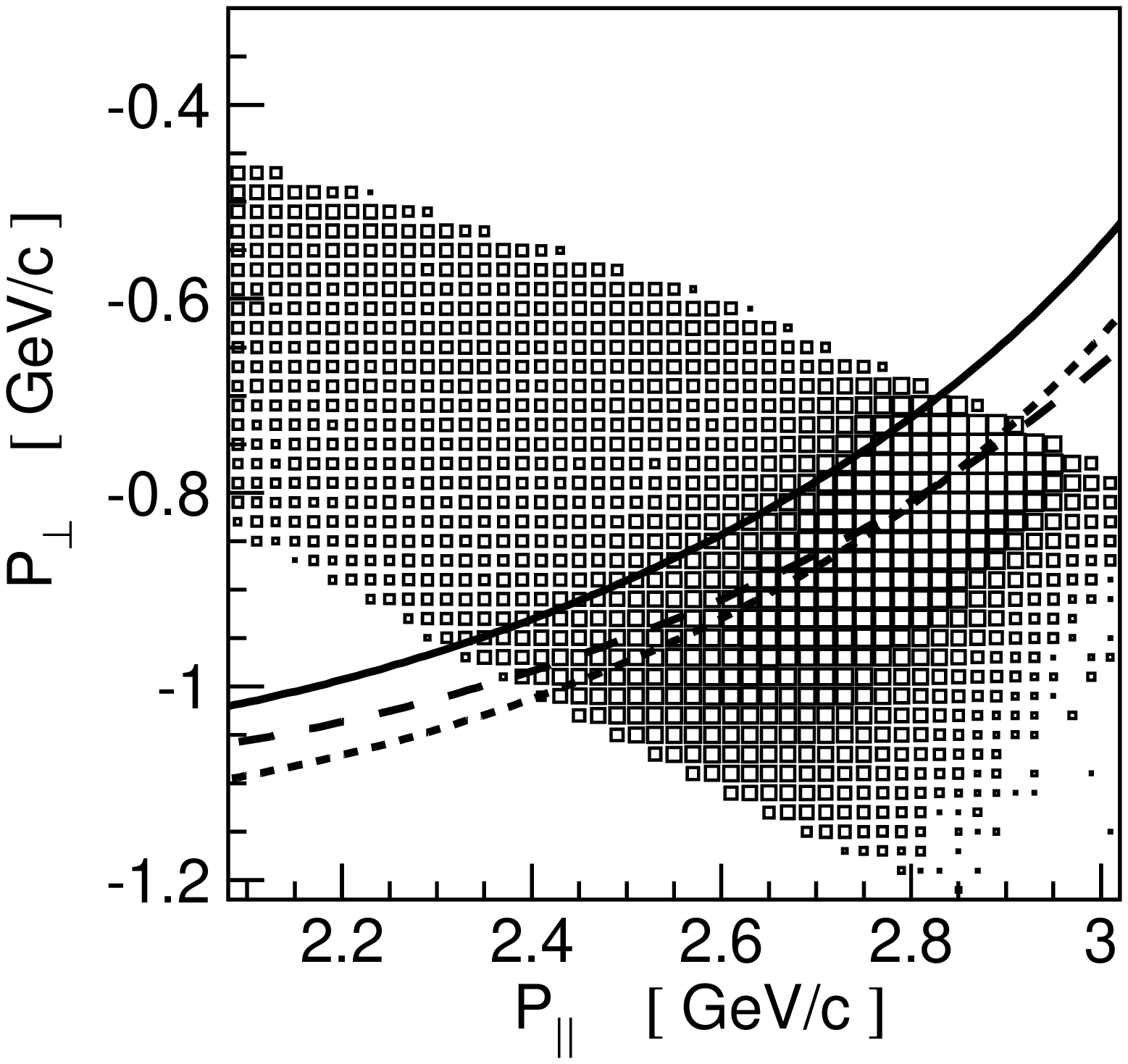}}
  \parbox{0.33\textwidth}{
    \includegraphics[width=0.409\textwidth]{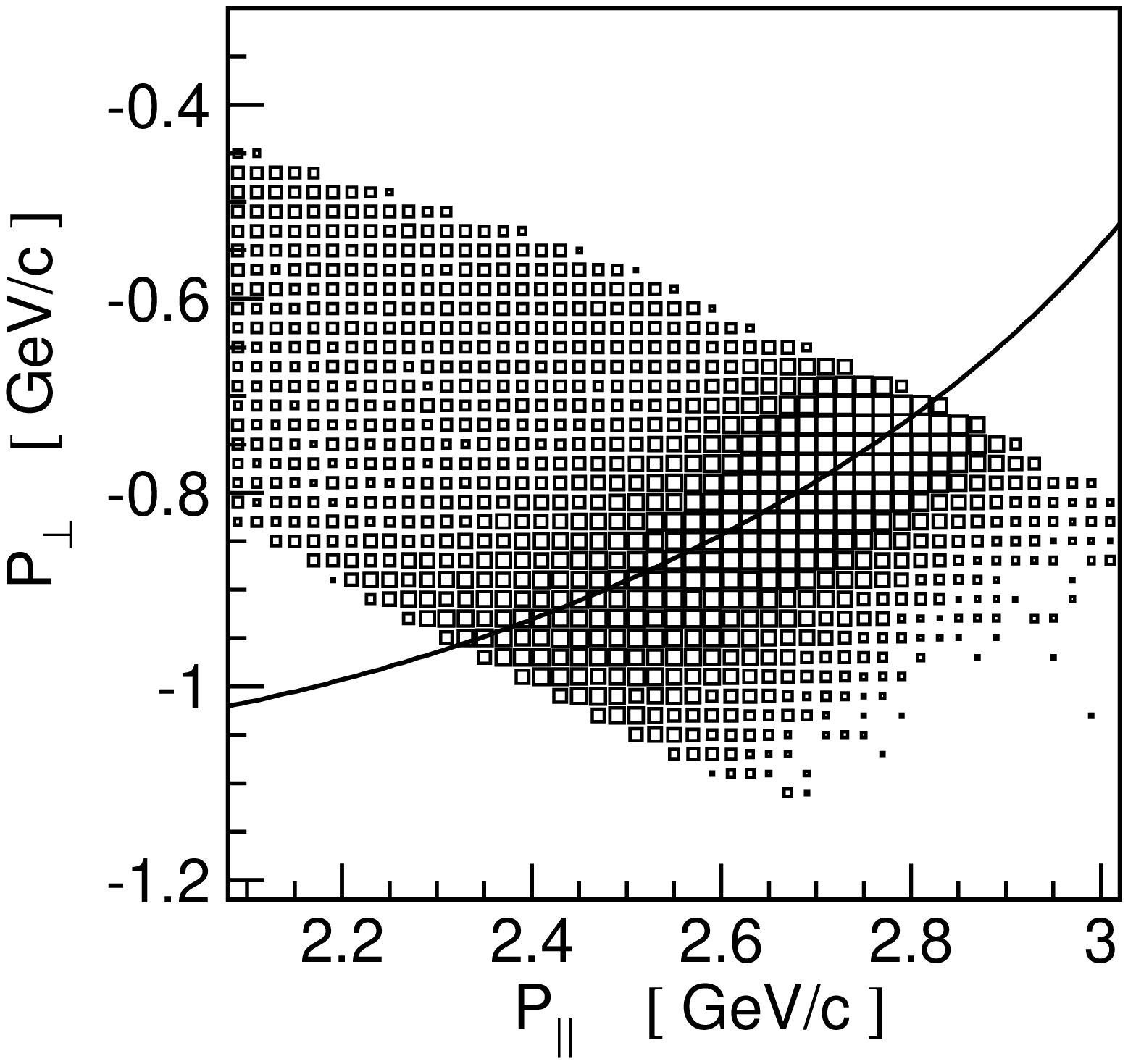}}

     \vspace{-0.4cm}
    \parbox{0.29\textwidth}{\raisebox{0ex}[0ex][0ex]{\mbox{}}} \hfill
\parbox{0.310\textwidth}{\raisebox{0ex}[0ex][0ex]{\large a)}} \hfill
\parbox{0.31\textwidth}{\raisebox{0ex}[0ex][0ex]{\large b)}} \hfill
\parbox{0.011\textwidth}{\raisebox{0ex}[0ex][0ex]{\large c)}}
\vspace{0.0cm}
\caption{  
          (a) \ \  Identification of elastically scattered protons from the correlation
            of hits in the silicon detector Si and the S1 scintillator hodoscope.
            Note that the number of entries per bin is given in
            a logarithmic scale, 
            ranging from 1 (smallest box) to 19000 (largest box).\protect\\
           (b) \ \ Perpendicular versus parallel 
           (with respect to the beam direction)
           momentum components 
           of particles registered at a beam
           momentum of 3.227~GeV/c. The number of entries per bin is shown
           logarithmically. The solid line corresponds to the momentum ellipse
           expected for elastically scattered protons at a beam momentum of
           3.227~GeV/c, the dashed line refers to a beam momentum of
           3.350~GeV/c, and the dotted line shows the momentum ellipse
           obtained for a proton beam inclined by 40~mrad. \protect\\
           (c) \ \ \ The same data as shown in b) but analyzed
           with the target point shifted by -0.2~cm perpendicularly to the beam
           direction (along the X-axis in figure~\protect\ref{elas_princip}b). The
           solid line shows the momentum ellipse at a beam momentum of
           3.227~GeV/c.
        }
\label{correl}
\end{figure}
In figure~\ref{correl}b,  similarly as in figure~\ref{correl}a, it is obvious
that data of elastically scattered events arise clearly over a certain low
level
background. However, what is important here is that the mean of the
elastically scattered data is significantly shifted from the expected line,
indicating that the
reconstructed momenta are on average larger than expected.
This discrepancy cannot be explained by an alternative assumption of either
the proton beam momentum or the proton beam angle because of the following
reason:
Trying to reach an acceptable agreement between data and expectation
the beam momentum must be
changed by more than 120~MeV/c (dashed line in figure~\ref{correl}b),
which is 40~times more than the conservatively estimated error of the
absolute beam momentum~($\pm$~3~MeV/c)~\cite{prasuhn167,moskalabsolut1,moskalabsolut2}.
Similarly, the effect could have been corrected by changing the
beam angle by 40~mrad~(see dotted line in figure~\ref{correl}b),
which also exceeds
the admissible deviation of the beam
angle~($\pm$~1~mrad~\cite{prasuhn167}) by at least a factor of~40.

However, the observed discrepancy can be explained by shifting  the
assumed reaction point relative to the nominal value by $-0.2$~cm
perpendicular to the beam axis towards the center of the COSY-ring along
the X-axis defined in figure~\ref{elas_princip}b.
The experimentally extracted momentum components obtained under this
assumption, shown in
figure~\ref{correl}c, agree with the expectation depicted by the
solid line, and the data are now spread symmetrically around the
ellipse.
   This spread is essentially due to
   the finite extensions of the cluster target and the proton beam overlap,
   which corresponds to about $\pm 0.2$~cm
   as can easily be inferred from the value of the target shift required.
Other contributions as the spread of the beam momentum and still some multiple
scattering events appear to be negligible and therefore we consider further the
influence of the effective target dimensions and the spread of the COSY beam.

Assuming both the target beam to be described by a cylindrical pipe,
with a diameter of 9~mm,
homogeneously filled with protons~\cite{dombrowski228,khoukazepj},
and the COSY proton beam density
distribution to be described by Gaussian functions with standard deviations
$\sigma_{X}$ and $\sigma_{Y}$ for the horizontal and vertical
directions, respectively~\cite{prasuhn167,schulzphd}, we performed
Monte-Carlo calculations varying the 
distance between
the target  and the beam centres ($\Delta_{X}$), and the
horizontal proton beam extension
($\sigma_{X}$)~(see figure~\ref{elas_princip}a).

In order to account for the angular distribution of the elastically scattered
protons,
to each generated event an appropriate weight $w$ was assigned according
to the  differential distributions of the cross sections
measured by the EDDA collaboration~\cite{albers1652}.
The generated events were  analyzed in the same way as the
experimental data.
A comparison of the data from figure~\ref{correl}b (data analyzed by
using the nominal interaction point) with the
corresponding simulated histograms allows to determine both
$\sigma_{X}$  and $\Delta_{X}$.
 For finding an estimate of the parameters $\sigma_{X}$ and $\Delta_{X}$
we construct the $\chi^{2}$ statistic according to the {\em method of least
squares}:
\begin{equation}
\chi^2 =  \sum_i \, \frac{\left(  \alpha N_i^s + b_i - N_i^e \right)^{2} }
                                { \alpha^{2} \displaystyle \sum_{i^{th}bin} w^2
+ N_i^e + b_i},
\label{eqchi2_sl}
\end{equation}
where  $N_i^e$ and
${\displaystyle N_i^s = \sum_{i^{th}bin} w }$
denote the content of the $i^{th}$ bin of
the P$_{\perp}$-versus-P$_{\parallel}$ spectrum determined from
experiment (figure~\ref{correl}b) and simulations, respectively. The
background
events $b_i$ in the i$^{th}$ bin amount to less than one per cent of the
data~\cite{moskalphd}
and were estimated by linear interpolations between the inner and outer part
of a broad distribution surrounding the expected ellipse originating from
 elastically
scattered protons. The free parameter $\alpha$ allows to adjust the overall
scale of
the fitted Monte-Carlo histograms. Thus, varying the $\alpha$ parameter the
$\chi^{2}_{min}$
for each pair of $\sigma_{X}$ and $\Delta_{X}$ was established
as a minimum of the $\chi^{2}(\alpha)$ distribution.
\begin{figure}[H]
\vspace{-0.8cm}
\hspace{0.0cm}
\parbox{0.49\textwidth}{\epsfig{file=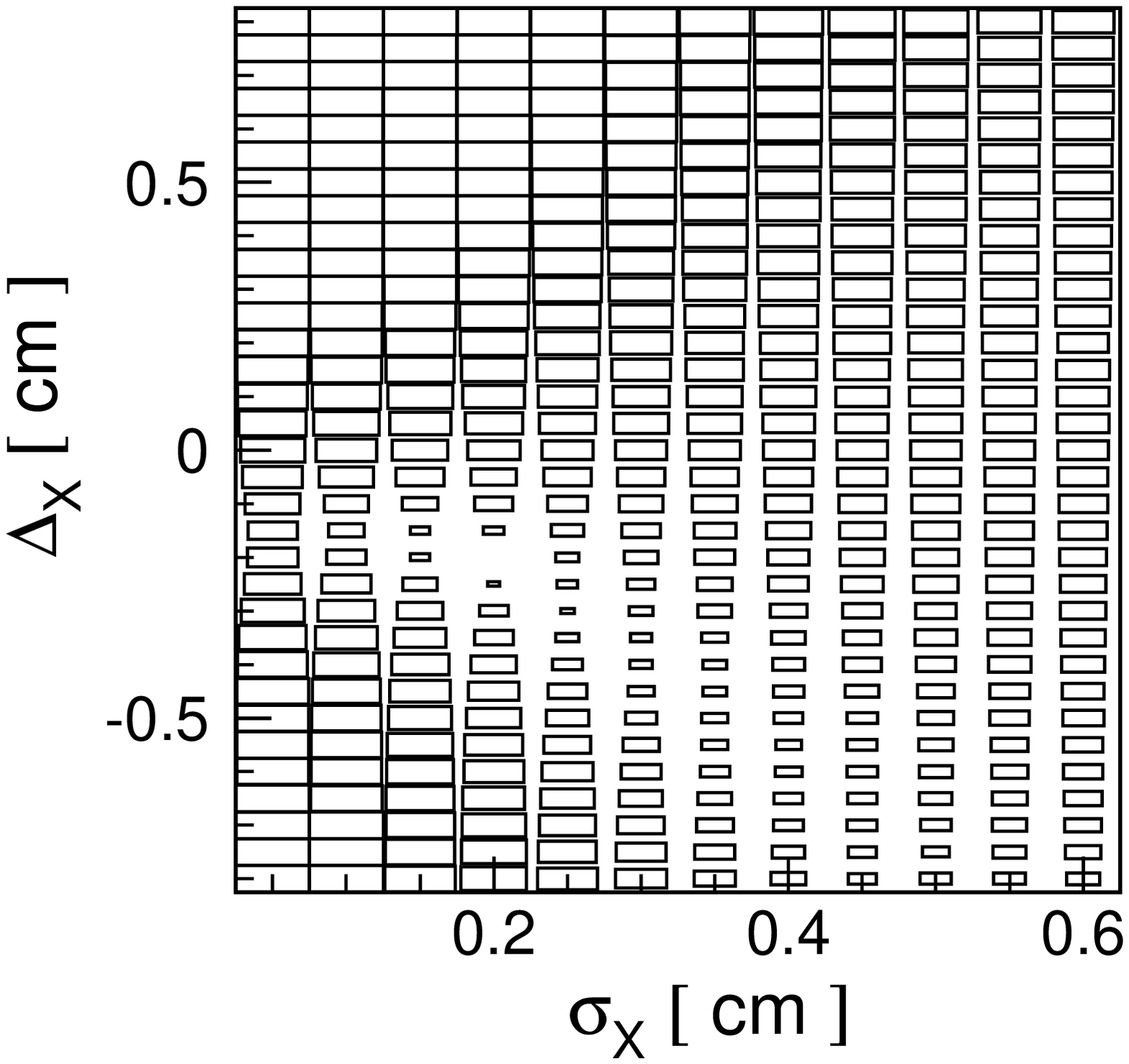,width=0.52\textwidth}}
\hfill
\parbox{0.49\textwidth}{\epsfig{file=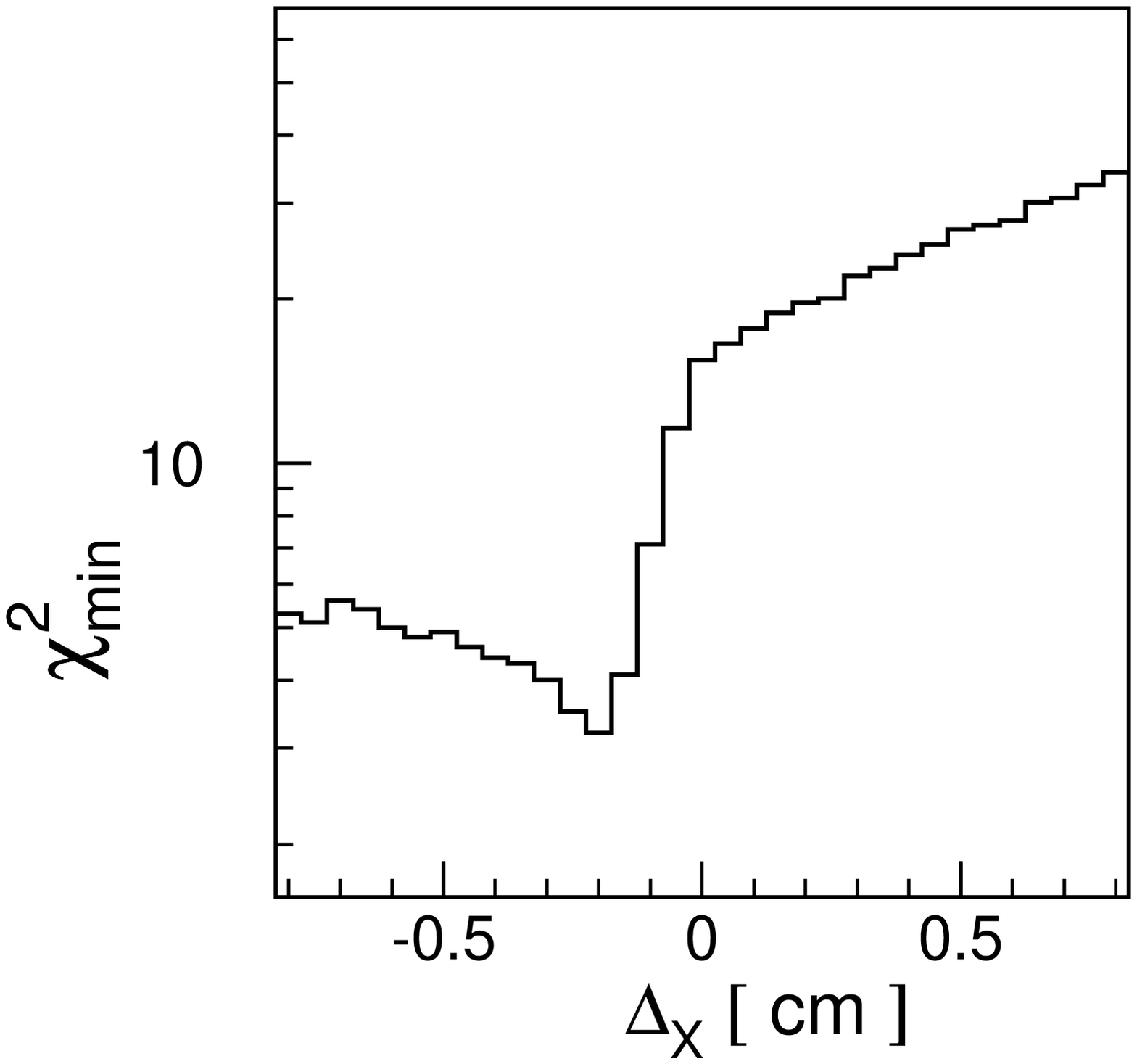,width=0.52\textwidth}}
 
\vspace{-0.4cm}
\parbox{0.43\textwidth}{\raisebox{1ex}[0ex][0ex]{\mbox{}}}\hfill
\parbox{0.50\textwidth}{\raisebox{1ex}[0ex][0ex]{\large a)}}\hfill
\parbox{0.03\textwidth}{\raisebox{1ex}[0ex][0ex]{\large b)}}
 
\vspace{-1.0cm}
\hspace{0.0cm}
\parbox{0.49\textwidth}{\epsfig{file=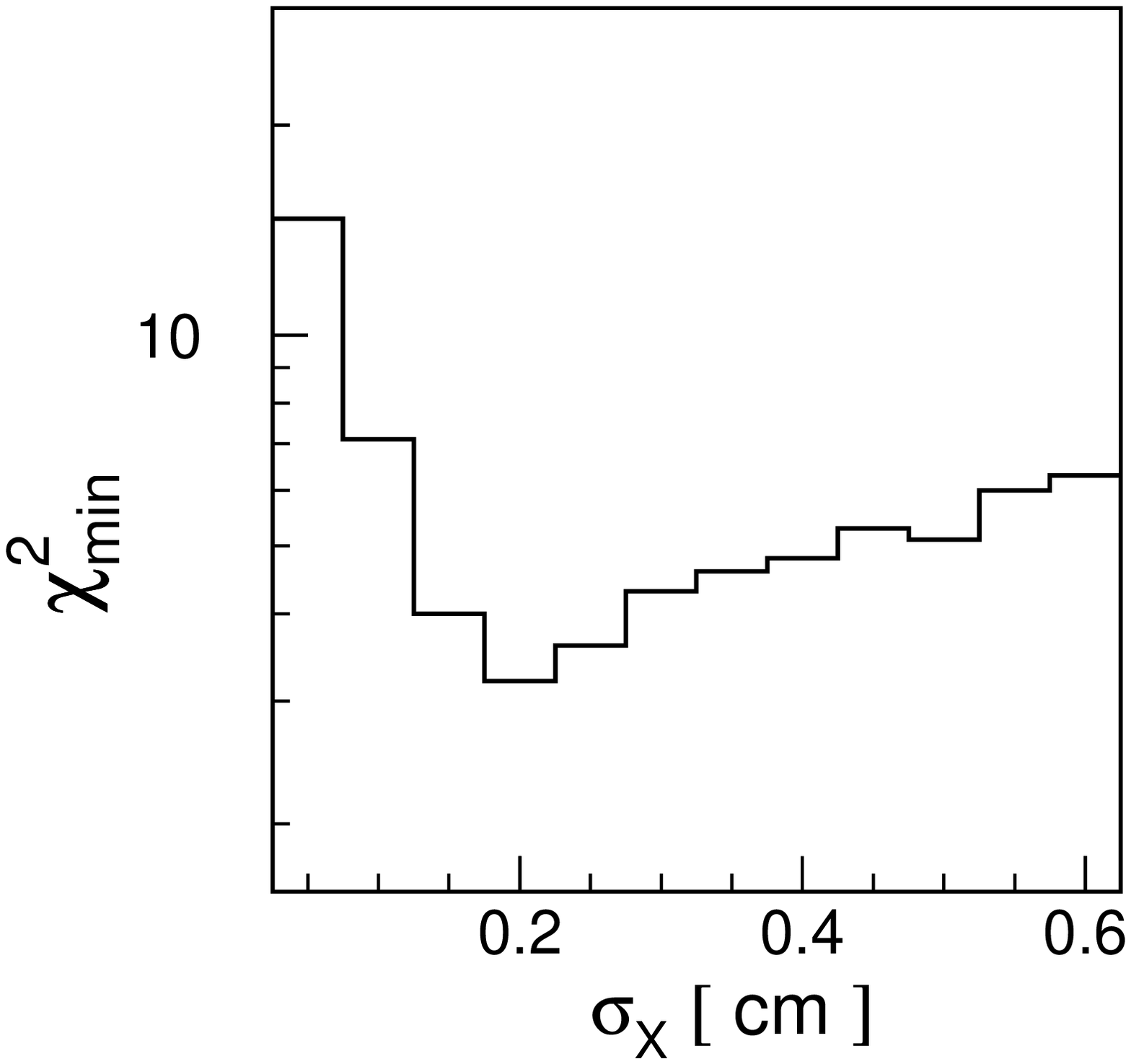,width=0.52\textwidth}}
\hfill
\parbox{0.49\textwidth}{\epsfig{file=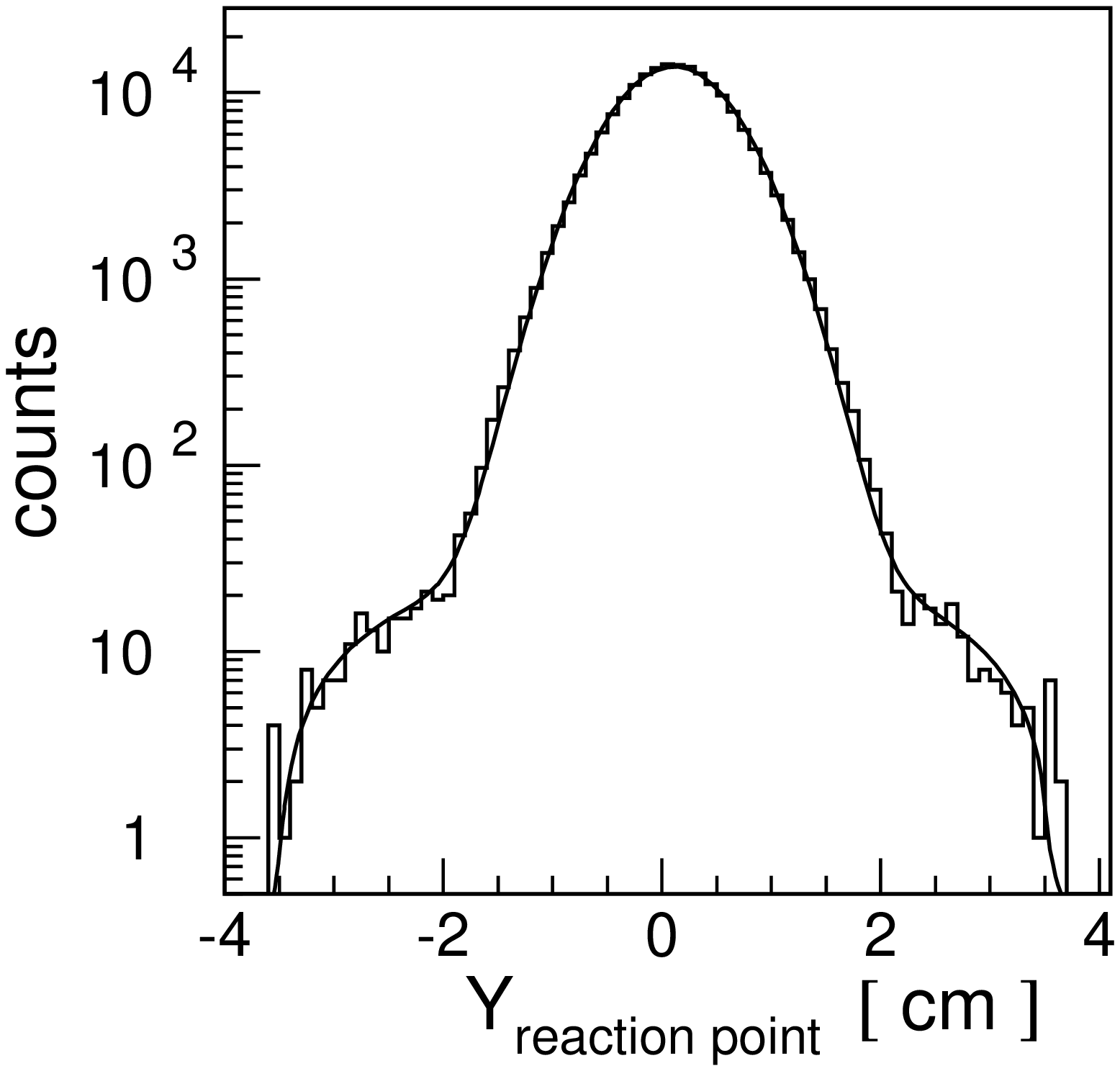,width=0.52\textwidth}}
\vspace{-0.3cm}
\parbox{0.43\textwidth}{\raisebox{1ex}[0ex][0ex]{\mbox{}}}\hfill
\parbox{0.50\textwidth}{\raisebox{1ex}[0ex][0ex]{\large c)}}
\parbox{0.03\textwidth}{\raisebox{1ex}[0ex][0ex]{\large d)}}
        \caption{
           \label{dx1_vs_sx1}
         (a) \ \ $\chi^{2}_{min}$ as a function of $\sigma_{X}$ and $\Delta_{X}$. 
             The number of entries is shown in  a logarithmic scale.  \ \ 
         (b) \ \ $\chi^{2}_{min}$ as a function of $\Delta_{X}$. \ \ \
         (c) \ \ $\chi^{2}_{min}$ as a function of $\sigma_{X}$. \
         The minimum value of  $\chi^{2}_{min}$ is larger than two. 
         However, it decreases when only a part of the 60 minutes long COSY cycle
         is taken into account (see subsection~\ref{stochastic}).
         This is due to the fact that the beam changes
         during the cycle and  its shape, when integrated over the whole cycle time,
         does not suit perfectly to the shape assumed in Monte-Carlo simulations.
        (d) \ \  Distribution of the vertical component of the reaction
         points determined by tracing back trajectories from the drift
         chambers through the dipole magnetic field to the centre of the
         target (in the horizontal plane). ``Tails'' are due to secondary
         scattering on the vacuum chamber and were parametrized by a
         polynomial of second order. The solid line shows the simultaneous fit
         of the Gaussian distribution and the polynomial of second order.
        }
\end{figure}
Figure~\ref{dx1_vs_sx1}a shows the logarithm of the obtained $\chi^{2}_{min}$
as a function of the values of $\sigma_{X}$ and $\Delta_{X}$ used
in the Monte-Carlo calculations:
A valley identifying a minimum range is clearly recognizable, which gives the
unique possibility to determine the varied parameters~$\sigma_{X}$ and
$\Delta_{X}$.
The overall minimum of the
$\chi^{2}_{min}(\sigma_{X},\Delta_{X})$-distribution, at
$\sigma_{X}$~=~0.2~cm and $\Delta_{X}$~=~-0.2~cm, is obviously and
better seen in
figures~\ref{dx1_vs_sx1}b~and~\ref{dx1_vs_sx1}c, which show the projections of
the valley line onto the respective axis.
The same results were obtained when employing
 the Poisson likelihood $\chi^{2}$ derived from the
maximum likelihood method~\cite{bakernim,feldmanpr}
\begin{equation}
\chi^2 = 2 \cdot \sum_i \, [\alpha N_i^s + b_i - N_i^e +  N_i^e \,
ln(\frac{N_i^e}{\alpha N_i^s + b_i})],
\label{eqchi2_mh}
\end{equation}

The vertical beam extension of $\sigma_{Y}$~=~0.51~cm was established
directly from the distribution of the vertical component of the
particle trajectories at the centre of the target.
This is possible, since for the momentum reconstruction only the
origin of the track in the horizontal plane is used, while the
vertical component remains a free parameter.
As shown in figure~\ref{dx1_vs_sx1}d, the reconstructed distribution
of the vertical component of the reaction points indeed can well be
described by a Gaussian distribution with the $\sigma$~=~0.53~cm.
 The width of this distribution
is primarily due to the  vertical spread of the proton beam.
The spread caused by
 multiple scattering and  drift chambers resolution was determined to be
about 0.13~cm (standard deviation)~\cite{moskalphd}.
Therefore, $\sigma_{Y}~=~\displaystyle\sqrt{\displaystyle 0.53^{2} -
0.13^{2}}~\approx$~0.51~cm.

The effect of a possible drift chamber misalignment, i.\ e.\ an
inexactness of the angle $\beta$ in figure~\ref{elas_princip}b, was
estimated to cause a shift in the momentum plane corresponding to a
value of 0.15~cm for $\Delta_{X}$.
This gives a rather large systematic bias to the estimation of the
absolute value of $\Delta_{X}$, but it does not influence the
implications concerning the parameter~$\sigma_{X}$, and still allows
for the determination of relative beam shifts $\Delta_{X}$ during
the measurement cycle.

The absolute value of $\Delta_{X}$ can be found if one additionally takes
into account the width of the mass spectrum of the studied meson.
Clearly, using in the analysis a wrong value of $\Delta_{X}$ broadens
a missing mass distribution. Therefore, the real value of $\Delta_{X}$
can be determined as the one at which the width of the mass spectrum 
of the investigated meson is the least.
\subsection{Influence of the stochastic cooling on the proton beam quality }
\label{stochastic}
\begin{flushright}
\parbox{0.65\textwidth}{
 {\em
   The cooling of a single particle circulating in a ring
   is particularly simple~\cite{meer}. \protect\\
 }
 \protect \mbox{} \hfill Simon van der Meer  \protect\\
 }
\end{flushright}
\vspace{-0.3cm}
Meson production cross sections in the proton-proton
collisions increase rapidly with the beam momentum,
and change by more than one order of magnitude when the
excess energy grows from 1~MeV to 10 MeV~\cite{smyrski182,moskal416,swave}.
 Due to the rapid change of the total cross section,
studies of the close-to-threshold meson production require
a precise beam with both small momentum and geometrical spread,
and an accurate determination of the momenta of the reaction products.
The use of a stochastically cooled beam~\cite{meer,mohl,stockhorst,prasuhn167}
and of internal cluster target
facilities allows to realize these conditions.
Very low target
densities ($\approx 10^{~14}$~atoms~cm$^{-2}$)~\cite{dombrowski228}
minimize changes
in the ejectiles' momentum vectors
due to secondary scattering in the target,
and simultaneously allow the beam to
circulate through the target
without  significant intensity losses
and changes of the momentum.
An improvement of the experimental conditions caused by the stochastic cooling
can be deduced already from a time dependence of the trigger frequency presented 
in figure~\ref{cooling}.
\begin{figure}[H]
\leavevmode
\centerline{\epsfig{file=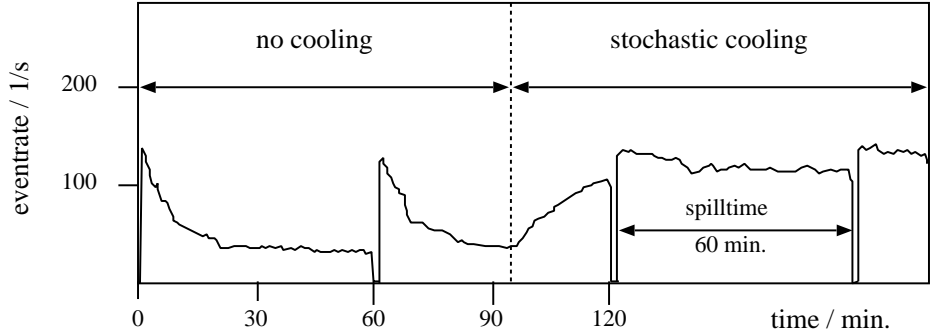,width=0.96\textwidth}}
\vspace{-0.5cm}
\caption{ \small{ 
          Rate of coincidences between the detectors S1 and S4
          shown in figure~\ref{elas_princip}b. In the middle of the second
          spill  the cooling system was switched on.
         }
        }
\label{cooling}
\end{figure}
 This figure presents the rate of the coincidence between the
scintillator detectors S4 and S1.
The observed event rate is proportional to the number of reactions
occuring in the region of the beam and target overlap.
 A steep decrease in the beginning of the first and second shown cycles
is caused by the outward movement of the uncooled beam from the target center.
Switching on the longitudinal and horizontal stochastic cooling,
 during the second cycle,  clearly
increases the luminosity and in the subsequent spills 
prevents the beam from spreading and shifting out of 
the target location. Consequently, a constant counting rate 
is obtained during the
cycles with a cooled beam.
In the following the beam monitoring technique introduced in the previous section
will be applied to inspect in more details an influence of the stochastic cooling 
on the proton beam quality.

The horizontal stochastic cooling~\cite{prasuhn167} is intended to squeeze
the proton beam in the horizontal direction until the beam reaches the
equilibrium between the cooling and the heating due to the target.
\vspace{-0.0cm}
\begin{figure}[H]
\vspace{-1.0cm}
  \parbox{0.31\textwidth}{\vspace{0.1cm}
    \includegraphics[width=0.40\textwidth]{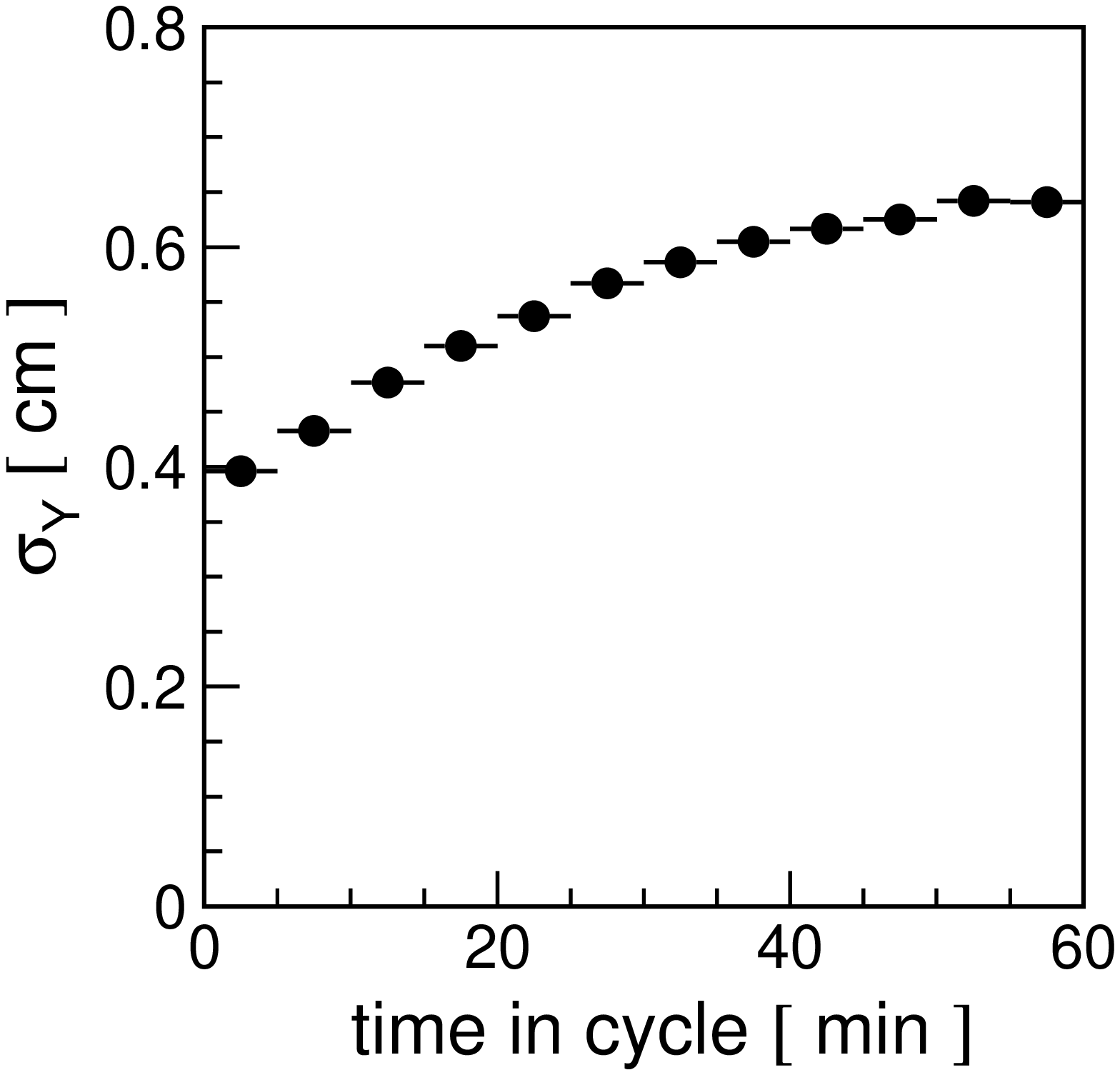}}
  \parbox{0.33\textwidth}{
    \includegraphics[width=0.41\textwidth]{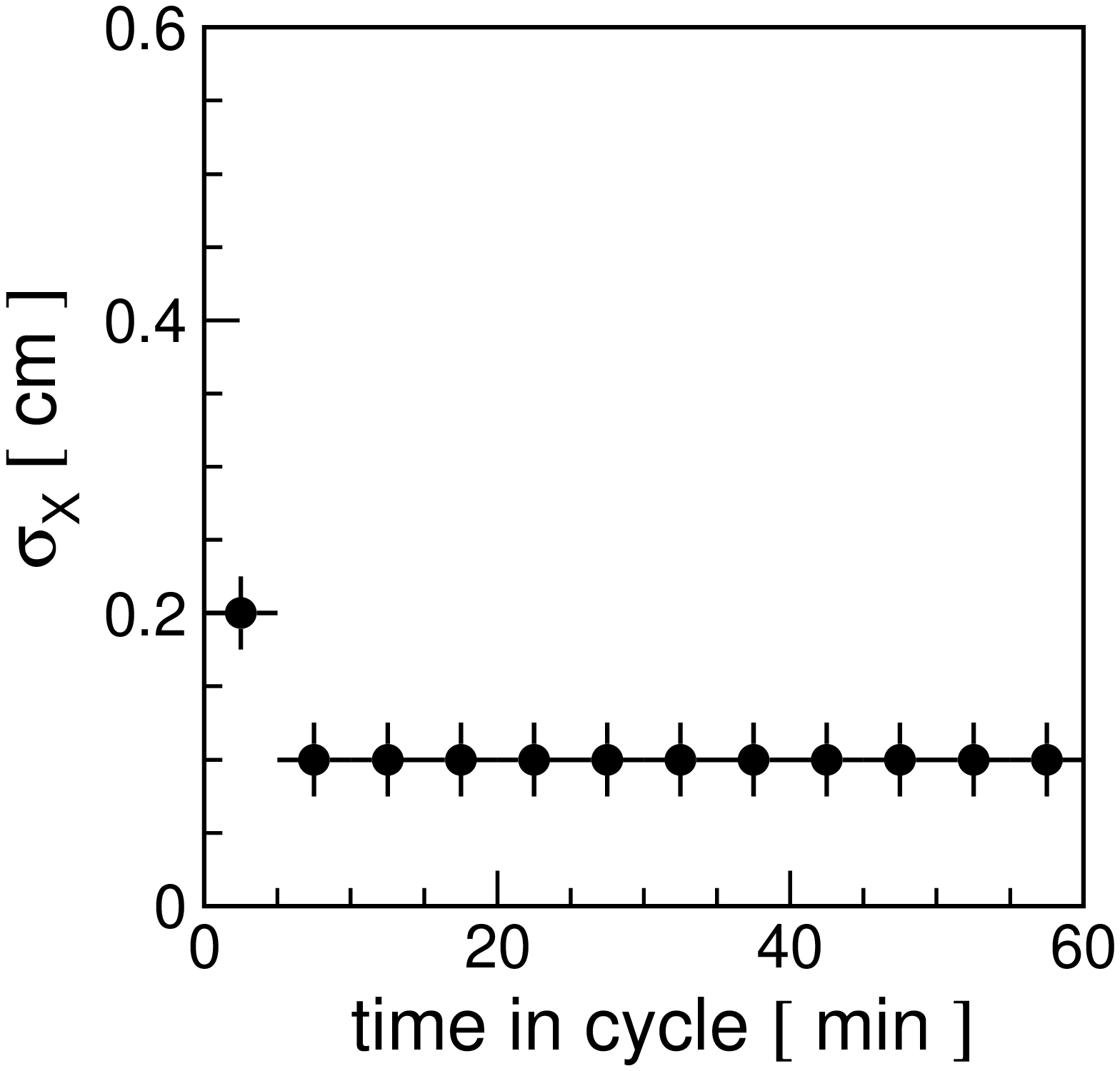}}
  \parbox{0.33\textwidth}{
    \includegraphics[width=0.41\textwidth]{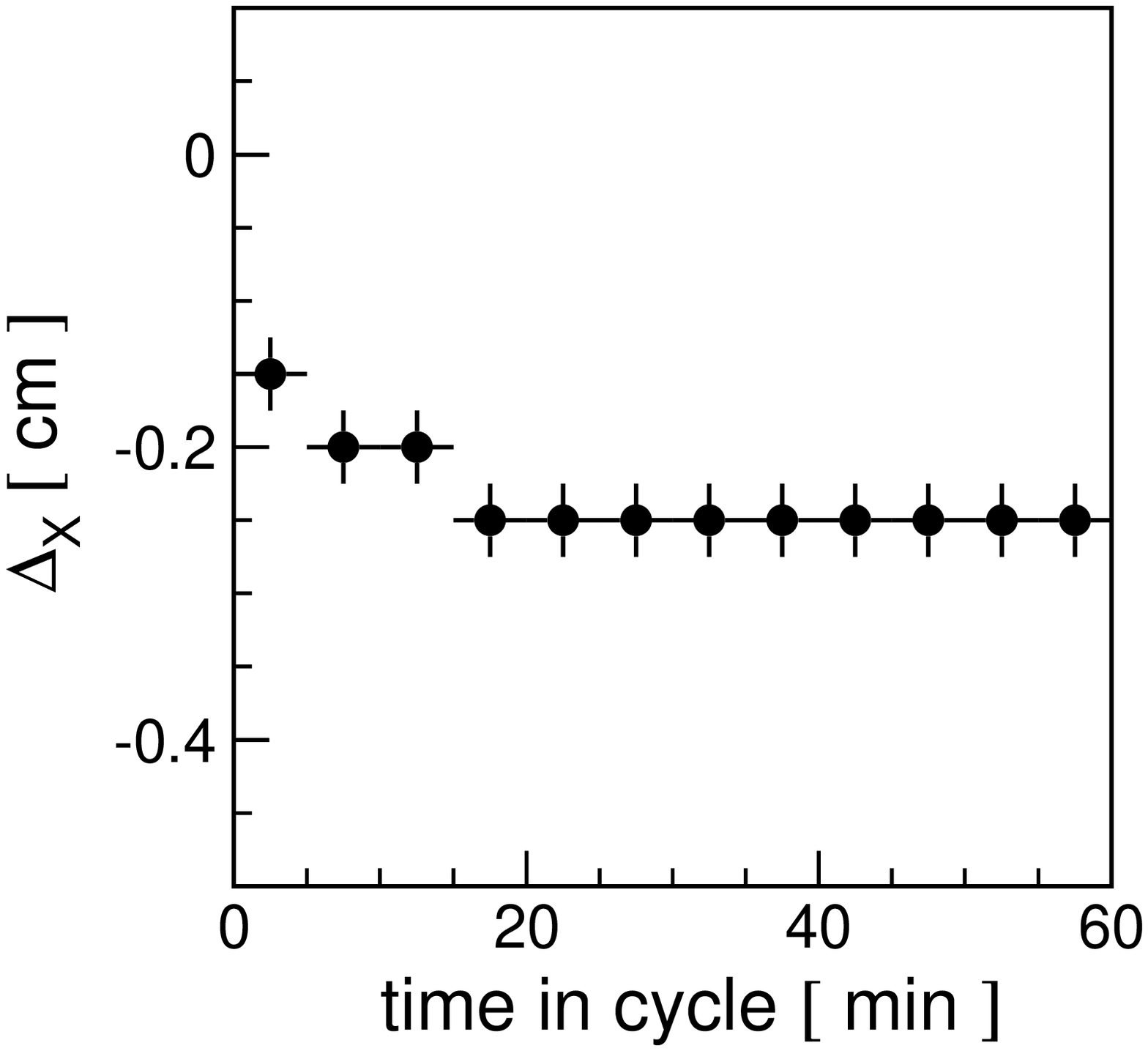}}

     \vspace{-0.3cm}
    \parbox{0.33\textwidth}{\raisebox{0ex}[0ex][0ex]{\mbox{}}} \hfill
\parbox{0.305\textwidth}{\raisebox{0ex}[0ex][0ex]{\large a)}} \hfill
\parbox{0.31\textwidth}{\raisebox{0ex}[0ex][0ex]{\large b)}} \hfill
\parbox{0.011\textwidth}{\raisebox{0ex}[0ex][0ex]{\large c)}}
\vspace{-0.3cm}
\caption{  The vertical a) and horizontal b) 
             beam width (one standard deviation)
          determined for  each five minutes partition of the COSY cycle. \protect\\
          Please note that the beam width $\sigma_{X}$ 
          determined for each five minutes period is smaller than the 
          $\sigma_{X}$  averaged over the whole cycle (compare figure~\ref{dx1_vs_sx1}c).
          This is due to the beam shift relative to the target during 
          the experimental cycle as shown in the panel c). \protect\\
          c) Relative settings of the  COSY proton beam 
          and the target centre versus the time of the measurement cycle.\protect\\
          The vertical error bars in pictures b) and c)  denote the size of the 
          step used in the Monte-Carlo simulations 
          ($\pm$~0.025~cm ; the bin width of figures~\ref{dx1_vs_sx1}b and~\ref{dx1_vs_sx1}c).
        }
 \label{sx_sy_cykl}
\end{figure}
\vspace{-0.1cm}
\begin{figure}[H]
\vspace{-1.4cm}
\parbox{0.45\textwidth}{\epsfig{file=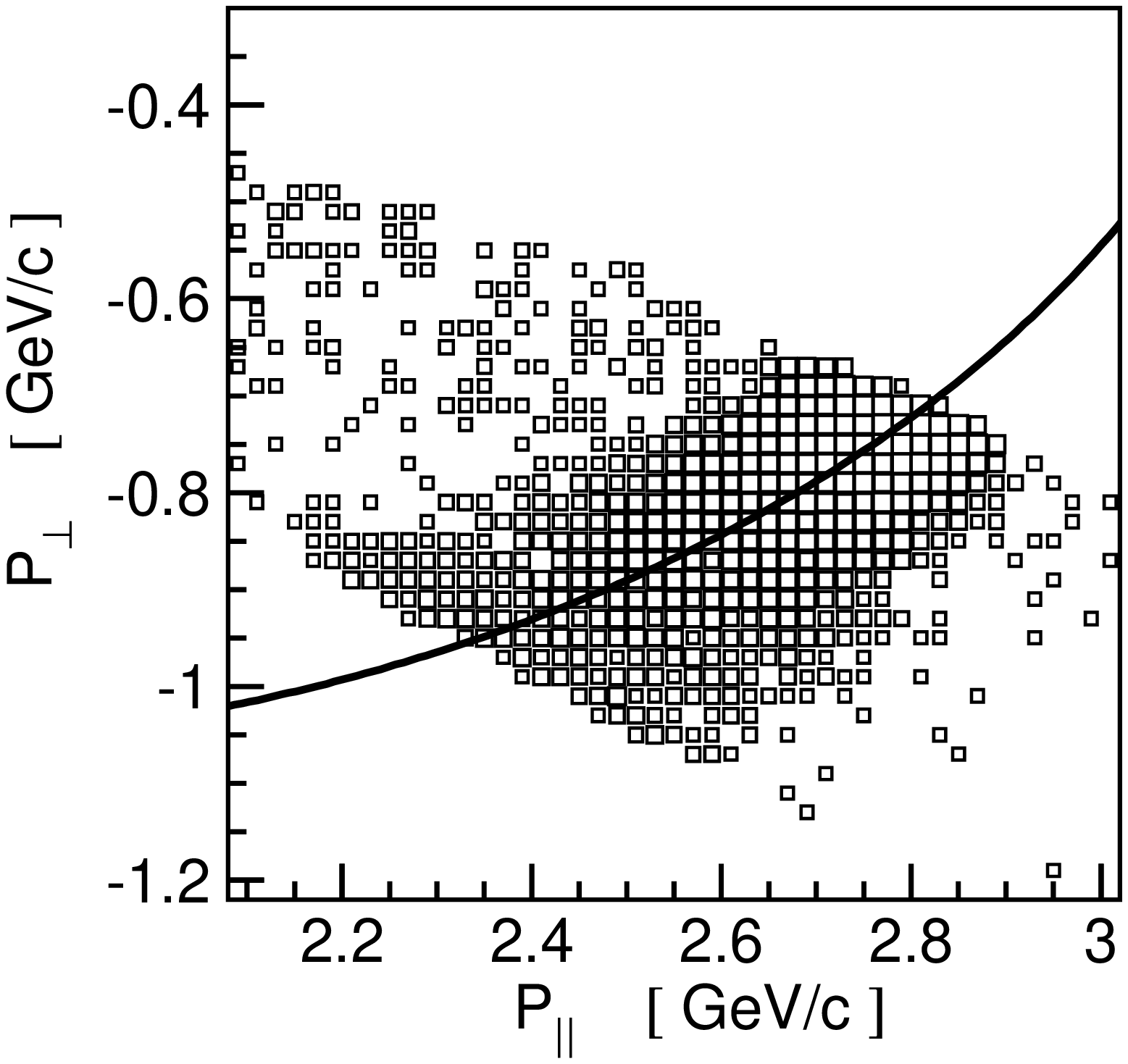,width=0.542\textwidth}} \hfill
\parbox{0.50\textwidth}
  {\epsfig{file=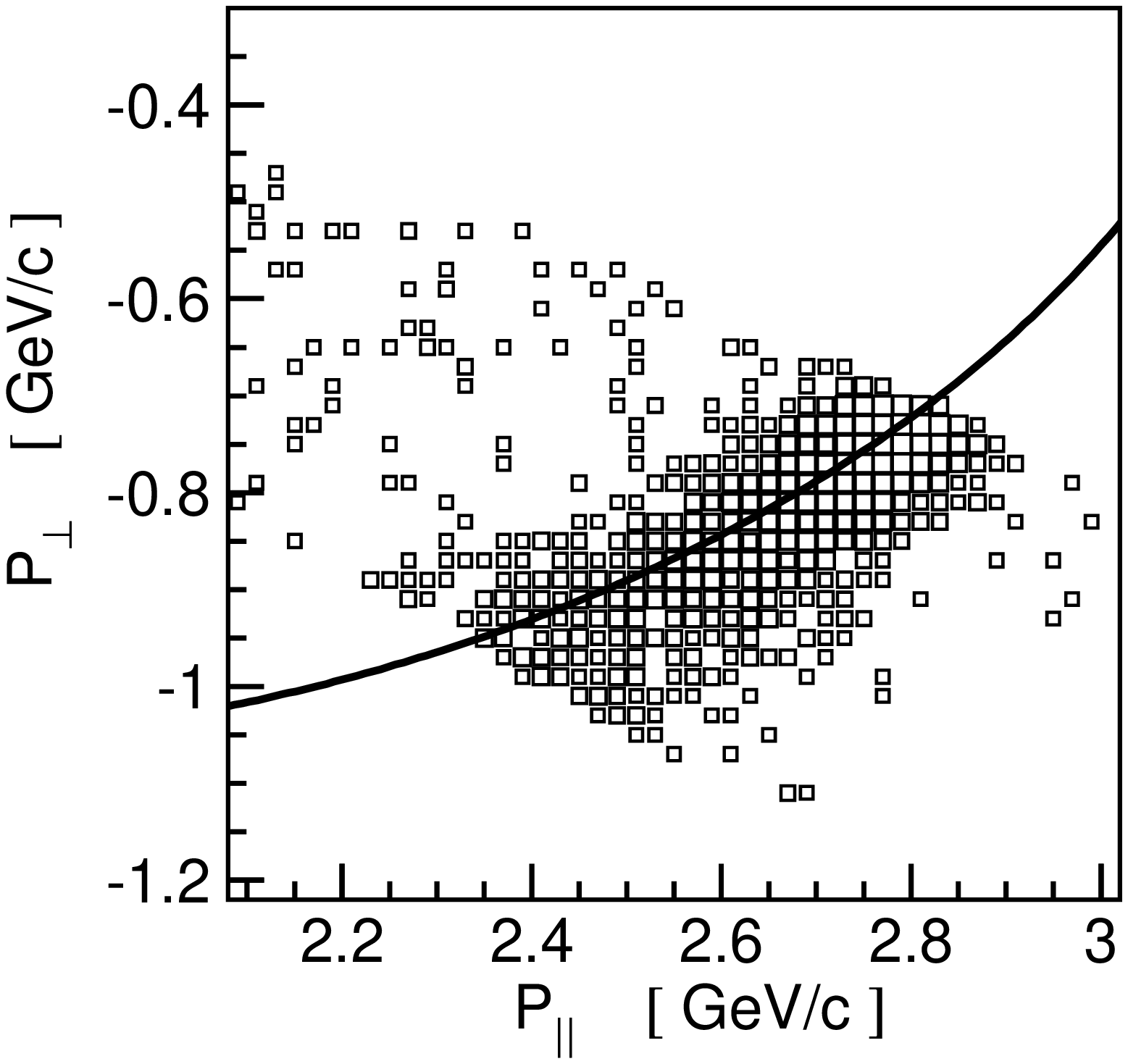,width=0.55\textwidth}}
 
\vspace{-0.6cm}
\parbox{0.38\textwidth}{\raisebox{1ex}[0ex][0ex]{\mbox{}}} \hfill
\parbox{0.49\textwidth}{\raisebox{1ex}[0ex][0ex]{\large a)}} \hfill
\parbox{0.03\textwidth}{\raisebox{1ex}[0ex][0ex]{\large b)}}

\vspace{-0.4cm}
\caption{  Perpendicular versus parallel momentum components with
           respect to the beam direction of particles registered at a beam
           momentum of 3.227~GeV/c, as measured during the first~(a),
           and the last minute (b) of the 60 minutes long COSY cycle.
           The data at the first minute were analyzed with $\Delta_{X}$~=~-0.15~cm,
           and the data of the 60's minute with $\Delta_{X}$~=~-0.25~cm,
           see figure~\ref{sx_sy_cykl}c.
           The number of entries per bin is shown
           logarithmically. The solid line corresponds to the momentum ellipse
           expected for protons scattered elastically at a beam momentum of
           3.227~GeV/c. 
        }
 \label{trans_paral_cykl}
\end{figure}
\vspace{-0.1cm}
In case of an uncooled beam its size would increase during the cycle
as can be seen in figure~\ref{sx_sy_cykl}a, which shows the
spreading of the beam in the vertical plane during the 60 minutes cycle.
This was expected since during the discussed experiment
the stochastic cooling in the vertical plane was not used.
 The influence of the applied cooling in the horizontal
 plane is clearly visible in figure~\ref{sx_sy_cykl}b.
During the first five minutes of the cycle the horizontal size of the beam 
of about  $2\cdot 10^{10}$ protons  was reduced by a factor of 2
reaching the equilibrium conditions, and remaining constant
for the rest of the COSY cycle.
 Figure~\ref{trans_paral_cykl} depicts the accuracy of the momentum reconstruction,
reflected in the spreading of the data, which is mainly due
to the finite horizontal size of the beam and target overlap.
The left and right panel correspond
to the first and the last minute of the measurement cycle, respectively.
The data were analyzed correcting for the relative target and beam shifts~$\Delta_{X}$.
The movement of the beam relative
to the target, during the cycle, is quantified in figure~\ref{sx_sy_cykl}c. The shift of the beam
denotes also changes of the average beam momentum,
due to the nonzero dispersion at the target place.
After some minutes the beam remains unchanged, this can be understood as 
reaching  the equlibrium between the energy losses
when crossing $1.6\cdot 10^{6}$ times per second
through the $H_{2}$ cluster target, and the power of the longitudinal
stochastic cooling
which cannot only diminish the  spread of the beam momentum but also shifts
it as a  whole.
\vspace{-0.3cm}
\section{Detailed data analysis on the example of the $pp\to pp\eta$ reaction} 
\label{detailedsection}
\vspace{-0.2cm}
\begin{flushright}
\parbox{0.73\textwidth}{
 {\em  
  Precision and exactness are not intellectual values in themselves,
  and we should never try to be more precise or exact than is 
  demanded by the problem in hand~\cite{popperbetter}.\\
 }
 \protect \mbox{} \hfill Karl Raimund  Popper  \protect\\
 }
\end{flushright}
\vspace{-0.4cm}
 Significant part of the differential distributions 
 of the cross sections used in this work
 was derived from a high
 statistics measurement of the $pp \rightarrow pp\eta$ reaction
 at a nominal beam momentum of 2.027~GeV/c. The experiment performed 
 at the COSY-11 facility 
 was based on the four-momentum registration
 of both outgoing protons, whereas the $\eta$ meson was identified
 via the missing mass technique.  The method allows for the extraction 
 of the kinematically complete information of the produced $pp\eta$ system, provided 
 that the reaction was identified. This is not feasible on the event-by-event basis
 due to the unavoidable physical background
 originating from  multi-pion production.
 However, with  a large number of identified events the background subtraction can
 be performed separately for each interesting phase space interval. In practice,
 the size of the studied phase space partitions 
 must be optimized between the statistical significance of 
 the signal-to-background ratio and the experimental resolution.
\begin{figure}[H]
  \vspace{-1.5cm}
  \parbox{0.5\textwidth}{\vspace{0.0cm}
     \includegraphics[width=0.54\textwidth]{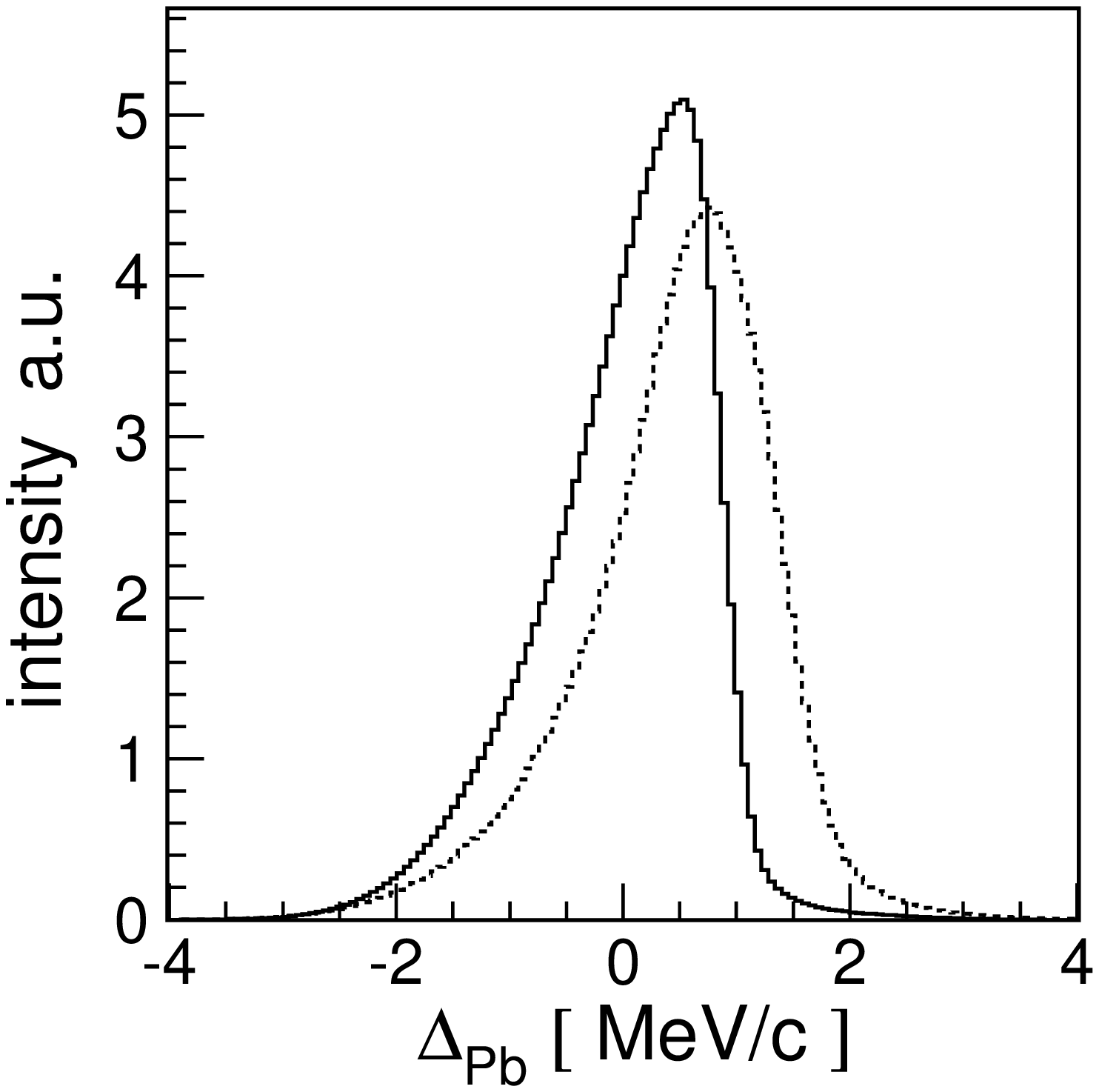}}
  \parbox{0.5\textwidth}{\vspace{0.0cm}
    \includegraphics[width=0.5\textwidth]{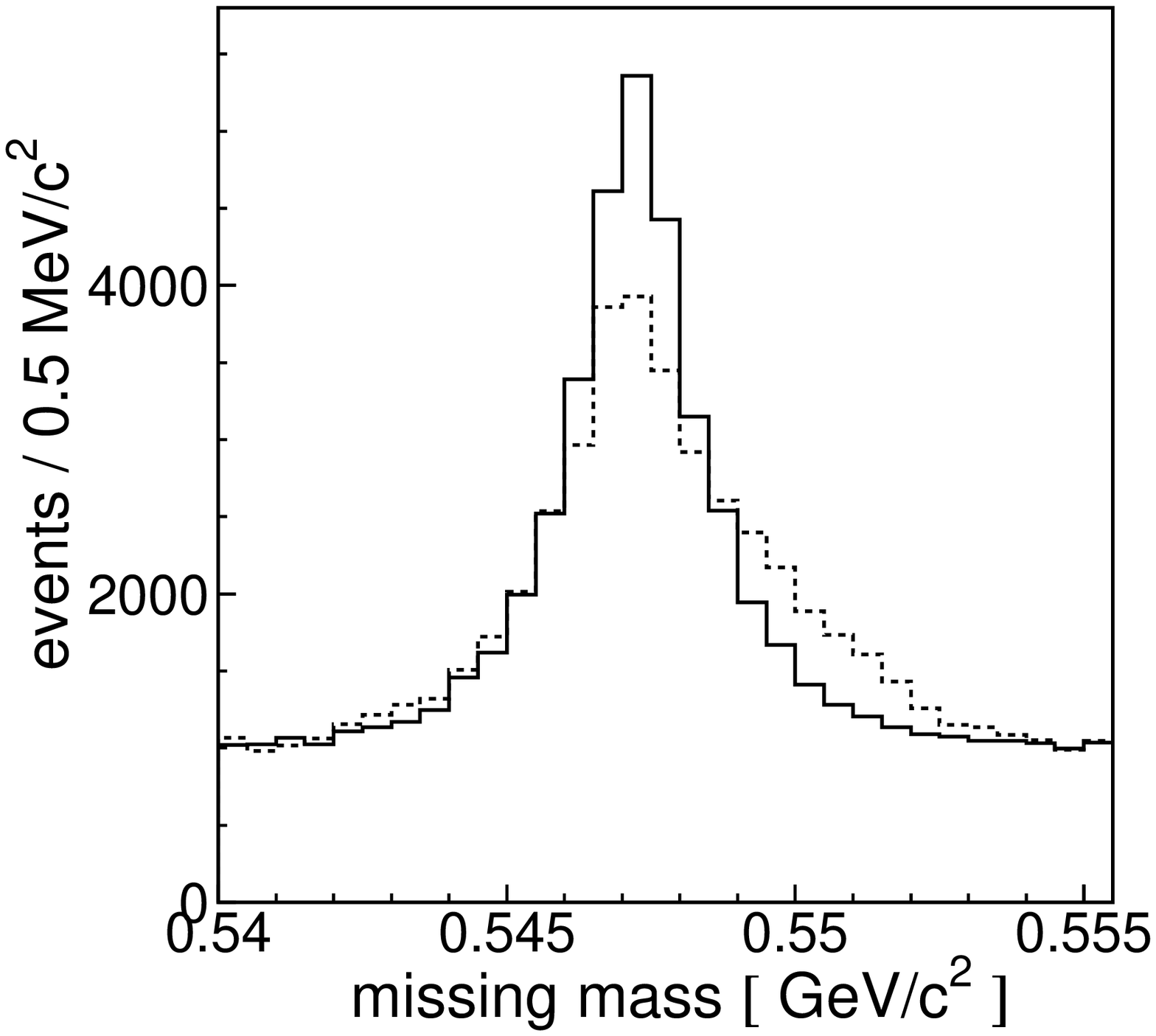}}

  \vspace{-0.4cm}
  \parbox{0.38\textwidth}{\raisebox{1ex}[0ex][0ex]{\mbox{}}} \hfill
  \parbox{0.49\textwidth}{\raisebox{1ex}[0ex][0ex]{\large a)}} \hfill
  \parbox{0.03\textwidth}{\raisebox{1ex}[0ex][0ex]{\large b)}}
  \vspace{-0.4cm}
  \caption{ 
          (a) \ \
          The dashed curve denotes the 
           proton beam momentum distribution 
           integrated over the whole measurement period.
           On the horizontal scale, the value of zero is set at a nominal beam momentum
           equal to 2.027~GeV/c.
           The solid line shows the
           beam momentum distribution after the correction for
           the mean value which was
           determined in  10 second intervals. 
          (b) \ \ Missing mass distribution for the $pp\rightarrow ppX$
           reaction determined by means of the COSY-11 detection system
           at a beam momentum of 2.027~GeV/c.
          The solid histogram presents the data corrected
          for effects of the time dependent relative
          shifts between the beam and
          the target using the method described in section~\ref{monitoring}.
          The dashed histogram shows the result before the correction.\protect\\
           \label{miss_all}
        }
\end{figure}
 The accuracy of the missing mass reconstruction
 depends on two parameters: the spread of the beam momentum as well as the 
 precision of the measured momenta of the outgoing protons. The latter is 
 predominantly due to the geometrical spread of the beam.  
 Since for the reconstruction of the momenta we assume that the reaction
 takes place in the middle of the target we cannot correct 
 on an event-by-event basis for the momentary spread of the beam.
 However, we can rectify the smearing due to the shifts of the centre of the beam relative
 to the target  as well as the average changes of the absolute beam momentum 
 during the experiment.
Therefore, after the selection of events with two registered protons,
as a first step of the more refined analysis
the data were corrected for the mean beam momentum
changes (see fig.~\ref{miss_all}a) determined from the
measured Schottky frequency spectra and the known beam optics.
In a next step, from the distributions of the elastically scattered protons,
the Schottky frequency spectrum, and the missing mass distribution
of the $pp \to pp X$ reaction,
we have estimated that 
the spread of the beam momentum,
and the spread of the reaction points in horizontal and
vertical direction
amount to
$\sigma(p_{beam})~=~0.63~\pm~0.03$~MeV/c,
$\sigma(x)~=~0.22~\pm~0.02$~cm,
and $\sigma(y)~=~0.38~\pm~0.04$~cm, respectively.
Details of this procedure can be found in section~\ref{monitoring}.
\vspace{0.4cm}
\begin{figure}[H]
  \parbox{0.55\textwidth}{\vspace{-0.3cm}
    \includegraphics[width=0.50\textwidth]{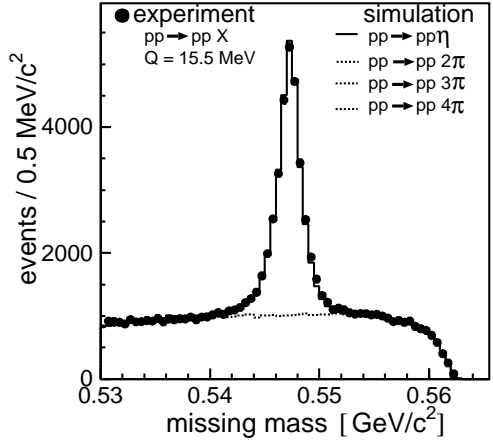}}
  \parbox{0.45\textwidth}{\vspace{-1.1cm}
  \caption{ \label{missall} 
         Missing mass spectrum for the $pp\to ppX$ reaction
         determined in the experiment at a beam momentum of 2.0259~GeV/c.
         The mass resolution amounts to 1~MeV/c$^2(\sigma)$.
         The superimposed histograms present the simulation
         for $1.5\cdot 10^8$ events of the $pp\to pp\eta$ reaction,
         and $10^{10}$ events for the reactions $pp\to pp 2\pi$,
         $pp\to pp 3\pi$, and $pp\to pp 4\pi$.
         The simulated histograms were fitted to the data varying only the magnitude.
         The fit resulted in $24009\pm 210$ events for the production
         of the $\eta$ meson.
 }} 
\end{figure}
\vspace{0.1cm}
Further on, a comparison of the experimentally determined
momentum spectra of the elastically scattered protons
with the distributions simulated with different beam and target
conditions allows us to establish the position at which the
centre of the 
beam crosses the target with an accuracy of 0.25~mm.
Accounting for the movement of the beam
relative to the target 
we improved the missing mass resolution
as demonstrated in figure~\ref{miss_all}b.

After this betterment, the peak originating from the $pp\to pp\eta$ reaction became 
more symmetric and the signal-to-background ratio increased significantly.
The background was simulated taking into account $pp\to pp X$ reactions with
$X~=~2\pi,~3\pi$, and~$4\pi$.
Since we consider here only
the very edge of the phase space distribution where the protons are
produced predominantly in the S-wave the shape of the background can
be reproduced assuming that the homogeneous phase space distribution
is modified only by the interaction between protons. Indeed, as can be observed
in figure~\ref{missall} the simulation describes the data very well.
The calculated spectrum is hardly distinguishable from the experimental points.
The position of the peak on the missing mass spectrum
and the known mass of the $\eta$ meson~\cite{PDG}
enabled to determine the actual absolute beam
momentum to be $p_{beam}~=~2.0259$~GeV/c~$\pm$~0.0013~GeV/c,
which agrees within error limits with the 
nominal value of $p_{beam}^{nominal}~=~2.027$~GeV/c.
The real beam momentum corresponds to an excess energy of the pp$\eta$ system 
equal to Q~=~15.5~$\pm$~0.4~MeV.

\subsection{Covariance matrix and kinematical fitting}
\begin{flushright}
\parbox{0.70\textwidth}{
 {\em  
 Beobachtungen, welche sich auf Gr{\"o}ssen\-besti\-mmungen aus der Sinnenwelt beziehen,
 werden immer, so sorgf{\"a}ltig man auch verfahren mag, gr{\"o}sseren oder kleineren Fehlern
 unterworfen bleiben~\cite{gauss}. \protect\\
 }
 \protect \mbox{} \hfill Carl Friedrich Gauss \protect\\
 }
\end{flushright}
\vspace{-0.3cm}

 As already mentioned in the previous section,
 at the COSY-11 facility the identification of the $pp \to pp \eta$
reaction is based on the measurement of the momentum vectors of
the outgoing protons and the utilisation of the missing mass technique.
Inaccuracy of the momentum determination manifests itself in the
population of  kinematically forbidden regions of the phase space,
preventing a precise comparison of the theoretically derived
and experimentally determined differential cross sections. 
\vspace{-0.3cm}
\begin{figure}[H]
  \parbox{0.55\textwidth}{\vspace{-0.3cm}
    \includegraphics[width=0.52\textwidth]{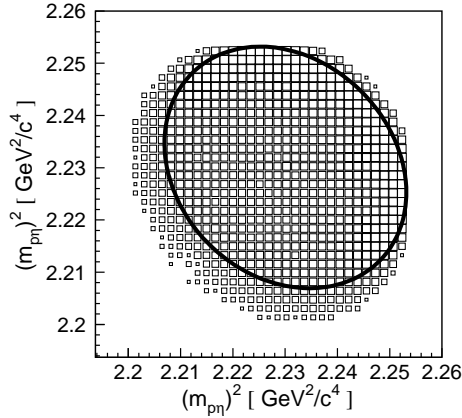}}
  \parbox{0.45\textwidth}{\vspace{-1.3cm}
  \caption{Dalitz plot distribution of the $pp\to pp\eta$ reaction
         simulated at Q~=~15.5~MeV.
         The number of entries is shown
          in a logarithmic scale.
         The solid line gives the kinematically allowed area.
          The  result was obtained
          taking into account the experimental conditions
          as described in the text.
         \label{dalitzsym}
  }}
\end{figure}
\vspace{-0.1cm}
Figure~\ref{dalitzsym}
visualizes this effect and
clearly demonstrates that the data
scatter significantly
outside the kinematically allowed region~(solid line),
in spite of the fact that the precision of the fractional momentum determination
in the laboratory system ($\sigma(p_{lab})/p_{lab}\approx 7\cdot 10^{-3}$) is quite high.
Therefore, when seeking for  small effects
like for example the influence of the proton-$\eta$ interaction
on the population density of the phase space,
one needs either to fold the theoretical calculations with the experimental resolution,
or to perform  a kinematical fitting of the data.  Both procedures require the knowledge
of the covariance matrix, and thus its determination constitutes a necessary step
in the differential analysis and interpretation of the data.

In order to derive the covariance matrix we need to recognize and quantify
all possible sources of errors
in the reconstruction of the two proton momenta $\vec{p}_1$ and $\vec{p}_2$.
The four dominant effects are:
i)   finite distributions of the beam momentum and of the reaction points,   
ii)  multiple scattering in the dipole chamber exit foil, air, and detectors,
iii) finite resolution of the position determination of the drift chambers, and 
iv) a possible inadequate assignment of hits to the particle tracks in drift chambers in the case
    of very close tracks.
Some of these, like the multiple scattering, depend on the outgoing protons' momenta,
others, like the beam momentum distribution, depend on the specific
run conditions and therefore must be determined for each run separately.
 
In order to estimate the variances and covariances for all possible
combinations of the momentum components of two registered protons
we have generated $1.5\cdot 10^8$ $pp\to pp\eta$ events and simulated the response
of the COSY-11 detection setup taking into account the above listed
factors and the known resolutions of the detector components.
Next, we  analysed the signals by means of the same reconstruction procedure
as used in case of the experimental data.
Covariances between the $i^{th}$ and the $j^{th}$ components of the
event vector ${\displaystyle{ \left( P=[p_{1x},p_{1y},p_{1z},p_{2x},p_{2y},p_{2z}] \right)}}$
were established as the average of the product of the deviations between the reconstruced
and generated values.
The explicit formula for the sample of $N$ reconstructed events reads:
 
\vspace{-0.4cm}
 
\begin{equation}
   \nonumber
   cov(i,j)~=~\frac{1}{N} \sum_{k=1}^{N}(P_{i,gen}^k - P_{i,recon}^k)(P_{j,gen}^k - P_{j,recon}^k),
\end{equation}

\vspace{-0.0cm}

\hspace{-0.5cm} where $P_{i,gen}^k$ and $P_{i,recon}^k$ denote the generated and reconstructed
values for the $i^{th}$ component of the vector $P$ describing the $k^{th}$ event.
 
Because of the inherent symmetries of the covariance matrix 
$(cov(i,j)=cov(j,i))$
and the indistinguishability of the registered 
protons\footnote{\mbox{} The symmetry of all observables 
   under the exchange of the two protons 
  ($\vec{p}_1 \leftrightarrow \vec{p}_2$) implies that 
   cov(i,j)~=~cov(i$\pm$3,j$\pm$3), where the '+' has to be taken
   for i,j~=~1,2,3 and the '-' for i,j~=~4,5,6. Thus for example
   cov(2,4)~=~cov(5,1).}
there are only
12 independent values which determine the $6~{\mbox{x}}~6$
error matrix $V$ unambiguously.
 
Since inaccuracies of the momentum determination depend on the particle momentum
itself (eg.~multiple scattering) and on the relative momentum between protons
(eg. trajectories reconstruction from signals in drift chambers),
we have determined the covariance matrix as a function of the  absolute momentum
of both protons: $cov(i,j,|\vec{p_{1}}|,|\vec{p_{2}}|)$.
As an example we present
the covariance matrix for the mean values of $|\vec{p_{1}}|$
and  $|\vec{p_{2}}|$ in units of MeV$^2$/c$^2$,
as established in the laboratory system with  the $z$-coordinate parallel
to the beam axis and $y$-coordinate corresponding to the vertical direction.
 
\vspace{-0.3cm}
\hspace{-1cm}\begin{equation}
 \mbox{}\hspace{-0.7cm}V~=~\begin{array}{c}
 \begin{array}{cccccc}
       p_{1x} \ \  &   p_{1y} \ \ &  p_{1z} \ \ & p_{2x} \ \ &  p_{2y} \ \ &  p_{2z}
 \end{array}
 \begin{array}{c}
  \mbox{\ \hspace{0.3cm}\ }
 \end{array}\\
 \left[
   \begin{array}{cccccc}
       5.6   &   0.0 &  -13.7 & 1.7 & 0.1  &  -3.0 \\
       -     &   7.1 &  0.1 &  - & -0.2  &  -0.2 \\
       -     &   -   &  37.0 & - & -  &  5.4 \\
       -   &   - &  - & - & -  &  - \\
       -   &   - &  - & - & -  &  - \\
       -   &   - &  - & - & -  &  -
   \end{array}
 \right]
 \begin{array}{c}
      p_{1x} \ \\\
      p_{1y}\\
      p_{1z}\\
      p_{2x}\\
      p_{2y}\\
      p_{2z}
 \end{array}
 \end{array}
\end{equation}
 
Since the measurements have been performed close to the kinematical threshold
the ejectile momentum component parallel to the beam is by far the largest one
and
its variance ($var(p_{z})~=~37~MeV^2/c^2$)
determines in first order the error of the momentum measurement.
The second largest contribution stems from
an anti-correlation
between the $x-$ and $z-$ momentum components ($cov(p_{x},p_{z})~= -~13.7~MeV^2/c^2$),
which is due to the bending
of the proton trajectory --~mainly in the horizontal direction~--
inside the COSY-11 dipole magnet~(see fig.~\ref{detector}).
There is also a significant correlation between the $z$ components of different
protons which is due to the smearing of the reaction points,
namely, if in the analysis the assumed reaction point differs from the
actual one, a mistake made in the reconstruction affects both protons similarly.

\vspace{-0.7cm}
\begin{figure}[H]
  \parbox{0.33\textwidth}{\vspace{0.4cm}
    \includegraphics[width=0.345\textwidth]{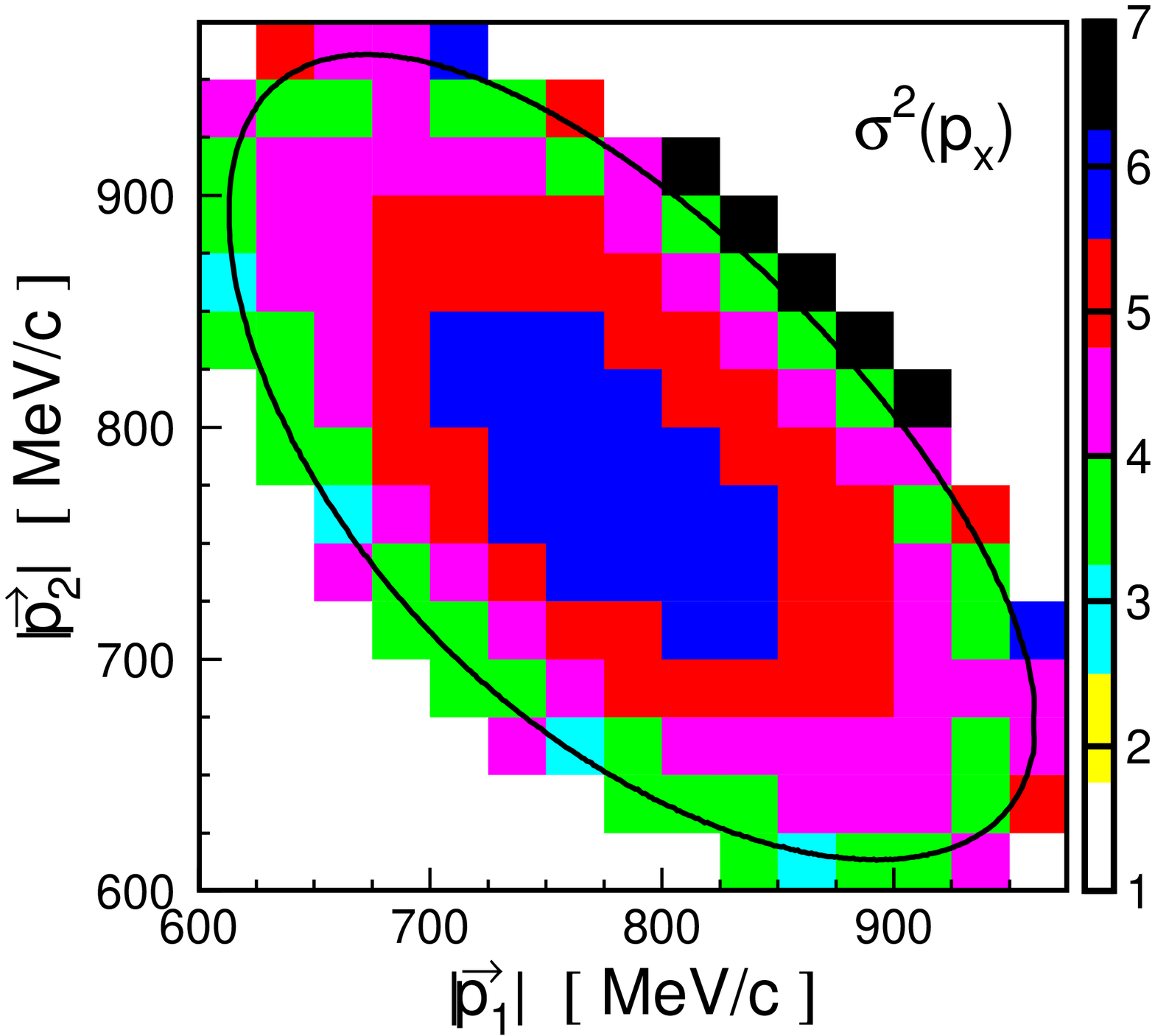}}
  \parbox{0.32\textwidth}{\vspace{0.0cm}
    \includegraphics[width=0.345\textwidth]{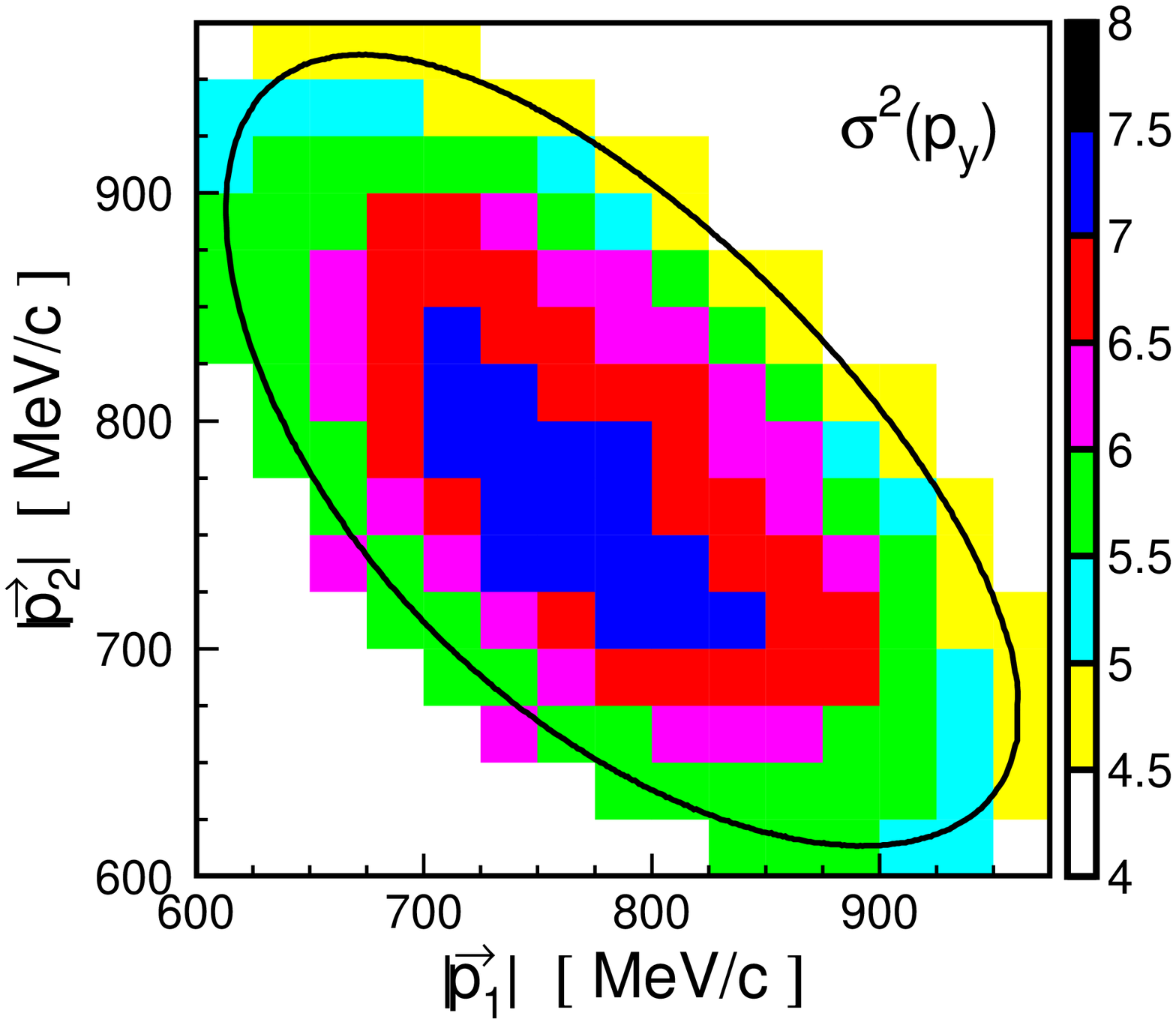}}
  \parbox{0.33\textwidth}{\vspace{0.0cm}
    \includegraphics[width=0.345\textwidth]{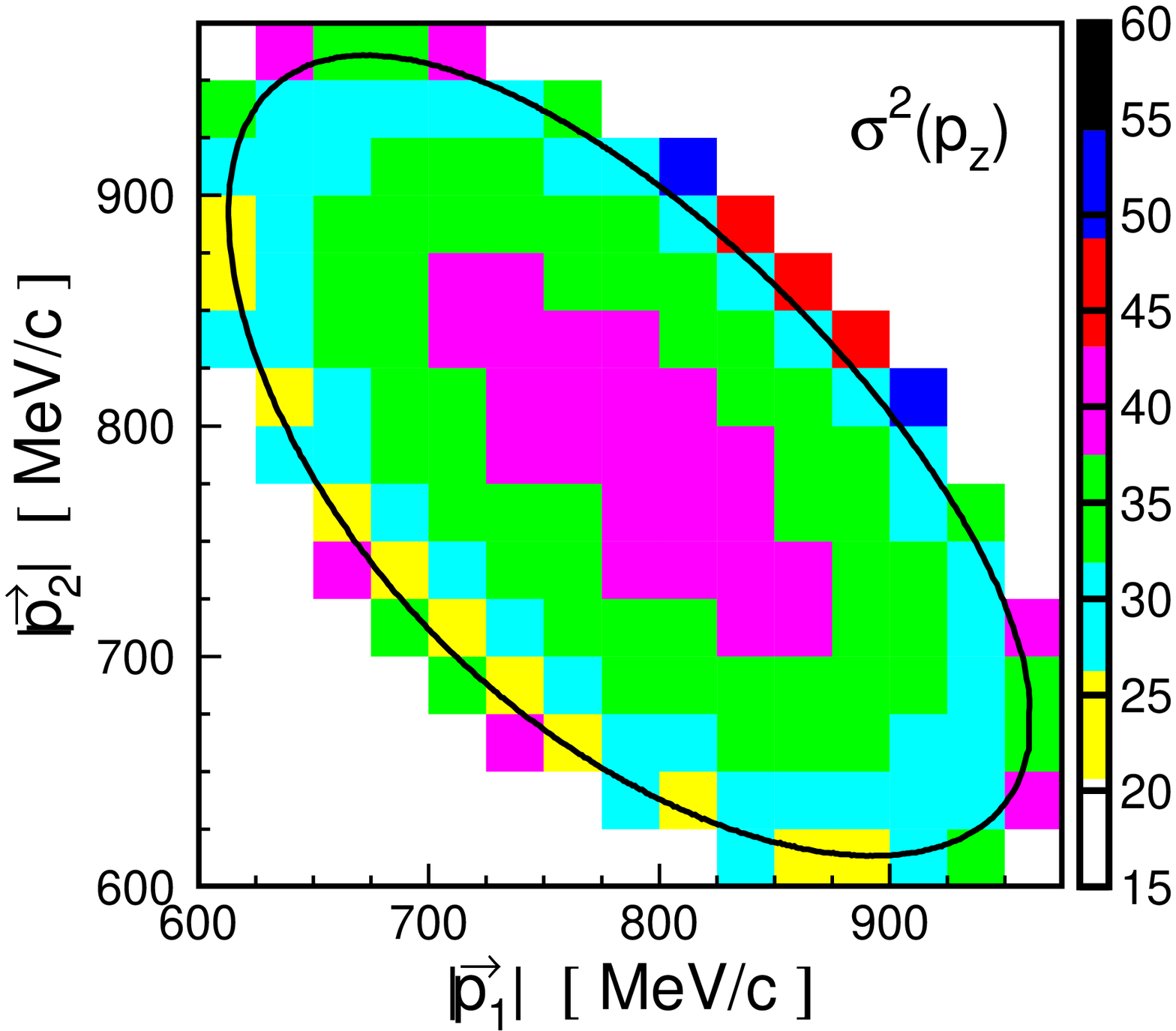}}
   \vspace{-0.2cm}
  \caption{ \label{covariance}
     Variances of the protons's momentum components 
     $\sigma^2(p_x)$, $\sigma^2(p_z)$, and $\sigma^2(p_y)$ shown
     as a function of the absolute values of the measured momenta.
  }
\end{figure}
\vspace{-0.0cm}
 
Figure~\ref{covariance} depicts the variation of $var(p_x)$, $var(p_y)$, and $var(p_z)$
over the momentum plane ($|\vec{p_{1}}|$, $|\vec{p_{2}}|$).
Taking into account  components of the covariance matrices V($|\vec{p_{1}}|$, $|\vec{p_{2}}|$)
and the distribution of the
proton momenta for the $pp\to pp \eta$ reaction at Q~=~15.5~MeV results in an
average error for the measurement of the proton momentum of about 6~MeV/c.
This can be also deduced from the distribution of the difference between the generated
and reconstructed absolute momenta of the protons.
The corresponding spectrum is plotted as a dashed line in figure~\ref{deltafit}.
\vspace{-0.4cm}
\begin{figure}[H]
  \parbox{0.55\textwidth}{\vspace{-0.3cm}
    \includegraphics[width=0.54\textwidth]{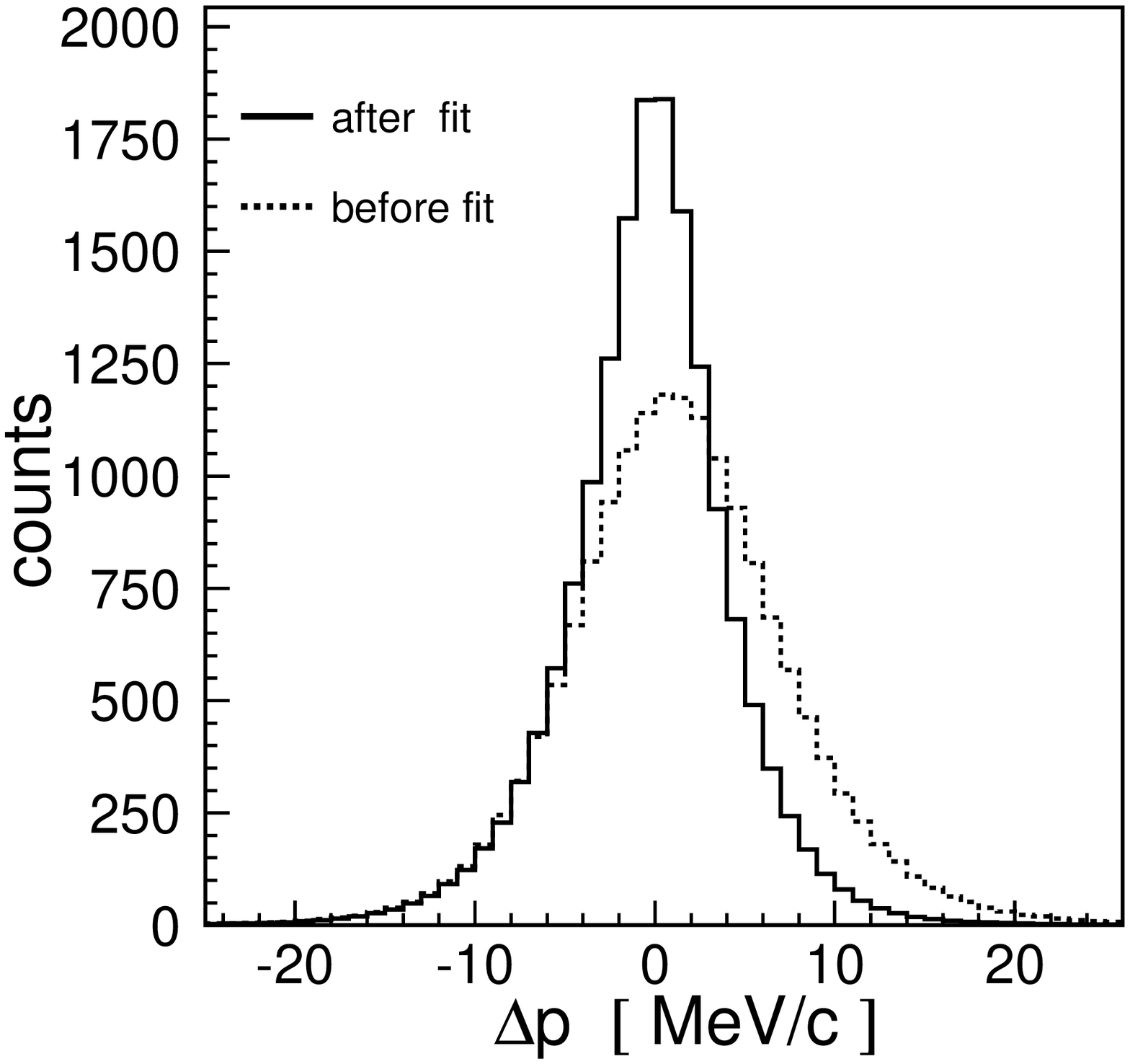}}
  \parbox{0.45\textwidth}{\vspace{-1.0cm}
  \caption{\label{deltafit}
          Spectrum of  differences between generated
          and --~after simulation of the detector response~--
          reconstructed
          absolute momenta of  protons, 
          as determined before (dashed line) and after
          the kinematical fit (solid line). \protect\\
          The picture  shows results obtained
          taking into account the experimental conditions
          as described in the text.
  }}
\end{figure}

In the experiment we have measured 6 variables and once we assumed that the event
corresponds to the $pp\to pp\eta$ reaction only
5 of them are independent. Thus we have varied the values of the event components
demanding that the missing mass is equal to the mass of the $\eta$ meson
and we have chosen that vector which was the closest to the experimental one
in the sense of the Mahalanobis distance.
The inverse of the covariance matrix was used as a metric for the distance
calculation.
The kinematical fit improves the  resolution
by a factor of about 1.5
as can be concluded
from comparing the dashed and solid lines in figure~\ref{deltafit}.
 The finally resulting error of the momentum determination  amounts to 4~MeV/c.

\subsection{Multidimensional acceptance corrections}
\label{multiaccept}
\begin{flushright}
\parbox{0.73\textwidth}{
 {\em Unfehlbar h{\"a}tte ich umkommen m{\"u}ssen, 
      wenn ich mich nicht mit der St{\"a}rke meiner
      Arme an meinem eigenen  Haarzopf samt meinem Pferd, 
      das ich fest zwischen meine
      Knie schlo{\small{$\beta$}},
      wieder herausgezogen h{\"a}tte~\cite{baron}.\protect\\
 }
 \protect \mbox{} \hfill Baron M{\"u}nchhausen
 }
\end{flushright}
At the excess energy of Q~=~15.5~MeV
the COSY-11 detection system does not cover the full 4$\pi$ solid angle
in the centre-of-mass  system of the $pp\to pp\eta$ reaction.
Therefore, the detailed study of differential cross sections
requires corrections for the acceptance. Generally, the acceptance
should be expressed as a function of the full set of mutually
orthogonal variables which describe the studied 
reaction unambiguously.
As introduced in section~\ref{choiceofobservables}, 
to define the relative movement of the particles in the reaction plane
we have chosen two squares of the invariant masses: $s_{pp}$  and $s_{p\eta}$,
and to define the orientation of this plane in the center-of-mass frame
we have taken the three Euler angles:
The first two are simply  the polar $\phi_{\eta}^{*}$ and azimuthal $\theta_{\eta}^{*}$ angles
of the momentum of the $\eta$ meson and the third angle $\psi$
describes the rotation of the reaction plane around the axis defined by the momentum vector
of the $\eta$ meson.

In the data evaluation we considerably benefit from the basic geometrical symmetries
satisfied by the $pp\to pp\eta$ reaction.
 Due to the axial symmetry of  the initial channel of the two
unpolarized colliding protons the event distribution over
$\phi_{\eta}^{*}$ must be isotropic. Thus, we can safely integrate
over $\phi^{*}_{\eta}$, ignoring
that variable in the analysis.
 Furthermore, taking advantage of the symmetry
due to the two identical particles in the initial channel,
without losing the generality, we can express
the acceptance as a function of $s_{pp}$,$s_{p\eta}$,$|cos(\theta_{\eta}^{*})|$, and $\psi$.
To facilitate the calculations we have divided the range of $|cos(\theta_{\eta}^{*})|$ and $\psi$
into 10 bins and both $s_{pp}$ and $s_{p\eta}$ into 40 bins each.
In the case of  $s_{pp}$ and $s_{p\eta}$ the choice was made such
that the width of the interval corresponds to the standard deviation of the experimental
accuracy. For $|cos(\theta_{\eta}^{*})|$ and $\psi$  we have taken only ten partitions
since from the previous experiments we expect only a small variation of the cross section
over these variables~\cite{TOFeta,calen190,jim}.
In this representation, however, the COSY-11 detection system covers only 50\%
of the phase space for the $pp\to pp\eta$ reaction at Q~=~15.5~MeV.
To proceed with the analysis we assumed that the distribution over the angle $\psi$
is isotropic as it was for example experimentally determined
for the $pp\to pp\omega$, $pp\to pp\rho$,
or $pp\to pp\phi$ reactions~\cite{balestra092001,jim}.
Please note that this is the only assumption
of  the reaction dynamics performed in the present evaluation.
The validity of this supposition in the case of
the $pp\to pp\eta$ reaction will be discussed later.

In the calculations we exploit also the symmetry of the cross sections
under the exchange of the two identical particles in the final state which reads:
$\sigma(s_{p_{1}\eta},\psi)~=~\sigma(s_{p_{2}\eta},\psi+\pi)$.
The resultant acceptance is shown in figure~\ref{acceptanceijk}.
One recognizes that only a small part (3\%) of the phase space is not covered by the detection system.
In further calculations these  holes were corrected according to the  assumption
of a homogeneous phase space distribution. Additionally it was checked that the corrections
under other suppositions e.g. regarding also the proton-proton FSI
lead to negligible differences.
\begin{figure}[H]
 \vspace{-4.2cm}
 
 \parbox{1.00\textwidth}{\centerline{\epsfig{file=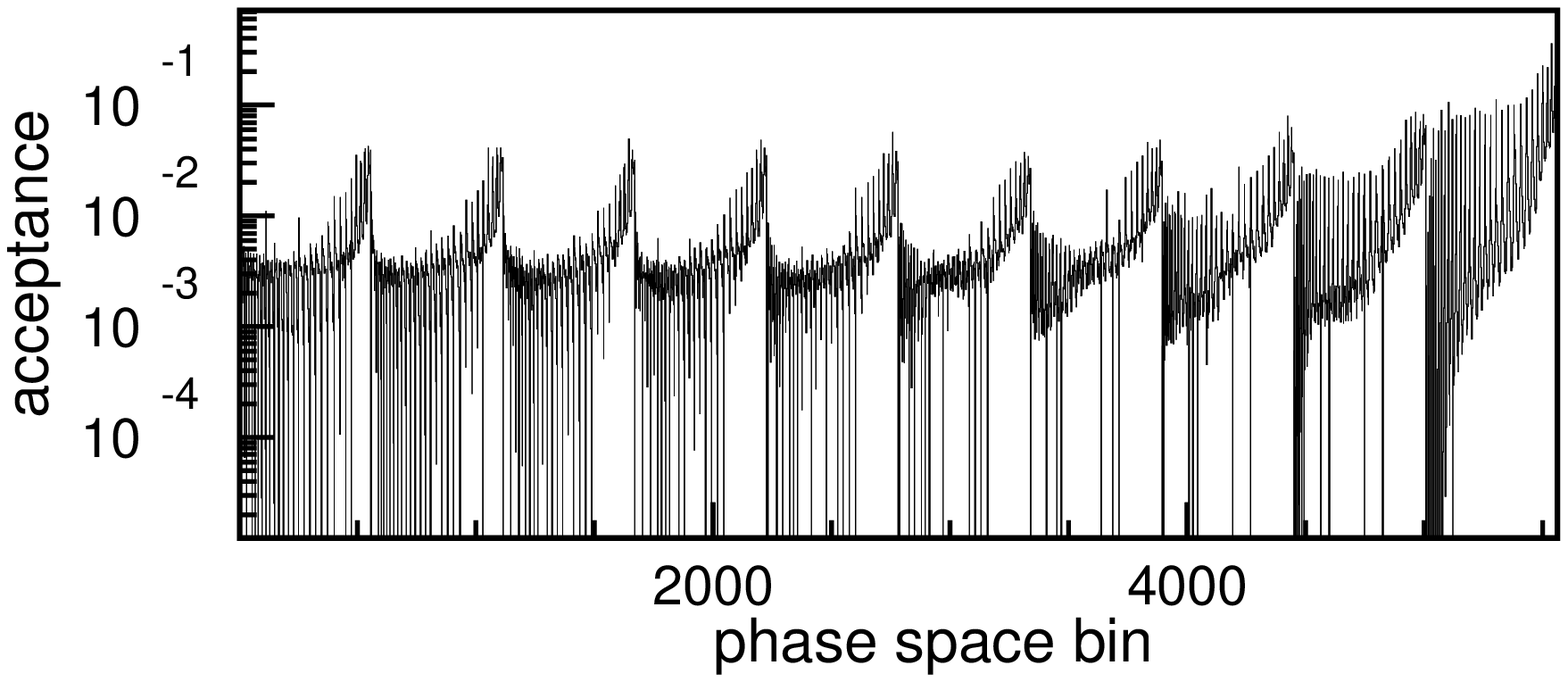,width=1.14\textwidth}}}
 
 \vspace{-8.1cm}
 
 \parbox{1.00\textwidth}{\centerline{\epsfig{file=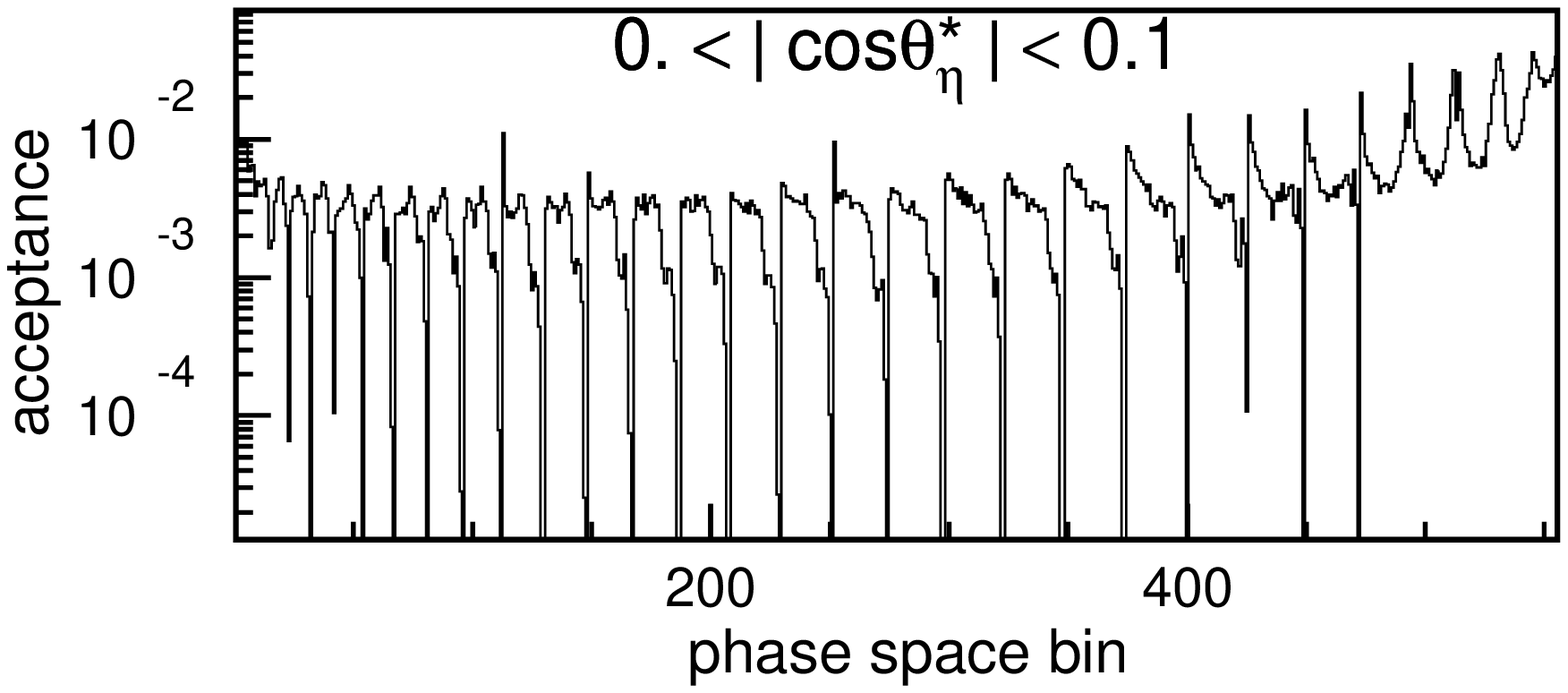,width=1.14\textwidth}}}
 
 \vspace{-8.1cm}
 
 \parbox{1.00\textwidth}{\centerline{\epsfig{file=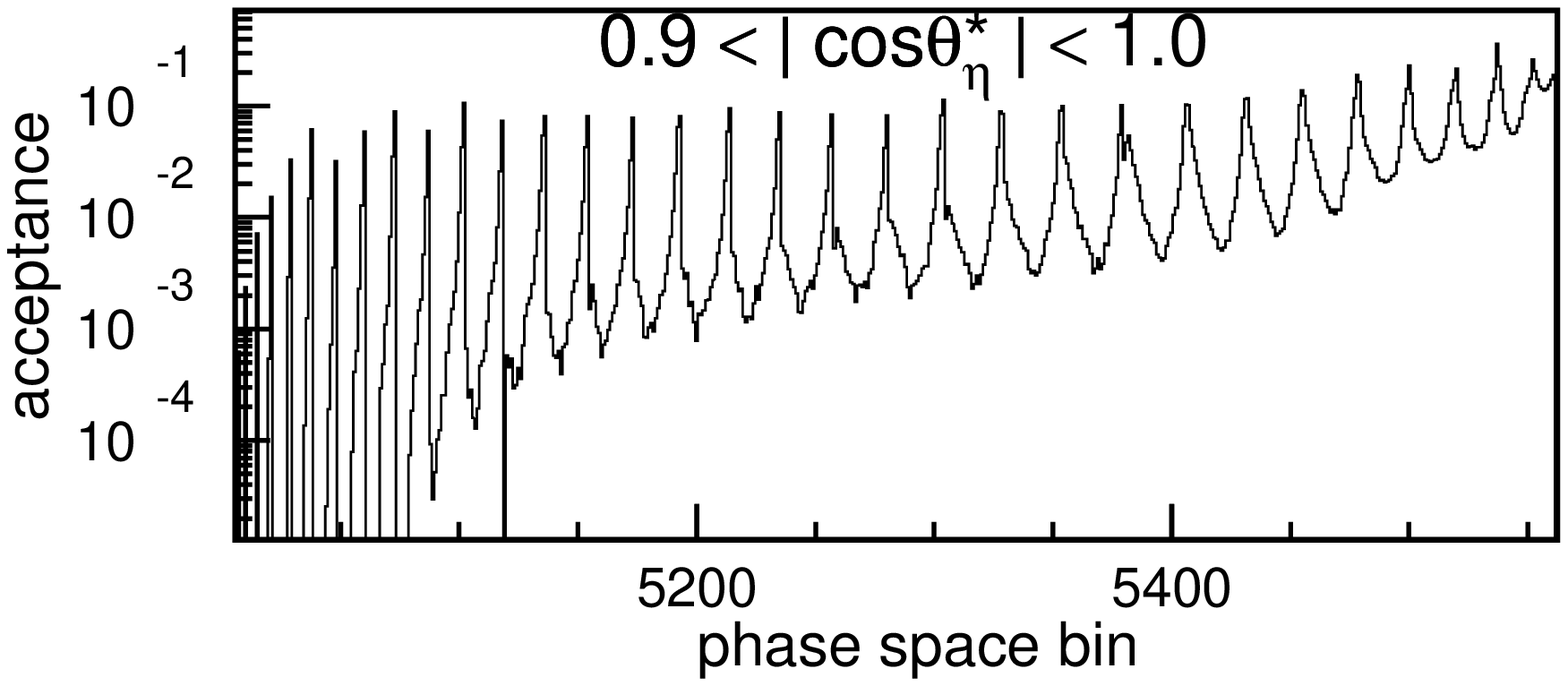,width=1.14\textwidth}}}
  \vspace{-3.8cm}
  \caption{ \label{acceptanceijk} 
    Acceptance of the COSY-11 detection system for the $pp\to pp\eta$ reaction
    at an excess energy of Q~=~15.5~MeV presented as a function of $s_{pp}$,$s_{p\eta}$,
    and $|cos(\theta_{\eta}^{*})|$.
    The numbers were assigned to the bins in the three dimensional space
    $s_{pp}-s_{p\eta}-|cos(\theta_{\eta}^{*})|$
    by first incrementing the index of $s_{pp}$ next of $s_{p\eta}$ and in the end that
    of $|cos(\theta_{\eta}^{*})|$.
    Partitioning of $|cos(\theta_{\eta}^{*})|$ 
    into ten bins 
    is easily recognizable. 
    The two lower pictures show the acceptance for the first and the last
    bin of $|cos(\theta_{\eta}^{*})|$.
  }
\end{figure}

\vspace{-0.4cm}
\begin{figure}[H]
  \parbox{0.45\textwidth}{\vspace{0.0cm}
    \includegraphics[width=0.49\textwidth]{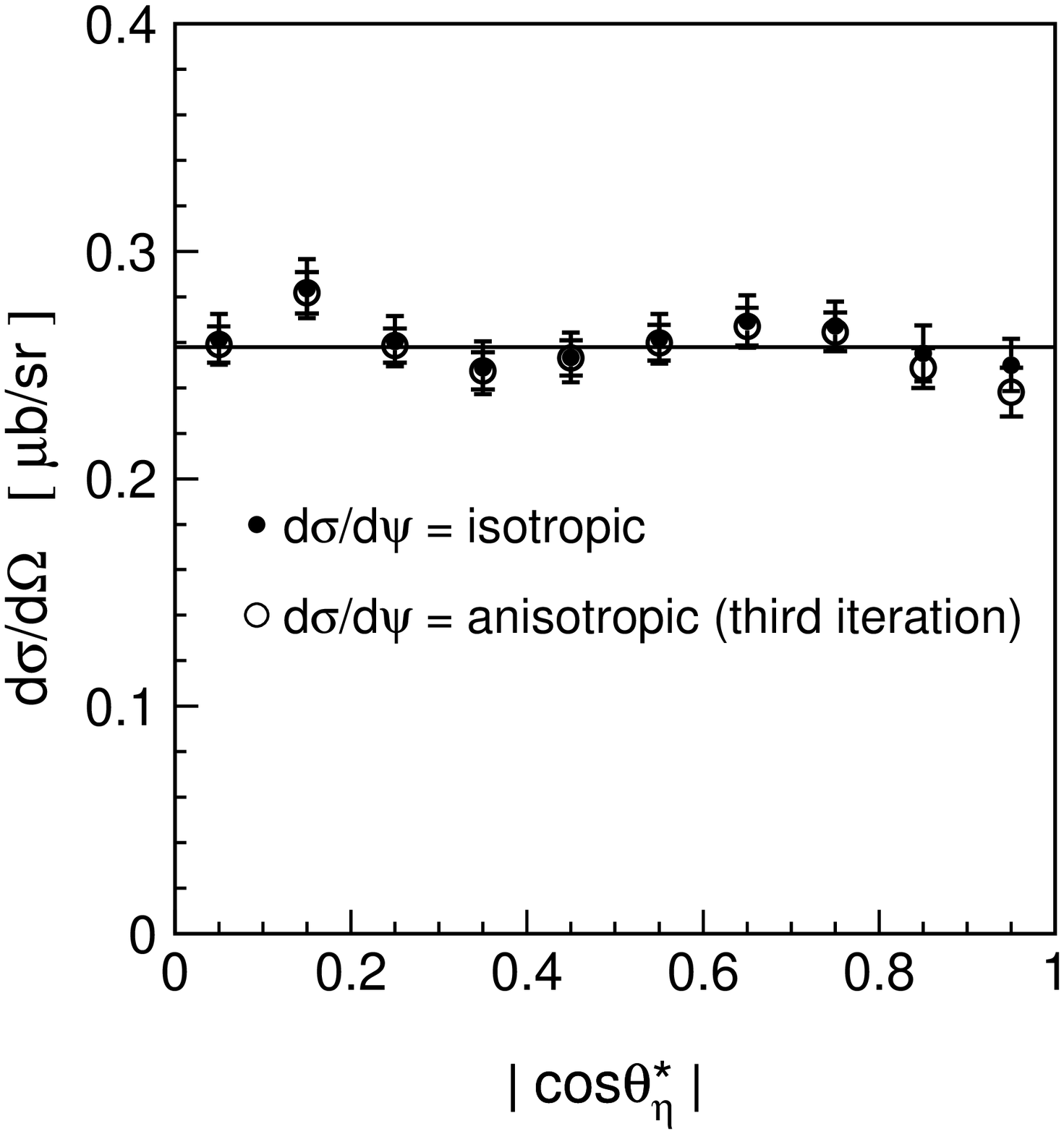}}
  \parbox{0.55\textwidth}{\vspace{-0.2cm}
    \includegraphics[width=0.58\textwidth]{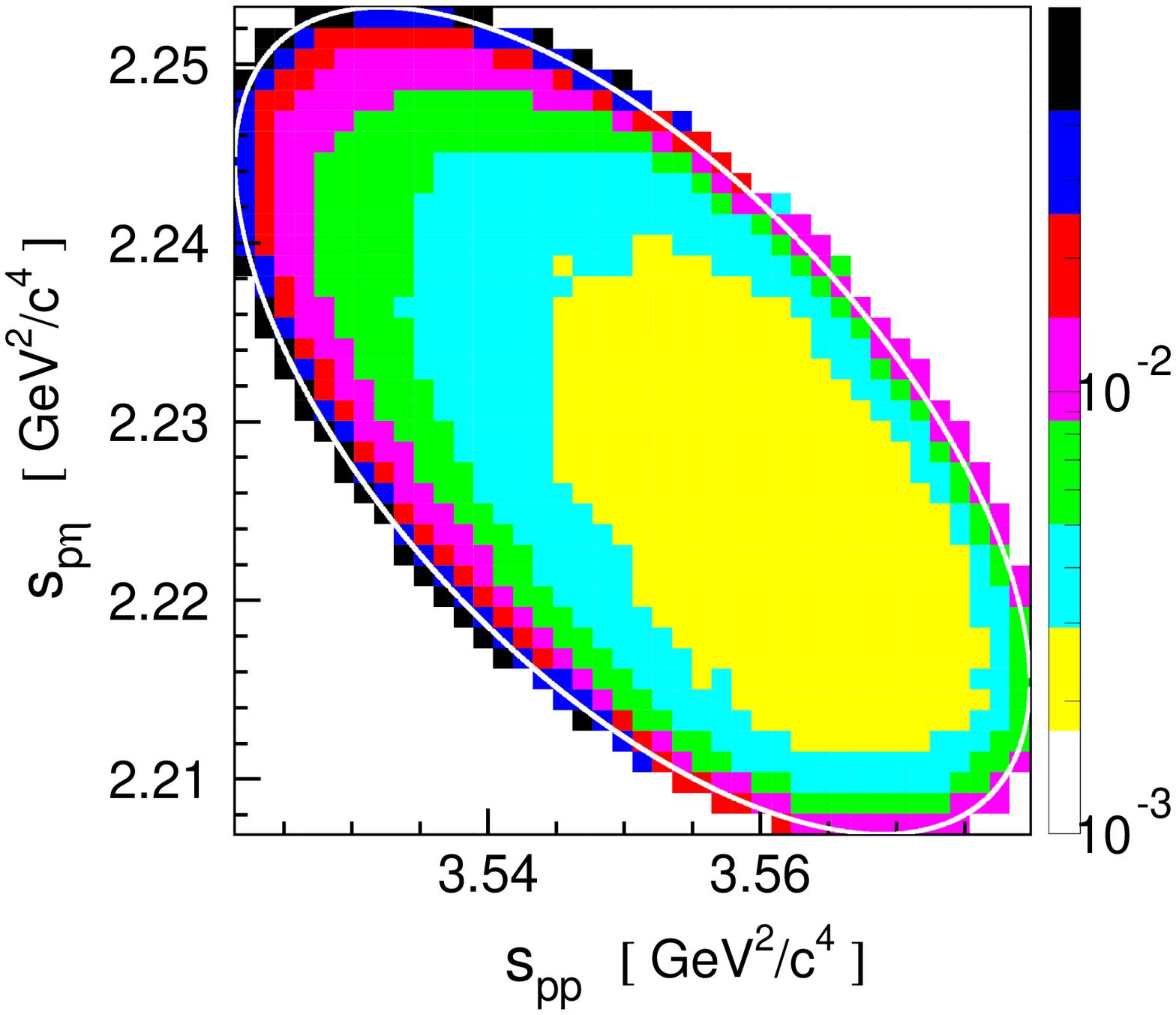}}

  \vspace{-0.9cm}
  \parbox{0.38\textwidth}{\raisebox{1ex}[0ex][0ex]{\mbox{}}} \hfill
  \parbox{0.49\textwidth}{\raisebox{1ex}[0ex][0ex]{\large a)}} \hfill
  \parbox{0.03\textwidth}{\raisebox{1ex}[0ex][0ex]{\large b)}}
  \vspace{0.0cm}

  \caption{ 
    \label{rozkladthetaeta} 
    (a) \ \ Distribution of the polar angle of the emission of the $\eta$ meson in the 
    centre-of-mass system.
    Experimental data were corrected for the acceptance
    in the three dimensional space~($s_{pp},s_{p\eta},|cos(\theta_{\eta}^{*})|)$.
    Full circles show the result with the assumption that the distribution of the $\psi$ angle
    is isotropic, and the open circles are extracted under the assumption that
    $\frac{d\sigma}{d\psi}$ is as derived from the data (see text).
    Both results have been normalized to each other in magnitude.\protect\\
   (b) \ \ COSY-11 detection acceptance as a function of $s_{pp}$ and $s_{p\eta}$,
   calculated under the assumption that the differential cross sections
   $\frac{d\sigma}{d\cos(\theta_{\eta}^{*})}$ and $\frac{d\sigma}{d\psi}$
   are isotropic.
   }
\end{figure}
\begin{figure}[H]
    \vspace{-0.2cm}
    \parbox{0.5\textwidth}{\vspace{-1.0cm} 
     \includegraphics[width=0.53\textwidth]{nnewdalitz_mc_gen_ppcfs_hab.eps}} 
    \parbox{0.5\textwidth}{\vspace{-1.0cm}
     \includegraphics[width=0.53\textwidth]{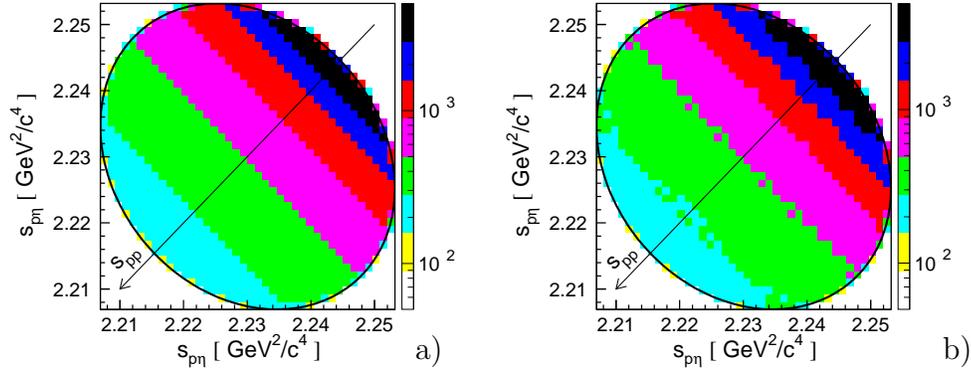}} 

\vspace{-0.6cm}
\parbox{0.38\textwidth}{\raisebox{1ex}[0ex][0ex]{\mbox{}}} \hfill
\parbox{0.49\textwidth}{\raisebox{1ex}[0ex][0ex]{\large a)}} \hfill
\parbox{0.03\textwidth}{\raisebox{1ex}[0ex][0ex]{\large b)}}

  \vspace{-0.5cm}
  \caption{ \label{dalitzporownanie}
   {\bf (a)} Dalitz plot distribution simulated
    for the $pp\to pp\eta$ reaction at Q~=~15.5~MeV. In the calculations the interaction between
    protons was taken into account.\protect\\
    {\bf (b)}
    Dalitz plot distribution reconstructed from the 
    response of the COSY-11 detectors simulated for events from figure (a) taking into account the 
    smearing of beam and target, multiple scattering in the materials,
    and the detectors' resolution. The evaluation
    included momentum reconstruction, kinematical fitting, and the acceptance
    correction exactly in the same way as performed for the experimental data.
    The lines surrounding the Dalitz plots depict the kinematical limits.
  } 
\end{figure}
Full points in figure~\ref{rozkladthetaeta}a present the distribution of the polar angle of the $\eta$
meson as derived from the data after the acceptance correction. Within the statistical accuracy
it is isotropic.  Taking into account this angular distribution of the cross section
we can calculate the acceptance
as a function of $s_{pp}$ and $s_{p\eta}$ only.
 This  is shown in figure~\ref{rozkladthetaeta}b, where
one sees that now the  full phase space is covered.
This allows us to determine the distributions of $s_{pp}$ and $s_{p\eta}$.
The correctness of the performed  procedures for the simulation of the
detectors response, the  event reconstruction programs, the kinematical fitting,
and acceptance correction can be confirmed by comparing
the distribution generated
(figure~\ref{dalitzporownanie}a) with the ones which underwent the complete 
analysis chain  described in this section
(figure~\ref{dalitzporownanie}b).
 
\vspace{-0.1cm}
\begin{figure}[H]
    \parbox{0.55\textwidth}{\vspace{0.0cm}
    \includegraphics[width=0.52\textwidth]{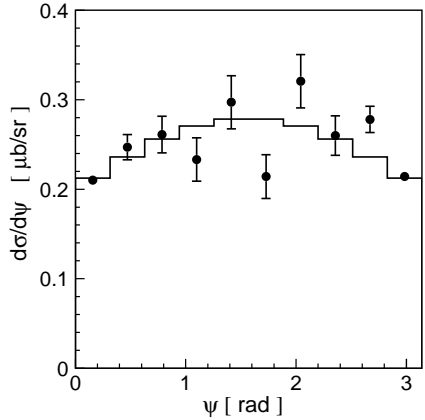}}
    \parbox{0.45\textwidth}{\vspace{-1.0cm}
  \caption{ \label{rozkladpsi1}
       Distribution of the cross section as a function of the angle $\psi$ as determined in 
       the first iteration. The superimposed histogram corresponds to the fit 
       of the function  $\frac{d\sigma}{d\psi}~=~a~+~b~\cdot~|sin(\psi)|$.
       The range of the $\psi$ angle is shown from 0 to $\pi$ only, 
       since in the analysis we take advantage of the symmetry 
      $\frac{d\sigma}{d\psi}(\psi)~=~\frac{d\sigma}{d\psi}(\psi+\pi)$.
  }
   } 
\end{figure}
\vspace{-0.1cm}
Knowing the distribution
of the polar angle of the $\eta$ meson $\theta_{\eta}^{*}$
and those for the invariant masses $s_{pp}$ and $s_{p\eta}$
we can check whether the assumption of the isotropy of the cross section
distribution versus the third Euler's angle $\psi$ is corroborated by the data.
For that purpose we calculated the acceptance as a function of
$\psi$ and $s_{p\eta}$ assuming the shape of the differential cross sections
of $\frac{d\sigma}{ds_{pp}}$ and $\frac{d\sigma}{dcos(\theta_{\eta}^{*})}$
as determined experimentally.
The obtained $\frac{d\sigma}{d\psi}$ distribution is shown
in figure~\ref{rozkladpsi1} and is not isotropic as assumed at the 
beginning.  
A fit of the  function of the form $\frac{d\sigma}{d\psi}~=~a~+~b~\cdot~|sin(\psi)|$ 
gives the value of 
 b~=~0.079~$\pm$~0.014~$\mu b / sr$, 
indeed significantly different 
from the isotropic solution.
This deviation cannot be assigned to any unknown behaviour
of the background since the
 obtained distribution can be regarded
as background free. This is because the number of $pp\to pp\eta$ events was
elaborated  for each invariant mass interval separately.
The exemplarily missing mass spectra 
for the first, fourth, seventh,
and tenth interval of $\psi$ values,
corrected for the acceptance,
are presented in figure~\ref{misspsi}.
From this figure one can infer that the shape of the background is well
reproduced not only for the overall missing mass spectrum as shown
previously in figure~\ref{missall} but also locally  in each
region of the phase space.
Since the experimental data are quite well described by the simulations 
one can 
rather exclude the possibility of a significant systematic error
which could cause the observed anisotropy 
of the differential cross section $\frac{d\sigma}{d\psi}$.
\vspace{-0.1cm}
\begin{figure}[H]
\hspace{0.0cm}
\vspace{-0.6cm}
\parbox{0.49\textwidth}{\epsfig{file=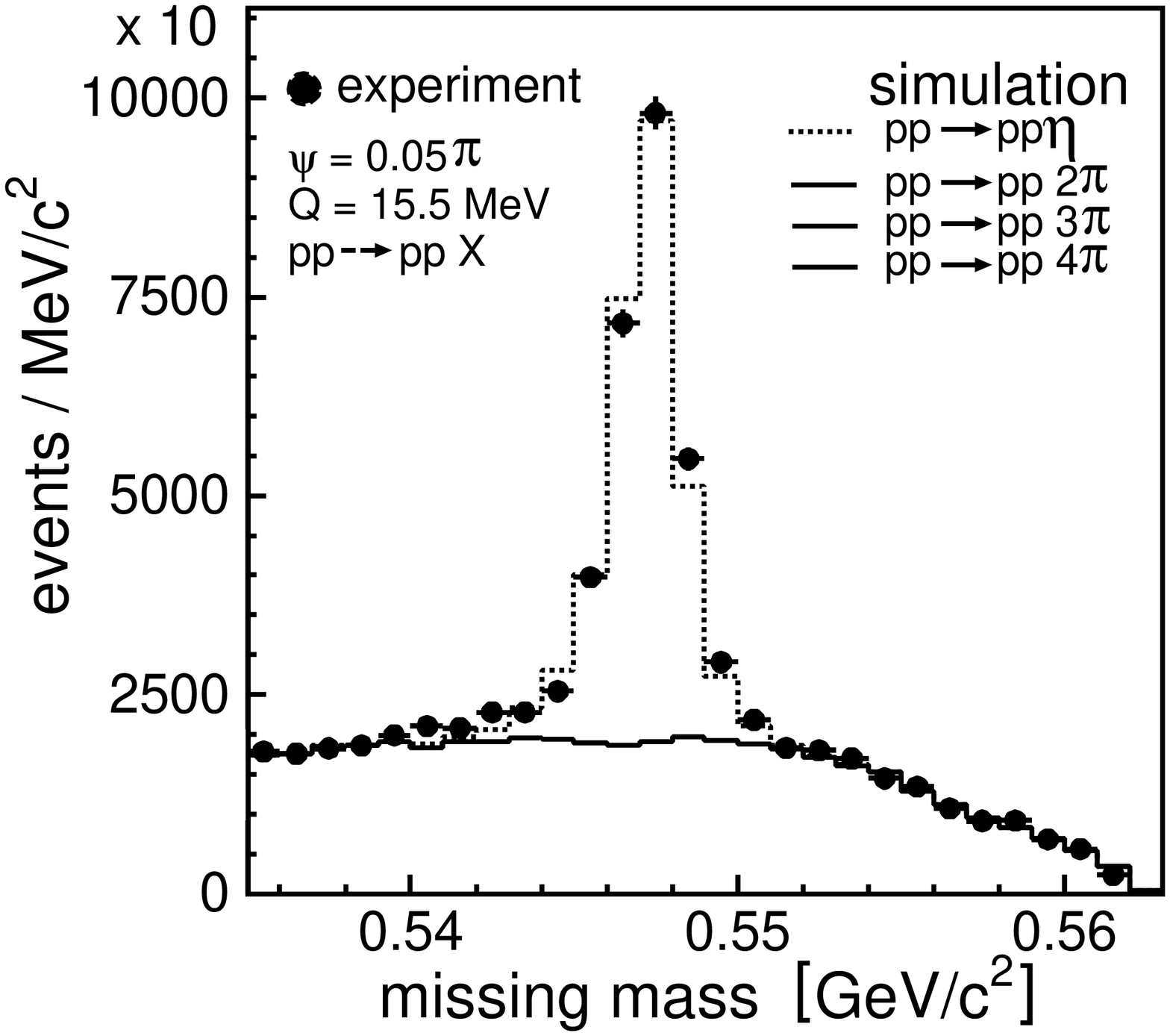,width=0.47\textwidth}}
\hfill
\parbox{0.49\textwidth}{\epsfig{file=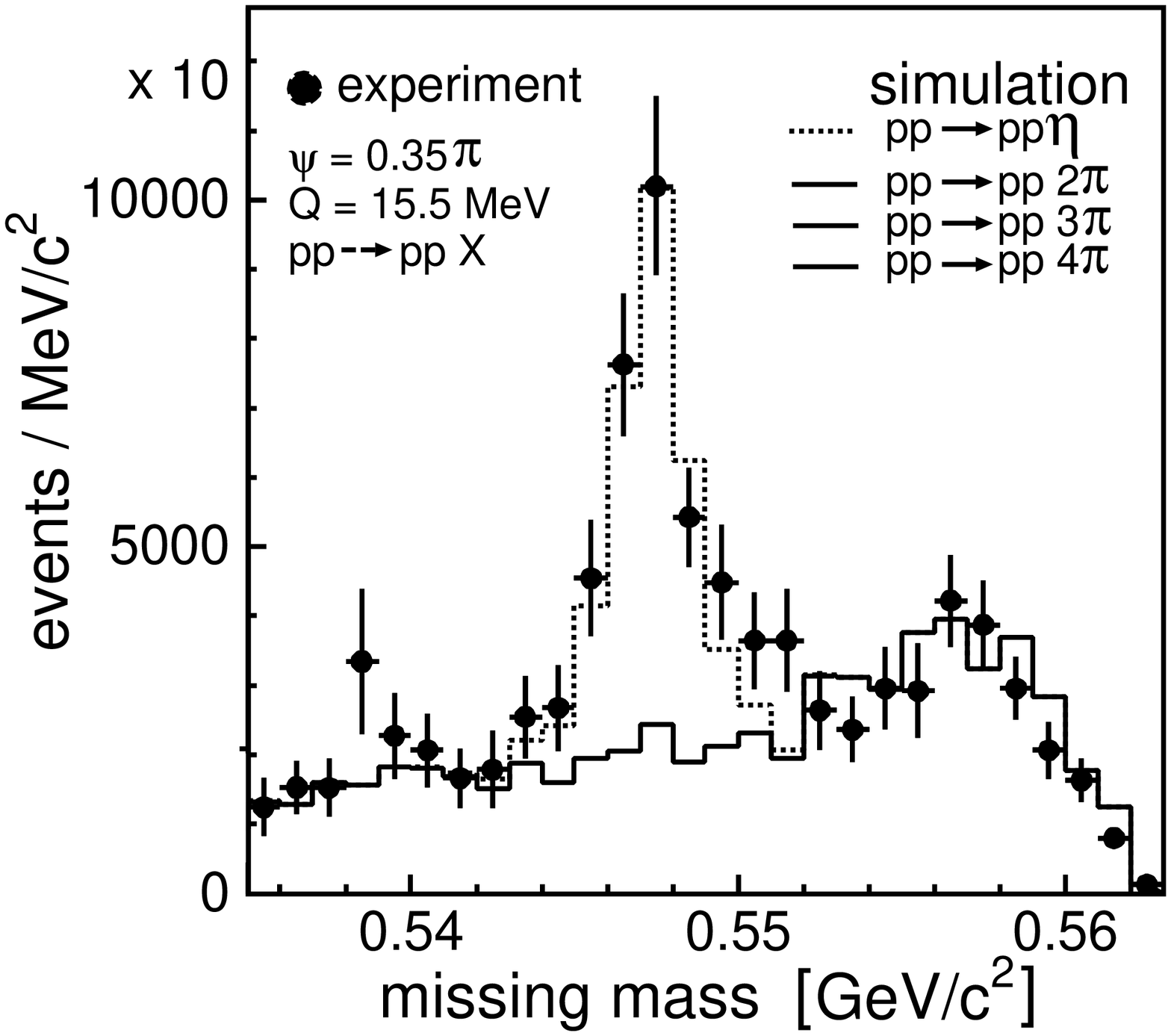,width=0.47\textwidth}}

\vspace{0.6cm}
\parbox{0.49\textwidth}{\epsfig{file=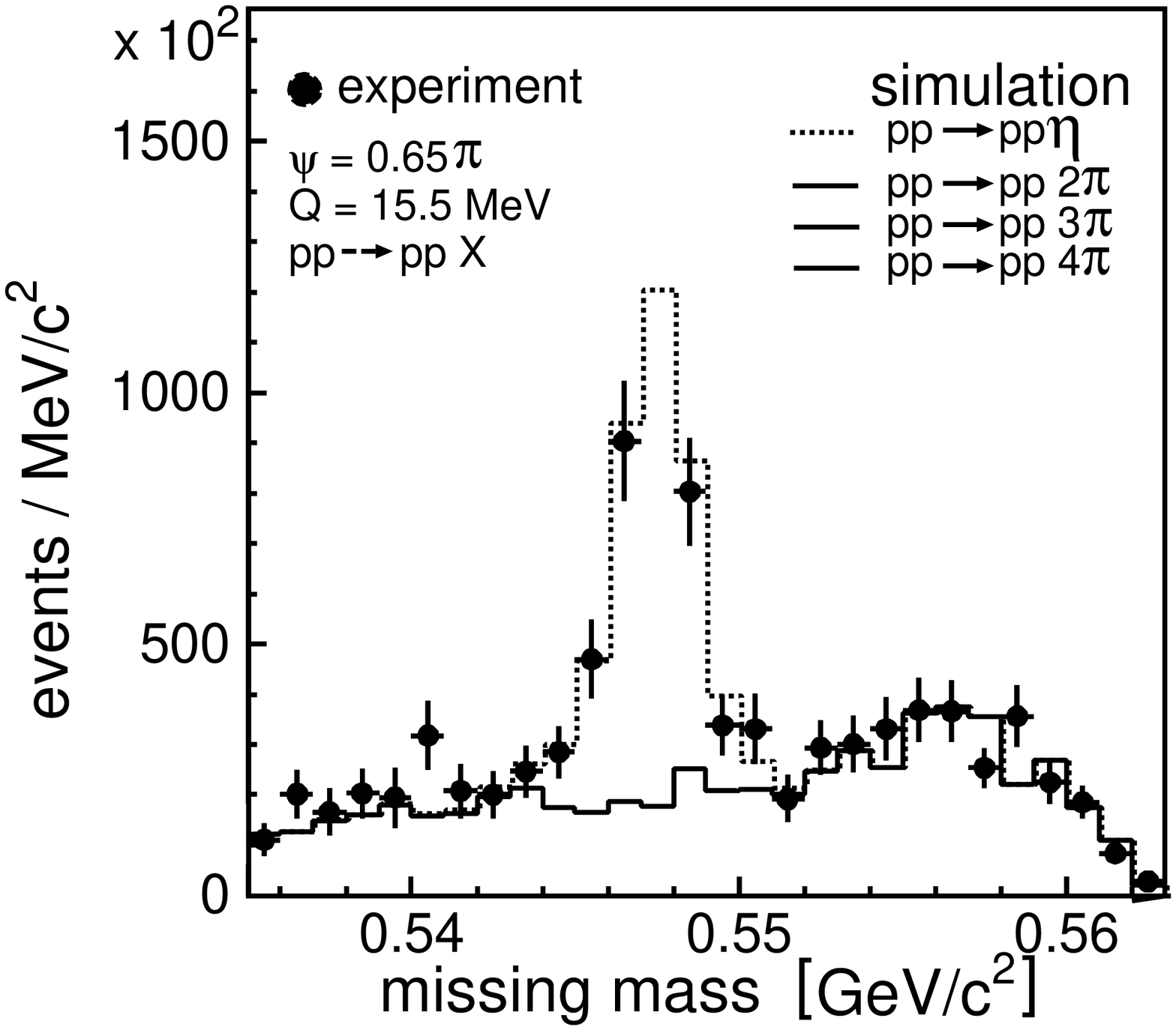,width=0.47\textwidth}}
\hfill
\parbox{0.49\textwidth}{\epsfig{file=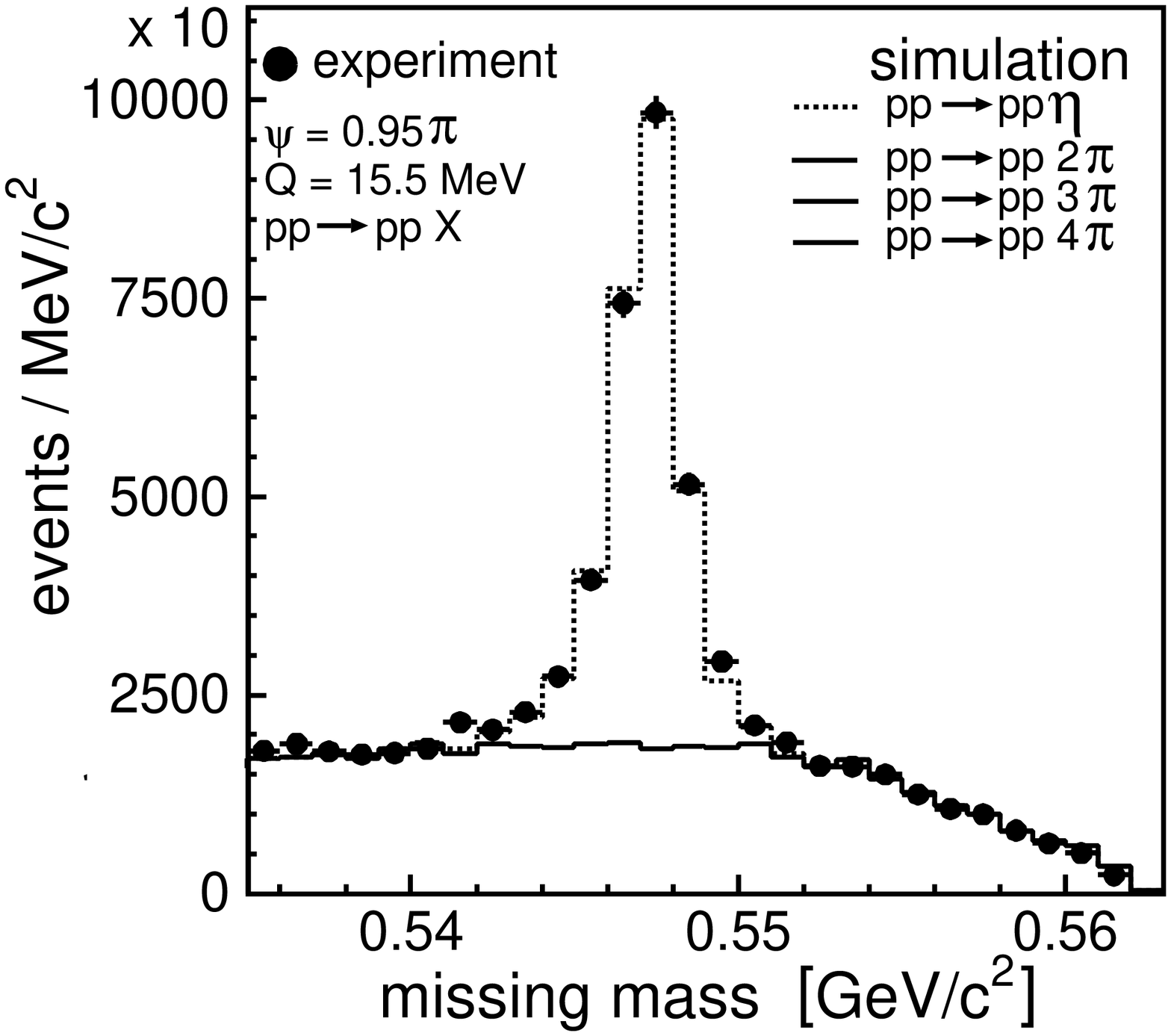,width=0.47\textwidth}}

  \vspace{-0.0cm}
  \caption{ \label{misspsi}
     Missing mass distributions for the first, fourth, seventh, and tenth  bin of $\psi$
     with the superimposed lines from the simulation 
     of $pp\to pp\eta$ and the multi-pion background $pp\to pp (m\pi)$ reactions. 
     Amplitudes of simulated distributions were fitted to the experimental points.
  }
\end{figure}
The evaluated distribution is however  
in disagreement with the working assumption that  $\sigma(\psi)$ is isotropic.
Therefore we performed a full acceptance correction procedure 
from the very beginning assuming that the distribution of $\frac{d\sigma}{d\psi}$ is as 
determined from the data. After repeating the procedure three times
we observed that the input and resultant  distributions are in good agreement. 
The result after the third iteration is shown in figure~\ref{rozkladpsi} by the full circles.
It is only slightly different from the one obtained after the first iteration
as shown in figure~\ref{rozkladpsi1}.
\vspace{0.2cm}
\begin{figure}[H]
    \centerline{
    \parbox{0.55\textwidth}{\vspace{-0.4cm}
    \includegraphics[width=0.55\textwidth]{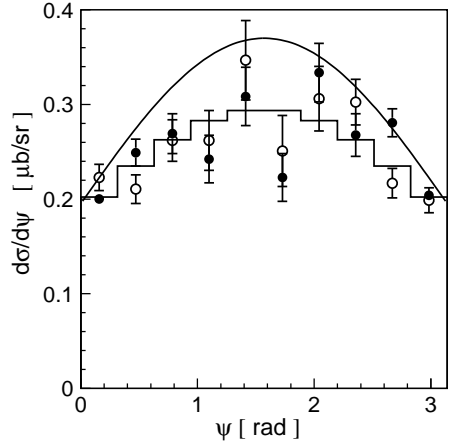}}}

 \vspace{-0.6cm}
  \caption{ \label{rozkladpsi}
       Distribution of the cross section as a function 
       of the angle $\psi$.
       Full circles stand for the final results 
       of the $\frac{d\sigma}{d\psi}$
      obtained after three iterations. 
       The superimposed histogram (solid line) corresponds to the fit 
       of the function  $\frac{d\sigma}{d\psi}~=~a~+~b~\cdot~|sin(\psi)|$
       which  resulted in 
       $a~=~0.186~\pm~0.004~\mu b/sr$ and $b~=~0.110~\pm~0.014~\mu b/sr$. 
       The dashed line shows the entry distribution used for the second series of iterations
       as described in the text.
     Open circles represent the 
     data from the left upper corner of the Dalitz plot 
     (see for example figure~\ref{rozkladthetaeta}b).
     At that region of the Dalitz plot due to the non-zero four dimensional acceptance
     over ($s_{pp}$, $s_{p\eta}$, $|cos(\theta_{\eta}^{*})|$, $\psi$) bins
     the spectrum (open circles) was corrected
     without a necessity of any assumptions concerning the reaction cross section.
  } 
\end{figure}
\vspace{-0.1cm}
To raise the confidence  of the convergence of the performed iteration
we accomplished  the full procedure once more, but now assuming that the distribution 
of $\frac{d\sigma}{d\psi}$ is much more anisotropic 
than determined from the data. As an entry distribution we took the 
dashed line shown in figure~\ref{rozkladpsi}.
Again after two iterations we have got the same result as before.
To corroborate this observation
we have evaluated the distribution over the $\psi$ angle (see figure~\ref{rozkladpsi})
from the phase space region which
has no holes in the acceptance expressed as a four-dimensional function of
the variables $s_{pp}$, $s_{p\eta}$, $|cos(\theta_{\eta}^{*})|$, and $\psi$,
this is for the values of $s_{pp}$ and $s_{p\eta}$ corresponding to the 
upper left corner of figure~\ref{rozkladthetaeta}b.
 Again the obtained distribution 
presented as open circles in figure~\ref{rozkladpsi} is anisotropic,
and moreover agrees
with the spectrum determined from all events.
It is important to note that the shape of the $s_{pp}$, $s_{p\eta}$, and 
$cos(\theta_{\eta}^{*})$ distributions keeps unchanged during the whole 
iteration procedure.  
Figure~\ref{invporownanie}
shows the distributions of the square of the proton-proton and proton-$\eta$
invariant masses. The spectra after the second and third
iterations are shown. One recognizes that the form of the spectra remains unaltered.
The same 
conclusion can be drawn for the $\frac{d\sigma}{d|cos(\theta^{*}_\eta)|}$ distribution
as demonstrated in figure~\ref{rozkladthetaeta}a. 
\vspace{-0.5cm}
\begin{figure}[H]
  \parbox{0.5\textwidth}{\vspace{0.0cm}
    \includegraphics[width=0.52\textwidth]{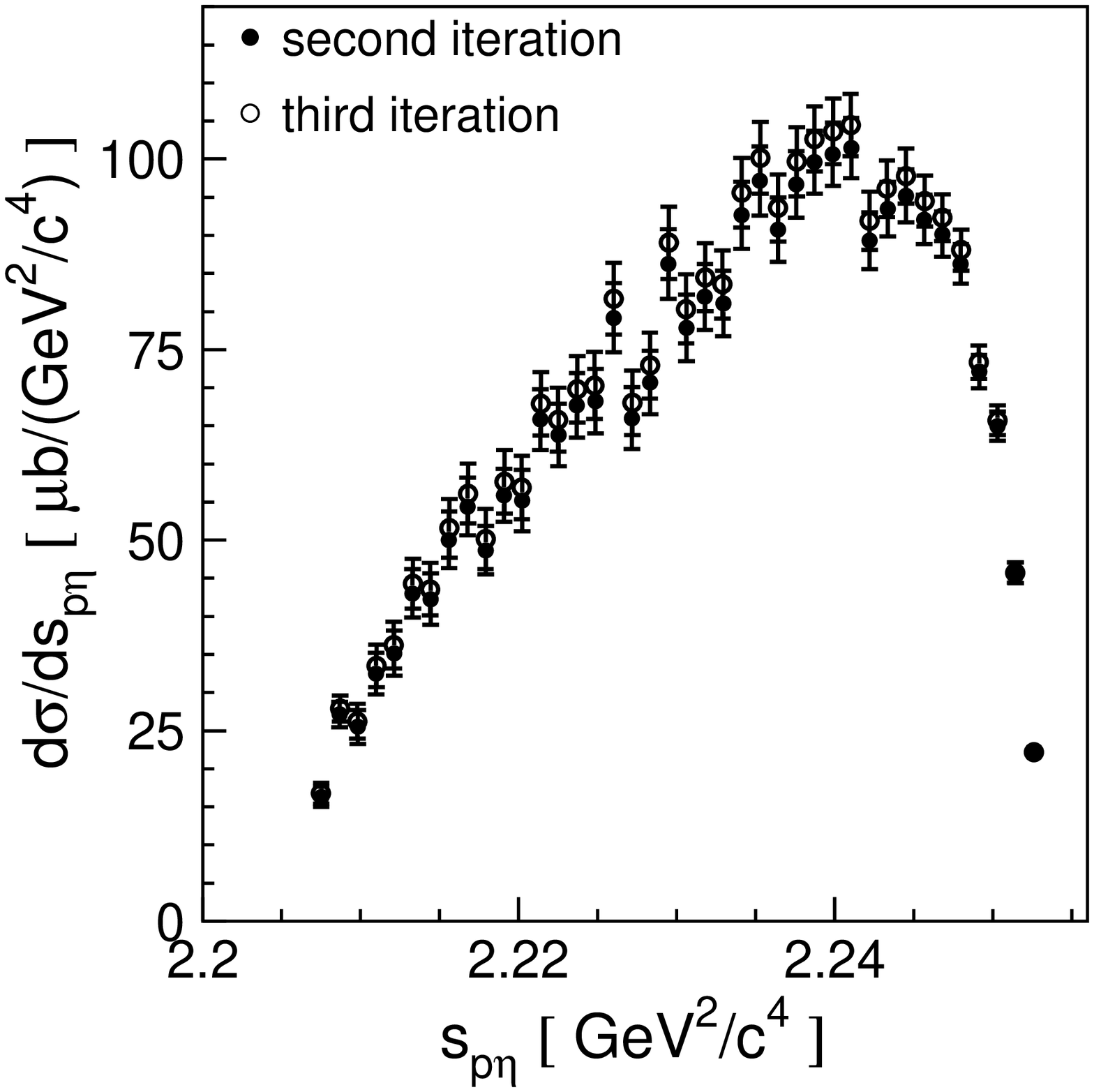}}
  \parbox{0.5\textwidth}{\vspace{0.0cm}
    \includegraphics[width=0.52\textwidth]{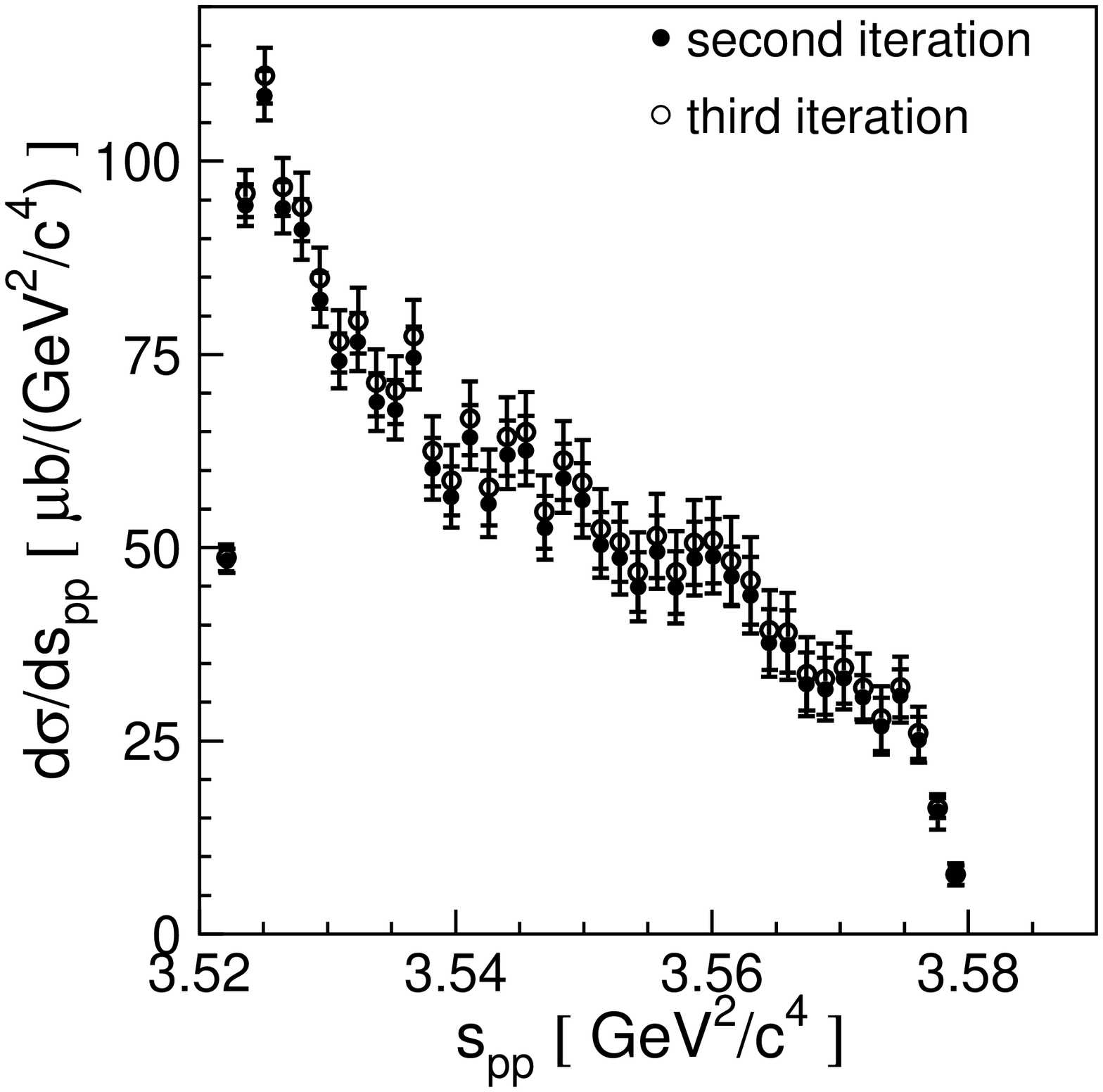}}
  \vspace{-0.6cm}
  \caption{ \label{invporownanie} 
      Distributions of the invariant masses $s_{pp}$ and $s_{p\eta}$ determined for the
      two different assumptions about the cross section dependence of the $\psi$ angle.
  }
\end{figure}
\vspace{-0.4cm}
From that comparison one can conclude
that the shapes of the determined distributions are
--~within the statistical accuracy~-- independent
of the shape of the $\frac{d\sigma}{d\psi}$ cross section
and can be treated as derived in a completely model independent manner.
Similarly, as in the case of the $\frac{d\sigma}{d\psi}$ distribution,
the differential cross sections 
in all other variables reported in this article
are not deteriorated  by the background. This is because the 
number of $pp\to pp\eta$ events was
determined  for each investigated  phase space interval separately.
\vspace{-0.3cm}
\begin{figure}[H]
    \parbox{0.5\textwidth}{
     \includegraphics[width=0.47\textwidth]{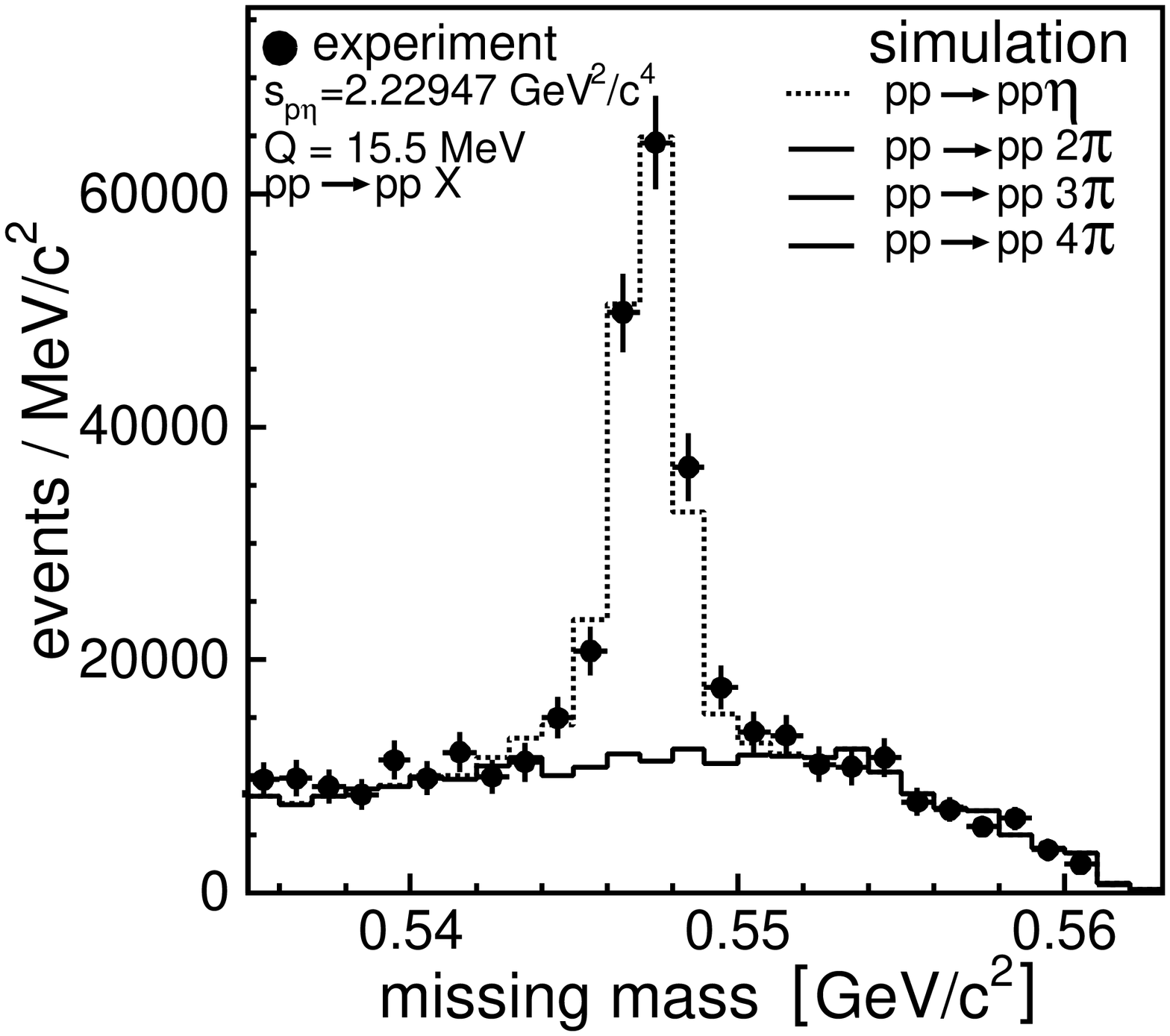}}
    \parbox{0.5\textwidth}{
     \includegraphics[width=0.47\textwidth]{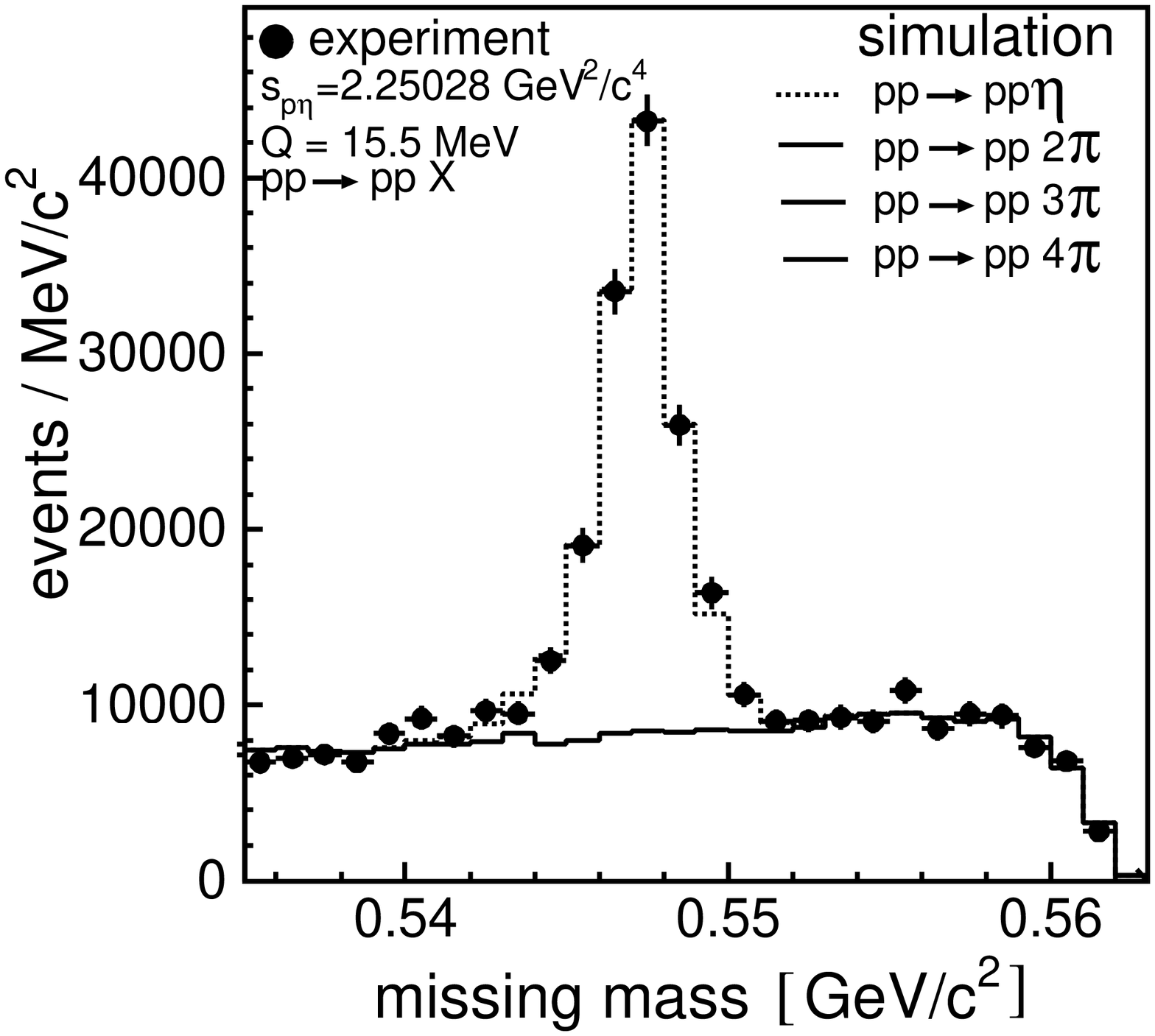}}
    \vspace{-0.4cm}
  \caption{ \label{invmiss}
     Missing mass distributions determined for the invariant mass bins as depicted inside
     the figures.  The spectra were corrected for the acceptance.
     The  histograms show the simulations for the
     multi-pion background  $pp\to pp (m\pi)$ and the $pp\to pp \eta$ reactions 
     fitted to the data with the amplitudes
     as the only free parameters.
  }
\end{figure}
\vspace{-0.2cm}
As an example 
the missing mass spectra 
for two bins of the proton-$\eta$ invariant mass 
are presented in figure~\ref{invmiss}. 
As already noticed for $\frac{d\sigma}{d\psi}$ 
the shape of the background is well
reproduced locally  in each region of the phase space.

\subsection{Angular distributions}
\label{angularsection}
\begin{flushright}
\parbox{0.87\textwidth}{
 {\em
   Before we declare our consent we must carefully examine\\
   the shape of the architecture the rhythm of the drums and pipes\\
   ...~\cite{herbert}.
 }
 \protect \mbox{} \hfill Zbigniew Herbert \protect\\
 }
\end{flushright}
\vspace{-0.3cm}
In previous subsections we derived the distributions of the cross section
in the invariant masses $s_{pp}$ and $s_{p\eta}$ (figure~\ref{invporownanie}),
and  in angle $\psi$ (figure~\ref{rozkladpsi}).
For the sake of completeness we will also present the distribution in the
$cos(\theta_{\eta}^{*})$, $cos(\theta_N^*)$, and $\psi_{N}$. 
This will allow us to have insight into 
distributions in all nontrivial variables describing the $pp\eta$ system 
in the case when the polarization
of nucleons is ignored. 
Here, as introduced in section~\ref{choiceofobservables} we consider two different sets of 
orthogonal variables, namely ($s_{pp}$,$s_{p\eta}$,$|cos(\theta_{\eta}^{*})|$,$\psi$) and
($s_{pp}$,$s_{p\eta}$,$|cos(\theta_{N}^{*})|$,$\psi_{N}$).
  If possible the data will be compared to the 
results of measurements performed at the non-magnetic spectrometer COSY-TOF~\cite{TOFeta}.

In the previous subsection we have shown that even  close-to-threshold at Q~=~15.5~MeV
the production of the $pp\eta$ system is not fully isotropic. In particular,
we found that the cosine of angle $\psi$ is not uniformly populated.
The anisotropy of the cross section 
reflects itself in an anisotropy of the orientation of the emission plane,
and the latter has a simple physical interpretation.

The determined cross section distribution in function 
of the polar angle $\theta^*_{N}$ of the vector normal to that plane
is shown in figure~\ref{costhetaetatof}a.
The distribution is not isotropic,
which is particularly visible for the low values of $|cos(\theta^*_{N})|$
burdened with  small errors.
As depicted in figure~\ref{figvariables} the $|cos(\theta^*_{N})|$~=~0
denotes such configurations of the ejectiles momenta in which the emission plane
comprises the beam axis. In that case the acceptance of the COSY-11 detection
system is much larger than for the configuration where the 
emission plane  is perpendicular to the beam.
Due to this reason the error bars  in figure~\ref{costhetaetatof}a
increase with growth of  $|cos(\theta^*_{N})|$.
It is worth to stress that the tendency 
of the $pp\eta$ system  to be produced 
preferentially
if the emission plane is perpendicular to the beam 
is in line with the preliminary analysis of the experiment performed 
by the TOF collaboration~\cite{roderburg2299}. Elucidation of that non-trivial 
behaviour can reveal interesting features of the dynamics of the 
production process. 
\vspace{-0.6cm}
\begin{figure}[t]
  \parbox{0.5\textwidth}{
    \includegraphics[width=0.5\textwidth]{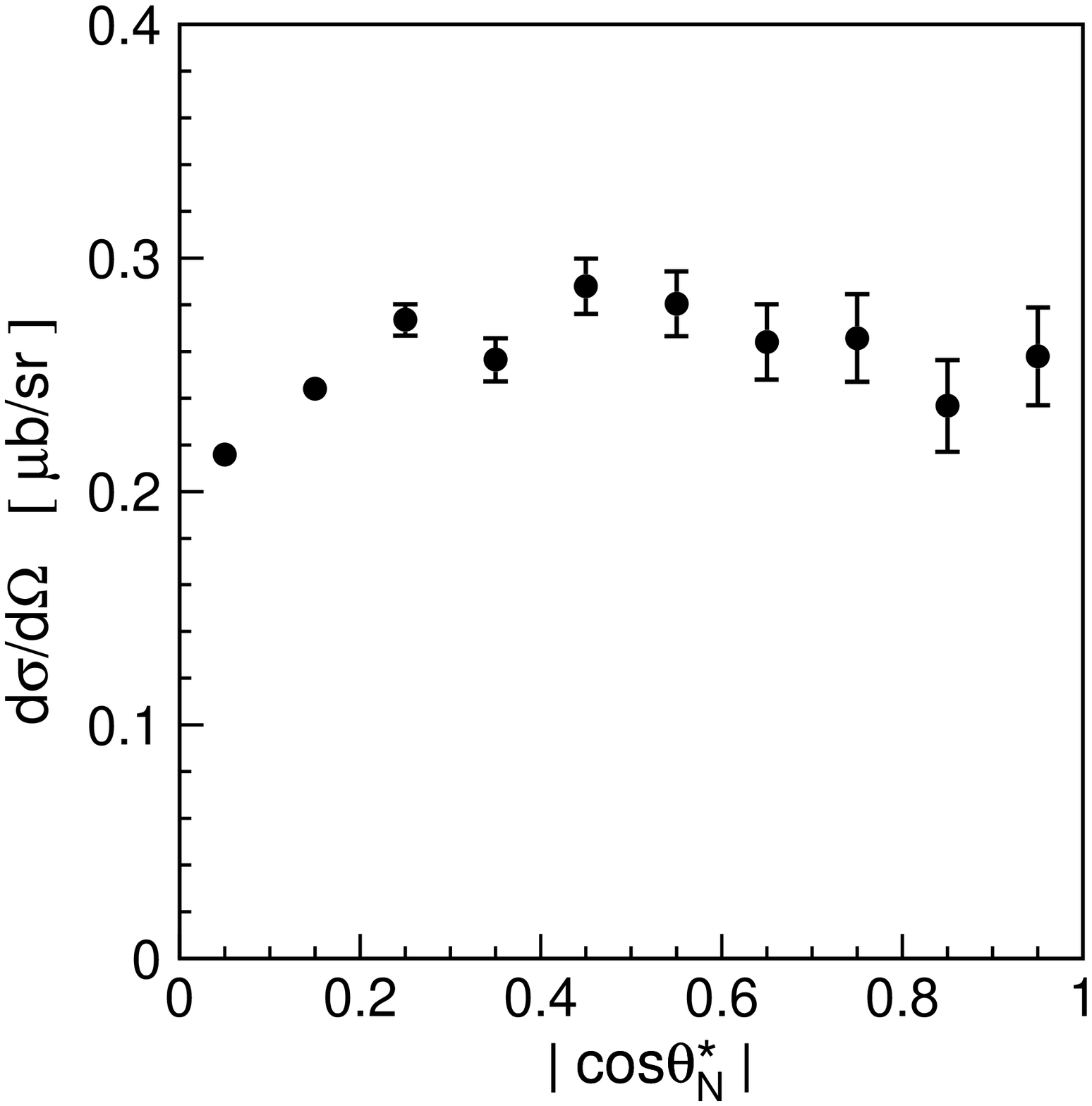}}
  \parbox{0.5\textwidth}{
    \includegraphics[width=0.5\textwidth]{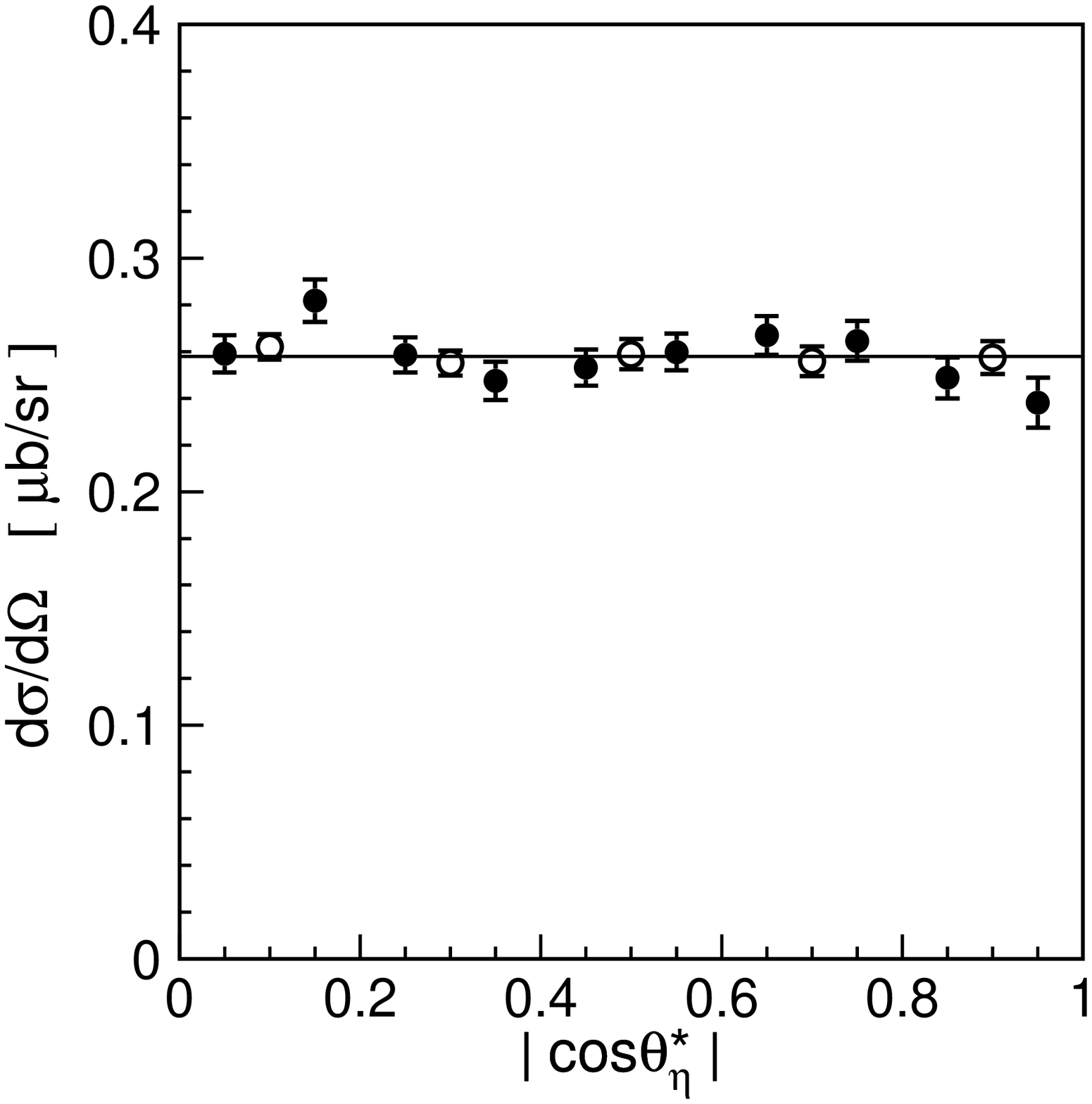}}
     
  \vspace{-0.5cm}
  \parbox{0.40\textwidth}{\raisebox{1ex}[0ex][0ex]{\mbox{}}} \hfill
  \parbox{0.47\textwidth}{\raisebox{1ex}[0ex][0ex]{\large a)}} \hfill
  \parbox{0.03\textwidth}{\raisebox{1ex}[0ex][0ex]{\large b)}}
  \vspace{-0.3cm}

    \caption{\label{costhetaetatof}
     (a) \ \ Differential cross section as a function of 
     the polar angle of the vector normal to the emission plane.\protect\\
     (b) Differential cross section of the $pp\to pp\eta$ reaction as a function of the
     $\eta$ meson centre-of-mass polar angle. 
     Full circles depict experimental results
     for the $pp\to pp\eta$ reaction measured at Q~=~15.5~MeV by the COSY-11 collaboration
     (this article) and the open circles were determined by the TOF collaboration
     at Q~=~15~MeV~\cite{TOFeta}. The TOF points were normalized in amplitude to our result,
     since for that data the absolute scale is not evaluated.
    }
\end{figure}
\vspace{-0.1cm}
\begin{figure}[H]
 \parbox{0.58\textwidth}{ \vspace{-0.5cm}
\includegraphics[width=0.54\textwidth]{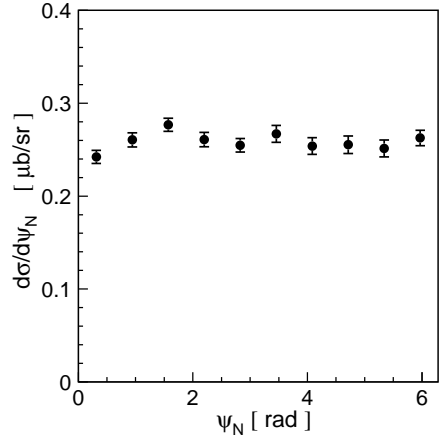}}
 \parbox{0.4\textwidth}{ \vspace{-1.5cm}
\caption{\label{figurepsiN} 
 Differential cross section in $\psi_{N}$ for the $pp\to pp\eta$ reaction
 measured at Q~=~15.5~MeV. The variable $\psi_{N}$ is defined 
 in section~\ref{choiceofobservables}. 
}}
\end{figure}
\vspace{-0.5cm}
The distribution of the polar angle 
of the $\eta$ meson  emission in the centre-of-mass system 
is shown in figure~\ref{costhetaetatof}b.
Clearly, the COSY-11 data agree very well with the angular dependence determined 
by the TOF collaboration.
As already mentioned, the two identical particles in the initial state of the $pp\to pp\eta$ reaction
imply that the angular distribution of either ejectile must be symmetric around
90 degree in the centre-of-mass frame. 
Here we use that reaction characteristic, yet
in figure~\ref{differentialeta}a (see also reference~\cite{moskal367} ) 
we presented the differential cross section
of the $\eta$ meson centre-of-mass polar angle for the full range of $cos(\theta^{*}_{\eta})$
and found that this is 
completely symmetric around $cos(\theta^{*}_{\eta})~=~0$.
These observations can be regarded as a check of the correctness of the acceptance calculation.
The cross section distribution in the angle $\psi_{N}$ --~defining the orientation of the 
$pp\eta$ system within the emission plane~--  
is shown in figure~\ref{figurepsiN}. 

Up to now we presented invariant mass spectra of the proton-proton
and proton-$\eta$ systems and angular distribution for two sets of 
non-trivial angles which describe the orientation of ejectiles within the 
emission plane and the alignment of the plane itself,
namely ($cos(\theta^{*}_{\eta})$, $\psi$) and ($cos(\theta^{*}_{N})$, $\psi_{N}$).
However, since one of the important issues discussed in this work is the 
contribution from  higher partial waves 
we evaluated also an angular distribution
of the relative momentum of two  protons seen from the proton-proton 
centre-of-mass subsystem~(see figure~\ref{figureqpppp}a).
The distribution of that angle should deliver  information 
about the partial waves of the proton-proton system 
in the exit channel (see section~\ref{partialwaves}). 
\vspace{-0.6cm}
\begin{figure}[H]
    \parbox{0.5\textwidth}{
    \includegraphics[width=0.5\textwidth]{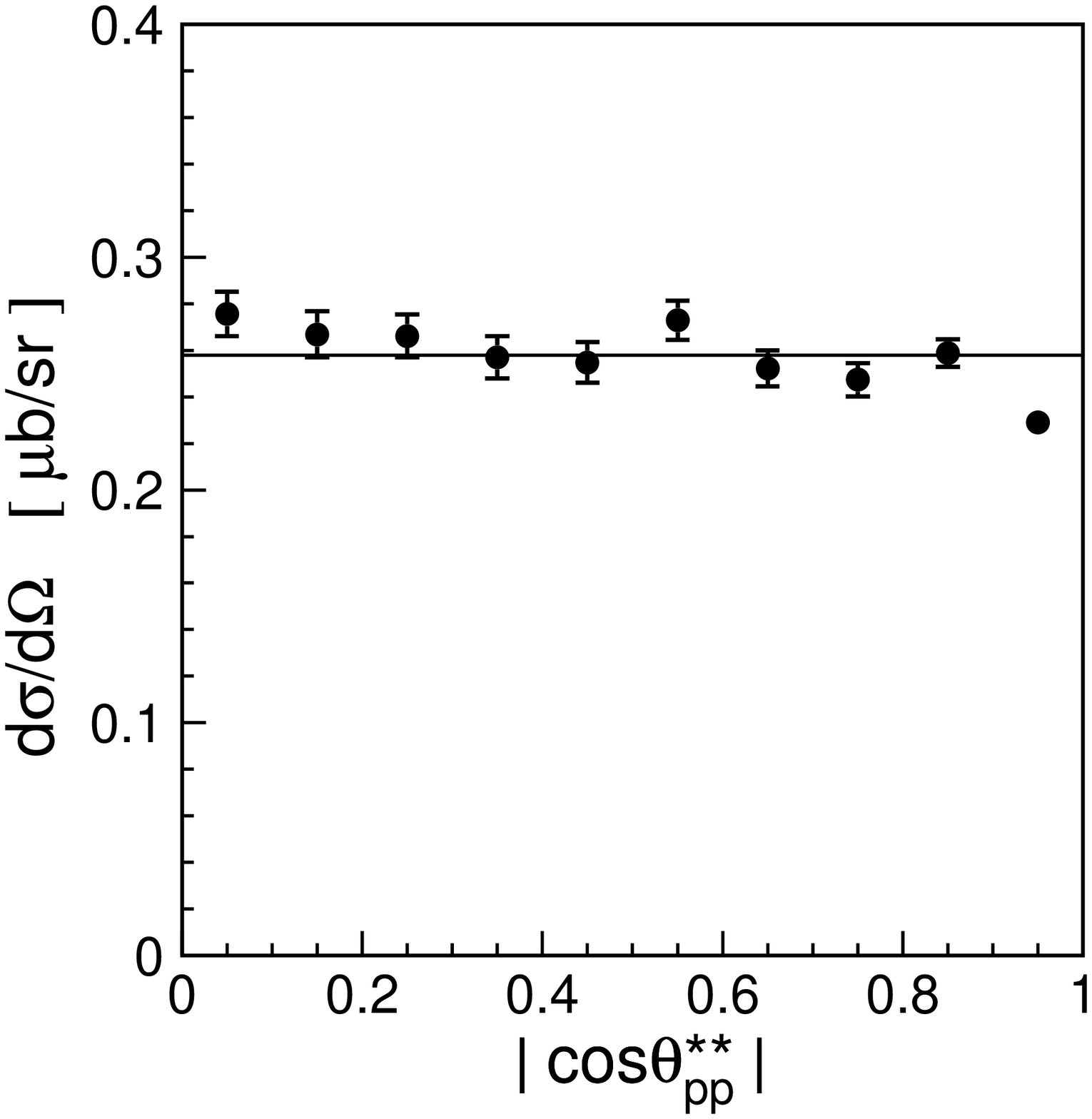}}
    \parbox{0.5\textwidth}{ 
    \includegraphics[width=0.50\textwidth]{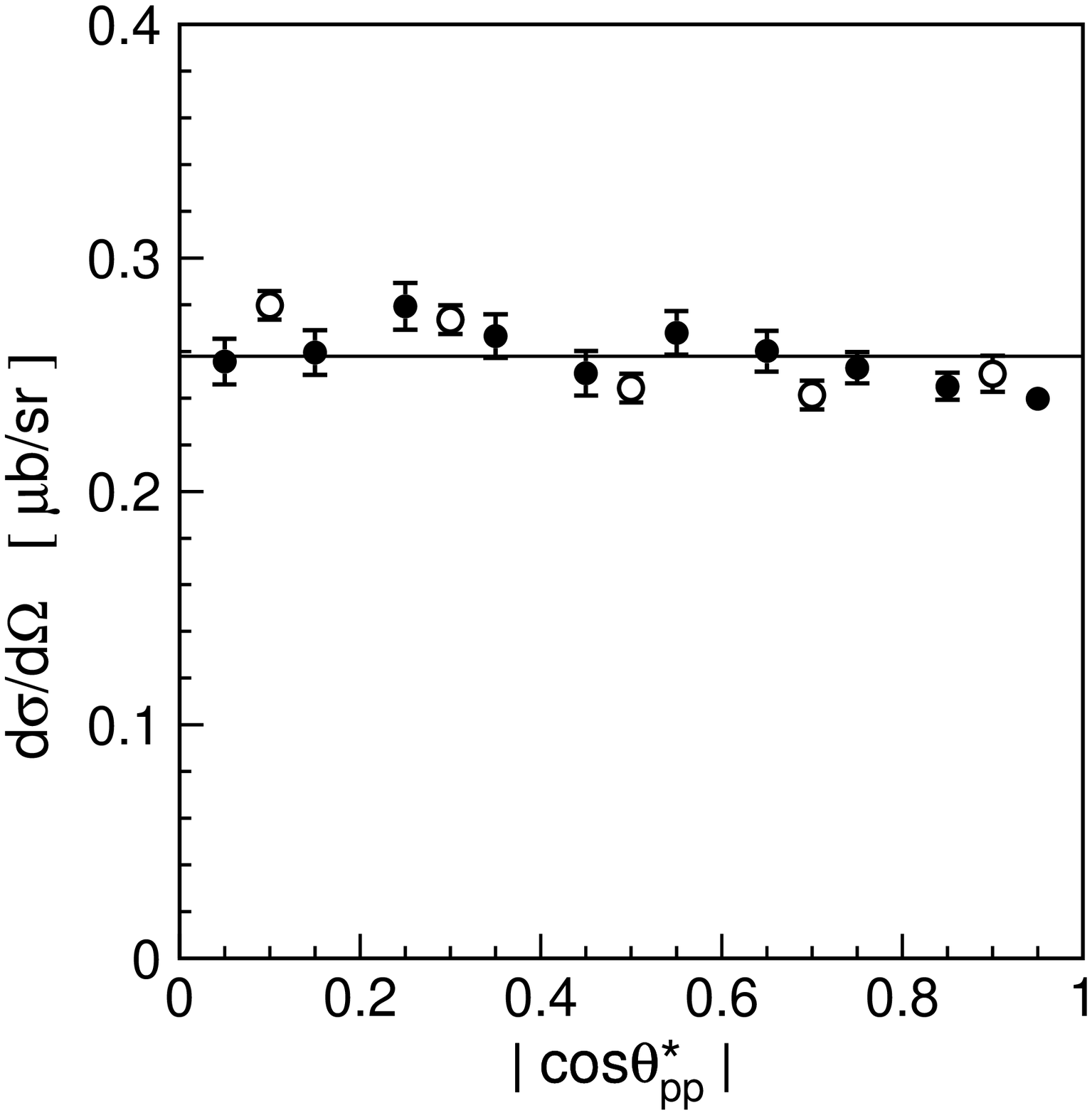}}

  \vspace{-0.4cm}
  \parbox{0.41\textwidth}{\raisebox{1ex}[0ex][0ex]{\mbox{}}} \hfill
  \parbox{0.46\textwidth}{\raisebox{1ex}[0ex][0ex]{\large a)}} \hfill
  \parbox{0.03\textwidth}{\raisebox{1ex}[0ex][0ex]{\large b)}}
  \vspace{-0.5cm}
    \caption{\label{figureqpppp}
     (a) \ \ Differential cross section in $\theta^{**}_{pp}$
     as determined for the $pp\to pp\eta$ reaction at Q~=~15.5~MeV. \protect\\
    (b) \ \ Distribution of the centre-of-mass polar angle of the relative protons momentum
    with respect to the beam direction
    determined  for the $pp\to pp\eta$ reaction at Q~=~15.5~MeV.
    The COSY-11 result (closed circles) is compared to the data points determined 
    at Q~=~15~MeV by the TOF collaboration (open circles)~\cite{TOFeta}.
    }
\end{figure}
\vspace{-0.2cm}
Figure~\ref{figureqpppp}b presents the
differential cross section  as a function of angle $\theta_{pp}^{*}$
of the relative proton
momentum seen from the overall centre-of-mass frame,
as this is often considered in the theoretical works.
This figure compares COSY-11 results to the angular distribution
extracted by the TOF collaboration. Both experiments agree very well within the 
statistical accuracy, and both indicate a slight decrease of the cross section 
with increasing $|cos(\theta^{*}_{pp})|$.

\clearpage
\section{Usage of the spectator model for the study of the
           $\eta$ and $\eta^{\prime}$ mesons
            via the proton-neutron interaction}
\label{Sitd}
\begin{flushright}
       \parbox{0.73\textwidth}{
          {\em
            As regards any subject we propose to investigate, we must
            inquire not what other people have thought, or what we ourselves
            conjecture, but what we can clearly and manifestly perceive by intuition
            or deduce with certains. For there is
            no other way of acquiring knowledge~\cite{descartes2}.\\
          }
          \protect \mbox{} \hfill Ren\'{e} Descartes  \protect\\
       }
\end{flushright}
In order to measure the $pn \to pn Meson$ reactions by means of the proton beam
it is necessary to use a nuclear target, since a pure neutron target does not exist.
Naturally, least complications in the interpretation of the experiment
will be encountered when using the simplest nuclei.
Therefore, here  deuterons will be considered as a source of neutrons,
and for the evaluation of the data an impulse approximation 
will be exploited. The main conjecture of this approach is that the 
bombarding proton interacts exclusively with one nucleon in the target nucleus 
and that the other nucleons affect the reaction by providing a momentum 
distribution to the struck constituent only (fig.~\ref{qfree}).
This assumption is justified if the kinetic energy of a projectile is large 
compared to the binding energy of the hit nucleus.
In fact, as noticed by Slobodrian~\cite{slobodrian175}, also the scattering of 
protons on a hydrogen target, where the protons are bound by molecular forces, 
may serve as an extreme example of the quasi-free reaction. 
In that case, although the hydrogen atoms rotate or vibrate in the molecule, 
their velocities and binding forces are totally negligible with respect to the 
velocity and nuclear forces operating the scattering of the relativistic 
proton~\cite{slobodrian175}.
\vspace{-0.8cm}
\begin{figure}[H]
\begin{center}
\epsfig{file=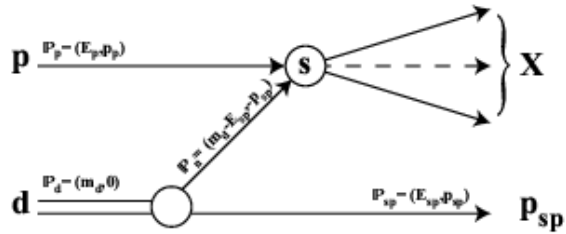,scale=0.70}
\end{center}
\vspace{-0.8cm}
\caption{\label{qfree} Spectator model for a particle production reaction via 
$p d \rightarrow p_{sp} X$.}
\end{figure}
\vspace{-0.0cm}
The deuteron is also relatively weakly bound with a binding energy of 
$\mbox{E}_B \approx 2.2\,\mbox{MeV}$, which is by far smaller -- more than two 
orders of magnitude in the case of pion and already more than three orders of 
magnitude in the case of $\phi$ meson -- than the kinetic energy of 
the bombarding protons needed for the creation of mesons in the proton-neutron 
interaction.
However, it has to be considered that even the low binding energy of 
$\mbox{E}_B \approx 2.2\,\mbox{MeV}$ results in large Fermi momenta of the 
nucleons which can not be neglected. 
The momentum and kinetic energy distributions of the nucleons in the deuteron are 
shown in figure~\ref{fermi_mom_and_kin}.
In the considered approximation 
the internucleon force manifests itself only as the Fermi motion of the nucleons and hence
the struck neutron is treated as a free particle 
in the sense that the matrix element for quasi-free meson production off a 
bound neutron is identical to that for the free $pn \rightarrow p n\,Meson$ 
reaction.  The reaction may be symbolically presented as:
\be
\begin{array}{c}
  p \; \\
  \\
\end{array}
\left(\begin{array}{c}
  n \\
  p \\
\end{array}\right) 
\begin{array}{c}
  \longrightarrow \\
  \\
\end{array}
\begin{array}{c}
  p \; n \; Meson \\
  p_{sp} \\
\end{array}
\end{equation}
where $p_{sp}$ denotes  the proton from the deuteron  regarded as a spectator which does not 
interact with the bombarding particle, but rather escapes untouched and hits 
the detectors carrying the Fermi momentum possessed at the moment of the 
collision. 
\vspace{-0.1cm}
\begin{figure}[H]
\centerline{
\parbox{0.49\textwidth}
  {\epsfig{file=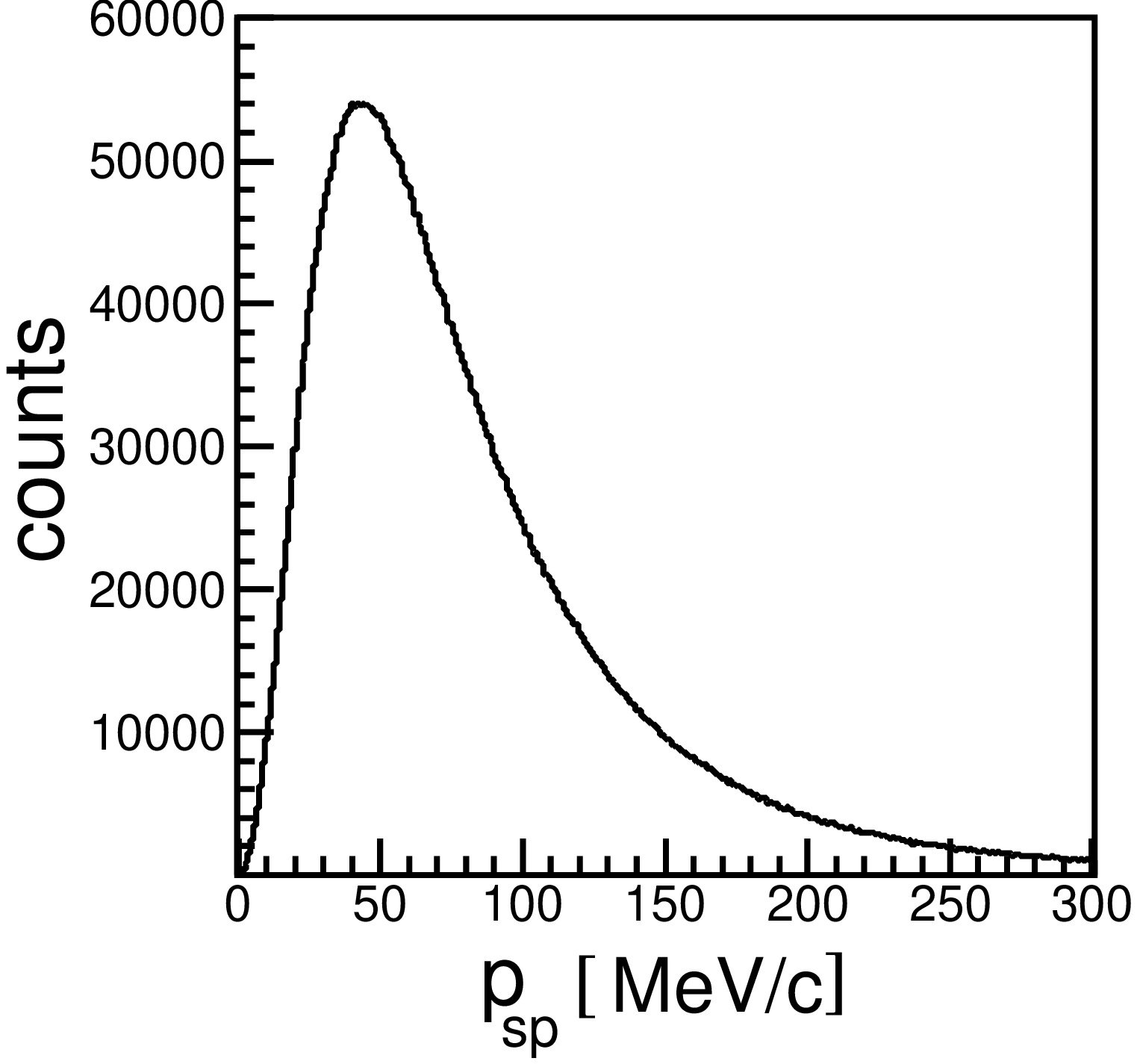,width=0.45\textwidth}}\hfill
\parbox{0.49\textwidth}
  {\epsfig{file=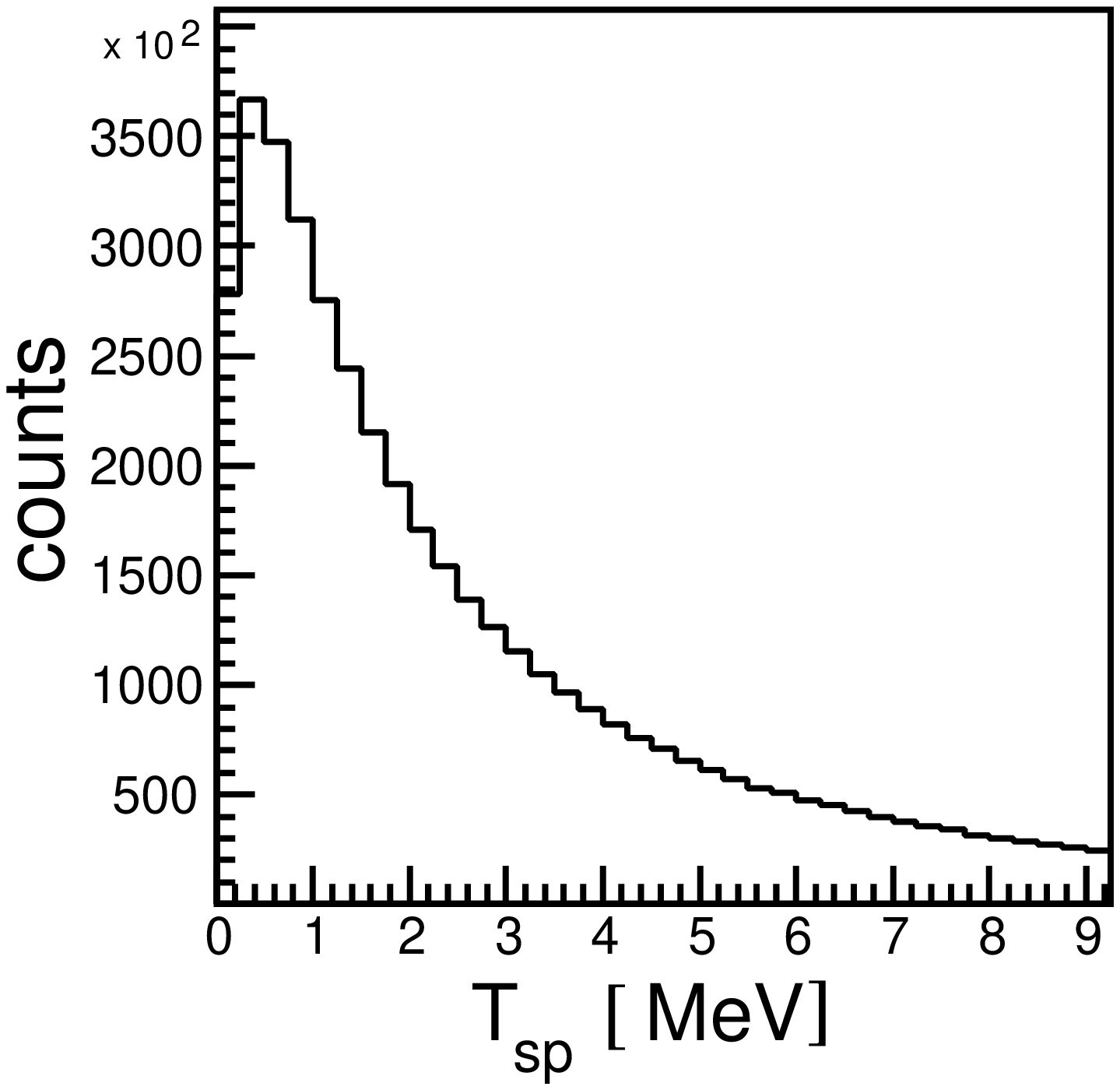,width=0.45\textwidth}}}

\vspace{0.4cm}
\parbox{.44\textwidth}{\raisebox{4ex}[0ex][0ex]{\mbox{}}} \hfill
\parbox{.48\textwidth}{\raisebox{4ex}[0ex][0ex]{\large a)}} \hfill
\parbox{.04\textwidth}{\raisebox{4ex}[0ex][0ex]{\large b)}}
\vspace{-0.9cm}
\caption{\label{fermi_mom_and_kin} (a) Momentum and (b) kinetic energy 
distribution of the nucleons in the deuteron, generated according to an 
analytical parametrization of the deuteron wave 
function~\cite{lacombe139,stina} calculated from the Paris 
potential~\cite{lacombe861}.}
\end{figure}
From the measurement of the momentum vector of the spectator proton 
$\vec{\mbox{p}}_{sp}$ one can infer the momentum vector of the neutron 
$\vec{\mbox{p}}_n = - \vec{\mbox{p}}_{sp}$ 
and hence 
calculate the excess energy Q for each event, provided that the beam momentum 
is known. 
As an example, a distribution of the excess energy in the quasi-free $pn 
\rightarrow pn \eta^{\prime}$ reaction is presented in 
figure~\ref{Q_and_mass_off}a~\cite{moskal0110001}. 
Due to the large centre-of-mass velocity ($\beta \approx 0.75$) with respect 
to the colliding nucleons, a few MeV wide spectrum of the neutron kinetic 
energy inside a deuteron (fig.~\ref{fermi_mom_and_kin}b) is broadened by more than 
a factor of thirty.\\[-0.7cm]
\begin{figure}[H]
\centerline{
\parbox{0.49\textwidth}{\epsfig{file=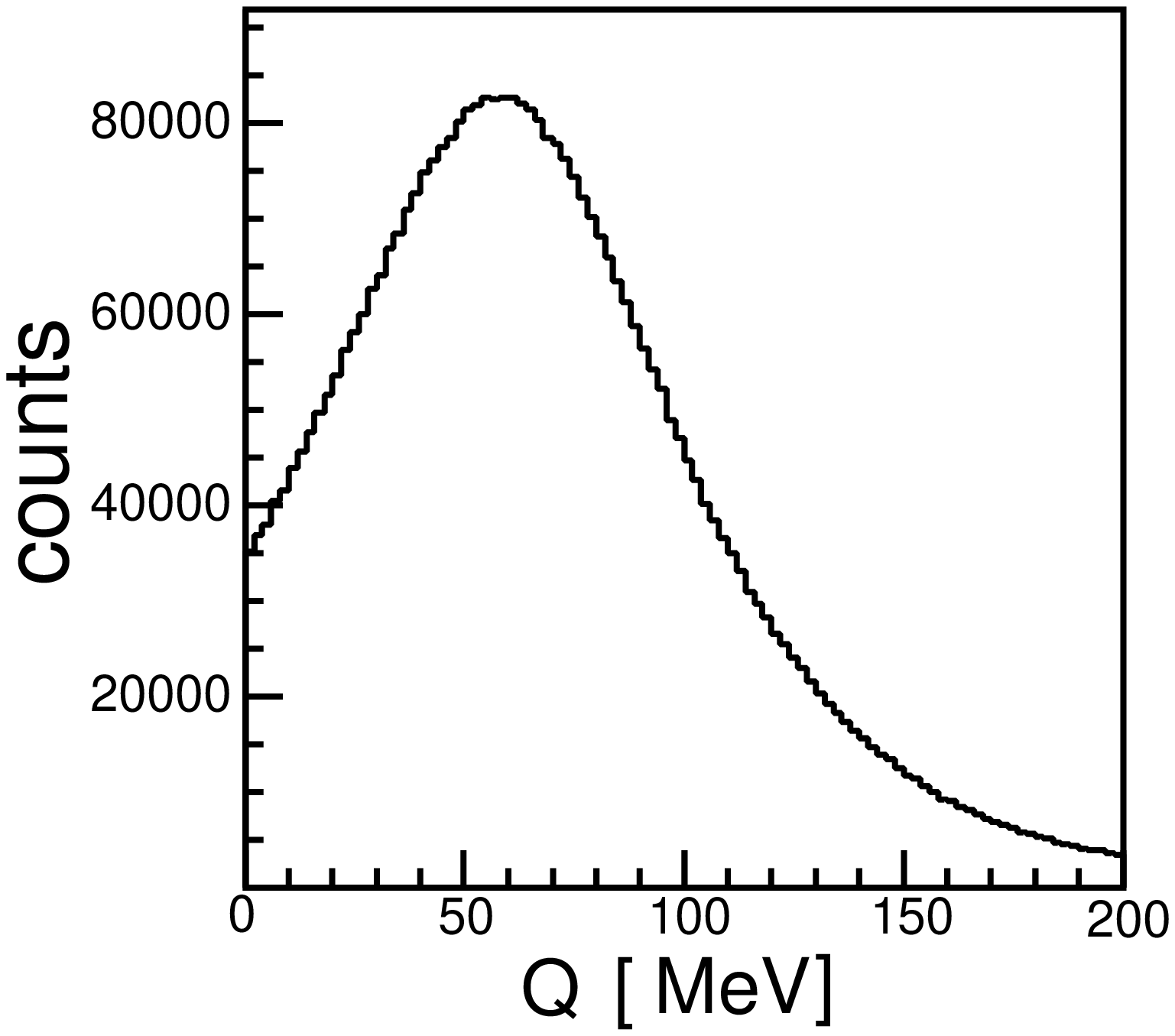,width=0.45\textwidth}}
\parbox{0.49\textwidth}{\epsfig{file=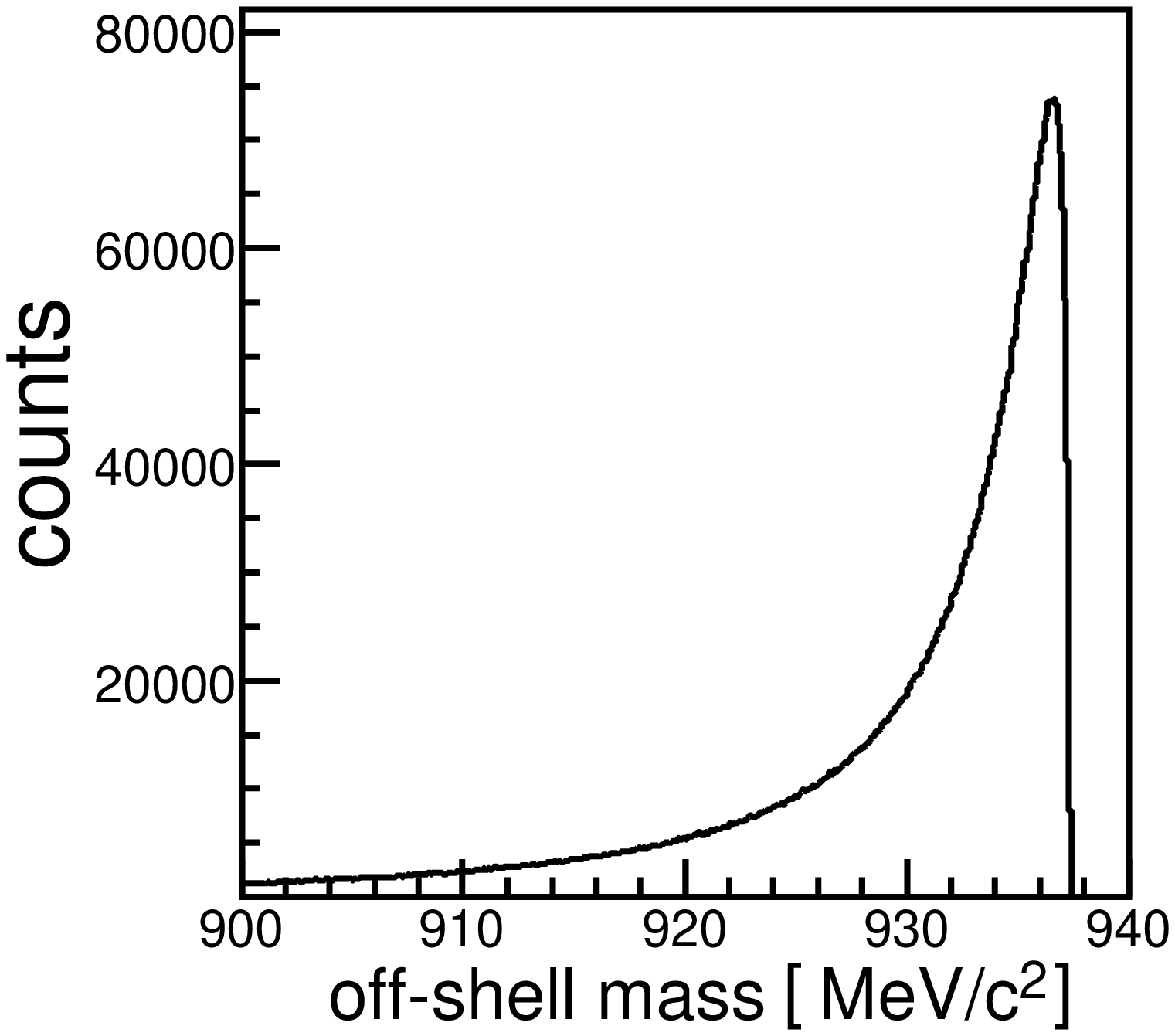,width=0.45\textwidth}}}

\vspace{0.8cm}
\parbox{.41\textwidth}{\raisebox{5ex}[0ex][0ex]{\mbox{}}}\hfill
\parbox{.47\textwidth}{\raisebox{5ex}[0ex][0ex]{\large a)}}\hfill
\parbox{.020\textwidth}{\raisebox{5ex}[0ex][0ex]{\large b)}}\hfill
\vspace{-1.1cm}
\caption{\label{Q_and_mass_off} (a) Distribution of the excess energy Q for the 
$pn \eta^{\prime}$ system originating from the reaction $pd \rightarrow 
p_{sp} pn \eta^{\prime}$ calculated with a proton beam momentum of 
$3.350\,\mbox{GeV/c}$ and the neutron momentum smeared out according to the 
Fermi distribution shown in figure~\ref{fermi_mom_and_kin}a. 
(b) Spectrum of the off-shell mass of the interacting neutron, as calculated 
under the assumption of the impulse approximation.}
\end{figure}
\vspace{-0.1cm}
Therefore, especially in the case of near-threshold measurements,
where the 
cross section grows rapidly with increasing excess energy~(see e.g.\ 
fig.~\ref{cross_eta_etap}), the total centre-of-mass energy 
$\sqrt{\mbox{s}}$ has to be determined on an event-by-event level. 
For this purpose, the spectator protons are usually registered 
by means of silicon pad- or $\mu$-strip 
detectors~\cite{bilger64,moskal0110001,ankespec} which allow to determine their kinetic 
energy ($\mbox{T}_{sp}$) and polar emission angle ($\theta$).
Thus, it is useful to express the total energy  squared $s$ as function of these variables:
\be
\label{eq:sqf}
\mbox{s} = |\mathbb{P}_p + \mathbb{P}_n|^2 =
 \mbox{s}_0 - 2\,\mbox{T}_{sp}\,(\mbox{m}_d + \mbox{E}_{p} ) + 
 2\,\mbox{p}_p\,\sqrt{\mbox{T}_{sp}^2 + 2\mbox{m}_p\mbox{T}_{sp}}\,\cos(\theta) 
\ee
with $\mbox{s}_0$ denoting the squared centre-of-mass energy, assuming a vanishing 
Fermi motion. 
Measuring both the energy and the emission angle of the spectator protons it is 
possible to study the energy dependence of a meson production cross section 
from data taken at only one fixed beam momentum. 

It must be noted, however, that in the framework of the impulse approximation, 
illustrated in figure~\ref{qfree}, the measured spectator proton is a physical 
particle, yet the reacting neutron is off its mass shell, where the explicit 
expression for its four-momentum vector $\mathbb{P}_n$, in the rest frame of 
the deuteron, reads:
\be
\label{neutronfourvector}
\mathbb{P}_n = (\mbox{m}_d - \mbox{m}_p - \mbox{T}_{sp},\,-\vec{\mbox{p}}_{sp}),
\ee
with $\mbox{T}_{sp}$ and $\vec{\mbox{p}}_{sp}$ denoting the kinetic energy and 
the momentum vector of a spectator proton, respectively.
The  mass spectrum of the interacting neutron ($\mbox{m}_n^2 = 
|\mathbb{P}_n|^2$) resulting from the distribution of Fermi momentum is shown 
in figure~\ref{Q_and_mass_off}b.
It can be seen that the maximum of this spectrum differs only by about 
$3\,\mbox{MeV}/\mbox{c}^2$ from the free neutron mass ($\mbox{m}_n = 
939.57\,\mbox{MeV}/\mbox{c}^2$), however on the average it is off by about 
$9\,\mbox{MeV}/\mbox{c}^2$. 
In the framework of the discussed approximation, the struck neutron is never on its 
mass shell and the minimum deviation from the real mass occurs for vanishing 
Fermi-momentum and --~as can be inferred from 
equation~\eqref{neutronfourvector}~-- is equal to the binding energy 
$\mbox{E}_B = \mbox{m}_d - \mbox{m}_n - \mbox{m}_p$. 
Measurements performed at the CELSIUS and TRIUMF accelerators for the 
$pp \rightarrow pp \eta$~\cite{calen2642} and $pp \rightarrow 
d \pi^+$~\cite{duncan4390} reactions, respectively, have shown that within the 
statistical errors there is no difference between the total cross section of 
the free and quasi-free processes. This conjecture is confirmed also by the
theoretical investigations~\cite{kaptari0212066}.
\vspace{-0.2cm}
\begin{figure}[H]
\parbox{0.59\textwidth}
  {\epsfig{file=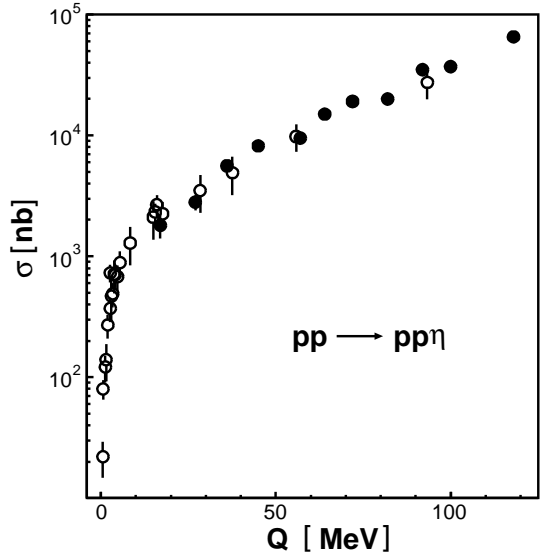,width=0.55\textwidth}}
\parbox{0.40\textwidth}{
\caption{\label{free_vs_quasifree} (a) Total cross sections for the $pp 
\rightarrow pp \eta$ reaction as a function of excess energy measured for the 
scattering of protons in vacuum (open 
symbols)~\cite{smyrski182,hibou41,calen39,chiavassa270,bergdoltR2969,moskal367}
and inside a deuteron (filled symbols)~\cite{calen2642}.\protect\\
}}
\end{figure}
\vspace{-0.1cm}
In figure~\ref{free_vs_quasifree} the production of the $\eta$ meson in free
proton-proton collisions is compared to the production inside a deuteron and 
in the overlapping regions the data agree within the statistical errors. 
These observations allow to anticipate that indeed the assumption of the 
identity for the transition matrix element for the meson production off free 
and quasi-free nucleons bound in the deuteron is correct, 
at least on the few per cent level.
In case of the meson production off the deuteron, one can also 
justify the assumption of the quasi-free scattering with a geometrical argument,
since the average 
distance between the proton and the neutron is 
in the order of~\footnote{\mbox{} The matter radius of the deuteron amounts to 
$\approx 2\,\mbox{fm}$~\cite{garcon049}.} $3\,\mbox{fm}$.
Of course, the other nucleon may scatter the incoming proton and the outgoing 
meson.
However, these nuclear phenomena are rather of minor importance in case of the 
production on the neutron bound in the deuteron, but should be taken into 
account for derivations of total cross sections from experimental data.
Indeed, calculating an influence on the cross section 
from  a possible rescattering of the $\omega$ meson
in the final state of the $pd\to d \omega p_{sp}$ reaction,
authors of reference~\cite{golubeva} found that 
for the spectator-proton momentum below 100-150~MeV/c 
the rescattering effects play a minor role.
The reduction of the beam flux on a neutron, due to the presence of the proton, 
referred to as a shadow effect, decreases for example the total cross section 
by about $4.5\,\%$~\cite{chiavassa192} for the $\eta$-meson production.
Similarly, the reduction of the total cross section due to the reabsorption of 
the outgoing $\eta$ meson on the spectator proton was found to be only about 
$3\,\%$~\cite{chiavassa192}. 
The appraisals were performed according to a formula derived in 
reference~\cite{smith647} which shows that the cross section for the deuteron 
reduces by a factor of:
\be
R = 1 - \sigma_{\eta N}^{inel} <\mbox{r}^{-2}> / 4\pi
\ee
compared to the free nucleon cross sections. 
Here $\sigma_{\eta N}^{inel}$ denotes the $\eta N$ inelastic cross section and 
$<\mbox{r}^{-2}>$ stands for the average of the inverse square nucleon 
separation in the deuteron taking the nucleon size into 
account~\cite{chiavassa192}.
The latter effect for the production of mesons like $(\pi,\omega,\eta^{\prime},
\phi)$ is expected to be much smaller, since 
the s-wave interaction of the $\eta$-meson with nucleons is by far 
stronger than for any of the mentioned ones.

\section{Test measurement of the $pn \to pn \eta$ reaction }
\label{testpnpneta}
\begin{flushright}
\parbox{0.63\textwidth}{
 {\em
    Nothing is so trivial as treating serious subject in a trivial manner~\cite{erasmus}.\\
 }
 \protect \mbox{} \hfill Erasmus of Rotterdam \protect\\
 }
\end{flushright}
\vspace{-0.3cm}
As a general commissioning of the extended COSY-11 facility
to investigate quasi-free $pn\to pn X$ reactions,
we have performed a measurement of the $pn \to pn\eta$ process
at a beam momentum of 2.075~GeV/c~\cite{hadronmoskal}. 
The experiment, carried out in June 2002,
had been preceded  by the  installation of a spectator~\cite{bilger64,spectatorraporty,spectatorraporty1} 
and  neutron detectors~\cite{neutronraporty,neutronraporty1},
and by a series of thorough simulations performed
in order to determine the best conditions for measuring quasi-free
$pn\to pn \eta$ and $pn\to pn \eta^{\prime}$ reactions~\cite{moskal0110001,proposalc11,rafalmgr}.
Figure~\ref{detectionsystem} presents the COSY-11 detection facility with  superimposed tracks
of protons and neutron originating from the quasi-free $pn\to pn X$ reaction induced  
by a proton beam~\cite{prasuhn167} impinging on a deuteron target~\cite{dombrowski228}. 
The  identification of the $pn\to pn\eta$ reaction is based on the measurement of the four-momentum vectors
of the outgoing nucleons and the $\eta$ meson is identified via the missing mass technique.
The slow proton  stopped in the first layer of the position sensitive silicon detector (Si$_{spec}$)
is, in the analysis, considered 
as a spectator non-interacting
with the bombarding particle 
and moving with the  Fermi momentum as possessed 
at the moment of the collision.
\vspace{-0.4cm}
\begin{figure}[H]
\parbox{1.0\textwidth}
{\hspace{4.8cm}\epsfig{file=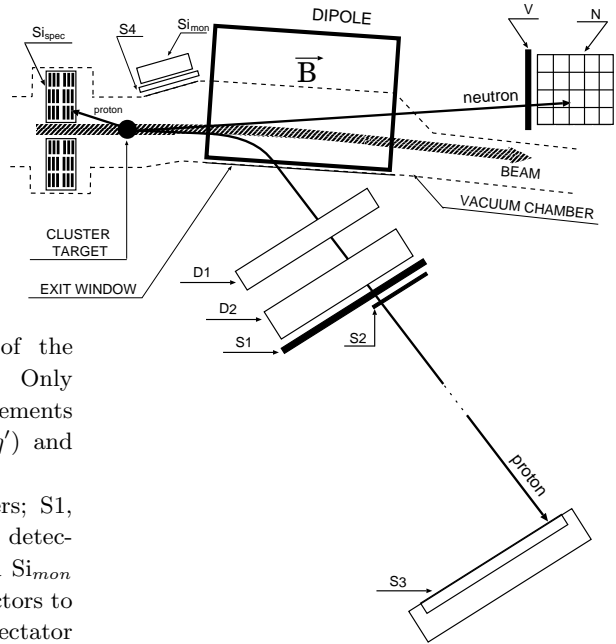,width=0.70\textwidth,angle=0}}\\

\vspace{-4.5cm}

\parbox{0.44\textwidth}{
            \caption{\small  Schematic view of the COSY-11 detection setup~\protect\cite{brauksiepe397}.
             Only detectors needed for the measurements of the  reactions
             $pd\rightarrow p_{sp}pn\eta(\eta^{\prime})$ and $pd\to pp n_{sp}$
             are shown. \protect\\
             D1, D2 denote the drift chambers; S1, S2, S3, S4 and V the scintillation detectors;
             N the neutron detector and Si$_{mon}$ and Si$_{spec}$~\cite{bilger64} silicon strip detectors
             to detect elastically scattered and spectator protons, respectively.
             \label{detectionsystem}
            }
          }
\end{figure}
\vspace{0.5cm}
As we described in subsection~\ref{Sitd}, 
from the measurement of the momentum vector of the spectator proton
$\vec{\mbox{p}}_{sp}$ one can infer the momentum vector of the struck neutron
$\vec{\mbox{p}}_n = - \vec{\mbox{p}}_{sp}$ at the time of the reaction and hence
calculate the total energy of the colliding nucleons for each event.
In figures~\ref{exp1}a and~\ref{exp1}b the measured and expected distribution of the kinetic energy of the spectator proton 
is presented. 
Though still a very rough energy calibration of the detector units  was performed,
one recognizes a substantial similarity in the shape 
of both distributions. 
\vspace{-0.1cm}
  \begin{figure}[H]
       \hspace{-0.4cm} 
       \parbox{0.5\textwidth}{\centerline{
       \epsfig{file=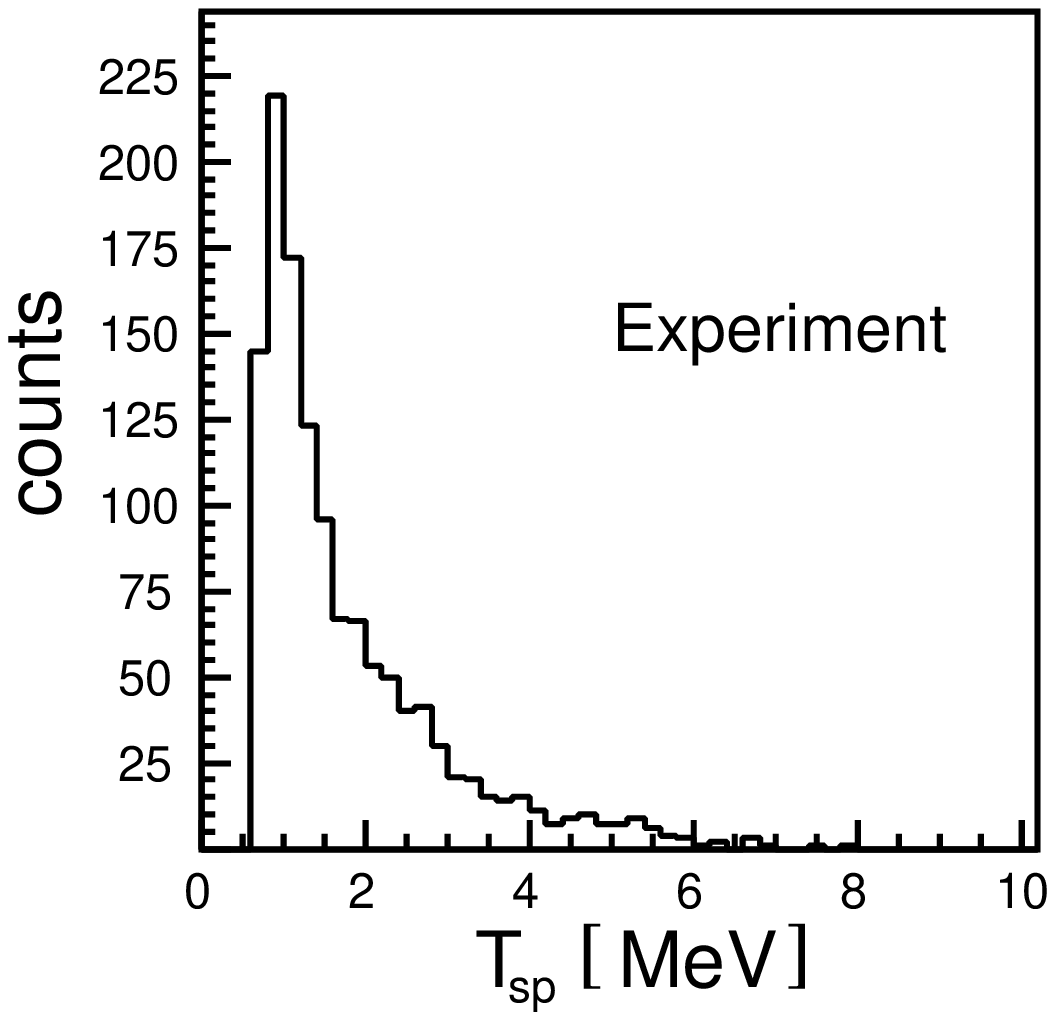,width=0.445\textwidth,angle=0}}}
       \parbox{0.5\textwidth}{\centerline{
       \epsfig{file=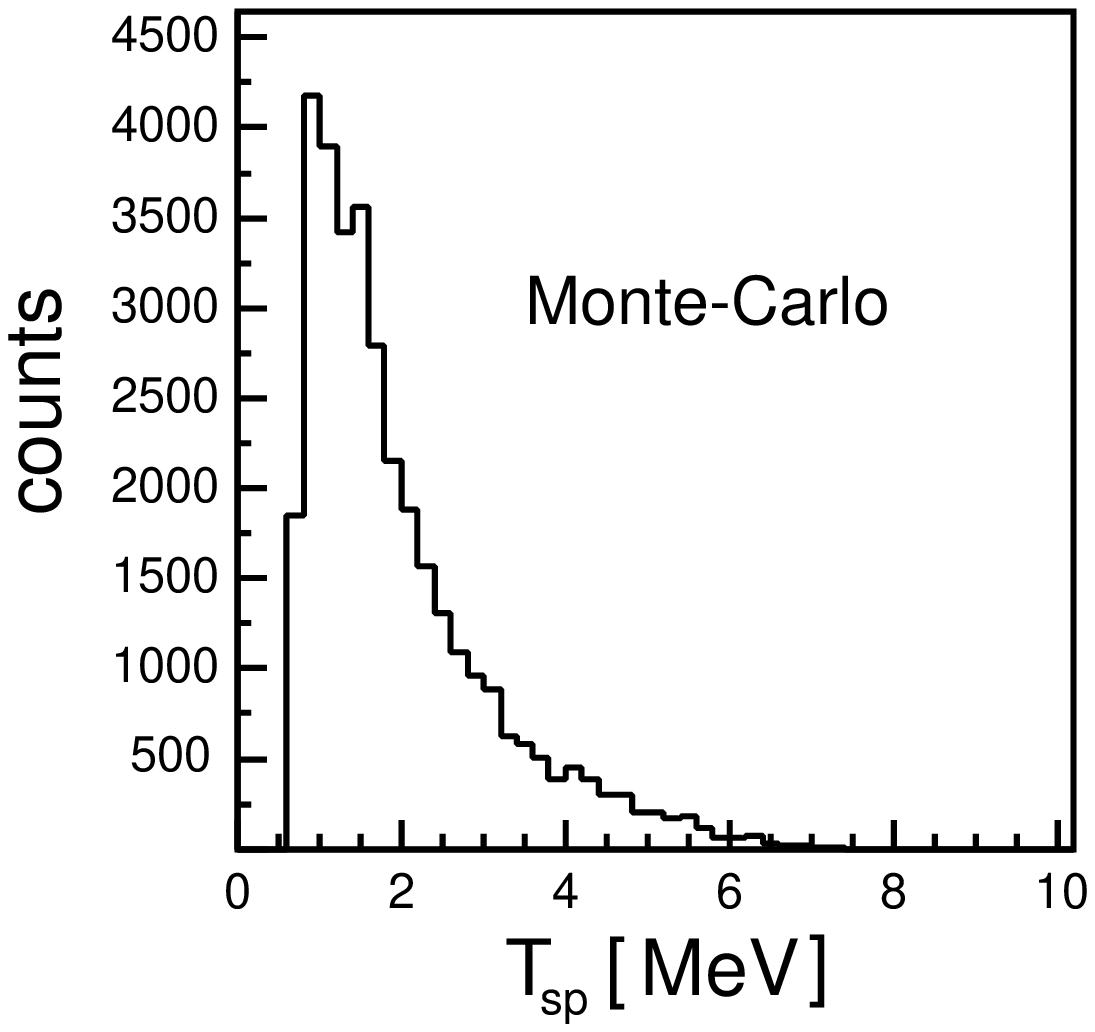,width=0.45\textwidth,angle=0}}}\\
       \hspace{-1.0\textwidth}
       \parbox{0.45\textwidth}{\mbox{} }\hfill
       \parbox{0.51\textwidth}{ \vspace{-0.7cm} a) }\hfill
       \parbox{0.04\textwidth}{ \vspace{-0.7cm} b) }\\
   \vspace{-0.7cm}
   \caption{ \small
          Distributions of the kinetic energy of the spectator protons.\protect\\
         (a) Experiment. \ \ \
         (b) Monte-Carlo simulations taking into account the acceptance
             of the COSY-11 detection system and 
           an analytical parametrization of the
           deuteron wave function~\protect\cite{lacombe139}
           calculated from the PARIS potential~\protect\cite{lacombe861}.
             \label{exp1}
         }
\end{figure}
\vspace{-0cm}
Figure~\ref{exp2} shows spectra of the excess energy in respect to the $pn\eta$ system
as obtained in the experiment~(\ref{exp2}a) and
the simulation~(\ref{exp2}b) for the $pn \rightarrow p n\eta$ reaction. The remarkable difference between the distributions comes from the
fact that in reality additionally to the $pn\to pn\eta$ reaction 
also the multi-pion production is registered.
The $\eta$ and multi-pion production cannot be distinguished from each other 
on the event-by-event basis by means of the missing mass technique.
\vspace{-0.1cm}
\begin{figure}[H]
       \hspace{-0.5cm}
       \parbox{0.5\textwidth}{\centerline{
       \epsfig{file=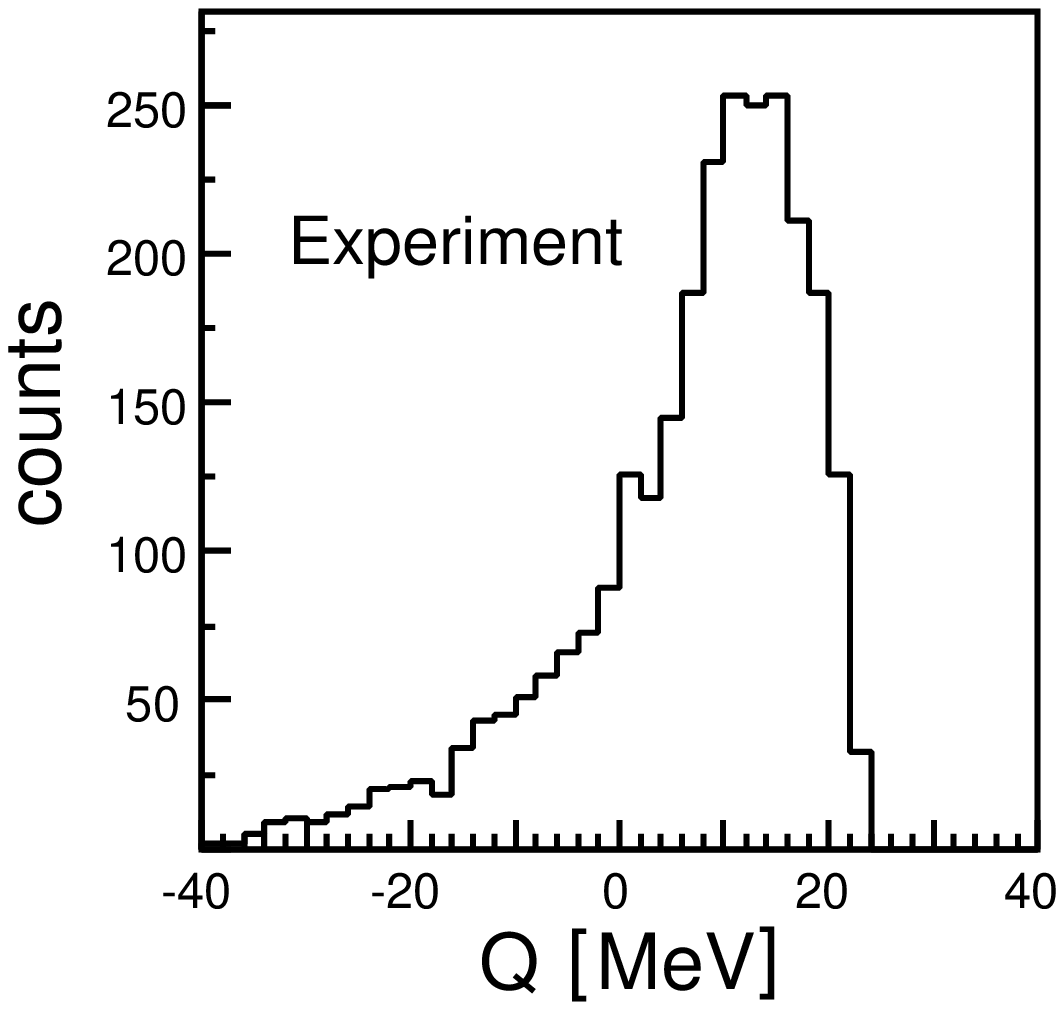,width=0.45\textwidth,angle=0}}}
       \parbox{0.5\textwidth}{\centerline{
       \epsfig{file=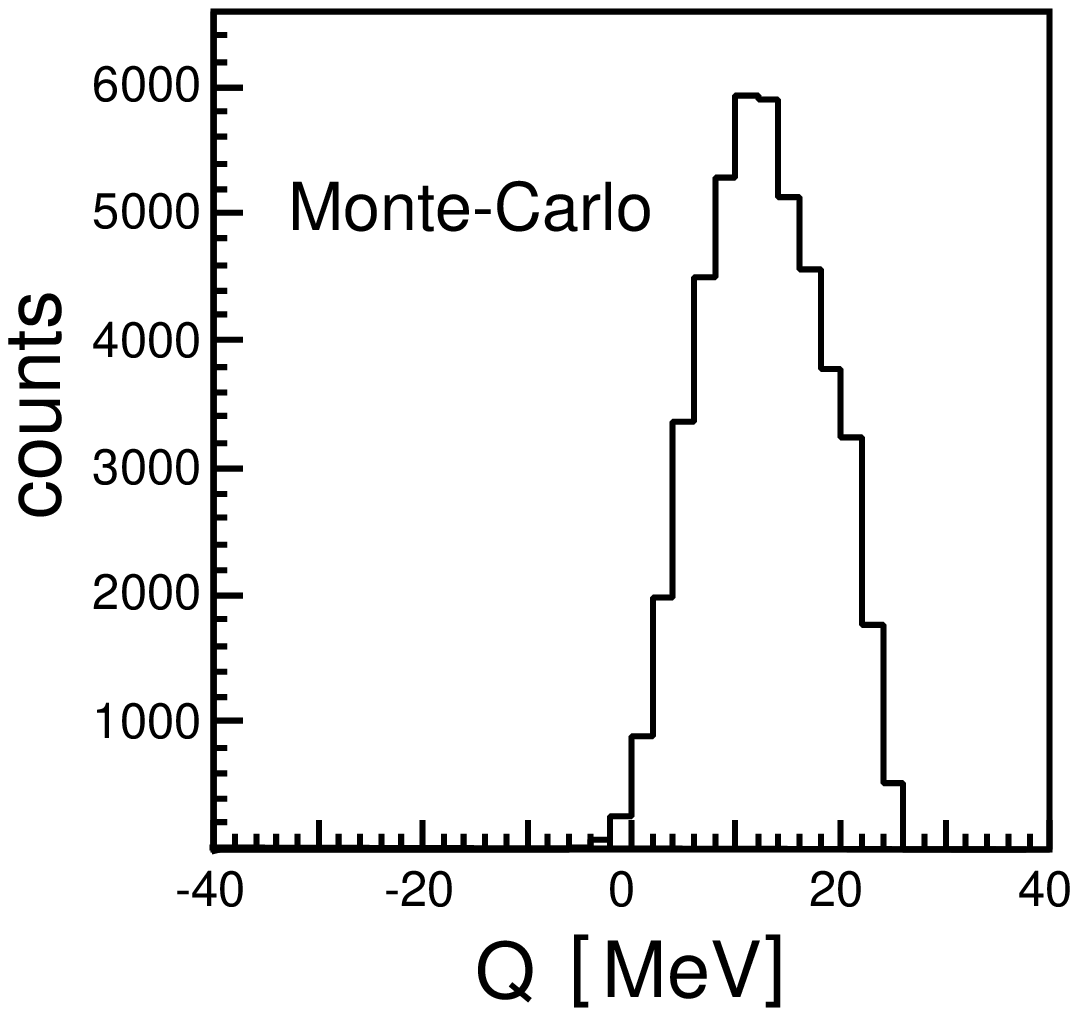,width=0.45\textwidth,angle=0}}}\\
       \hspace{-1.0\textwidth}
       \parbox{0.45\textwidth}{\mbox{} }\hfill
       \parbox{0.51\textwidth}{ \vspace{-0.7cm} a) }\hfill
       \parbox{0.04\textwidth}{ \vspace{-0.7cm} b) }
   \vspace{-1.0cm}
   \caption{ \small
          Distributions of the excess energy $Q$
           for the quasi-free $pn\rightarrow pnX$ reaction,
           determined with respect to the $pn\eta$ threshold.
           (a) Experiment. (b) Simulation.
             \label{exp2}
         }
\end{figure}
\vspace{-0.2cm}
However, we can determine the number of the registered $pn\to pn\eta$ reactions from the multi-pion background
comparing the missing mass distributions for Q values larger and smaller than zero. Knowing that negative values of Q can only be
assigned to the multi-pion events we can derive the shape of the missing mass distribution corresponding to these events.
This is shown as the dashed line in figure~\ref{exp6}. 
  \vspace{-0.1cm}
  \begin{figure}[H]
       \hspace{-0.5cm}
       \parbox{0.5\textwidth}{\centerline{
       \epsfig{file=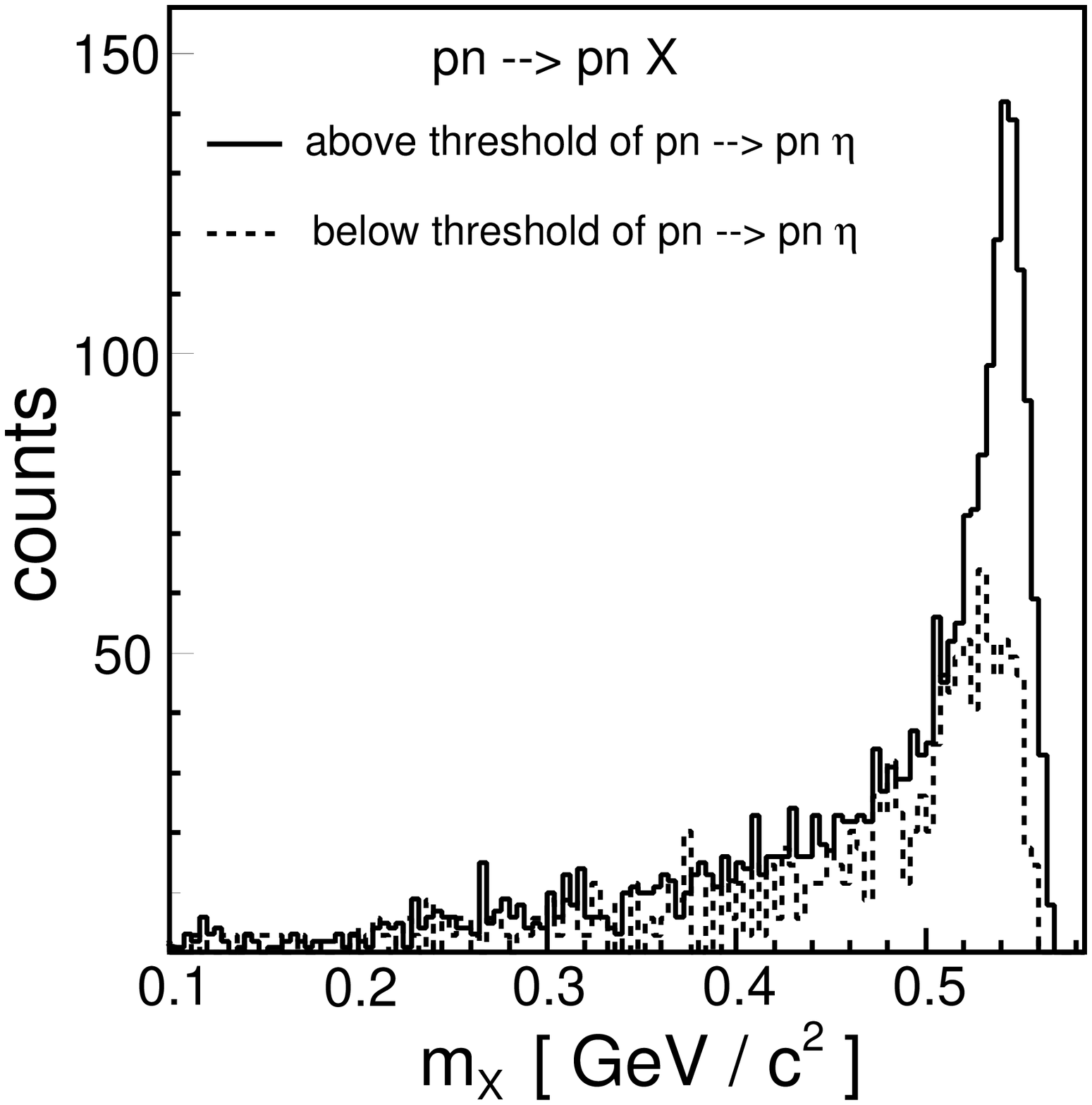,width=0.47\textwidth,angle=0}}}
       \parbox{0.5\textwidth}{\centerline{
       \epsfig{file=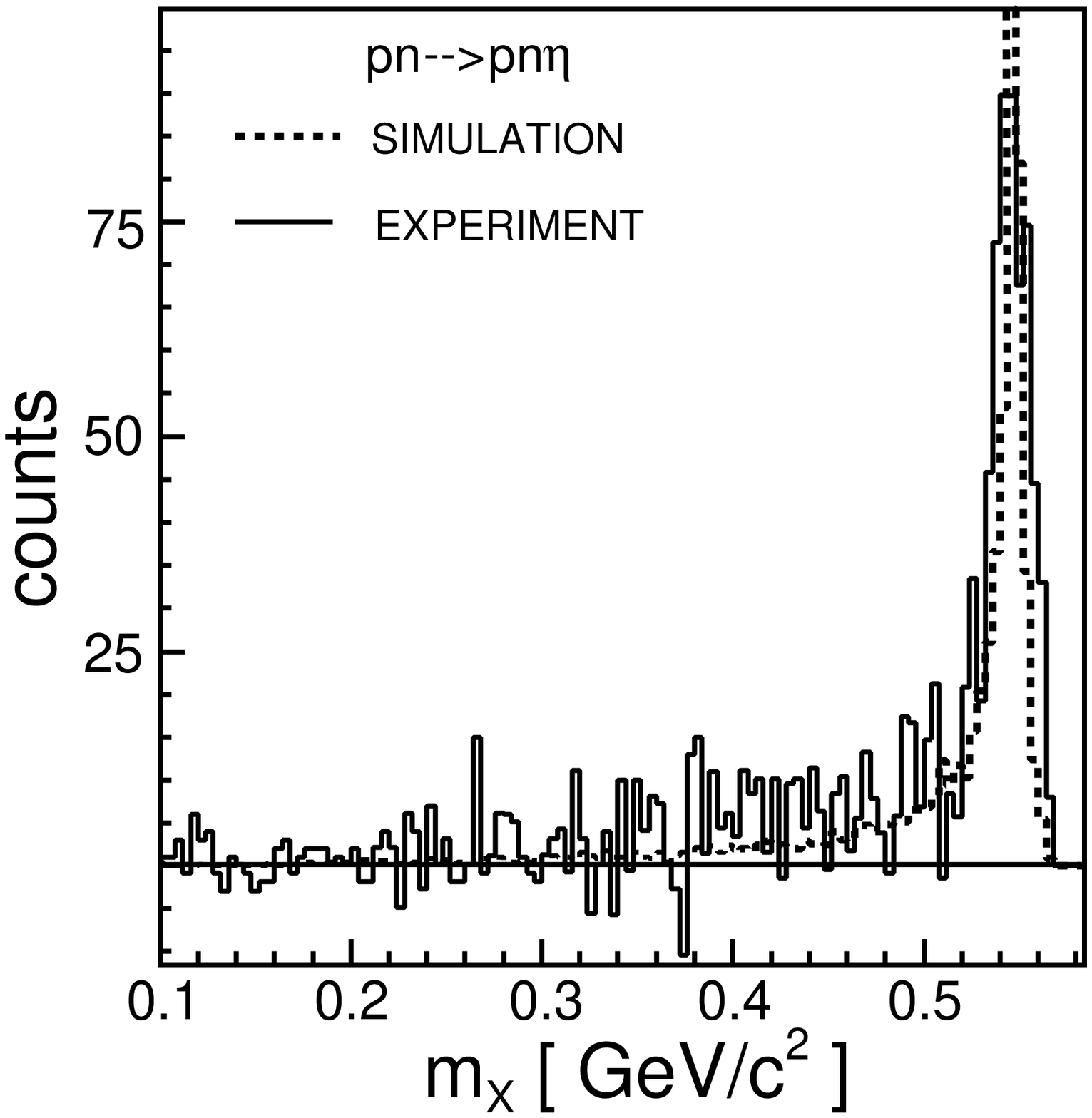,width=0.47\textwidth,angle=0}}}\\
       \hspace{0.2\textwidth}
       \parbox{0.38\textwidth}{\mbox{} }\hfill
       \parbox{0.45\textwidth}{ \vspace{-1.2cm} a) }\hfill
       \parbox{0.04\textwidth}{ \vspace{-1.2cm} b) }
   \vspace{-0.2cm}
   \caption{ \small
          Missing mass spectra as obtained during the June'02 run: \protect\\
          a) Event distribution for $Q < 0$ ( dashed line) and for $Q > 0$ (solid line).\protect\\
          b) Solid line represents the difference between number of events above and below threshold
             for the $pn\to pn\eta$ reaction, and the dashed line
              corresponds to the Monte-Carlo simulation.
          \label{exp6}
         }
\end{figure}
A thorough evaluation of the background is in progress, however, rough 
comparison of events for positive and negative Q yields the promising results with a clear signal 
from the  $pn\to pn\eta$ reactions,
as can be deduced by inspection of figures~\ref{exp6}a and ~\ref{exp6}b.

\section{Double quasi-free production}
\label{nnnnn}

\begin{flushright}
\parbox{0.6\textwidth}{
 {\em
   There is need of a method  for investigating the truth about things~\cite{descartes2}. \\
 }
 \protect \mbox{} \hfill Ren\'{e} Descartes \protect\\
 }
\end{flushright}
     In article~\cite{moskal295} we described a method of measuring the close-to-threshold
     meson production in neutron-neutron collisions,
     where the momenta of the colliding neutrons could be
     determined with the accuracy obtainable for 
     the proton-proton reaction. The technique
     is based on the double quasi-free $nn\rightarrow nn X^{0}$ reaction,
     where deuterons are used as a source of neutrons.

Close-to-threshold meson production in proton-neutron
collisions were investigated by means of a technique
based on a  quasi-free scattering
of the proton off the  neutron bound in the deuteron. 
Thin windowless internal deuterium cluster targets  
($\sim$~$10^{14}$~atoms/cm$^{2}$) make
a detection of an undisturbed spectator proton and 
a precise determination of the reacting neutron momentum
--~and hence of the excess energy~-- possible.\\
Pioneering experiments of the $\pi^{0}$ meson creation in the proton-neutron 
reaction
with the simultaneous
tagging of the spectator proton resulted in a resolution of the excess energy 
equal to $\sigma$(Q) = 1.8~MeV~\cite{bilger64}.
Similar studies including  the production  of heavier mesons 
are carried out at COSY-11 and ANKE facilities~\cite{proposalankephiomega,proposalc11},
and we intend to continue them at the forthcoming facility WASA at COSY.\\
\vspace{-0.5cm}
\begin{figure}[H]
\hspace{-0.1cm}
\centerline{
\resizebox{0.85\textwidth}{!}{%
\includegraphics{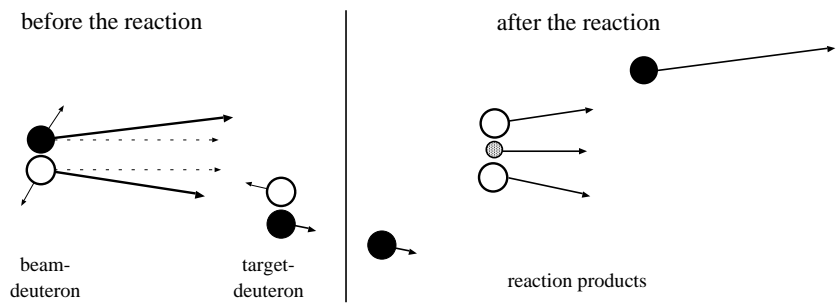}}}
\vspace{0cm}
\caption{Schematic depiction of the double quasi-free 
         {\mbox{$ nn \rightarrow nnX$}} reaction.
         During the collisions of deuterons (left hand side of the figure, 
	 with the total momentum (solid arrow) resulting from the sum of the
	 beam momentum (dotted arrow) plus the Fermi momentum
	 (short arrow)) a double quasi-free neutron-neutron
         reaction may lead to the creation of mesons (small gray circle).
         The spectator protons~(black circles) leave the reaction region
         with their initial momentum plus the Fermi momentum, 
         which they possessed at the moment of the reaction.
         Neutrons are plotted as open circles.
         Due to the large relative momenta between spectators and the outgoing neutrons
         ($\sim$~1~GeV/c close-to-threshold for the $\eta$ meson production)
         a distortion of the nnX system by the accompanied protons can be neglected.
        }
\label{dd_ppnneta}     
\end{figure}
Experimental investigations of the close-to-threshold
production in neutron-neutron collisions, however, have not yet been 
carried out. A realisation of such studies 
--~which are characterised by typical cross sections of $\le \mu$b~--
with high quality neutron beams bombarding a deuterium target is not 
feasible due to the low neutron beam intensities forcing to use liquid or 
solid deuterium targets which make the precise determination of the momentum
of the spectator proton impossible.
In this section a unique possibility of the precise measurement of the 
close-to-threshold meson production in neutron-neutron collisions is 
pointed out. The technique is based on the double quasi-free interaction of 
neutrons originating from colliding deuterons as depicted in 
Figure~\ref{dd_ppnneta}.
Utilizing this method, a precision of $\sim$1~MeV can be obtained for 
determining the excess energy, since it depends only on the 
accuracy of the momentum or angle reconstruction for the registered
spectator protons. At present cooled deuteron beams --~available at 
the facilities CELSIUS and COSY~-- give the possibility
of using this method for the studies of neutron-neutron scattering.
Moreover, the usage of a stored beam circulating through an internal
cluster target permits the study with high luminosities (~10$^{31}$cm$^{-2}$s$^{-1}$)
in spite of very low target densities.
These investigations can  be realized much more effectively once the WASA detector
is installed at COSY. 
In the double quasi-free interaction,
due to the small binding energy of the deuteron~($E_{B} = 2.2$~MeV),
the colliding neutrons may be approximately treated as 
free particles in the sense that the matrix element for quasi-free meson
production from bound neutrons is identical to that for the free
nn $\rightarrow $ NNX reaction at the same excess energy 
available in the NNX system.
The measurements at CELSIUS~\cite{calen2642,calen2667}
 and TRIUMF~\cite{duncan4390,hahn2258} have  proven 
that the offshellness of the reacting neutron
can be neglected and that the spectator proton influences the interaction
only in terms of the associated Fermi motion~\cite{duncan4390}.
 The registration of both spectator protons will allow for a
precise determination of the excess energy. 
A possible internal target facility based on the COSY-11 setup~\cite{brauksiepe397}
is presented in figure~\ref{detectors}. The energy and the
emission angle of the "slow" spectator can 
be measured by an appropriately segmented silicon detector,
whereas the momentum of the "fast" spectator proton can be analysed
by the magnetic spectrometer. By means of the detection system
shown in figure~\ref{detectors}, a resolution of the excess energy
of 2~MeV can be achieved for excess energies lower than 
30~MeV as demonstrated in reference~\cite{proposalc11}. 
The double quasi-free $nn\rightarrow nn X^{0}$ reaction
can be identified by the registration of both outgoing neutrons.
For example, in order to measure the production of the $\eta$ meson,
a seven meter distance for the time-of-flight measurement would be enough
to obtain  8~MeV (FWHM) missing mass resolution~\cite{proposalc11}
with a calorimeter segmented 
by 10~cm~$\ast$~10~cm 
and providing a 0.5~ns~($\sigma$) time resolution, which was obtained in test
runs using a scintillator/lead sandwich type of detector.
The suggested meson production via a double 
quasi-free neutron-neutron reaction with precisions achievable
for the proton-proton and proton-neutron reactions,
opens the possibility of 
studying for example the charge symmetry breaking by comparing cross sections 
for the $pp\rightarrow pp\eta$ and $nn\rightarrow nn\eta$ reactions,
similarly to investigations performed via the $\pi$-deuteron reactions~\cite{tippens052001}. 
The Dalitz-plot analysis of the $ nn \rightarrow nn$ Meson would allow for the study
of the neutron-neutron and neutron-Meson~\cite{kudriavtsev} scattering lengths,
the first being still not well established~\cite{machleidtR69} and the second being unknown.
In principle when studying the meson production in proton-proton and in 
proton-neutron collisions one has access to all possible isospin combinations, 
which can be derived after the correction for the electromagnetic interaction.  
\vspace{-0.0cm}
\begin{figure}[H]
\hspace{0.1cm}
\centerline{
\resizebox{0.8\textwidth}{!}{%
\includegraphics{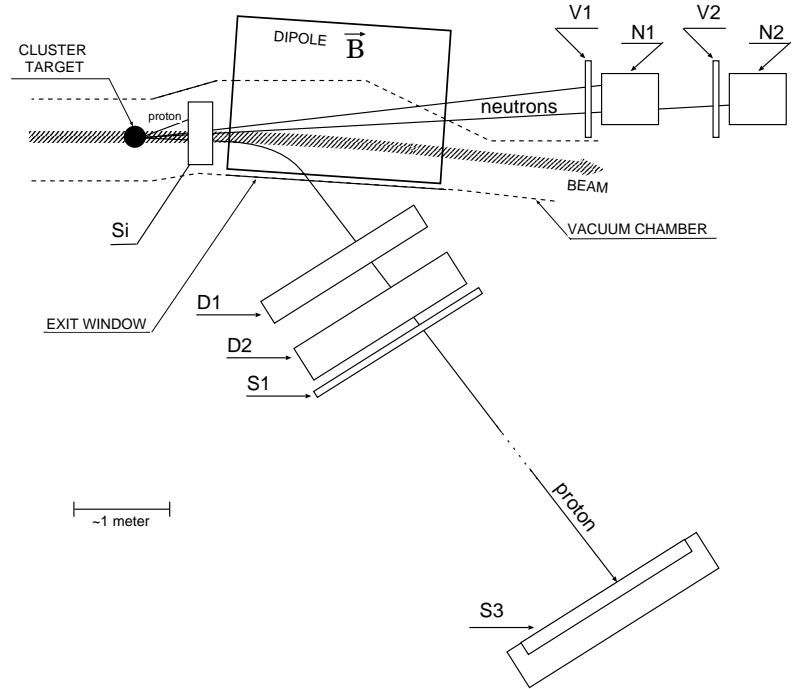}}}
\vspace{0cm}
\caption{
    Schematic view of the extended COSY-11 detection setup~\protect\cite{brauksiepe397}.
             Only detectors needed for the measurements of the 
	     $dd\rightarrow nn p_{sp}p_{sp} X$
             reactions are shown. \protect\\
             D1,D2 denote the drift chambers used for the track 
             reconstruction of the fast spectator proton;
             S1,S3 and V1,V2 are the scintillation detectors 
             used as  time-of-flight and veto counters, respectively,
             N1,N2  the neutron detectors, and 
	     Si~\protect\cite{bilger64} the silicon strip detectors.
 }
\label{detectors}       
\end{figure}
Exceptionally, close-to-threshold meson production via the
neutron-neutron scattering represents a pure $T=1$ isospin channel without 
accompanying Coulomb interaction and consequently no need for its correction.
Investigations of neutron-neutron scattering allow also for the production 
of  $K^{+}K^{-}$ pairs in a system with only two charged particles in the final
state ($nn\rightarrow nnK^{+}K^{-}$), simplifying the theoretical
calculations drastically,
which in case of the $pp\rightarrow ppK^{+}K^{-}$ are not feasible due to the difficulty
of treating the  electromagnetic forces in the system of four charged particles~\cite{hanhart}.

At present the COSY synchrotron can accelerate deu\-te\-rons up to 3.5~GeV/c~\cite{prasuhn167}
which, utilizing the Fermi momentum, allows for the $\pi$ and $\eta$ meson production
in the $nn\rightarrow nn X^{0}$ reaction. 
To investigate the neutron-neutron interaction
with the production of heavier mesons like $\omega$, $\eta^{\prime}$, or $\phi$,
a deuteron beam of $\sim 7$~GeV/c would be required.\\

\newpage
\clearpage
\thispagestyle{empty}
\pagestyle{plain}
\chapter{Conclusion}
\thispagestyle{empty}
\pagestyle{myheadings}
\markboth{Hadronic interaction of $\eta$ and $\eta'$ mesons with protons}
         {7. Conclusion}

Using the stochastically cooled proton beam of the cooler synchrotron COSY
and the COSY-11 facility we performed measurements of the $pp\to pp\eta$
and $pp\to pp\eta^\prime$ reaction close to the kinematical threshold.
We have succeeded in establishing the energy dependence of the total cross section
for  both reactions and the probability density of the phase space population
for the $pp\to pp\eta$ reaction.  At present we are pursuing the study by elaborating 
differential cross sections of the $pp\to pp\eta^{\prime}$ reaction based on the
high statistics data from a recently performed experiment~\cite{joannarap}.
Measuring four-momentum vectors of protons in the entrance and exit channels 
we determined complete kinematics for each registered event and hence
derived all in principle possible empirical information  about the studied processes.
Thus, the interpretation of the results is limited only by the experimental accuracy and 
ambiguities of the theoretical approaches. 

Inferences of the interaction taking place within the $pp\eta$ and $pp\eta^{\prime}$ systems
were based on the comparison between 
the experimental results and the predictions obtained assuming that
all kinematically permitted momentum combinations of the outgoing particles
are equally probable. 
Indeed, we have discovered  
statistically incontestable deformations of both the excitation function 
and the phase space abundance. 
The discrepancies can plausibly be
assigned to the influence of the hadronic interaction occuring among ejectiles
since the range of excess energy  was chosen such that other possible explanations 
--~like variations resulting from the dynamics of the primary production
or contributions from larger than zero angular momenta~-- can be excluded.

A unique precision achieved due to the low emittance 
of the cooled beam and the high resolution mass spectrometer enabled us to
distiguish effects resulting from the $\eta$-proton interaction from effects  caused by the almost
two orders of magnitude stronger proton-proton force. 
This conclusion was drawn from the phenomenological analysis
based on the assumption that the primary production can be separated
from the final state interaction and also that the three-body final state 
can be treated as an incoherent system of pairwise interacting objects.
We must admit that these simplifications enabled us to estimate the effects from the proton-$\eta$ and
proton-$\eta^{\prime}$ interaction only qualitatively. 
Nevertheless, a remarkable difference 
between the shape of the excitation functions 
of the $pp\to pp\eta$ and $pp\to pp\eta^{\prime}$ reactions allowed us to conclude that the 
interaction between the $\eta^{\prime}$ meson and the proton is significantly weaker 
than the analogous interaction
between the $\eta$ meson and the proton.
To raise the confidence in the 
concluded distinction between proton-proton and meson-proton interaction 
we compared the shapes of the  excitation functions of the reactions $pp\to pp\eta$ 
and $pp\to pp\eta^{\prime}$ with the one of $pp\to pp\pi^0$. 
We could plausibly assert that the differences in shapes are due to the proton-$\eta$,
or correspondingly, to the proton-$\eta^{\prime}$ interaction since
the well known proton-$\pi^0$ interaction is too weak to manifest itself within the
current accuracy. Therefore
the excitation function for the $pp\to pp\pi^0$ reaction 
reflects the influence of the interaction between protons only.
This comparison confirmed 
--~model independently~-- that 
the hadronic interaction between the $\eta^{\prime}$ meson and the proton 
is indeed much weaker than the $\eta$-proton one. 
This is the first ever empirical appraisal of this hitherto entirely unknown interaction
occuring between the $\eta^{\prime}$ meson and the proton.

We  minimized a systematic bias of the data performing a multidimensional 
acceptance correction and the bin-by-bin subtraction of the multi-pion background.
In particular, we elaborated invariant mass distributions of two-particle subsystems of the $pp\eta$ final state
independently of the reaction model used in the simulations. 
We must confess, however, that 
with the  measurement method used and within the achieved statistics it was not possible 
to determine a background-free occupation density over the
Dalitz-plot unless a small number of  bins was used, which in turn caused drastic losses in the experimental accuracy.
Therefore as a next step of our investigation we proposed a pertinent experiment at a WASA@COSY facility~\cite{LoImoskal}.

The experimental technique employed and the developed method of analysis 
allowed us to achieve an unique 
precision for the simultaneous determination of the absolute excess energy ($\sigma(Q)~=~0.4$~MeV),
missing mass  ($\sigma(m)~\approx~0.3$~MeV/c$^2$), and momentum of the outgoing particles 
($\sigma(p)~=~4~MeV/c$). This unprecedented accuracy  enabled us not only to find out that 
the density of the phase space population is strongly anisotropic but also to discern 
distortions originating from the proton-proton and proton-$\eta$ interactions.
An utterly satisfactory description of the deviations observed at low relative proton-proton momenta 
was achieved by taking into account the well known hadronic and electromagnetic interactions between the protons.
On the contrary, the enhancement at low relative proton-$\eta$ momenta is a few times larger than expected
under the assumption that the mutual interaction among particles in the pp$\eta$ system 
can be described by the incoherent pairwise interaction of proton-proton and proton-$\eta$ subsystems.
At present many theoretical groups are seeking for an explanation of this 
effect~\cite{deloff,Fixprivate,nakayama0302061,vadim024002,garcilazopriv}.
Undoubtedly, this and other results presented in this treatise 
accelerated the development of the formalism of the description of the three-body system 
in the complex hadronic potential~\cite{Fixprivate1,garcilazopriv} 
and led to significant progress in the understanding of the
production mechanism on both hadronic~\cite{christoph,kanzoheber,vadim024002} 
and quark-gluon levels~\cite{dillig050}.

Since the interaction between hadrons, their structure, and production dynamics are inseparably connected with each other,
in this treatise we have discussed not only the hadronic interaction between the $\eta$ and $\eta^{\prime}$  mesons and the protons
but also  presented our endeavour to understand the reaction dynamics and the structure of the studied mesons.
The observed  large difference of the total cross sections
between the $\eta$ and $\eta^{\prime}$ meson production
shows that these mesons are created via different mechanisms, since comparable
coupling constants are expected for both of them,  at least in the SU(3) limit.
It is well established that close-to-threshold the $\eta$ meson is produced predominantly
through the excitation of the $S_{11}(1535)$ resonance, 
and hence the large difference in the observed cross sections suggests that 
the $\eta^{\prime}$ primary production process is nonresonant. 
Different production mechanisms 
reflect 
differences in the structure of these mesons, however, at the present stage we cannot 
draw any quantitative conclusion.

As far as the meson structure is concerned we focused on the  $\eta^{\prime}$ meson 
demonstrating a method which will allow to investigate
a gluonic component of its wave function. The study of this very interesting aspect 
--~connected to the search of  matter
built exclusively out of gluons~-- 
will be based on the comparison of the $\eta^{\prime}$ meson production
via the $pp\to pp\eta^{\prime}$ and $pn\to pn\eta^{\prime}$ reaction.
The measurement of the excitation function of the $pp\to pp\eta^{\prime}$ reaction is finished,
and in order to accomplish the measurement of the $pn\to pn\eta^{\prime}$ reaction we have 
extended the COSY-11 detection setup by a neutron and a spectator detectors.
The neutron detector was designed and  built by us in Cracow~\cite{neutronraporty,neutronraporty1},
and for the registration of the spectator protons we adapted~\cite{proposalc11,spectatorraporty,spectatorraporty1}
to the COSY-11 facility
a silicon pad detector~\cite{bilger64}
which was previously used by the WASA/PROMICE collaboration at the CELSIUS accelerator.
A successful measurement of the $pn\to pn\eta$ reaction  has proven that it is possible
to study quasi-free $pn\to pnX$ reactions using the newly accomplished experimental setup.

An exploration of the isospin degrees of freedom in the study of the creation of mesons 
led us to the entirely novel idea of using  quasi-free reactions for the investigations
of meson production cross sections
as a function of the virtuality of the interacting nucleons. We demonstrated also a method 
for measuring the close-to-threshold production
of mesons in  quasi-free neutron-neutron collisions.
It is our great hope to conduct these investigations in the near future
at the facility WASA@COSY~\cite{LoImoskal}.

In parallel we began also to explore  spin degrees of freedom. In particular,
we have demonstrated that by measuring spin observables it is possible to learn about the very details
of the production dynamics of the $\eta$ meson. Predictions  for the analysing power differ so much
depending on the assumed mechanism that the determination of this quantity alone will allow
to establish whether vector or pseudoscalar meson exchange dominates the production process.

In this treatise we presented results of investigations aiming to determine the interaction
of $\eta$ and $\eta^{\prime}$ mesons with nucleons, the structure of these mesons and the mechanism which
governs their creation in the collision of nucleons. 
We have not arrived at the very aim yet.
But we have gained a qualitative understanding of the reaction mechanism, estimated the hithertho entirely unknown 
$\eta^{\prime}$-proton interaction, observed a signal from the $\eta$-proton interaction,
delivered a sample of high quality data on both total and differential cross sections,
and developed novel methods of measurements which should lead  to more quantitative statements in the near future.



\begin{flushright}
\parbox{0.7\textwidth}{
 {\em
So even though my speculations pleased me very much, I believed that other persons had
their own speculations which perhaps pleased them even more~\cite{descartes1}.\\
 }
 \protect \mbox{} \hfill  Ren\'{e} Descartes \protect\\
 }
\end{flushright}
\begin{flushright}
\parbox{0.7\textwidth}{
 {\em
 The approbation of the public I consider as the greatest reward of my
 labours; but am determin'd to regard its judgment, whatever it be, 
 as my best instruction~\cite{humetreatise}.\\
 }
 \protect \mbox{} \hfill  David Hume \protect\\
}
\end{flushright}

\clearpage
\newpage
\thispagestyle{empty}

  \cleardoublepage
        \def\bibname{References}
        \addcontentsline{toc}{chapter}{\protect\numberline{}{References}}

\newpage
\clearpage
\thispagestyle{empty}
\pagestyle{plain}
\pagestyle{myheadings}

\end{document}